\documentclass[12pt,twoside]{book}
\usepackage{graphicx,epstopdf}
\usepackage{a4wide}
\usepackage{fancyhdr}
\usepackage{palatino,amsmath,float,multicol,subfigure,shadow,fancybox,amsfonts,amsthm,bm}
\usepackage{amssymb}
\usepackage{stmaryrd}
\usepackage{mathrsfs}
\usepackage[nottoc]{tocbibind}
\usepackage{makeidx}
\usepackage{relsize}
\usepackage{bm}
\usepackage{latexsym}
\usepackage{dcolumn}
\usepackage{lipsum}
\usepackage{graphicx,epsfig,float}
\usepackage{psfrag}
\usepackage{bm}
\usepackage{bbold}
\usepackage{epstopdf}
\usepackage{latexsym}
\usepackage{multirow}
\usepackage{amssymb,amsfonts}
\usepackage{enumerate}
\usepackage{hyperref}
\usepackage{multirow}
\usepackage{dcolumn}
\usepackage{array}
\usepackage{color,wasysym}
\usepackage[titletoc]{appendix}
\usepackage[dvipsnames]{xcolor}
\usepackage{tikz}
\usetikzlibrary{shapes.geometric, arrows}
\tikzstyle{st} = [rectangle, rounded corners, text width = 3cm, text centered, draw = black ]
\tikzstyle{arrow} = [->,>=stealth]

\def\beq {\begin{equation}}
\def\eeq {\end{equation}}
\def\bea {\begin{eqnarray}}
\def\eea {\end{eqnarray}}
\def\br{\begin{eqnarray}}
\def\er{\end{eqnarray}}

\def\bc {\begin{center}}
	\def\ec {\end{center}}
\def\bi {\begin{itemize}}
	\def\ei {\end{itemize}}
\newcommand{\eel}[1] {\label{#1}\end{equation}}

\newcolumntype{L}[1]{>{\raggedright\arraybackslash}p{#1}}
\newcolumntype{C}[1]{>{\centering\arraybackslash}p{#1}}
\newcolumntype{R}[1]{>{\raggedleft\arraybackslash}p{#1}}

\allowdisplaybreaks
\definecolor{cpgreen}{rgb}{0.55, 0.0, 0.55}
\renewcommand{\headrulewidth}{.1pt}
\def\baselinestretch{1.1}
\date{\today}
\newcommand{\doublespace}{\par\renewcommand{\baselinestretch}{1.5}\small\normalsize\par}

\hypersetup{
    unicode=false,          
    pdftoolbar=true,        
    pdfmenubar=true,        
    pdffitwindow=false,     
    pdfstartview={FitH},    
    pdftitle={My title},    
    pdfauthor={Author},     
    pdfsubject={Subject},   
    pdfcreator={Creator},   
    pdfproducer={Producer}, 
    pdfkeywords={keywords}, 
    pdfnewwindow=true,      
    colorlinks=true,        
    linkcolor=Cerulean, 
    citecolor=RedOrange,	
    filecolor=magenta,      
    urlcolor=cpgreen          
}

\newcommand {\apgt} {\ {\raise-.5ex\hbox{$\buildrel>\over\sim$}}\ }
\newcommand {\aplt} {\ {\raise-.5ex\hbox{$\buildrel<\over\sim$}}\ }

\makeatletter
\def\myfnt{\ifx\protect\@typeset@protect\expandafter\footnote\else\expandafter\@gobble\fi}
\makeatother

\begin{document}
\doublespace

\thispagestyle{empty}
\begin{center}{
		\textbf{\LARGE{Hamiltonian formalism of cosmological \\perturbations and higher derivative theories}}
	}
\end{center}{ \par}

\vspace{.5cm}

\begin{center}{
		{\it {\large A thesis submitted for the degree of}}\\
		\vspace {0.5cm}}
\end{center}
%
\begin{center}{\textbf{\Large {\bf Doctor of Philosophy}}}\end{center}{\Large \par}
\begin{center}{\textit{\large {in the  }}}\end{center}{\Large \par}
\begin{center}{\textbf{\Large {\bf School of Physics }}}\end{center} 
\vspace {0.5cm}
\begin{center}{\textit{\large {by  }}}\end{center}{\Large \par}
\begin{center}{\textbf{\Large {\bf Debottam Nandi}}}\end{center} 
\begin{center}{\textbf{ {( Roll No: PHD12104) }}}\end{center} 


\vspace*{\fill}
\begin{center}
	\vspace{.5cm}
	\includegraphics[scale=.4]{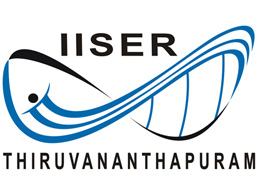}{\small}\\
	{\large {\bf School of Physics,}}\\
	{\large \bf Indian Institute of Science Education and Research Thiruvananthapuram (IISER-TVM),}\\
	{\large \bf Kerala- 695016, India}\\
\end{center}{\small
\par}
\begin{center}{\large May 2017}\end{center}{\small \par}
\clearpage
\newpage
\thispagestyle{empty}
\pagestyle{fancy}
\thispagestyle{empty}
\bigskip{}
\vspace{1cm} \cleardoublepage

\thispagestyle{plain}
\pagenumbering{roman} \setcounter{page}{3}
\begin{center}
	{\large{\bf{AUTHOR'S DECLARATION}}}
\end{center}
\vspace{1cm}
\noindent
I hereby declare that the work reported in this thesis is original and has been carried out by me during my tenure as a PhD student at { \bf School of Physics, Indian Institute of Science Education and Research Thiruvananthapuram (IISER-TVM)} under the supervision  of { \bf Dr. S.~ Shankaranarayanan}. This work has not been submitted earlier as a whole or in part for a degree/diploma at this or any other Institution/ University. Material  obtained from other sources has been duly acknowledged in the thesis.
\vspace{4cm}

\noindent Place: Thiruvananthapuram \hfill   Debottam Nandi\\
\noindent Date:\hfill  \\
\clearpage
\thispagestyle{empty}
\phantom{a}
\vfill
\vfill

\newpage
\thispagestyle{plain}
\pagenumbering{roman} \setcounter{page}{5}
\begin{center}
	{\large{\bf{CERTIFICATE}}}
\end{center}\vspace{1cm}

\noindent
This is to certify that the work contained in this thesis 
entitled \textbf{``Hamiltonian formalism of cosmological perturbations and higher derivative theories''}  submitted
by \textbf{Debottam Nandi} (\textbf{Roll No: PHD12104}) to { \bf School of Physics, Indian Institute of Science Education and Research Thiruvananthapuram }
towards the requirement of {\bf Doctor of Philosophy} in {\bf Physics }  has been carried out
by him under my supervision and that it has not been submitted elsewhere
for the award of any degree.

\vspace{4cm}

\noindent Place: Thiruvananthapuram \hfill Dr. S.~ Shankaranarayanan\\
\noindent Date:\hfill Doctoral Advisor \\

\clearpage
\thispagestyle{empty}
\phantom{a}
\vfill
\vfill
\newpage
\thispagestyle{plain}
\pagenumbering{roman} \setcounter{page}{7}
\clearpage\vspace*{8cm}
\begin{center}
	{\Large{\textit{\textbf {Dedicated to my beloved parents and my sister\\
					for their endless love, support and  encouragement.}}}}
\end{center}\vspace{1cm}
\clearpage
\thispagestyle{empty}
\phantom{a}
\vfill
\vfill
\newpage
\thispagestyle{plain}
\pagenumbering{roman} \setcounter{page}{9}
\begin{center}
	{\large{\bf{ACKNOWLEDGMENTS}}}
\end{center}\vspace{1cm}

It is a great honour and immense pleasure to pursue PhD at  School of Physics, IISER-TVM. Over the last five years, many have helped  to get to this stage. 
First of all, the credit goes to my advisor Dr. S.~ Shankaranarayanan (Shanki) who not only guided me but also inspired me for doing research. I would like to thank all my colleagues who have helped me during my PhD life. In Shanki's group, I have had many conversations about my research with Dr. Sanil Unnikrishnan, Dr. Swastik Bhattacharya, Dr. Abhishek Basak  and Dr. Oph\'elia Fabre and I am indebted to all the members of the group  for their valuable comments and criticism regrading my presentation and research  during the weekly group meeting. Also, I would like to sincerely thank my colleagues, Dr. Santhosh Kumar S., Mr. Jose Mathew and Mr. Soham Bhattacharyya, for the support they have given me during these years.    

I am very grateful to my doctoral committee members Dr. Sreedhar B Dutta and Dr. Archana Pai. I would like to take this opportunity to thank  all the  faculty members of School of Physics, IISER-TVM for their enormous support  during these years. I wish to acknowledge Council of Scientific \& Industrial Research (CSIR) for the JRF-SRF fellowships, IISER-TVM for providing  research facilities and Max Planck Partner Group on Cosmology and Gravity, IISER-TVM for international travel support throughout these years. I also thank Kasper Peeters for his useful program Cadabra \cite{Peeters:2007wn,DBLP:journals/corr/abs-cs-0608005} for doing algebraic calculations.

PhD life is not all about study and research. During these years, I  have enjoyed my life with all of my friends. I thank all those who  made me feel home. Special regards to Gayathri V., Saleem M., Sumanta Banerjee, Rajesh Ghosh, Swapnil Bhowmick, Balvinder Bindra and Akhileshwar Mishra. Also, I   take this opportunity to thank all the members of the 'GEM building' PhD room and SBT hostel where I stayed during this tenure. Also, I owe a great deal to the countless  friends  I enjoyed being with during my IISER days.
\markboth{Acknowledgements}{Acknowledgements}

Last, but not least, I want to thank my family, specially my parents and my sister,  for their constant  effort, sacrifice, and love. I also thank my all childhood and college friends for their unconditional support. None of this would have been possible without them. 
\noindent
\clearpage

\thispagestyle{plain}
\pagenumbering{roman} \setcounter{page}{11}
\begin{center}
	{\large{\bf{ABSTRACT}}}
\end{center}\vspace{1cm}

Inflation, an accelerated phase of the early universe, has now become
an integral part of the standard model of cosmology as it can explain
homogeneity of the Universe at large scales, the origin of primordial
perturbations and observation of anisotropic temperature fluctuations
in the Cosmic Microwave Background Radiation (CMBR). Einstein's
equations are highly non-linear, hence it is impossible to solve these
equations exactly. To evaluate the physical quantities during the
inflationary epoch, several approximation techniques have been
developed in the literature. The focus of the thesis is to obtain an
universal formalism to evaluate the perturbations at all orders that
can be applied to any theory of gravity and matter source in the
early universe. We first look at the equivalence of two approaches ---
action and order-by-order gravity equation approach --- for
cosmological perturbation theory, and establish that both lead to
equivalent results for any gravity models at any order of
perturbations. We then focus on Hamiltonian formalism that has not
been studied extensively in the literature. We provide a generalized
Hamiltonian approach for cosmological perturbations which is
equivalent to the Lagrangian approach. We show that, the approach can
be applied to any model at any order of perturbations. Using this
approach, we show that evaluating interaction Hamiltonian is simpler
and efficient than earlier approach. In the next work, we concentrate
on generalized non-canonical scalar field and by introducing a new
variable provide a technique to write Hamiltonian for generalized
non-canonical scalar field. We then implement our Hamiltonian approach
for non-canonical scalar field and evaluate interaction Hamiltonian
without slow-roll approximations. Finally, for the first time, we
construct a vector Galileon model in curved space-time in which, the
field equations do not contain any higher-derivative terms, yet,
preserving U(1) gauge-invariance. Conformal invariance is broken in
this model which leads to primordial magnetogenesis and we compare the
predictions of our model with observations.
\clearpage
\thispagestyle{empty}
\phantom{a}
\vfill
\vfill \pagenumbering{arabic}
\setcounter{page}1
\newpage
\let\cleardoublepage\clearpage

\setcounter{secnumdepth}{3} 
\tableofcontents
\thispagestyle{fancy}
\listoffigures
\clearpage
\thispagestyle{empty}
\phantom{a}
\vfill
\vfill
\listoftables
\let\cleardoublepage\clearpage
\clearpage

\newpage
\clearpage
\thispagestyle{empty}
\phantom{a}
\vfill
\vfill
\pagestyle{fancy}
\chapter*{Preface}
\markboth{Preface}{Preface}
\addcontentsline{toc}{chapter}{Preface} 
\label{generalintro}
 Einstein's general theory of relativity provides a compelling and
 testable theory of the Universe. However, this brings us a very
 crucial question that needs to be resolved: \emph{what started all
 	this?} The standard model of cosmology (Friedman-Robertson-Walker
 model) is intrinsically incomplete as it cannot provide answers to the
 questions like: Why the Universe is homogeneous at large scales? Why
 the Cosmic Microwave Background (CMB) temperature is same in all
 directions? How the primordial perturbations were generated in the
 early universe? The quest to answer these questions leads to a new
 realm of physics --- inflationary paradigm.  It is a period of
 accelerated expansion in the very early universe around
 $10^{14}$ GeV which is much remote as compared to the terrestrial
 experiments. Inflationary cosmology has two key theoretical
 aspects. One is the approximation schemes employed in solving gravity
 equations.  The other is the inflationary model building inspired by
 particle physics or a fundamental theory of quantum gravity. The
 problem with any theory of gravity is that it is highly non-linear, so
 one has to rely on approximation schemes to compare with observations.
 Primarily, two formalisms exist to solve the non-linear equations:
 \begin{itemize}
 	\item The separate universe approximations with either gradient
 	expansion theory or $\Delta N$ formalism.
 	\item Gauge invariant cosmological perturbation theory.
 \end{itemize}
 
 The temperature fluctuations as observed in CMB is $\sim 10^{-5}$,
 hence order-by-order perturbation theory can be considered to match
 with observations. With respect to inflationary model building, the
 proposed theories are primarily preferred through simplicity. In the
 case of canonical scalar field, the simplest, 60 e-foldings of
 inflation require the potential to be flat. However, in the case of non-canonical scalar field model, dependence on the potential is reduced  and the kinetic part of the action leads to different inflationary scenarios. In addition to this, non-canonical scalar field leads to the time dependence
 of the speed of perturbations and makes it difficult to be compared
 with CMB observations. In order to seek more generalized fields,
 scalar fields with higher time derivatives in action are
 considered. Beside these, modified gravity models, specifically $f(R)$
 lead to accelerated expansion in the early universe.
 
 There are two mathematical procedures that are currently used in the
 literature to study gauge invariant cosmological perturbation theory:
 Hamiltonian formulation and Lagrangian
 formulation. While the Lagrangian formulation has been widely studied,
 Hamiltonian framework in the context of cosmological perturbations is
 not common. This thesis focuses on the different formalisms and
 techniques that are used to obtain cosmological perturbation equations
 during early universe. The analysis/comparison of these
 approaches are much needed to enhance our theoretical
 predictions. First, we look into Lagrangian formulation and establish
 the equivalence of action approach and order-by-order Einstein's
 equation approach for generalized gravity models. Next we concentrate
 on Hamiltonian formalism and provide a generalized Hamiltonian
 approach for cosmological perturbations at all orders. We also provide
 a new technique to obtain Hamiltonian for generalized non-canonical
 scalar fields and apply our Hamiltonian approach for the same. We also
 apply our new approach to Galileon fields (fields whose action has
 higher derivative terms, however, the equations of motion is still
 second order).  Like Galileons, the Einstein-Hilbert action contains
 higher order derivatives, however Einstein's equations are still
 second order. We are led to ask: Can there exist a corresponding
 action for spin-1 or vector fields? While constructing such an action
 that preserves gauge-invariance is not possible in flat space-time, we
 explicitly show that it is possible in curved space-time. In other
 words, for the first time, we construct a vector Galileon model in
 curved space-time in which the field equations do not contain any
 higher derivative terms, yet, it preserves U(1)
 gauge-invariance. Conformal invariance automatically breaks down for
 the model which leads to primordial magnetogenesis and we compare the
 predictions of our model with observations.
 
 
 The thesis is divided into four parts. In the introductory part of the
 thesis, we discuss the motivation of our work and provide necessary
 background material. Part 2 contains three chapters focusing on
 different formalisms and their consistencies. Part 3 contains a single
 chapter about primordial magnetogenesis from vector Galileon in curved
 space-time. The concluding part is made up of chapter 6 giving the
 summary of the thesis and presenting the future outlook.
 
 A chapter wise summary of the thesis is given below:
 
 In {\bf chapter} $\mathbf 1$, the brief history of the Universe is
 presented followed by the Einstein's equations and the solutions of
 those equations. Then we focus on early universe and discuss the
 problems in understanding early universe within
 Friedman-Robertson-Walker model and the need for inflationary
 paradigm. We mainly concentrate on cosmological perturbation theory
 and review this in the context of Lagrangian formulation. We also
 briefly discuss about different formalisms and approaches which we
 discuss in detail in chapters $2, 3$ and $4$.
 
 In {\bf chapter} $\mathbf 2$, we, first, concentrate on Lagrangian
 formulation. In Lagrangian formulation, there are two approaches to
 obtain perturbed equations: (i) Order-by-order gravity equations and
 (ii) Action approach. Because of non-linear nature of gravity,
 equivalence of these two approaches is not established properly. In
 this work, we show that, in case of canonical scalar field minimally
 coupled with gravity, these two approaches are equivalent for any
 order of perturbations. We extend our method and show that the
 equivalence is generally true for any kind of gravity models at any
 order of perturbations. We also find out that, in order to obtain
 reduced equation that can lead to evolution of density fluctuations,
 `constraint consistency' has to be satisfied, i.e., Lapse function and
 shift vector should behave as constraints in the system. The crucial
 result we obtain is that all `constraint inconsistent' models have
 \emph{Ostrogradski's instability} but the reverse is not true. In
 other words, one can have models with constrained Lapse function and
 Shift vector, though it may have Ostrogradski's instabilities.

 In {\bf chapter} $\mathbf 3$, after focusing on Lagrangian
 formulation, we shift our concentration on Hamiltonian formulation. In
 order to match observations with theory, we need a proper approach of
 quantization and for canonical quantization, we need to know the
 Hamiltonian of the system. In the case of cosmological perturbations,
 by using order-by-order approximation, perturbed interaction
 Hamiltonian is obtained from the reduced perturbed Lagrangian. This
 method is not straightforward, difficult to implement for constrained
 system and also suffers other difficulties as well as restrictions. In
 this work, we present a consistent Hamiltonian approach of
 cosmological perturbation theory for any kind of particle as well as
 gravity models at any order of perturbation, which is simple, robust
 and provide much more information of the system than the earlier
 work. We show that, if Hamiltonian of any system is known to us, our
 approach can be applied to all those systems. We also show that, this
 approach is not only simple and straightforward to implement, but also
 provide extra information of the system, e.g., the true degrees of
 freedom of the system.

 In {\bf Chapter} $\mathbf 4$, we concentrate on the Hamiltonian
 formulation of generalized non-canonical scalar field. In the case of
 general non-canonical scalar fields, since Legendre transformation is
 non-invertible, Hamiltonian of the system cannot be obtained. However,
 in this work, we show that, by introducing a new phase-space variable,
 it is, in fact, possible to define the Hamiltonian also for a
 generalized non-canonical scalar field. Then, we apply our method:
 Hamiltonian approach of cosmological perturbation to non-canonical
 scalar field and show that our approach is consistent. We obtain a new
 expression of speed of sound in terms of phase-space variables for
 generalized non-canonical scalar field. Then we provide a general
 inversion relation between phase-space variable (and its derivatives)
 and configuration-space variable (and its derivatives) to compare and
 contrast general perturbed Hamilton's equations and Euler-Lagrange
 equations. Finally, we extend our approach to general
 higher-derivative gravity models like Horndeski's model.
 
 Both in case of Lagrangian and Hamiltonian approach, we encounter several
 higher derivative models that are interesting to study. Like Galileons 
 the Einstein-Hilbert action contains higher order derivatives, however
 Einstein's equations are still second order. We are led to ask: can there
 exist a corresponding action for spin-1 or vector fields? While 
 	constructing such an action that preserves gauge-invariance is not possible
 	in flat space-time, in {\bf chapter} $\mathbf 5$, we explicitly show that it
 	is possible in curved space-time by considering different non-minimal couplings between the vector field and the Riemann tensor. Thus, in flat space-time, the action explicitly vanishes and only contributes in curved space-time. Conformal invariance is broken in this
 model, hence, providing a new model for primordial magnetogenesis.

 Finally, in {\bf chapter} $\mathbf 6$, we list the key results of the
 thesis and future direction that arises from these results. In case of
 Hamiltonian formulation, for the first time, we provide the
 generalized approach, however, the application of this is still wide
 open. Also, in case of higher derivative models, we find that, we can
 extend the search for other higher derivative vector Galileon since
 whatever we have provided is the first of its kind. This can even be
 extended for higher order scalar-tensor theories. Not only we can
 search for those models but the applications of those models can be
 studied for different era of the universe at any length scales.
 \clearpage
 \thispagestyle{empty}
 \phantom{a}
 \vfill
 \vfill
 
 \newpage
 \thispagestyle{fancy}  
 \section*{List of publications}
 \markboth{List of publications}{List of publications} 
 \begin{itemize}
 	\item[1.] D. Nandi and S. Shankaranarayanan, "`Constraint
 		consistency' at all orders in cosmological perturbation theory", \href{https://arxiv.org/abs/1502.04036}{{\it 
 		JCAP} \textbf{1508} (2015) 050
 	[arXiv:1502.04036]} ({\bf
 		Chapter} $(\ref{Chap.2})$ is based on this work).
 	\item[2.] D. Nandi and S. Shankaranarayanan, "Complete
 		Hamiltonian analysis of cosmological perturbations at all orders", \href{https://arxiv.org/abs/1512.02539}{{\it 
 		JCAP} \textbf{1606} (2016) 038
 	[arXiv:1512.02539]} ({\bf
 		Chapter} $(\ref{Chap.3})$ is based on this work).
 	\item[3.] D. Nandi and S. Shankaranarayanan, "Complete
 		Hamiltonian analysis of cosmological perturbations at all orders II:
 		Non-canonical scalar field", \href{https://arxiv.org/abs/1606.05747}{{\it JCAP} \textbf{1610} (2016) 008
 	[arXiv:1606.05747]} ({\bf
 		Chapter} ($\ref{chap.4}$) is based on this work).
 	\item[4.] D. Nandi and S. Shankaranarayanan, "Primordial
 		magnetogenesis from vector Galileon", \href{https://arxiv.org/abs/1704.06897}{[arXiv:1704.06897]}
 		({\bf Chapter} $(\ref{Chap.5})$ is based on this
 	work).
 \end{itemize}
\clearpage
\clearpage
\thispagestyle{empty}
\phantom{a}
\vfill
\vfill

\newpage
\thispagestyle{plain}

\pagestyle{fancy}
\chapter{Introduction} \label{Chap.Introduction}

\begin{figure}
\centering
\includegraphics[width=0.7\linewidth]{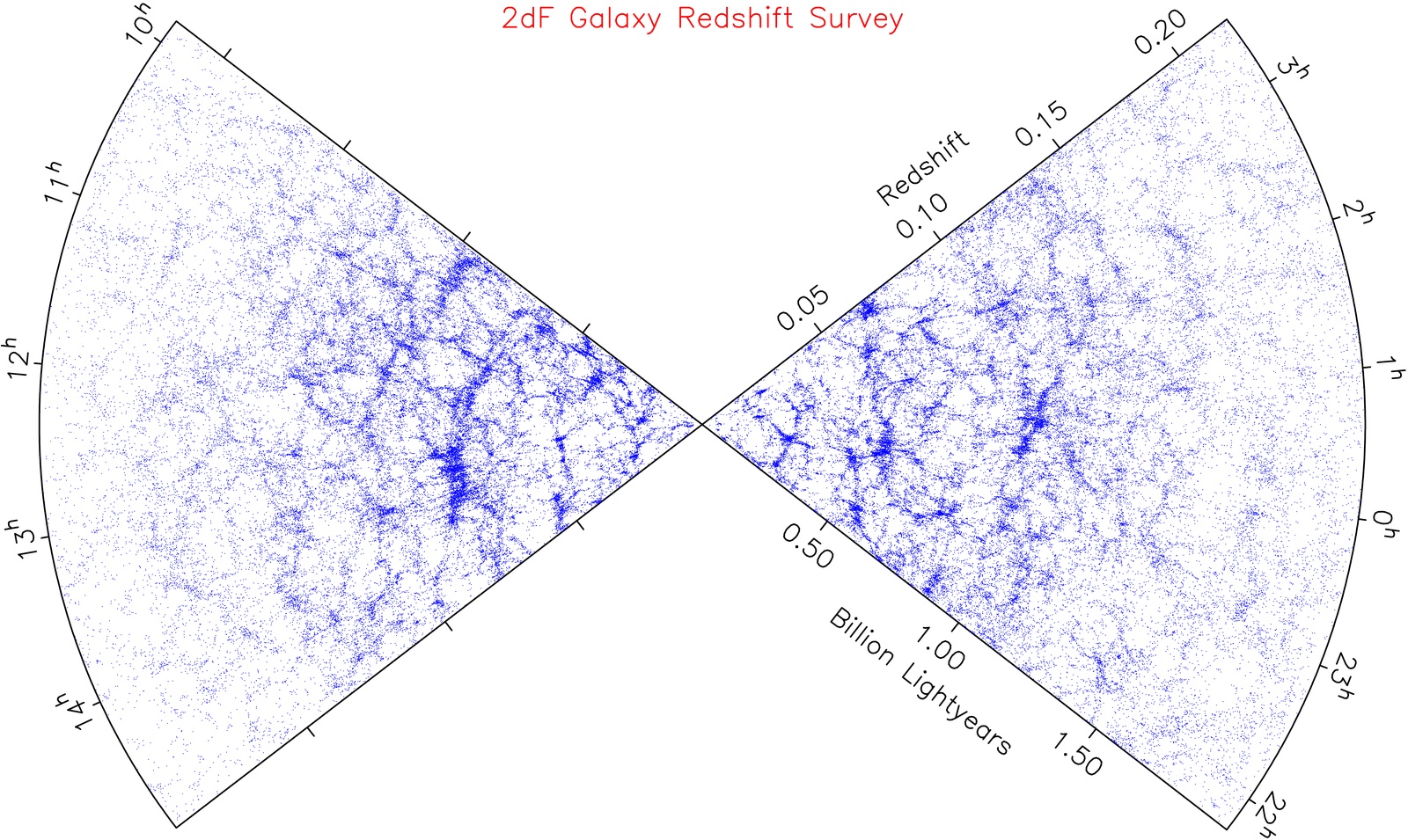}
\caption{2dF Galaxy Redshift Survey}
\label{fig:2dFzcone_big}
\end{figure}

Einstein's General Theory of Relativity has provided us a compelling, testable mathematical framework of the Universe.
It makes precise prediction about the evolution of the Universe starting from $10^{-34}$ seconds till now. The standard model of cosmology is based on the cosmological principle that states that on the largest cosmological scales, the Universe is homogeneous and isotropic \cite{1994ApJ...420..445F, 2009MNRAS.399.1755C}. Fig. (\ref{fig:2dFzcone_big}) shows the distribution of the large scale structures as a function of redshift. From the figure it is clear that at high redshifts less structures were formed and it was approximately homogeneous and isotropic. The success of the standard model of cosmology is based on three observational pillars: expanding Universe (Hubble's Law) \cite{1990-Kolb.Turner-Book}, light element abundances \cite{1990-Kolb.Turner-Book} and Cosmic Microwave Radiation Background (CMBR) \cite{1994ApJ...420..445F, Ade:2015xua, Komatsu:2008hk}. In this chapter, we briefly discuss the salient features of the standard model of cosmology, colloquially referred to as \emph{Big Bang cosmology}, by highlighting these three aspects.

\section{Observables in cosmology}
 We now have a good evidence that the universe is expanding --- at earlier time,  distance between us and distant galaxies was smaller than it is today. It is convenient to describe the expansion by introducing the scale factor $a(t)$, whose present value is set to one and physical distance between two points at a time $t$ is $a(t)$ times the comoving distance. At earlier times $a(t)$ was smaller than it is today, hence $\frac{da(t)}{dt} > 0 $ corresponds to expanding Universe \cite{1990-Kolb.Turner-Book}. In addition to the scale factor and its evolution, the smooth universe is characterized by other parameter --- the geometry of the 3-space \cite{1990-Kolb.Turner-Book}. There are three possibilities: flat (3-curvature is zero), open (3-curvature is negative), or closed (3-curvature is positive). The current observations indicate that we live in nearly flat Universe \cite{Ade:2015xua, Komatsu:2008hk}.

To understand the history of the universe, we need to determine the evolution of the scale factor $a(t)$. To quantify the change in the scale factor and its relation to the energy density of the matter in the Universe, it is useful to define Hubble rate 
\begin{equation}
H(t) = \frac{1}{a(t)}\frac{da(t)}{dt}
\end{equation}
 which measures how rapidly the Universe changes. Here and throughout, subscript $0$ denotes the value of $a(t),\, H(t)$ today. The Hubble constant ($H_0$)  is
parameterized by $h$ defined via 
\begin{eqnarray}
H_0 = 100h~\mbox{km sec}{}^{-1} \mbox{Mpc}{}^{-1} = \frac{h}{0.98 \times 10^{10}\mbox{years}} = 2.133 \times 10^{-33} \,h\, \frac{\mbox{eV}}{\hslash} \nonumber 
\end{eqnarray}
where the current value of $h \simeq 0.72$ and $\hslash$ is the Planck's constant. Scale factor $a(t)$ as well as the Hubble constant $H(t)$ determines the evolution of the geometry of the Universe. More specifically, Hubble rate is a measurable quantity and, hence, it is independent of the model. Relating Hubble parameter to physical quantities like energy density, pressure and temperature provide different stages the Universe underwent. Another model independent quantity is the redshift ($z$) which is defined as 
\begin{eqnarray}\label{Eq.Redshift}
1 + z \equiv \frac{\lambda_{\text obs}}{\lambda_{\text emit}} = \frac{1}{a}.
\end{eqnarray}

In the rest of this chapter, using Einstein's equations, we establish the equations relating the scale factor to the matter. We then briefly discuss the problems of standard model of cosmology and the need for inflationary paradigm. We discuss in detail the cosmological perturbation theory including the gauge issue in the linear and higher order. Finally, in the last section, we discuss the need for higher order perturbation theory by discussing different approaches to evaluate the physical quantities and the problem of generation of primordial magnetic fields during inflation which forms the motivation of the thesis \cite{1990-Kolb.Turner-Book, Baumann:2009ds,  2003-05-14-DodelsonScott-Moderncosmology, 2005-05-14-MukhanovViatcheslav-Physicalfoundationsofcosmology}.

In this thesis, we use natural units $\hslash = c = k_B = 1/(4\pi\epsilon_0) = 1, \kappa = 8\pi G, G$ is 4-dimensional Newton's constant. We use the following metric convention $(-, +, +, +)$. The Greek letters run from 0 to 3 and the lower case Latin letters run from 1 to 3.

\section{Dynamics of expansion}\label{Sec.Prelim}

A general 4-dimensional space-time line element is given by

\begin{eqnarray}
ds^2 = g_{\mu \nu} dx^\mu dx^\nu.
\end{eqnarray}

As mentioned earlier, the cosmological principle states that the Universe is spatially homogeneous and isotropic. The 4-dimensional line element consistent with the cosmological principle is given by

\begin{equation}\label{FRW k metric}
ds^2  = - dt^2 + a^2(t) \left(\frac{dr^2}{1- k r^2} + r^2\, d\Omega^2\right)
\end{equation}
where $k$ is called the curvature of the 3-space and can take three discrete values -1, 0, 1. -1 corresponds to open universe, +1 relates to closed universe and 0 corresponds to flat spatial curvature. $a(t)$ is the scale factor and $t$ is referred to as the cosmic time. The above line element is referred to as Friedman-Robertson-Walker metric. Since the spatial part of the line-element is multiplied by the scale factor $a(t)$, even if the comoving distance between two points are constant, the physical distance changes as the Universe evolves. The current observations indicate that the Universe is nearly spatially flat \cite{1994ApJ...420..445F,Ade:2015xua,Komatsu:2008hk} and the metric is

\begin{equation}\label{FRW metric}
g_{\mu \nu} = \left({\begin{array}{cccc}
-1 &0&0&0 \\
0&a(t)&0&0 \\
0&0&a(t)&0 \\
0&0&0&a(t)
\end{array}}\right)
\end{equation}

The evolution of scale factor $a(t)$ is obtained from the Einstein's equations:
\begin{eqnarray}\label{Eq.EinsteinEq}
G_{\mu \nu} &=& \kappa \,T_{\mu \nu},  \\
\text{where}\quad G_{\mu \nu} &\equiv & R_{\mu \nu} - \frac{1}{2} g_{\mu \nu} R,\qquad \kappa \equiv 8\pi G~ \mbox{is a constant}.
\end{eqnarray}
$T_{\mu \nu}$ is the stress-tensor corresponding to the matter, $R_{\mu \nu}$ is the Ricci tensor given by

\begin{eqnarray}\label{RicciTensor}
R_{\mu \nu} = \partial_{\gamma}{\Gamma^{\gamma}_{\mu \nu}} - \partial_{\nu}{\Gamma^{\gamma}_{\mu \gamma}} + \Gamma^{\gamma}_{\mu \nu} \Gamma_{\alpha}{\gamma \alpha} - \Gamma^{\gamma}_{\mu \alpha} \Gamma^{\alpha}_{\nu \gamma}
\end{eqnarray}
and $R$ is Ricci scalar given by
\begin{eqnarray}
\label{RicciScalar}
R  &=& g^{\mu \nu} R_{\mu \nu}.
\end{eqnarray}

For the Friedman-Robertson-Walker line element (\ref{FRW metric}), the Ricci tensor, Ricci scalar and the Einstein's tensor are given by
\begin{eqnarray}
R_{0 0} &=& -3 \frac{\ddot{a}}{a},~~R_{i j} = \delta_{i j} \left(2\dot{a}^2 + a\ddot{a} \right),~~R =  6\left[\frac{\ddot{a}}{a} + \left(\frac{\dot{a}}{a}\right)^2\right] \\
G_{0 0} &=& 3 \left(\frac{\dot{a}}{a}\right)^2,\quad G_{i j} = - \left(\dot{a}^2 + 2 a \ddot{a}\right) \delta_{i j}
\end{eqnarray}
where dot denotes derivative with respect to the cosmic time. Homogeneity of the FRW line element demands that the stress-tensor is independent of the spatial coordinates and depends only on the cosmic time. Isotropy demands that the stress tensor takes the following form
\begin{equation}
T^\mu_\nu = \left({\begin{array}{cccc}
-\rho(t) &0&0&0 \\
0&p(t)&0&0 \\
0&0&p(t)&0 \\
0&0&0&p(t)
\end{array}}\right)
\end{equation}
where $\rho$ is the energy density and $p$ is the pressure. It is important to note that it is not necessary to assume the matter content is a perfect fluid; it is a consequence of the symmetry of the FRW line element.  Then, 0-0  and i-j component of the Einstein's equations (\ref{Eq.EinsteinEq}) become

\begin{eqnarray}
\label{Friedman Eq}
 && H^2 = \frac{\kappa}{3} \rho \\
 \label{Acceleration Eq}
 && \frac{\ddot{a}}{a} \equiv \dot{H} + H^2  = -\frac{\kappa}{6} \left(\rho + 3p\right) 
\end{eqnarray}
where $H$ is the Hubble parameter. These are called Friedman equation and acceleration equation, respectively. Along with these, the continuity equation

\begin{eqnarray}\label{Continuity Eq}
\nabla_{\mu} T^{\mu}_0 = 0.
\end{eqnarray}
leads to 
 
 \begin{eqnarray}\label{Conserv Eq}
 \dot{\rho} + 3 H \left(\rho + p\right) = 0.
 \end{eqnarray}
 
It is important to note that the three equations (\ref{Friedman Eq}), (\ref{Acceleration Eq}) and (\ref{Conserv Eq}) are not independent equations; any one of the three equations can be obtained from the other two equations. Since there are three unknown functions --- $a(t), \rho(t)$ and $p(t)$ --- and only two independent differential equations, we need to assume one of these functions. One natural assumption is to consider that at all times, pressure is proportional to energy density, i.e., $ p = w \rho$ where $w$ is the equation of state parameter. Since the equations (\ref{Friedman Eq}), (\ref{Acceleration Eq}) and (\ref{Conserv Eq}) are highly non-linear, it is not possible to analytically solve the equations for multiple fields. Below, we obtain exact analytical solutions for three simple cases --- Universe filled with non-relativistic matter ($w = 0$), Universe filled with radiation or relativistic matter ($w = 1/3$) and cosmological constant ($w = -1$).

 In case of matter-dominated Universe, i.e., the Universe filled with dust matter $(w = 0)$, the continuity equation (\ref{Conserv Eq}) reduces to
 
 \begin{eqnarray}
  \partial_t\left(\rho_m a^3\right) = 0 \quad\Rightarrow \rho_m \propto \frac{1}{a^3}
 \end{eqnarray}
 
 Using the above form of the energy density in (\ref{Friedman Eq}), we get
 
 \begin{eqnarray}
 a(t) \propto t^{2/3}.
 \end{eqnarray}
 
 Similarly, in the case of radiation-dominated universe ($w = 1/3$), the energy conservation equation and Friedman equation lead to
 
 \begin{eqnarray}
 \rho_r \propto \frac{1}{a^4} \quad \mbox{and}\quad a(t) \propto t^{1/2}.
 \end{eqnarray}
 
 In the case of cosmological constant $\Lambda$ ($w = -1$), using the energy conservation and Friedman equations, we have
 
 \begin{eqnarray}
  \rho_\Lambda = \mbox{Const.} \quad \mbox{and} \quad a(t) \propto e^{H t}.
 \end{eqnarray}

 In general, $\rho$ in the Friedman equation (\ref{Friedman Eq}) consists of all types of matter. Dividing the equation with $H_0^2 = \frac{\kappa}{3}\rho_{cr}$, where $\rho_{cr}$ is the critical density, 
 
 \begin{eqnarray}\label{Eq.CriticalEn}
 \rho_{cr} = \frac{3H_0^2}{8\pi G} = 1.88 h^2 \times 10^{-29} \mbox{gm cm}{}^{-3}
 \end{eqnarray}
  and using the relation (\ref{Eq.Redshift}) we have
 
 \begin{eqnarray}
 \frac{H^2}{H_0^2} = \Omega_r (1 + z)^4 + \Omega_m (1 + z)^3 + \Omega_k (1 + z)^2 + \Omega_\Lambda
 \end{eqnarray}
 where $\Omega$ is defined as $\frac{\rho}{\rho_{cr}}$. It is important to note that we have also considered the energy contribution from the curvature, though it is negligible.

 For the matter and radiation dominated epoch, we see that as the Universe expands the energy density decreases. This implies that in the past the energy density of the fluids was very high. To understand the physics as we go back in time, we need to obtain relation between the energy density, pressure to the temperature. In the expanding Universe, the second law of thermodynamics as applied to comoving volume can be written as \cite{1990-Kolb.Turner-Book}
 
 \begin{eqnarray}
 T dS = d\,[(p + \rho) V] - V\,dp 
 \end{eqnarray}
 
 Using the integrability condition for the entropy, we get,
 
 \begin{eqnarray}
 dS \equiv d\left[ (p + \rho) \frac{V}{T}\right] = 0 \Rightarrow (p + \rho) \frac{V}{T} = {\text{Constant}} 
 \end{eqnarray}

This implies that in thermal equilibrium the entropy per comoving volume is constant. For a relativistic particle, the above expression implies that

\begin{eqnarray}
T \propto \frac{1}{a(t)}
\end{eqnarray}

This implies that, at earlier times, universe was much hotter than it is today \cite{1994ApJ...420..445F, Ade:2015xua, Komatsu:2008hk}. When the universe was much hotter and denser, $(T \sim$ MeV), there were no bound atoms or nuclei and all particles were highly energetic traveling close to the speed of light. Hence, at early times, all the particles can be treated as relativistic  and hence, the Universe was radiation dominated. As the Universe expanded and cooled down, and the interaction between radiation and particles reduced, the Universe became matter dominated. Also, when the Universe cooled down below the binding energy of nuclei, light elements were formed. This epoch is commonly referred to as {\em Big Bang Nucleosynthesis}. Knowing the conditions of the early universe and the relevant nuclear cross-sections, it is possible to  calculate the expected primordial abundances of all the elements. These are consistent with the predictions and this consistency test is another confirmation of the standard model \cite{1990-Kolb.Turner-Book}. The table below lists the key events in the Universe history along with the approximate time and the temperature.
\begin{table}[h!]
\begin{center}
	\begin{tabular}{|m{4cm}|m{3cm}|m{9cm}|}
		\hline
		{\bf Epoch} & {\bf Temperature} & {\bf Characteristics} \\
		\hline
		{\bf Big Bang} & $\infty$ K & singularity (vacuum fluctuations?) \\ 
		$0$ s & $\infty$ eV & \\
		\hline
		{\bf Planck} & $> 10^{40} $ K & quantum gravity\\
		$0 < t \leq 10^{-43}$ s & $ >10^{36}$ eV & \\
		\hline
		{\bf Grand Unification} & $10^{36} - 10^{40}$ K & Gravity freezes out. \\
		$10^{-43} \leq t \leq 10^{-36}$ s & $10^{36} - 10^{32}$ eV & the `grand unified force'(GUT).\\
		\hline
		{\bf Inflationary}& $10^{33} - 10^{36}$ K& Inflation begins \\
		$10^{-36} - 10^{-32}$ s & $10^{29} - 10^{32}$ eV & strong forces freezes out.\\
		\hline
		{\bf Electroweak}& $10^{20} - 10^{33}$ K& weak force freezes out \\
		$10^{-32} - 10^{-12}$ s&$10^{16} - 10^{29}$ eV & 4-distinct forces (EM dominates)\\
		&& {\bf baryogenesis:} baryons and anti-baryons annihilate.\\
		\hline
		{\bf Quark}&$10^{16} - 10^{20}$ K& Universe contains hot quark-gluon plasma:\\
		$10^{-12} - 10^{-6}$ s& $10^{12} - 10^{16}$ eV& quarks, gluons and leptons.\\
		\hline
		{\bf Hadron}& $ 10^{12} - 10^{16}$ K & quarks and gluons bind into hadrons.\\
		$10^{-6} - 1$ s & $10^8 - 10^{12}$ eV&\\
		\hline
		{\bf Lepton} & $10^{10} - 10^{12}$ K& Universe contains photons ($\gamma$), muons ($\mu^{\pm}$),\\
		1 s - 3 mins & $10^6 - 10^8$ eV & electrons/positrons ($e^{\pm}$) and neutrinos ($\nu, \bar{\nu}$);\\
		&& nucleons $n$ and $p$ in equal numbers \\
		\hline
		1 s& $\leq 10^{12}$ K & $\mu^+$ and $\mu^-$ annihilate; $\nu$ and $\bar{\nu}$ decouple;\\
		& $\leq 10^{8}$ eV& $e^{\pm}, \gamma$ and nucleons remain.\\
		\hline
		 100 s& $10^{10}$ K, $10^{6}$ eV & $e^+, e^-$ annihilate\\
		 \hline 
	\end{tabular}
\end{center}
\caption{Brief history of Universe}
\end{table}

As mentioned above, when the Universe was much hotter and denser, then a typical photon would scatter many times on a trip across the Universe. This would lead to occasional absorption and reemission, leading to thermodynamic equilibrium between radiation and matter and hence the radiation would have blackbody spectrum Fig. (\ref{fig:BackgroundCMB}) \cite{1994ApJ...420..445F,Ade:2015xua,Komatsu:2008hk}

\begin{eqnarray}
I_\nu = \frac{4\pi \hslash \nu^3/c^2}{e^{2\pi \hslash \nu/k_B T} - 1}
\end{eqnarray}

As Universe expanded, the temperature and frequency of the radiation redshifts in the same way and hence, preserving blackbody spectrum. Therefore, when the radiation and matter decoupled at a redshift of 1100 (commonly referred to as last scattering surface) the radiation streams across the Universe and reach us from all the directions \cite{1994ApJ...420..445F, 1990-Kolb.Turner-Book, Ade:2015xua,Komatsu:2008hk}. The current temperature of the background radiation is $T_{\text{CMB}} = 2.73$ K \cite{1994ApJ...420..445F,Ade:2015xua,Komatsu:2008hk}, and the energy in this background is greater than the energy in all other photons in the Universe combined. CMB temperature is same in all directions, to roughly a part in $10^5$ \cite{1994ApJ...420..445F, Ade:2015xua,Komatsu:2008hk}. 

In short, CMB is the source of the most precise information about the Universe and about its early state. However, CMB also possesses several questions which the standard model of cosmology cannot address like: Why the CMB photons from different patches of the sky have the same temperature? Why the Universe is isotropic at the largest possible scales? What is the source of CMB anisotropies and primordial density perturbations (Fig. \ref{Fig.FullCMB})? Why the Universe is nearly flat? What is the origin of the primordial magnetic fields?

Inflationary paradigm \cite{1990-Kolb.Turner-Book, Baumann:2009ds,  2003-05-14-DodelsonScott-Moderncosmology, 2005-05-14-MukhanovViatcheslav-Physicalfoundationsofcosmology} can solve many of the above listed problems of standard model of cosmology. In the following section, we discuss some of the following problems and how inflationary paradigm provides solution.

\section{Very Early Universe}

The CMB photon carries the information of the earliest Universe and by carefully analyzing these photons, we can extract information about the early Universe. It tells us that the Universe was nearly homogeneous and isotropic at scales larger than 100 Mpc. In fact, the origin of fluctuations in CMB can also answer the origin of the large scale structures. But, unfortunately, it also leads us a fundamental problem of early universe: What is the mechanism for the generation of primordial temperature perturbations?

\begin{figure*}[h!]
\centering
\includegraphics[width=0.7\linewidth]{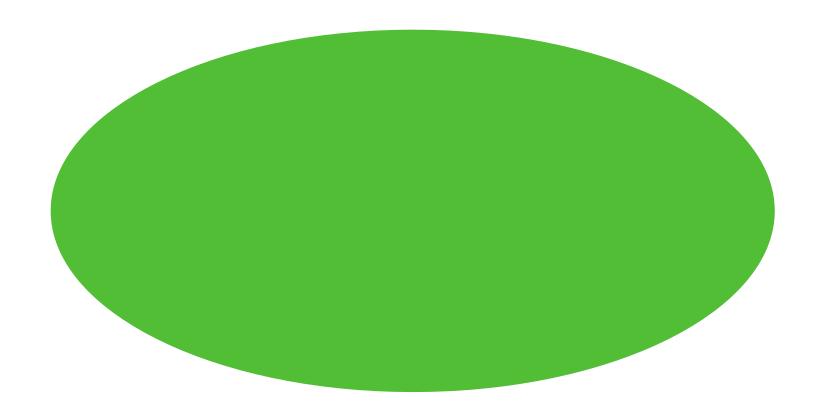}
\caption{Homogeneous and isotropic CMB}
\label{fig:BackgroundCMB}
\end{figure*}


Interestingly, the solution to the above problem also solves the problem of isotropy of the CMB temperature. In other words, how the two regions that were too far apart for them to be in contact know to start expanding in synchronization with each other? (Fig. \ref{fig:BackgroundCMB}).

\begin{figure}
\begin{center}
	\label{Fig.FullCMB}
\includegraphics[scale=.4]{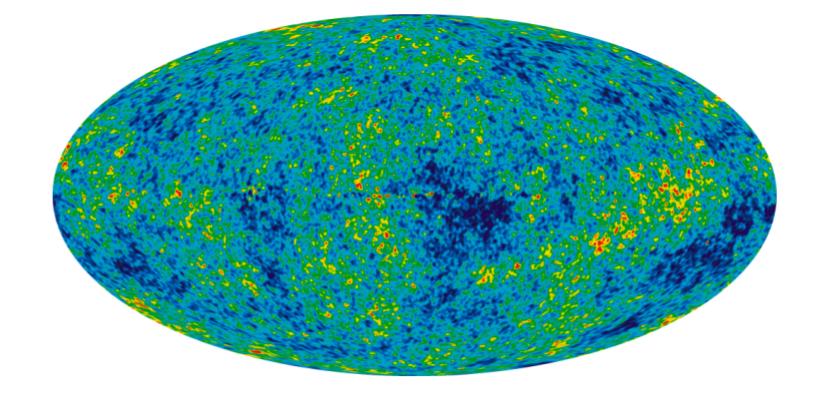}
\caption{Temperature fluctuations in CMB}
\end{center}
\end{figure}

In the rest of this section, we discuss the Horizon problem and the flatness problem in the standard cosmology.  We then discuss the elegant solution provided by the inflationary paradigm to overcome these problems. We also discuss in detail how inflation generates primordial density perturbations leading to temperature fluctuations in the CMB (Fig. \ref{Fig.FullCMB}).

\subsection{Horizon problem}

The maximum distance a photon can travel between an initial time $t_i$ and later time $t$, for the flat FRW metric (\ref{FRW metric}) is determined by the null geodesic $ds^2  = 0,$ or

\begin{eqnarray}
\Delta r = \int_{ t_i}^{t} \frac{dt^\prime}{a(t^\prime)}.
\end{eqnarray}

Therefore, the maximal distance a photon, hence any particle can travel since the beginning of the Universe is

\begin{eqnarray}\label{CoHorizon1}
\eta = \int_{0}^{t} \frac{dt^\prime}{a(t^\prime)}.
\end{eqnarray}

$\eta$ is referred to as \emph{comoving horizon} or \emph{particle horizon}. This can be re-written in a different form

\begin{eqnarray}\label{CoHorizonFinal}
\eta \equiv \int_{0}^{t} \frac{dt^\prime}{a(t^\prime)}  = \int (a H)^{-1} d\ln a.
\end{eqnarray}

$(a H)^{-1}$ is referred to as the \emph{comoving Hubble radius} which is a way of measuring causally connected region. If particles are separated with a distance larger than the Hubble radius, then, they are causally disconnected and cannot communicate with each other. However, later they may communicate and hence be in a causally connected region. 	In a universe with a perfect fluid with equation of state $w \equiv p/\rho$,  using equations (\ref{Conserv Eq}) and (\ref{Friedman Eq}), we find that

\begin{eqnarray}
(a H)^{-1} \propto a^{\frac{1}{2}\left(1 + w\right)},
\end{eqnarray}
which when plugged back in the equation (\ref{CoHorizonFinal}) leads to

\begin{eqnarray}
\eta \propto a^{\frac{1}{2}\left(1 + w\right)}.
\end{eqnarray}

All known matter sources satisfy $1 + 3 w > 0$ which leads to the fact that comoving Hubble radius as well as comoving horizon increase as the Universe expands. Hence, the initial singularity, in this case is $\eta_i = 0$ for $1+ 3 w > 0$ and the comoving horizon is finite.

Hence, this leads to the following conclusion: what we see today in the CMB with very large wavelength mode enters to the current horizon at late times and at very early times, they were well outside the horizon. This means that, at earlier times, these modes were in causally disconnected regions, yet, they were synchronized.

\subsection{Flatness problem}

Consider the Friedman equation (\ref{Friedman Eq})	 with the line element (\ref{FRW k metric})

\begin{eqnarray}
H^2  = \frac{\kappa}{3} \rho - \frac{k}{a^2}
\end{eqnarray} 

Dividing both sides with the Hubble factor leads to the equation

\begin{eqnarray}
1 - \Omega(a) = \frac{k}{(aH)^2}
\end{eqnarray}
where $\Omega$ is the ratio of energy density to critical energy density (\ref{Eq.CriticalEn}). The left hand side is the measure of the curvature of our Universe. In other words, the deviation of $\Omega$ from unity is the measurement of the curvature of the 3-space. From observations, we know that, $|1 - \Omega_0| \sim 0.01 $ \cite{Ade:2015xua, Komatsu:2008hk}. However, even for a small value of $k$ at earlier times, the right hand side will grow with time. Thus, starting with a large curvature, it is difficult to maintain the small flatness of $0.01$ at present time. On the other side, if we want to maintain flatness of $0.01$ at present time, then at very earlier times, the Universe had to be extremely flat. Within the standard cosmology, there is no reason why the 3-curvature of the early Universe need to be flat.

\subsection{Solution: Inflationary paradigm}
One important feature to observe in the case of matter or radiation dominated Universe is that the particle horizon as well as Hubble radius grow with time. To understand this, let us transform to conformal time coordinate ($\eta$), i.e., 

\begin{eqnarray}
d\eta = \frac{dt}{a}.
\end{eqnarray}

and the line element (\ref{FRW k metric}) for $k = 0$ becomes

\begin{eqnarray}
ds^2  = - dt^2 + a^2 (t) d{\bf x}^2 = a^2 (\eta) \left(-d\eta^2 + d{\bf x}^2\right).
\end{eqnarray}

In the new coordinate, FRW metric is conformally flat and the conformal time ($\eta$) is same as the comoving horizon or the particle horizon i.e., $\eta = \int \frac{dt}{a(t)}$. For the equation of state $p = w \rho$, as discussed in section (\ref{Sec.Prelim}), the scale factor is given by $a(t) \propto t^n$ where $n = 2/3 (1 + w)$, the particle horizon exists only if $w > -1/3$ or $(\rho + 3 p) > 0$ and grows with time. In the conformal time, Hubble radius $\mathcal{H}$ is $\mathcal{H} \equiv \frac{a(\eta)^\prime}{a(\eta)} = a(t) H(t)$ where $^\prime$ denotes derivative with respect to conformal time. Hence, we can see that $d\mathcal{H}/dt = (1 + 3w)/2 \,t^{-n} > 0$. We infer that the perfect fluid for which $w > -1/3$, particle horizon and comoving Hubble radius increases. In the case of cosmological constant $w = -1$, (as discussed in section (\ref{Sec.Prelim}), we also infer that the comoving Hubble radius decreases. One immediate consequence of this is 

\begin{eqnarray}
\frac{d}{dt}(aH)^{-1} < 0 \quad \Rightarrow \frac{d^2 a}{dt^2} > 0
\end{eqnarray}
which corresponds to \emph{accelerated expansion}. This is called \emph{Inflation}. The above relation can be written in a different form as \cite{2000-05-14-LiddleAndrewR.-Cosmologicalinflationandlarge-scalestructure}
 
 \begin{eqnarray}\label{LinkAccEq}
 \frac{d^2 a}{dt^2} > 0 \quad \Rightarrow \dot{H} + H^2 > 0
\end{eqnarray}

\begin{figure}
\centering
\includegraphics[width=1 \linewidth]{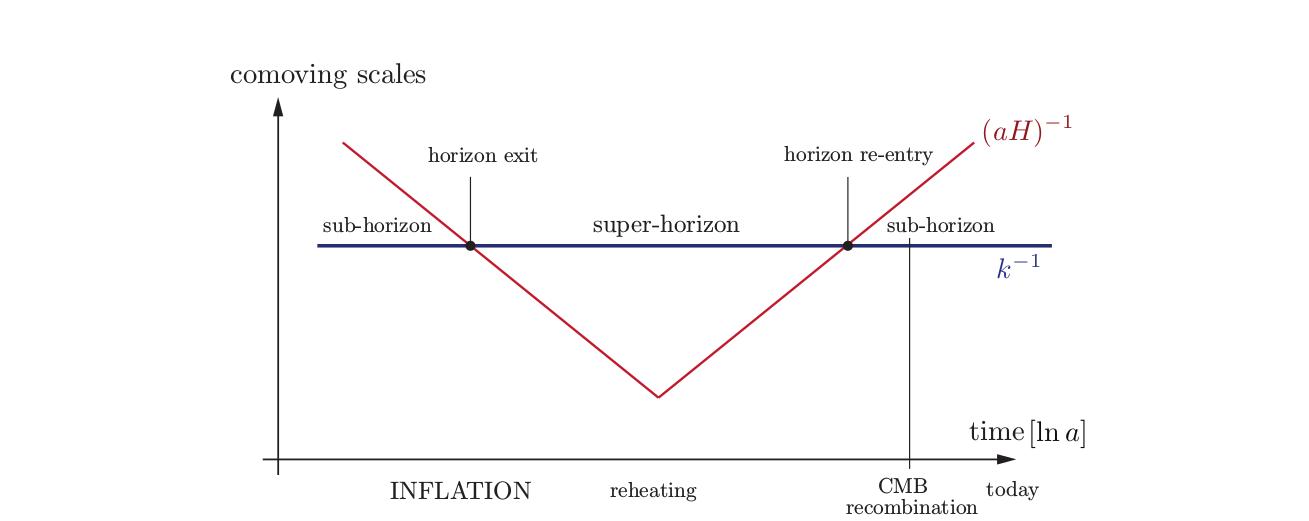}
\caption{Comoving Hubble radius (red) during and after inflation}
\label{fig:Horizon-comoving1}
\end{figure}

This has to be contrasted with matter or radiation dominated Universe where $\ddot{a} < 0$. Left hand side of the equation (\ref{LinkAccEq}) is the same as the left hand side of the acceleration equation (\ref{Acceleration Eq}) which again implies that $\rho + 3p < 0\, \Rightarrow\, w < -1/3$. This directly solves the horizon problem as well as flatness problem. Since $w < -1/3$, 

\begin{eqnarray}
\eta_i \sim a_i^{\frac{1}{2}(1 + 3w)} \rightarrow -\infty
\end{eqnarray} 
and hence, at very early time, horizon was so large that all modes were inside the horizon and causally connected (see Fig. \ref{fig:Horizon-comoving1}). Later, these modes cross the horizon and re-enter the horizon. Since comoving Hubble radius is shrinking, all initial fluctuations in the curvature $|1 - \Omega|$ decay and become flat, which also solves the flatness problem.

 Equation (\ref{LinkAccEq}) can be re-written as 
\begin{eqnarray}\label{DefSlow-roll}
 1 + \frac{\dot{H}}{H^2} > 0 \quad \Rightarrow \epsilon_1 < 1, \quad \epsilon_1 \equiv -\frac{\dot{H}}{H^2}.
 \end{eqnarray}

 \noindent $\epsilon_1$ is called the first slow-roll parameter and condition for inflation to occur is $\epsilon_1 < 1$. The condition for the exit of inflation is $\epsilon_1 = 1$. The slow-roll parameter can be written alternatively as \cite{2000-05-14-LiddleAndrewR.-Cosmologicalinflationandlarge-scalestructure}
 
 \begin{eqnarray}\label{Eq.SlowRoll1}
 \epsilon_1 \equiv -\frac{\dot{H}}{H^2} = -\frac{d \ln\, H}{dN}
 \end{eqnarray}
 
 Here, we have defined $dN  = d \ln a = H dt$, which measures the number of e-folds during the expansion of the Universe. It tells us that, fraction change of the Hubble parameter per e-fold is small. Moreover, to last inflation for a sufficient time ($N \sim 60$) to solve cosmological problems, we need $\epsilon_1$ to be small for sufficient time. This is measured by the second slow-roll parameter \cite{2000-05-14-LiddleAndrewR.-Cosmologicalinflationandlarge-scalestructure}
 
 \begin{eqnarray}\label{Eq.Slow-roll2}
 \eta_1 \equiv \frac{\dot{\epsilon_1}}{H\epsilon_1} = \frac{d\ln \epsilon_1}{dN} \ll 1.
\end{eqnarray}

Therefore, the necessary requirements for a suitable inflationary model are as follows:
\begin{itemize}
		\item Accelerated expansion, $\ddot{a} > 0$ with $H >0$,
		\item Negative pressure with $w < -1/3$,
		\item Slow-roll parameters $\epsilon_1 < 1,~\eta_1 \ll 1$.
	\end{itemize}

\section{Dynamics of inflation}\label{Sec.DynamicsInflation}
As discussed above, inflation is a dynamical theory and it starts when a component of the matter field with negative pressure starts dominating the stress tensor and after sufficient number of e-foldings should exit to radiation dominated Universe. This implies that the cosmological constant cannot be a possible candidate of inflation. In this case, the two slow-roll parameters, $\epsilon_1 = \eta_1 = 0$. Even it ends locally, post-inflationary Universe looks nothing like our Universe. The Universe is either empty or too much homogeneous. This problem is called \emph{the graceful exit problem} \cite{1990-Kolb.Turner-Book}. Since the two slow-roll parameters do not change in time, inflation once started with continue \emph{ad infinitum} and will never exit inflation. Since the pressure has to be negative, the fluid cannot be standard. 

Using the fact that the inflation takes place at very high energies \cite{2000-05-14-LiddleAndrewR.-Cosmologicalinflationandlarge-scalestructure}, at those energies, the natural description of matter is field theory. The simplest field theory --- relativistic self-interacting scalar field --- can lead to negative pressure that is necessary for the inflation \cite{Martin:2004um}.

 \subsection{Slow-roll inflation}\label{SubSec.SlowRoll}
 
Action for single canonical scalar field ($\varphi$) minimally coupled to gravity is given by

\begin{eqnarray}\label{ActionScalarMin}
\mathcal{S} = \int d^4 x\sqrt{-g}\left(\frac{1}{2\kappa}R - \frac{1}{2} g^{\mu \nu} \partial_{\mu}\varphi\partial_\nu \varphi - V(\varphi)\right).
\end{eqnarray}
where $R$ is the Ricci scalar, $V(\varphi)$ is the self-interacting potential, $g_{\mu \nu}$ is the metric and $\sqrt{-g}$ is the determinant of the metric. The stress-energy tensor which is defined by 

\begin{eqnarray}
T_{\mu \nu} = -\frac{2}{\sqrt{-g}}\frac{\delta\mathcal{S}}{\delta g^{\mu \nu}}
\end{eqnarray}
is given by
 \begin{eqnarray}\label{EMScalar}
 T_{\mu \nu} = \partial_\mu \varphi \partial_\nu \varphi - g_{\mu \nu} \left(\frac{1}{2} g^{\alpha\beta} \partial_\alpha\varphi \partial_\beta\varphi + V(\varphi)\right)
 \end{eqnarray}
 
From this expression, it is clear that the above stress-tensor can be written as perfect fluid with energy density and pressure given by
\begin{eqnarray}
 - T^0_0 &=& \rho = \frac{1}{2}\dot{\varphi}^2  + V(\varphi)  \\
 p &=& \frac{1}{2}\dot{\varphi}^2  - V(\varphi).
\end{eqnarray}

 Substituting the above $\rho$ and $p$ in the Friedman equation (\ref{Friedman Eq}) and acceleration equation (\ref{Acceleration Eq}) for the metric (\ref{FRW metric}), we get
 
 \begin{eqnarray}
 \label{FriedSc}
 H^2  &=& \frac{\kappa}{3}\left(\frac{1}{2} \dot{\varphi}^2 + V \right), \\
 \label{AccSc}
 \dot{H} &=& -\frac{\kappa}{2} \dot{\varphi}^2
 \end{eqnarray}
 
 Similarly, the continuity equation (\ref{Conserv Eq}) becomes
 
 \begin{eqnarray}
 \label{kleinSc}
 \ddot{\varphi} + 3 H \dot{\varphi} + V_\varphi = 0
 \end{eqnarray}

It is important to note again that only two of the above three equations are independent. To extract physics from these equations, let us obtain the slow-roll parameters defined in (\ref{Eq.SlowRoll1}) and (\ref{Eq.Slow-roll2}) for the canonical scalar field. Substituting (\ref{FriedSc}) and (\ref{AccSc}) in (\ref{Eq.SlowRoll1}), we get
\begin{eqnarray}
\epsilon_1 = \frac{3}{2} \frac{\dot{\varphi}^2}{\left(\frac{1}{2}\dot{\varphi}^2 + V\right)} .
\end{eqnarray}

As discussed above, condition for the inflation is $\epsilon_1 < 1$. From the above expression, we see that this condition can be satisfied when potential energy $V(\varphi)$ dominates over the kinetic energy ($\frac{1}{2}\dot{\varphi}^2$). In order to obtain sufficient amount of e-folds, we need the acceleration of the scalar field to be very small. Rewriting (\ref{kleinSc}), we have 

\begin{eqnarray}
3 H \dot{H} \left( 1 + \frac{1}{3}\delta\right) + V_\varphi = 0
\end{eqnarray}
where
\begin{eqnarray}
\delta_1 &\equiv& \frac{\ddot{\varphi}}{H\dot{\varphi}} = \frac{1}{2}\eta_1 - \epsilon_1
\end{eqnarray}  

As mentioned in the last section, $\eta_1 \ll 1$ for sufficiently long inflation. Till now, we have not made any approximations. Based on the above discussions, we can make approximations, which are referred to as slow-roll approximations, correspond to  $\{\epsilon_1, \eta_1, \delta_1\} \ll 1$. Using these approximations in (\ref{FriedSc}) and (\ref{AccSc}), we get

\begin{eqnarray}
 H^2 \approx \frac{\kappa}{3} V(\varphi).
\end{eqnarray}

Using $\delta_1 \ll 1 $, equation (\ref{kleinSc}) becomes

\begin{eqnarray}
3 H \dot{\varphi} + V_\varphi \approx 0
\end{eqnarray} 

During \emph{Slow-roll inflation}, the Hubble parameter is nearly constant, leading to quasi de-Sitter Universe and inflation stops when $\epsilon_1$ becomes unity. Since, scale factor $a(t)$ grows quasi-exponentially, the amount of inflation is measured by the number of e-folds of accelerated expansion as

\begin{eqnarray}
N \equiv \int\, d \ln a = \int_{t_i}^{t_f} H(t)\,dt.
\end{eqnarray} 

This can be re-written as

\begin{eqnarray}
\int H(t)\,dt = \int \frac{H}{\dot{\varphi}}\,d\varphi = \int \sqrt{\frac{\kappa}{2\epsilon}} d\varphi
\end{eqnarray}

To solve horizon and flatness problem require at least 50 e-foldings of inflation. PLANCK observations provide the upper limit of number of e-foldings to be around 65 \cite{Ade:2015xua}.

\section{Cosmological perturbation theory}

FRW line element (\ref{FRW metric}) was obtained from the assumption of cosmological principle that the Universe is homogeneous and isotropic. CMB observations also indicate that at the largest possible length scales, the Universe is indeed homogeneous and isotropic. However, as mentioned earlier, the CMB is the same in all directions up to 1 part in $10^5$ (see Fig.\ref{Fig.FullCMB}). 2dF/SDSS surveys \cite{2009MNRAS.399.1755C} show that the large scale structures were tiny at the early Universe and started growing due to gravitational attraction. To understand the evolution of the inhomogeneities, we need to solve Einstein's equations perturbatively by expanding around FRW line-element. The perturbation theory is valid only if the perturbations are small as in the CMB anisotropies \cite{Ade:2015xua,Komatsu:2008hk}.

Defining inhomogeneity over homogeneous FRW background leads to ambiguities in the choice of coordinates. General relativity is diffeomorphism invariant and hence, there is no preferred choice of coordinates. Choosing different coordinate system may lead to different descriptions of the same physical solution. This is referred as the \emph{gauge problem}. Therefore, it is important to describe the perturbations in terms of gauge-invariant variables \cite{Kodama01011984, Mukhanov:1990me,Bardeen:1980kt,Garriga1999,Lidsey:1995np,Sasaki1986}.

Any tensor can be decomposed into unperturbed and perturbed parts as

\begin{eqnarray}
{\bf T}(\eta, x^i) = {\bf T}_{0}(\eta) + \delta{\bf T}(\eta, x^i).
\end{eqnarray}
where any quantity with subscript 0 refers to those defined in the FRW background and $\delta$ refers to the perturbed quantity. The background part is independent of spatial dependence. The perturbation can further be written as 

\begin{eqnarray}
\delta{\bf T}_{0}(\eta, x^i) = \sum_{n = 1}^{\infty}\frac{\epsilon^n}{n!} \delta{\bf T}_{n}(\eta, x^i)
\end{eqnarray}
where $n$ denotes the order of perturbation and $\epsilon$ is the book keeping parameter that will be set to unity at the end. In the case of linear order perturbation, we consider only $\epsilon$ terms and neglect $\epsilon^2$ or higher order terms. FRW background line element in the conformal coordinate (\ref{FRW metric}) is given by

\begin{eqnarray}\label{ConformalMetric}
ds^2  = a^2(\eta) \left(-d\eta^2 + \delta_{i j} dx^i dx^j\right)
\end{eqnarray}

The first order perturbed part of the metric tensor can be written as \cite{Kodama01011984,Mukhanov:1990me}

\begin{eqnarray}\label{MetricPert}
\delta g_{0 0} &=& - 2 \epsilon\, a^2 \phi \nonumber \\
\delta g_{0 i} &=& a^2 \epsilon\,\left(B_{, i} - S_i\right)\nonumber \\
\delta g_{i j} &=& 2\,a^2 \epsilon\,\left(-\psi \delta_{i j} + E_{, ij} + F_{(i, j)} + \frac{1}{2} h_{i j}\right)
\end{eqnarray}
where we have used decomposition theorem and written tensor in terms of 3-scalars, 3-vectors and 3-tensor. Vector part is divergence-free, i.e., $S^i_{, i} = F^{i}_{, i} = 0$ and tensor part is traceless and divergence-less, i.e, $h^{i j}_{, i} = 0,~ h^{i}_{ i} = 0$.

The first order inverse perturbed metric takes the form

\begin{eqnarray}
\delta g^{0 0} &=& 2 a^{-2} \epsilon\,\phi \\
\delta g^{0 i} &=&  a^{-2} \epsilon\,\left(B^{, i} - S^i\right)\\
\delta g^{i j} &=& - 2 a^{-2} \epsilon\,\left( -\psi \delta^{i j} + E^{, ij} + F^{(i, j)} + \frac{1}{2} h^{i j}\right) 
\end{eqnarray}

Similarly scalar field ($\varphi$) and its energy density ($\rho$) can be decomposed in to background and perturbed parts as

\begin{eqnarray}
\varphi(\eta, x^i) &=& \varphi_0(\eta) + \epsilon\,\delta\varphi(\eta, x^i) \\
\rho(\eta, x^i) &=& \rho_0(\eta) + \epsilon\,\delta\rho(\eta, x^i) 
\end{eqnarray}

In the next subsection, we discuss the gauge problem and the way to go about resolving them by defining gauge-invariant variables.

\subsection{Gauge transformation and gauge-invariant variables}\label{SubSec.Gauge}

The gauge problem arises due to the decomposition of background and perturbed parts. However, the full GR is manifestly gauge-invariant. This gauge problem affects only the perturbations while the background remains the same in different coordinate systems. In our case, we assume the difference between coordinate systems is infinitesimal. There are two approaches to evaluate how perturbations behave under such transformation: active and passive approach \cite{Mukhanov:1990me, Malik2009}. Here, we use passive approach and evaluate how different types of perturbations change under such transformations and how to construct the gauge-invariant variables.

In case of first order perturbation, the relation between old and new coordinate systems at the same physical point $p$ is given by

\begin{eqnarray}\label{GaugeTrans}
\tilde{x}^\mu(p) = x^\mu(p) - \xi^\mu(p)
\end{eqnarray}
where $\xi^\mu$ is 4-vector corresponding to infinitesimal coordinate transformation. Under these transformations, the scalar field as well as energy density behave as

\begin{eqnarray}
\tilde{\varphi}(\tilde{x}^\mu) &=& \varphi(x^{\mu}),~~ \tilde{\rho}(\tilde{x}^\mu) = \rho(x^{\mu})\nonumber \\
\label{VarRho}
\Rightarrow \widetilde{\delta\varphi} &=& \delta\varphi + \varphi_0^\prime\,\xi^0, ~~\mbox{and}~~\widetilde{\delta\rho} = \delta\rho + \rho_0^\prime\,\xi^0
\end{eqnarray}

At linear order, the generating vector $\xi^{\mu}$ can again be decomposed in terms of scalar and vector parts as

\begin{eqnarray}\label{Generator}
\xi^{\mu} = \left(\alpha, \beta_{,}^{i} + \gamma^i\right), ~~\gamma_{, i}^{i} = 0
\end{eqnarray}
where $\alpha$ and $\beta$ are scalars, $\gamma_i$ is divergence-less vector. Substituting the above decomposition in (\ref{VarRho}), we get,

\begin{eqnarray}
\widetilde{\delta\varphi} = \delta\varphi + \varphi_0^\prime\,\alpha, ~~\mbox{and}~~\widetilde{\delta\rho} = \delta\rho + \rho_0^\prime\,\alpha
\end{eqnarray}

Since the metric $g_{\mu \nu}$ is a tensor, it follows

\begin{eqnarray}\label{MetricTrans}
\tilde{g}_{\mu \nu}(\tilde{x}^\eta) = g_{\alpha\beta}(x^\eta)\,\frac{\partial x^\alpha}{\partial\tilde{x}^\mu}\frac{\partial x^\beta}{\partial\tilde{x}^\nu}.
\end{eqnarray}

\noindent Using (\ref{MetricPert}), (\ref{MetricTrans}) and  (\ref{Generator}), the following relations for the 3-scalar perturbations can be established \cite{Kodama01011984, Mukhanov:1990me, Malik2009}:

\begin{eqnarray}
\widetilde{\phi} &=& \phi + \mathcal{H} \alpha + \alpha^\prime \\
\widetilde{\psi} &=& \psi - \mathcal{H} \alpha \\
\widetilde{B} &=& B - \alpha + \beta^\prime \\
\widetilde{E} &=& E + \beta.
\end{eqnarray}

The 3-vector perturbations transform as

\begin{eqnarray}
\widetilde{S}^i &=& S^i - \gamma^i{}^\prime \\
\widetilde{F}^i &=& F^i + \gamma^i
\end{eqnarray}

and, at linear order perturbation, the tensor perturbation remains invariant

\begin{eqnarray}
\widetilde{h}_{i j} = h_{i j}.
\end{eqnarray}

As one can see, only 3-tensor perturbations are invariant under gauge-transformation at linear order, whereas, all other perturbations are not invariant. By taking combinations of four scalar potentials defined in the metric perturbations, we can construct two such gauge-invariant quantities 
\begin{eqnarray}
\Phi = \phi + \mathcal{H}\left(B - E^\prime\right) + \left(B - E^\prime\right)^\prime, \quad\Psi = \psi - \mathcal{H} \left(B - E^\prime\right)
\end{eqnarray}

These variables were introduced first by Bardeen \cite{Bardeen:1980kt} and called as Bardeen potentials. Gauge-invariant variables are not limited to the above, we can construct infinite numbers of them. One easy way to construct such quantities is to choose a proper gauge, i.e., fix the value of $\xi^\mu$ and then identify and map those variables in generic gauge. For example, in \emph{longitudinal gauge}, we fix the gauge by choosing $B  = E = 0$ which leads to

\begin{eqnarray}
\beta = - E,\quad \alpha = B - E^\prime.  
\end{eqnarray}

Thus,
\begin{eqnarray}
\phi_l &=& \phi + \mathcal{H}\left(B - E^\prime\right) + \left(B - E^\prime\right)^\prime \equiv \Phi \\
\psi_l &=& \psi - \mathcal{H} \left(B - E^\prime\right) \equiv \Psi
\end{eqnarray}

Hence, in the \emph{longitudinal gauge}, metric perturbations $\phi$ and $\psi$ coincide with Bardeen potentials. Similarly, in \emph{flat-slicing gauge} with $\psi = E =0$, we can construct gauge-invariant quantity which is of interest during inflation. In this particular gauge, scalar field $\varphi$ coincides with Mukhanov-Sasaki variable which is gauge-invariant. Mukhanov-Sasaki variable is also related to \emph{comoving curvature perturbation} $\mathcal{R}$ as

\begin{eqnarray}\label{MukhSasaVar}
\delta\varphi_{\text{flat}} = \delta\varphi + \frac{\varphi_0^\prime}{\mathcal{H}}\,\psi = \frac{\varphi_0^\prime}{\mathcal{H}}\,\mathcal{R}
\end{eqnarray} 

This is also related to \emph{curvature perturbation} which is defined as

\begin{eqnarray}
-\zeta  = \psi + \frac{\mathcal{H}}{\rho_0^\prime}\,\delta\rho
\end{eqnarray}

At large scale, $\zeta = -\mathcal{R}$ which leads to $\zeta = \frac{\mathcal{H}}{\varphi_0^\prime} \delta\varphi_{\text{flat}}$. For details, see \cite{Malik2009}. For simplification, throughout the thesis, we use \emph{flat-slicing gauge}.

\subsection{Perturbations generated during inflation}

Since we know how to construct gauge-invariant variables by fixing a gauge, we can obtain the dynamics of those gauge-invariant variables from Einstein's equations. Let us consider the action of canonical scalar field minimally coupled to gravity (\ref{ActionScalarMin}). The stress-tensor for the canonical scalar field is given by (\ref{EMScalar}). The equation of motion of scalar field and the Einstein's equations in the FRW background are given in (\ref{kleinSc}), (\ref{FriedSc}) and (\ref{AccSc}).  

For the linear order scalar perturbations in the flat-slicing gauge, the Einstein's equations reduce to

\begin{eqnarray}\label{0-0FirstSc}
&&2a^2 V_0 \phi + \varphi_0^\prime\delta\varphi^\prime + a^2 V_{0, \varphi}\delta\varphi + \frac{2\mathcal{H}}{\kappa} \nabla^2 B  = 0 \\
\label{0-iFisrtSc}
&&\mathcal{H}\phi - \frac{\kappa}{2}\varphi_0^\prime\delta\varphi = 0 \\
&& B^\prime + 2\mathcal{H}B + \phi = 0.
\end{eqnarray}

The perturbed Klein-Gordon equation becomes
\begin{eqnarray}
\label{KGFirstSc}
\delta\varphi^{\prime\prime} + 2\mathcal{H}\delta\varphi^\prime + 2a^2 V_{0,\varphi}\phi - \nabla^2\delta\varphi - \varphi_0^\prime{}^2\nabla^2 B - \varphi_0^\prime\phi^\prime + a^2 V_{0,\varphi\varphi}\delta\varphi = 0
\end{eqnarray}

As one can see, $\phi$ as well as $B$ behave like constraints as they are not dynamical quantities. Using equations (\ref{0-0FirstSc}) and (\ref{0-iFisrtSc}), $\phi$ and $B$ can be written in terms of the background variables and $\delta\varphi$. Substituting these in (\ref{KGFirstSc}), we then have
\begin{eqnarray}
\delta\varphi^{\prime\prime} + 2\mathcal{H}\delta\varphi^\prime - \nabla^2\delta\varphi + \frac{a^2\kappa}{3}\left(2\varphi_0^\prime V_{0,\varphi} + \frac{\kappa\varphi_0^\prime{}^2}{\mathcal{H}} V_0\right)\delta\varphi + a^2 V_{0,\varphi\varphi}\delta\varphi = 0
\end{eqnarray}

Rewriting $\delta\varphi = v/a$ ($v$ is the Mukhanov-Sasaki variable), the above equation can be written in a simplified form: 

\begin{eqnarray}
\label{MukhSasaEq}
v^{\prime\prime} - \nabla^2 v - \frac{z^{\prime\prime}}{z}\, v = 0.
\end{eqnarray}

 Fourier decomposing $v$:

\begin{eqnarray}\label{FourierV0}
\hat{v}(\eta, {\bf x}) = \int \frac{d^3{\bf k}}{(2\pi)^{3/2}}\left(\hat{b}_{\bf k} u_{\bf k}(\eta)\,e^{i{\bf k.x}}+ \hat{b}^{\dagger}_{\bf k} u^*_{\bf k}(\eta)\,e^{-i{\bf k.x}}\right),
\end{eqnarray}
equation (\ref{MukhSasaEq}) becomes
\begin{eqnarray}\label{EqUk}
u_{\bf k}^{\prime\prime} + \left(k^2 - \frac{z^{\prime\prime}}{z}\right)\,u_{\bf k} = 0.
\end{eqnarray}

\noindent The above equation is referred to as Mukhanov-Sasaki equation \cite{Kodama01011984, Mukhanov:1990me}. Physically, Mukhanov-Sasaki variable is related to the comoving curvature perturbation $\mathcal{R}$ which coincides with curvature perturbation $\zeta$ at very large scales given in equation (\ref{MukhSasaVar}). Hence, solving the above equation, we can obtain the solution for $\zeta$ which is directly related to temperature fluctuations of the CMB ($\Delta T/T_0$) that we observe today. In case of de-Sitter Universe, the quantity $z$ vanishes, hence, $z^{\prime\prime}/z$ is undefined as pure de-Sitter cannot support density perturbations.

\subsection{Perturbations generated during slow-roll}

In case of leading order slow-roll approximations $\{\epsilon_1, \eta_1\} \ll 1$, as discussed in section (\ref{SubSec.SlowRoll}), we have
\begin{eqnarray}
z^2  &=& 2a^2 \epsilon_1 \\
\label{Z''}
\frac{z^{\prime\prime}}{z} &\approx& \left(aH\right)^2 \left(2 - \epsilon_1 + \frac{3}{2}\eta_1 \right)
\end{eqnarray}
where $\eta_1$ is the second slow-roll parameter defined in (\ref{Eq.Slow-roll2}). Also, in case of leading order slow-roll approximation

\begin{eqnarray}
aH \approx -\frac{1+ \epsilon_1}{\eta}.
\end{eqnarray}

Using this in the above equation, (\ref{Z''}) reduces to
\begin{eqnarray}
\frac{z^{\prime\prime}}{z} \approx \frac{\nu^2 - \frac{1}{4}}{\eta^2},\quad\text{where}\quad \nu \equiv \frac{3}{2} + \epsilon_1 + \frac{1}{2} \eta_1.
\end{eqnarray}
and Mukhanov-Sasaki equation ($\ref{EqUk}$) becomes

\begin{eqnarray}
\label{EqUKT}
u_{\bf k}^{\prime\prime} + \left(k^2 - \frac{\nu^2 - \frac{1}{4}}{\eta^2}\right)\,u_{\bf k} = 0.
\end{eqnarray}

There exists an exact solution for the above equation, which can be written in terms of Hankel functions of first and second kind as

\begin{eqnarray}
\label{EqUKsol}
u_{\bf k} = \sqrt{-\eta}\left[C_1\,H^{(1)}_\nu(-k \eta) + C_2\,H^{(2)}_\nu(-k \eta)\right].
\end{eqnarray}

As discussed earlier, inflation occurs at very high energies and natural description of matter is quantum field theory. We can fix the constants using the initial conditions for the Mukahnov-Sasaki variable at $\eta \rightarrow - \infty$. Using the fact that at very early times ($-k\eta \gg 1$), all modes are inside the horizon and are not affected by the gravity,  we set the modes to be as like in the flat space-time:

\begin{eqnarray}
\lim_{-k\eta\rightarrow \infty} u_{\bf k} = \frac{1}{\sqrt{2k}}\,e^{-ik\eta}.
\end{eqnarray}

This is referred to as \emph{Bunch-Davies vacuum} \cite{1978RSPSA.360..117B}. Using this condition, we find that $C_2 = 0$ and $C_1 = \sqrt{\pi/4}$. Therefore, we obtain

\begin{eqnarray}
u_{\bf k} = \sqrt{\frac{\pi}{4}}\sqrt{-\eta} H^{(1)}_\nu(-k \eta).
\end{eqnarray}

The 2-point correlation function is defined as

\begin{eqnarray}
<0|\hat{v}_{\bf k^\prime}\, \hat{v}_{\bf k}|0> &=& |u_{\bf k}|^2 \delta ({\bf k} + {\bf k}^\prime) \nonumber \\
&=& P_{v} \,\delta ({\bf k} + {\bf k}^\prime)
\end{eqnarray}  
where $P_{v}$ is called the power-spectrum of $v$. $\zeta$ is related to $v$ as $\zeta = z^{-1} \,v,\,z \sim \eta^{\frac{1}{2} - \nu}$. Hence, power-spectrum for $\zeta$ is

\begin{eqnarray}
P_\zeta = \frac{\pi}{4}  (-\eta)^{2\nu} \left|H^{(1)}_{\nu}(-k\eta)\right|^2
\end{eqnarray}

 Dimensionless power-spectrum is defined as \cite{2000-05-14-LiddleAndrewR.-Cosmologicalinflationandlarge-scalestructure}

\begin{eqnarray}
\Delta^2_{s} \equiv \frac{k^3}{2\pi^2} \,P_\zeta 
\end{eqnarray}

which, at large scale becomes
\begin{eqnarray}
\lim_{-k\tau \rightarrow0}\Delta^2_s \sim  k^{3 - 2\nu}.
\end{eqnarray}

Finally, spectral index is defined as

\begin{eqnarray}
n_s -1 \equiv \frac{d\ln \,\Delta^2_s}{d\ln \,k} &=& 3 - 2\nu \\
&=& - 2\epsilon_1 - \eta_1 . 
\end{eqnarray}

This implies that, the scalar power-spectrum is perfectly scale-invariant in de-Sitter Universe with $n_s  =1$. In case of slow-roll inflation, the Universe is quasi de-Sitter and deviation from scale-invariant spectra is measured via the slow-roll parameters.

In a similar manner, one can calculate vector and tensor power spectra. However, single scalar field driven inflationary models do not generate vector perturbations \cite{Kodama01011984}, hence it is insignificant. Repeating the above calculations for the tensor perturbations, we can get another important quantity, tensor-to-scalar ratio ($r$) which is given by \cite{2000-05-14-LiddleAndrewR.-Cosmologicalinflationandlarge-scalestructure, Mukhanov:1990me}

\begin{eqnarray}
 r  \equiv \frac{\Delta_s^2}{\Delta_t^2} = 16 \epsilon_1.
\end{eqnarray}

\section{Motivation of the thesis}\label{Sec.Motivation}

In the previous sections, we discussed how inflation solves many problems of standard model of cosmology  like horizon and flatness problem, and provides a casual mechanism for the generation of the primordial density fluctuations. However, inflation has several problems including it is still unclear what is the field that drives inflation (inflaton). 

Inflationary paradigm tells us that at early times, Universe has gone through near-exponential expansion phase. During this time, the temperature also drops down very fast, i.e., inflation is also a period of supercooled expansion. Hence, the Universe has to reheat rapidly which is referred to as \emph{reheating}. During this phase, inflaton energy decays to standard model particles which then lead to radiation-dominated era of the Universe. Inflation itself cannot solve all the problems of early Universe as \emph{reheating} phase is still not understood properly \cite{Bassett:2005xm}.

We have referred to the accelerated phase of the early Universe as inflationary paradigm indicating that there are no unique models of inflation. Many models lead to the near scale-invariance of the observed CMB power-spectra \cite{Ade:2015xua}. In the standard inflationary models, inflation is driven by canonical scalar field(s). The canonical scalar fields are the simplest and requires flat-potential to get sufficient e-foldings. However, there exist other scalar fields that can lead to inflation. Non-canonical scalar fields \cite{Armendariz-Picon1999, Garriga1999,
	Armendariz-Picon2001} are generalizations
of canonical scalar fields and reduces the dependence on the
potential. In case of non-canonical scalar field, even in the absence
of potential energy term, a general class of non-quadratic kinetic
terms can drive inflationary evolution. These models satisfy two crucial
requirements of inflationary scenarios: the scalar perturbations are
well-behaved during inflation and there exists a natural mechanism for
exiting inflation in a graceful manner. Non-canonical scalar field
also contains extra parameters than canonical scalar field such as
speed of sound. However, unlike canonical scalar field models, the
speed of propagation of the scalar perturbations in these inflationary
models can be time-dependent\cite{Armendariz-Picon1999, Garriga1999,
  Armendariz-Picon2001}. Recently, in order to seek more generalized
field, modified gravity models \cite{STAROBINSKY1982175}, scalar fields with higher time derivatives models like
Hordenski scalar fields, Kinetic Gravity Braiding models \cite{Horndeski1974,Kobayashi2010,Kobayashi2011,Nicolis2008,Deffayet2009,Deffayet:2010qz}
are considered. Besides these, there are plenty of other models including vector and dirac field that can lead to accelerated universe. Hence, it is still unknown what drives inflation.

Not only we do not know what drives inflation, inflationary paradigm depends on models which lead to `slow-roll' inflation. `Slow-roll' inflation, till now, fits best with the observation, however, is in contradiction with scalar field potential in standard model of particle physics \cite{Lyth:1998xn, Lidsey:1995np}.

As mentioned earlier, inflationary paradigm provides a natural explanation for primordial perturbations. However, the mechanism by which these perturbations are generated is not yet established. To be specific, the predictions of the inflationary models assume that Fourier modes generated are independent. Hence, the linear order perturbation theory development is sufficient. However, theory or observations do not rule out the possibility that the primordial perturbations generated during inflation are non-Gaussian. In other words, it is possible that the Fourier components of the fluctuations may not be independent --- non-Gaussian. This naturally raises two questions: How does one quantify primordial non-Gaussianity? What new information non-Gaussianity can provide about the inflationary model building?

If the fluctuations are indeed non-Gaussian, then the 3-point correlation of the temperature fluctuations of the CMB:
$$<\frac{\delta T ({\bf k_1})}{T_0}\, \frac{\delta T ({\bf k_2})}{T_0}\, \frac{\delta T({\bf k_3})}{T_0}>$$
must be non-zero. As mentioned earlier, temperature fluctuations are related to the 3-curvature perturbations. To compare the theoretical models with observations, we need to evaluate $$<\zeta({\bf k_1}) \zeta({\bf k_2}) \zeta({\bf k_3})>.$$

If we restrict to linear order perturbations, the above correlation function is identically zero. Hence, we need to go beyond  the linear order perturbation theory. To go about understanding this, let us treat $\zeta$ as an effective quantum field. Within canonical quantization, the above 3-point correlation of $\zeta$ can be written as

\begin{eqnarray}
<\zeta({\bf k_1}) \zeta({\bf k_2}) \zeta({\bf k_3})>\, \sim \int dt <0|\,[\zeta({\bf k_1}) \zeta({\bf k_2}) \zeta({\bf k_3}), \mathcal{H}^{\text int}_3]\,|0>
\end{eqnarray}
where $\mathcal{H}_3^{\text int}$ is the interaction Hamiltonian. Hence, to obtain higher order correlations, we need to include self-interacting terms and in the case of cosmological perturbations, this corresponds to higher order perturbed quantity. Therefore, we need to go beyond linear order perturbations and evaluate the interaction Hamiltonian. 

\emph{There are different approaches to evaluate higher order perturbations and one of the aim of this thesis is to obtain a unified formalism to obtain interaction Hamiltonian for different types of matter fields and gravity theories.}

The problem with any theory of gravity is that it is typically highly
non-linear, so one has to rely on approximation schemes to match the
observations. Primarily there exist two formalisms to deal with the
non-linear equations:
\begin{itemize}
\item The separate universe
  approximations\cite{Rigopoulos:2004gr,Rigopoulos:2004ba,Rigopoulos:2005xx}  with either gradient expansion theory or $\Delta N$
  formalism.\cite{PhysRevD.42.3936,Sasaki:1998ug,Lyth:2004gb,Lyth:2005fi,Langlois:2005ii,Langlois:2006vv,Sasaki:1995aw}
\item Gauge invariant cosmological perturbation
  theory\cite{Bardeen:1980kt,Bruni:1996im,Acquaviva:2002ud,Nakamura:2003wk,Maldacena2003,PhysRevD.69.104011,Bartolo:2001cw,Rigopoulos:2002mc,Bernardeau:2002jf,Bernardeau:2002jy,Malik:2003mv, Bartolo:2003bz,Finelli:2003bp,Bartolo:2004if,Enqvist:2004bk,Vernizzi:2004nc,Tomita:2005et,Lyth:2005du, Seery:2005gb, Malik:2005cy, Seery:2006vu,Langlois1994,BardeenSteinhardt1983}.
\end{itemize}


There are two mathematical procedures that are currently used in the
literature to study gauge invariant cosmological perturbation theory:
Hamiltonian formulation(ADM formulation)\cite{Langlois1994} and
Lagrangian
formulation\cite{Bardeen:1980kt,Bruni:1996im,Acquaviva:2002ud,Nakamura:2003wk,Maldacena2003,
  PhysRevD.69.104011,Bartolo:2001cw,Rigopoulos:2002mc,
  Bernardeau:2002jf,Bernardeau:2002jy,Malik:2003mv, Bartolo:2003bz,
  Finelli:2003bp,
  Bartolo:2004if,Enqvist:2004bk,Vernizzi:2004nc,Tomita:2005et,
  Lyth:2005du, Seery:2005gb, Malik:2005cy, Seery:2006vu}. The Lagrangian formulation can further be classified into two different approaches: 

\begin{enumerate}
\item Since gravity and matter are coupled to each other, one can write the full
action and vary the action with respect to metric and matter fields to
obtain general equations of motion (e.g., Einstein's equation in
General relativity). Those equations can be expanded in terms of
perturbed variables (metric and field variables) and one can write
down equations in the perturbation theory \cite{Lifshitz:1963ps}. This is order-by-order gravity equation approach.

\item In the action formalism \cite{Lukash:1980iv}, the action is expanded to
the required perturbed order in terms of the perturbed variables
(metric and field variables) and then varied the action with respect
to these variables. For example, to obtain perturbed equations in the
first order, we need to expand the action to second order and vary the
action with respect to the first order perturbed variables. This
formalism can be extended to obtain perturbed equations of motion up
to any order. In the reduced action formalism, constraint variables
are replaced in the action using constraint equations so that we can
rewrite the action only in terms of dynamical variables.
\end{enumerate}

While the Lagrangian formulation has been widely studied,
Hamiltonian framework in the context of cosmological perturbations is not common.
While calculating interaction Hamiltonian, in standard literature, perturbed Lagrangian is used and series of approximations are used to convert to into interaction Hamiltonian \cite{Chen:2006nt}. The thesis focuses on the different formalisms and techniques that are used to obtain
cosmological perturbation equations in the early universe. The analysis/comparison
of these approaches are much needed to enhance our theoretical predictions. First,
we look into Lagrangian formulation and establish the equivalence of action approach
and order-by-order gravity equation approach for generalized gravity models. Next
we concentrate on Hamiltonian formalism and provide a generalized Hamiltonian approach for cosmological perturbations at all orders. We also provide a new technique to
obtain Hamiltonian for generalized non-canonical scalar fields and apply our Hamiltonian approach for the same. We also apply our new approach to Galileon fields (fields
whose action has higher derivative terms, however, the equations of motion is still second order).

Observations indicate that Universe is filled with magnetic fields at all length scales \cite{Turner:1987bw, Dolgov:1993vg}. Like the growth of large scale structures, magnetic fields can be amplified by the dynamo mechanism. However, the dynamo mechanism requires primordial magnetic fields that are generated in the early Universe. Although, inflation provides a mechanism for the generation of primordial density perturbations, the same mechanism cannot provide an explanation for the magnetic field generation \cite{Birrell:1982ix}. Unlike scalar fields, the standard 4-dimensional electromagnetic action is conformally invariant.  Hence, inflation cannot generate the seed magnetic field with standard electromagnetism. The thesis focuses on modifications to electromagnetic theory leading to breaking of conformal invariance, and thus may able to produce primordial magnetic field \cite{Turner:1987bw}. Hence, we seek the modifications to the standard electromagnetic action without non-minimal coupling to any scalar. We are led to ask the following question: Can there exist a higher derivative action for spin-1 or vector fields that is free of ghosts? While constructing such an
action that preserves gauge-invariance is not possible in flat space-time, we explicitly
show that it is possible in curved space-time. In other words, for the first time, we
construct a vector Galileon model in curved space-time in which the field equations do
not contain any higher derivative terms, yet, it preserves U(1) gauge-invariance. Conformal invariance automatically breaks down for the model which leads to primordial
magnetogenesis and we compare the predictions of our model with observations.

A chapter wise summary of the thesis is given below:

In {\bf chapter 2,} we, first, concentrate on Lagrangian formulation. In Lagrangian formulation, there are two approaches to obtain perturbed equations: (i) Order-by-order
gravity equations and (ii) Action approach. Because of non-linear nature of gravity,
equivalence of these two approaches is not established properly. In Ref.\cite{Appignani2010a}, it was shown that when the metric perturbations are frozen then the two approaches
do not, in general, lead to the same expressions. In this work, we show
that, in case of canonical scalar field minimally coupled with gravity, these two approaches are equivalent for any order of perturbations. We extend our method and
show that the equivalence is generally true for any kind of gravity models at any order
of perturbations. We also find out that, in order to obtain reduced equation that can
lead to evolution of density fluctuations, `constraint consistency' has to be satisfied, i.e.,
Lapse function and shift vector should behave as constraints in the system. The crucial
result we obtain is that all `constraint inconsistent' models have Ostrogradski's instability
but the reverse is not true. In other words, one can have models with constrained Lapse
function and Shift vector, though it may have Ostrogradski's instabilities.

In {\bf chapter 3,} after focusing on Lagrangian formulation, we shift our concentration
on Hamiltonian formulation. In order to match observations with theory, we need a
proper approach of quantization and for canonical quantization, we need to know the
Hamiltonian of the system. In the case of cosmological perturbations, by using order-by-order approximation, perturbed interaction Hamiltonian is obtained from the reduced perturbed Lagrangian. This method is not straightforward, difficult to implement for constrained system and also suffers other difficulties as well as restrictions. In
this work, we present a consistent Hamiltonian approach of cosmological perturbation
theory for any kind of particle as well as gravity models at any order of perturbation,
which is simple, robust and provide much more information of the system than the earlier work. We show that, if Hamiltonian of any system is known to us, our approach can
be applied to all those systems. We also show that, this approach is not only simple and
straightforward to implement, but also provide extra information of the system, e.g.,
the true degrees of freedom of the system.

In {\bf Chapter 4,} we concentrate on the Hamiltonian formulation of generalized non-
canonical scalar field. In the case of general non-canonical scalar fields, since Legendre transformation is non-invertible, Hamiltonian of the system cannot be obtained.
However, in this work, we show that, by introducing a new phase-space variable, it is,
in fact, possible to define the Hamiltonian also for a generalized non-canonical scalar
field. Then, we apply our method: Hamiltonian approach of cosmological perturbation to non-canonical scalar field and show that our approach is consistent. We obtain a new expression of speed of sound in terms of phase-space variables for generalized non-canonical scalar field. Then we provide a general inversion relation between phase-space variable (and its derivatives) and configuration-space variable (and its derivatives) to compare and contrast general perturbed Hamilton’s equations and
Euler-Lagrange equations. Finally, we extend our approach to general higher-derivative
gravity models like Horndeski’s model.

Both in case of Lagrangian and Hamiltonian approach, we encounter several higher
derivative models that are interesting to study. Like Galileons the Einstein-Hilbert action, Lovelock, Hordenski model contain higher order derivatives, however equations of motion for those models are still second
order. We are led to ask: can there exist a corresponding action for spin-1 or vector
fields? While constructing such an action that preserves gauge-invariance is not possible in flat space-time, in {\bf chapter 5,} we explicitly show that it is possible in curved space-time. Conformal invariance is broken in this model, hence, providing a new model for primordial magnetogenesis.

Finally, in {\bf chapter 6,} we list the key results of the thesis and future direction that
arises from these results. In case of Hamiltonian formulation, for the first time, we
provide the generalized approach, however, the application of this is still wide open.
Also, in case of higher derivative models, we find that, we can extend the search for
other higher derivative vector Galileon since whatever we have provided is the first of
its kind. This can even be extended for higher order scalar-tensor theories. Not only
we can search for those models but the applications of those models can be studied for
different era of the universe at any length scales.




\pagestyle{fancy}
\chapter{Equivalence of two approaches and `constraint consistency'} \label{Chap.2}

\section{Introduction}\label{Sec.Intro.Chap2}
As mentioned in sections (\ref{Sec.DynamicsInflation}) and (\ref{Sec.Motivation}), inflation has now become an integral part of the standard model, that
can eliminate cosmological initial value problems, explain
homogeneities as well as inhomogeneities and observation of
anisotropic Cosmic Microwave Background Radiation
(CMBR)\cite{Komatsu:2008hk}. It is a period of accelerated expansion
in the very early universe and it occurs around $10^{14}$ GeV which is
much remote in time compared to the terrestrial
experiments. Inflationary cosmology has two key theoretical
aspects. One is the approximation schemes employed in solving gravity
equations. The other is the inflationary model building inspired by
particle physics or a fundamental theory of Quantum gravity. The
problem with any theory of gravity is that it is typically highly
non-linear, so one has to rely on approximation schemes to match the
observations. As mentioned in section (\ref{Sec.Motivation}), primarily there exist two formalisms to deal with the
non-linear equations:
\begin{itemize}
	\item The separate universe
	approximations\cite{Rigopoulos:2004gr,Rigopoulos:2004ba,Rigopoulos:2005xx}  with either gradient expansion theory or $\Delta N$
	formalism.\cite{PhysRevD.42.3936,Sasaki:1998ug,Lyth:2004gb,Lyth:2005fi,Langlois:2005ii,Langlois:2006vv,Sasaki:1995aw}
	\item Gauge invariant cosmological perturbation
	theory\cite{Bardeen:1980kt,Bruni:1996im,Acquaviva:2002ud,Nakamura:2003wk,Maldacena2003,PhysRevD.69.104011,Bartolo:2001cw,Rigopoulos:2002mc,Bernardeau:2002jf,Bernardeau:2002jy,Malik:2003mv, Bartolo:2003bz,Finelli:2003bp,Bartolo:2004if,Enqvist:2004bk,Vernizzi:2004nc,Tomita:2005et,Lyth:2005du, Seery:2005gb, Malik:2005cy, Seery:2006vu,Langlois1994,BardeenSteinhardt1983}.
\end{itemize}

The temperature fluctuations as observed in CMB is $\sim 10^{-5}$,
hence it is consistent to use order-by-order perturbation theory to
match with
observations \cite{Bardeen:1980kt,Mukhanov:1990me,Sasaki1986}. In the
first order, one assumes that the perturbed fields are linear. This
implies that the 3-point and higher order correlation functions are
zero. In the second order, the interactions of the first order need to
be included, hence, leading to non-zero 3-point functions. Also it is
widely believed that the detection of these 3-point correlation
functions can reduce the field space of inflationary
models \cite{Bruni:1996im,Maldacena2003,Seery:2005wm}.

With respect to inflationary model building, the proposed theories are
primarily preferred through simplicity. In the case of the canonical
scalar field, the simplest, $60$ e-foldings of inflation require the
potential to be flat, which is in contradiction with particle physics
models \cite{Lyth:1998xn,Lidsey:1995np}. Non-canonical scalar field
model \cite{Armendariz-Picon1999,Garriga1999,Armendariz-Picon2001}
removes the dependence of the potential, however it leads to time
dependence of the speed of perturbations and makes it difficult to be
compared with CMB observations \cite{Phys}. In order to seek more
generalized fields, scalar fields with higher time derivatives in
action are
considered \cite{Kobayashi2010,Kobayashi2011,Nicolis2008,Deffayet2009}. Beside
these, modified gravity models, specifically $f(R)$ lead to
accelerated expansion in the early universe.

As mentioned briefly in section (\ref{Sec.Motivation}), there are two mathematical procedures that are currently used in the
literature to study gauge invariant cosmological perturbation theory:
Hamiltonian formulation (ADM formulation) \cite{Langlois1994} and
Lagrangian
formulation \cite{Bardeen:1980kt,Bruni:1996im,Acquaviva:2002ud,Nakamura:2003wk,Maldacena2003,
	PhysRevD.69.104011,Bartolo:2001cw,Rigopoulos:2002mc,
	Bernardeau:2002jf,Bernardeau:2002jy,Malik:2003mv, Bartolo:2003bz,
	Finelli:2003bp,
	Bartolo:2004if,Enqvist:2004bk,Vernizzi:2004nc,Tomita:2005et,
	Lyth:2005du, Seery:2005gb, Malik:2005cy, Seery:2006vu}. Since
gravity and matter are coupled to each other, one can write the full
action and vary the action with respect to metric and matter fields to
obtain general equations of motion (e.g., Einstein's equation in
General relativity). Those equations can be expanded in terms of
perturbed variables (metric and field variables) and one can write
down equations in the perturbation theory \cite{Lifshitz:1963ps}. In
the action formalism \cite{Lukash:1980iv}, the action is expanded to
the required perturbed order in terms of the perturbed variables
(metric and field variables) and then varied the action with respect
to these variables. For example, to obtain perturbed equations in the
first order, we need to expand the action to second order and vary the
action with respect to the first order perturbed variables. This
formalism can be extended to obtain perturbed equations of motion up
to any order. In the reduced action formalism, constraint variables
are replaced in the action using constraint equations so that we can
rewrite the action only in terms of dynamical variables.

Since the matter fields (Non-canonical \& Galilean scalar field model)
and Gravity are highly non-linear, it is not clear whether the two
approaches, i.e., Einstein's equations writing in order-by-order
perturbation theory and action/reduced action formalism, lead to the
same equations of motion. In Ref. \cite{Appignani2010a}, it was shown
that when the metric perturbations are frozen then the two approaches
do not, in general, lead to the same expressions. In this work we
address the issue by including the metric perturbations in the theory.

In the next section, we study higher order cosmological perturbation
theory for a single scalar field minimally coupled to gravity and show
the equivalence of the two approaches at all orders. We point out a
crucial and novel consistency check which we refer to as `constraint
consistency' condition that needs to be verified. We also show that
this provides a fast and efficient way to check the consistency and
apply it to minimally coupled non-canonical scalar field.

In section (\ref{sec3}), we apply the `constraint consistency' condition
to many inflationary models that are proposed in the literature. First
we check the theory with higher derivative Lagrangian models minimally
coupled to gravity. Then we extend the procedure to other different types
of models like modified gravity models and modified gravity with
higher order matter Lagrangian. Appendix (\ref{App:Appendix A}) contains
some of the derived expressions used in section (\ref{sec2}) and in
Appendix (\ref{App:Appendix B}), we obtain a single variable equation of
motion for non-canonical scalar fields in terms of second order
perturbed variables.

In this chapter and other chapters, some of the 
long calculations have been performed using Cadabra \cite{Peeters:2007wn,DBLP:journals/corr/abs-cs-0608005}.

\section{Consistency of Higher order perturbations in two different
  approaches}\label{sec2}

The action for gravity sourced by a single, non-minimally coupled
scalar field ($\varphi$) is,
\begin{equation}
\label{act}
\mathcal{S} = \int d^4x\sqrt{-g} \left( \frac{R}{2 \kappa} + 
  \mathcal{L}_m \right)
\end{equation}
where  
\begin{equation}\label{boxx}
  \mathcal{L}_m = P(X, \varphi) + G(X, \varphi) \Box \varphi,~~~ ~X \equiv 
\frac{1}{2} g^{\mu \nu} \partial_{\mu}{\varphi} \partial_{\mu}{\varphi}, ~~~ 
\Box \equiv -\frac{1}{\sqrt{-g}}\partial_{\mu}\left(\sqrt{-g} g^{\mu \nu} \partial_{\nu}\right)
\end{equation}
is the Lagrangian for the Galilean field \cite{Kobayashi2010,Kobayashi2011,Unnikrishnan2014}. Varying the action with respect to metric leads to the Einstein's equations:
\begin{equation}
\label{eine}
R_{\mu \nu} - \frac{1}{2} g_{\mu \nu}~ R = \kappa~ T_{\mu \nu}, 
\end{equation}
where the stress tensor $T_{\mu \nu}$ is,
\begin{eqnarray}
\label{TG}
T_{\mu \nu} &=& g_{\mu \nu} \{ P + G_X g^{\alpha
  \beta} \partial_{\alpha}{X} \partial_{\beta}{\varphi} + G_\varphi
g^{\alpha
  \beta} \partial_{\alpha}{\varphi} \partial_{\beta}{\varphi}\} \nonumber\\
 && - \{P_X + 2 G_\varphi + G_X \Box
\varphi\} \partial_{\mu}{\varphi} \partial_{\nu}{\varphi} - 2
G_X \partial_{\mu}{X} \partial_{\nu}{\varphi}
\end{eqnarray}

For simplicity and to obtain the physical features, we consider only
single scalar field theory minimally coupled to the gravity. Variation of the
action (\ref{act}) with respect to the scalar field ($\varphi$) leads
to the following equation of motion,
\begin{equation}
\label{Eq.Eg}
\begin{split}
  &\{2 G_\varphi - 2 X G_{X \varphi} + P_X\} \Box \varphi - \{P_{XX} +
  2 G_{X \varphi}\}\partial_{\mu}{\varphi} \partial^{\mu}{\varphi} -
  2X
  \{G_{\varphi} + P_{X \varphi} \} + P_\varphi\\
  & - G_X \{ \varphi_{, \mu \nu} \varphi_{,}^{\mu \nu} - \{\Box
  \varphi\}^2 + R_{\mu
    \nu}\partial^{\mu}{\varphi} \partial^{\nu}{\varphi}\} - G_{XX}
  \{\partial_{\mu}X \partial^{\mu}X +
  \{\partial_{\mu}{\varphi} \partial^{\mu}X\}\Box \varphi\}\} = 0
\end{split}
\end{equation}

As one can see, although the Lagrangian is of the form, $\mathcal{L}_m
= \mathcal{L}_m ( \varphi,\partial{ \varphi}, \partial^2 {\varphi})
$, i.e., it contains higher time derivatives of the scalar field but
equations of motion are second order, thus does not suffer from
Ostrogradsky's instability \cite{Ostro,Woodard:2015zca}. With $G(X, \varphi) = 0$ the
field becomes non-canonical. Further fixing $ P = - X - V(\varphi)$,
where $V(\varphi)$ is the potential, the Lagrangian corresponds to
canonical scalar field.

The four-dimensional line element in the ADM form is given by,
\begin{eqnarray}
ds^2 &=& g_{\mu \nu} dx^{\mu} dx^{\nu} \nonumber \\
\label{line}
&=& -(N^2 - N_{i} N^{i} ) d\eta^2 + 2 N_{i} dx^{i} d\eta + \gamma_{i j} dx^{i} dx^{j},
\end{eqnarray}
where $N(x^{\mu})$ and $N_i(x^{\mu})$ are Lapse function and Shift
vector respectively, $\gamma_{i j}$ is the 3-D space metric. Note
that, in the case of Galilean model, $N(x^{\mu})$ and $N_i(x^{\mu})$
are the gauge constraints and variation of action (\ref{act}) with
respect to those lead to Hamiltonian and Momentum constraints,
respectively.

Action (\ref{act}) for the line element (\ref{line}) takes the form,
\begin{equation}
\label{Eq.AA}
\mathcal{S} = \int d^4x \sqrt{-g} \Big\{ \frac{1}{2 \kappa}
\left(^{(3)}R + K_{i j} K^{i j} - K^2\right) + \mathcal{L}_m \Big\}
\end{equation}
where $K_{i j}$ is extrinsic curvature tensor and is given by
\begin{eqnarray}
  && K_{i j} \equiv \frac{1}{2N} \Big[ \partial_{0}{\gamma_{i j}} - N_{i|j} - N_{j | i} \Big] \nonumber \\
  && K \equiv \gamma^{i j} K_{i j} \nonumber
\end{eqnarray}
  
`$|$' denotes covariant derivative in the 3-D spatial hypersurface which is characterized  by the metric $\gamma_{i j}$. As discussed in section (\ref{SubSec.Gauge}), in case of flat-slicing gauge at second order of perturbations and considering only scalar perturbations for simplicity, the metric and matter parts can be decomposed as
\begin{eqnarray}
  && g_{0 0} =- a^2(1 + 2 \epsilon \phi_1 + \epsilon^2 \phi_2 + ...) \\
  && g_{0 i} \equiv N_{i} = a^2 (\epsilon \partial_{i}{B_1} + \frac{1}{2} \epsilon^2 \partial_{i}{B_2} + ...) \\
  && g_{i j} =a^2 \delta^{i j}\\
  &&\varphi = \varphi_0 + \epsilon \varphi_1 + \frac{1}{2} \epsilon^2 \varphi_2+ ...
\end{eqnarray}
where $\epsilon$ is book keeping parameter which will be set to unity at the end. Now, we can use the above perturbed fields to evaluate equations of motion using both the approaches. Again, to confirm or infirm the result of Ref. \cite{Appignani2010a} that the
two approaches lead to different results we focus on non-canonical
scalar field, i.e., setting $G(X, \varphi) = 0$.

\subsection{Order-by-Order Einstein's equation approach}
 For the background, $g_{\mu \nu} = \mbox{diag } (- a^2, a^2, a^2, a^2)$,
equations (\ref{eine}) and (\ref{Eq.Eg}) lead to,

\begin{eqnarray}
\label{Back1}
 &&- \frac{\kappa}{3}  (P_X {\varphi^{\prime}_0}^{2} + P a^2)  = {\mathcal H}^{2} \\
 \label{Back2}
&&- 2 \frac{a^{\prime \prime}}{a} +   \mathcal{H}^2 = \kappa P a^2 \\
\label{Back3}
&& P_X {\varphi_0^{\prime \prime}} - P_{XX} {\varphi_0^{\prime \prime}}
{\varphi_0^{\prime}}^{2} a^{-2} + P_{X\varphi}
{\varphi_0^{\prime}}^{2} + 2 P_X {\varphi_0^{\prime}} \mathcal{H} +
P_{XX} \mathcal{H} {\varphi_0^{\prime}}^{3} a^{-2} + P_{\varphi}
a^{2} = 0
\end{eqnarray}

Equations (\ref{Back1}) and (\ref{Back2}) are zeroth order 0-0 and i-j
Einstein's equations where as equation (\ref{Back3}) is the zeroth
order equation of motion of the scalar field. Similarly, the first
order 0-0, 0-i Einstein's equations and equation of motion of the
perturbed scalar field are,

\begin{eqnarray}
\label{P1}
\mathcal{H} \nabla^2{B_1} &=& \frac{\kappa}{2} ( P_X \phi_1
{\varphi_0^{\prime}}^{2} + 2 P a^2 \phi_1 + 
P_X \varphi_0^{\prime} \varphi_1^{\prime} + P_{XX}
\phi_1 {\varphi_0^{\prime}}^{4} a^{-2} 
- P_{XX} \varphi_1^{\prime} {\varphi_0^{\prime}}^{3} a^{-2}
+ \nonumber \\
&& P_{X \varphi} {\varphi_0^{\prime}}^{2} \varphi_1 
+ P_\varphi \varphi_1 a^2 )\\
\label{P2}
\mathcal{H} \phi_1 &=& - \frac{\kappa}{2}P_X \varphi_0^{\prime} \varphi_1
\end{eqnarray}

\begin{eqnarray}
\label{P3}
&&- P_X {\varphi_1^{\prime \prime}} a^{2} - P_{XX}
{\phi_1^{\prime}}{\varphi_0^{\prime}}^{3} + P_{XX} {\varphi_1^{\prime
    \prime}} {\varphi_0^{\prime}}^{2} - P_{XX\varphi} \phi_1
{\varphi_0^{\prime}}^{4} 
+ P_{XX\varphi} {\varphi_1^{\prime}}
{\varphi_0^{\prime}}^{3} -P_{\varphi} \phi_1 a^{4} \nonumber \\
&&- P_{\varphi \varphi} a^{4} \varphi_1 + P_X \phi_1 {\varphi_0^{\prime \prime}}
a^{2} + P_X \nabla^2{\varphi_1} a^{2} + P_X {\phi_1^{\prime}}{\varphi_0^{\prime}} a^{2} - 2 P_X {\varphi_1^{\prime}} \mathcal{H}
a^2 - 4 P_{XX} \phi_1 {\varphi_0^{\prime
    \prime}}{\varphi_0^{\prime}}^{2} \nonumber\\
  &&+ 3 P_{XX} {\varphi_0^{\prime}}
{\varphi_1^{\prime}}{\varphi_0^{\prime \prime}} + P_{XX\varphi}
{\varphi_0^{\prime \prime}}{\varphi_0^{\prime}}^{2} \varphi_1 -
P_{X\varphi} {\varphi_0^{\prime}} {\varphi_1^{\prime}} a^{2} -
P_{X\varphi} {\varphi_0^{\prime \prime}} a^{2} \varphi_1 -P_{X\varphi\varphi} {\varphi_0^{\prime}}^{2} a^{2} \varphi_1\nonumber\\
&& + 2 P_X
\phi_1 {\varphi_0^{\prime}} \mathcal{H} a^2 + P_X {\varphi_0^{\prime}}
\nabla^2{B_1} a^{2} + P_{XX} \phi_1 \mathcal{H}
{\varphi_0^{\prime}}^{3} - P_{XX} {\varphi_1^{\prime}}\mathcal{H}
{\varphi_0^{\prime}}^{2} - P_{XXX} \phi_1 \mathcal{H}
{\varphi_0^{\prime}}^{5} a^{-2} \nonumber\\
&&+ P_{XXX} \phi_1 {\varphi_0^{\prime
    \prime}}{\varphi_0^{\prime}}^{4} a^{-2} + P_{XXX}
{\varphi_1^{\prime}}\mathcal{H}{\varphi_0^{\prime}}^{4} a^{-2} -
P_{XXX}{\varphi_1^{\prime}}{\varphi_0^{\prime
    \prime}}{\varphi_0^{\prime}}^{3} a^{-2} - P_{XX\varphi}
\mathcal{H} {\varphi_0^{\prime}}^{3} \varphi_1\nonumber\\
&& - 2 P_{X\varphi}
{\varphi_0^{\prime}}\mathcal{H} \varphi_1 a^2 = 0
\end{eqnarray}

Note that, there are no $\phi_1^{\prime \prime}$ and $B_1^{\prime
  \prime}$ terms in the above three equations and equations (\ref{P1})
and (\ref{P2}) are, as expected, the constraint equations
corresponding to Lapse function and Shift vector. Hence, $\phi_1$ and
$B_1$ are constraints and we can eliminate them from the first order
equation of motion of the scalar field (\ref{P3}). In first order,
single variable equation for non-canonical scalar field in terms of
Mukhanov-Sasaki variable ($v$) is,
\begin{equation}
\label{S1}
v^{\prime \prime} - c_s^2 \nabla^2 v - \frac{z^{\prime \prime}}{z} v = 0
\end{equation}
where,
\begin{equation}
\label{vzc}
v \equiv a \varphi_1,~~~ z \equiv \frac{a \varphi_0^\prime}{\mathcal{H}},~~~ 
c_s^2 \equiv \frac{P_X}{P_X + 2X P_{XX}}
\end{equation}

Similarly, perturbed second order 0-0 and 0-i Einstein's equations for
non-canonical scalar fields at second order are, {\small
\begin{eqnarray}
\label{G002}
&&- 4 \phi_1 \mathcal{H} \nabla^2{B_1} + \mathcal{H} \nabla^2{B_2} - 2
\delta^{i j} \mathcal{H} \partial_{i}{B_1} \partial_{j}{\phi_1} -
\frac{1}{2} \nabla^2{B_1} \nabla^2{B_1} + \frac{1}{2} {\delta}^{i j}
{\delta}^{k l} {\partial}_{i k}{B_1} {\partial}_{j l}{B_1} \nonumber\\
&& + \kappa (
- \frac{1}{2} P_X {\delta}^{i j} {\partial}_{i}{B_1}
{\partial}_{j}{B_1} {\varphi_0^{\prime}}^{2} - P {\delta}^{i j}
{\partial}_{i}{B_1} {\partial}_{j}{B_1} a^2 - \frac{1}{2} P_X \phi_2
{\varphi_0^{\prime}}^{2} - P \phi_2 a^2 + 2 P_X \phi_1^{2}
{\varphi_0^{\prime}}^{2} + 4 P \phi_1^{2} a^2 \nonumber\\
&&+ 2 P_X \phi_1
\varphi_0^{\prime} \varphi_1^{\prime} + \frac{7}{2} P_{XX} \phi_1^{2}
{\varphi_0^{\prime}}^{4} a^{-2} - 5 P_{XX} \phi_1 \varphi_1^{\prime}
{\varphi_0^{\prime}}^{3} a^{-2} + P_{X \varphi} \phi_1
{\varphi_0^{\prime}}^{2} \varphi_1 - \frac{1}{2} P_X
\varphi_0^{\prime} \varphi_2^{\prime}\nonumber\\
&& - \frac{1}{2} P_X
{\varphi_1^{\prime}}^{2} - \frac{1}{2} P_{XX} \phi_2
{\varphi_0^{\prime}}^{4} a^{-2} - \frac{1}{2} P_{XX} {\delta}^{i j}
{\partial}_{i}{B_1} {\partial}_{j}{B_1} {\varphi_0^{\prime}}^{4}
a^{-2} + 2 P_{XX} {\varphi_0^{\prime}}^{2} {\varphi_1^{\prime}}^{2}
a^{-2} + \frac{1}{2} P_{XX} \varphi_2^{\prime}
{\varphi_0^{\prime}}^{3} a^{-2} \nonumber\\
&& - P_{XX} {\delta}^{i j}
{\partial}_{i}{B_1} {\partial}_{j}{\varphi_1} {\varphi_0^{\prime}}^{3}
a^{-2} - \frac{1}{2} P_{XX} {\delta}^{i j} {\partial}_{i}{\varphi_1}
{\partial}_{j}{\varphi_1}{\varphi_0^{\prime}}^{2} a^{-2} - P_{X
  \varphi} {\varphi_0^{\prime}} {\varphi_1^{\prime}} \varphi_1 -
\frac{1}{2} P_{X \varphi} {\varphi_0^{\prime}}^{2} \varphi_2\nonumber\\
&& -
\frac{1}{2} P_{XXX} \phi_1^{2} {\varphi_0^{\prime}}^{6} a^{-4} 
 + P_{XXX} \phi_1 {\varphi_1^{\prime}} {\varphi_0^{\prime}}^{5} a^{-4} -
\frac{1}{2} P_{XXX}{\varphi_0^{\prime}}^{4} {\varphi_1^{\prime}}^{2}
a^{-4} - \frac{1}{2} P_{X\varphi\varphi} {\varphi_0^{\prime}}^{2}
\varphi_1^{2}  \nonumber\\
&&- P_{XX\varphi} \phi_1 {\varphi_0^{\prime}}^{4} a^{-2}
\varphi_1 + P_{XX\varphi} {\varphi_1^{\prime}}
{\varphi_0^{\prime}}^{3} a^{-2} \varphi_1 
- \frac{1}{2} P_X
{\delta}^{i j} {\partial}_{i}{\varphi_1} {\partial}_{j}{\varphi_1} -
\frac{1}{2} P_\varphi \varphi_2 a^2 - \frac{1}{2} P_{\varphi \varphi}
\varphi_1^{2} a^2) = 0 \\
\label{G0i2}
&&\mbox{and} \nonumber\\
 &&- 2 \delta^{j k} \mathcal{H} \partial_{j}{B_1} \partial_{i k}{B_1} +
 \partial_{i}{\Phi_1} \nabla^2{B_1} + 4 \phi_1 \mathcal{H}
 \partial_{i}{\phi_1} - \mathcal{H} \partial_{i}{\phi_2} - \delta^{j
   k} \partial_{j}{\phi_1} \partial_{i k}{B_1}\nonumber\\
   && +\kappa ( - \frac{1}{2}
 P_X {\varphi_0^{\prime}} \partial_{i}{\varphi_2} - P_X {\varphi_1^{\prime}}
 \partial_{i}{\varphi_1} - P_{XX} \phi_1
 \partial_{i}{\varphi_1}{\varphi_0^{\prime}}^{3} a^{-2} +
 P_{XX}{\varphi_1^{\prime}}
 \partial_{i}{\varphi_1} {\varphi_0^{\prime}}^{2} a^{-2} \nonumber \\
 && - P_{X\varphi}
 {\varphi_0^{\prime}} \partial_{i}{\varphi_1} \varphi_1) = 0
\end{eqnarray}}
 and the equation of motion of the scalar field is,
 
 \begin{equation}
\label{eos2}
\begin{split}
&C_X P_X + C_{XX} P_{XX} + C_{XXX} P_{XXX} + C_{XXXX} P_{XXXX} +
C_{XXX\varphi} P_{XXX\varphi}\\
& + C_{XX\varphi} P_{XX\varphi} +
C_{XX\varphi\varphi} P_{XX\varphi\varphi} + C_{X\varphi} P_{X\varphi}
+ C_{X\varphi\varphi} P_{X\varphi\varphi} + C_{X\varphi\varphi\varphi}
P_{X\varphi\varphi\varphi}\\
& + C_{\varphi} P_{\varphi} +
C_{\varphi\varphi} P_{\varphi\varphi} + C_{\varphi\varphi\varphi}
P_{\varphi\varphi\varphi} = 0
\end{split}
\end{equation}
where $C_X, C_{XX}, ...$ are all second order perturbed quantities and
$P_X, P_{XX}, P_{X \varphi}, ...$ are background quantities. The
explicit form of C's are given in Appendix (\ref{App:Appendix A}).

It is important to note that the second order equations also do not
contain $\phi_2^{\prime \prime}$ and/or $B_2^{\prime \prime}$. Hence
one can obtain a single variable equation of motion of non-canonical
scalar field at second order. Malik et al obtained the single variable
equation of motion in second order for canonical scalar
field \cite{Malik2007}. So far we obtain all unperturbed and perturbed equations using order-by-order gravity equations approach. In the next section, using reduced action approach, we obtain all relevant background and perturbed equations and then we compare the two approaches.

\subsection{Reduced Action approach}\label{reducedaction}
In the reduced action approach, one perturbs the field variables ($g_{\mu
  \nu}, \varphi$) in the action and expands the action to the required
order. In other words, one assumes a priori the form of the metric and
the matter variables in the lowest order and expands
order-by-order. For instance, in the case of FRW background, the
action (\ref{Eq.AA}) becomes,

\begin{equation}
  ^{(0)}\mathcal{S}^{NC} = \int d^4x \left( P a^{4} - 3 \frac{1}{\kappa} 
    {a^\prime}^{2} \right)
\end{equation}

Varying the above action with respect to metric variable $a(\eta)$ and
$\varphi_0(\eta)$ leads to the equations (\ref{Back2}) and
(\ref{Back3}). Note that, as expected, these two equations are
independent of each other since $a(\eta)$ and $\varphi_0(\eta)$ are
dynamical variables. To obtain first order (in $\epsilon$) equations,
one expands action (\ref{Eq.AA}) upto second order of
$\epsilon$. \textit{In general, varying the $n^{th}$ order action with
  respect to $m^{th}$ order perturbed variables leads to $(n -m)^{th}$
  order perturbed equations. It may be worth noting that a given order
  equations of motion can be obtained in several ways, e.g., varying
  first order action with respect to first order variables leads to
  zeroth order equations of motion}.

Expanding the action (\ref{Eq.AA}) to the second order, only in terms of
first order variables ($\varphi_1, \phi_1, B_1$), we get {\small
\begin{eqnarray}
  ^{(2)}\mathcal{S}^{NC} &= & \int d^4x\Big(\frac{1}{2} P_X {\delta}^{i
    j} {\partial}_{i}{B_1} {\partial}_{j}{B_1}
  {\varphi_0^{\prime}}^{2} a^{2} - P_X \phi_1^{2}
  {\varphi_0^{\prime}}^{2} a^{2} + P_X \phi_1 {\varphi_0^{\prime}}
  {\varphi_1^\prime} a^{2} - \frac{1}{2} P_X {\varphi_1^\prime}^2
  a^{2} + \nonumber\\
  &&P_X {\delta}^{i j} {\varphi_0^{\prime}} {\partial}_{i}{B_1}
  {\partial}_{j}{\varphi_1} a^{2} + \frac{1}{2} P_X {\delta}^{i j}
  {\partial}_{i}{\varphi_1} {\partial}_{j}{\varphi_1} a^{2} +
  \frac{1}{2} P_{XX} \phi_1^{2} {\varphi_0^{\prime}}^{4} - P_{XX}
  \phi_1 {\varphi_1^\prime} {\varphi_0^{\prime}}^{3} +  \nonumber\\
  &&\frac{1}{2}
  P_{XX} {\varphi_0^{\prime}}^{2} {\varphi_1^\prime}^{2} + \frac{1}{2}
  P_{\varphi\varphi} \varphi_1^{2} a^{4} + P_{X\varphi} \phi_1
  {\varphi_0^{\prime}}^{2} a^{2} \varphi_1 - P_{X\varphi}
  {\varphi_0^{\prime}} {\varphi_1^\prime} a^{2} \varphi_1 + P_\varphi
  \phi_1 a^{4} \varphi_1 + \nonumber\\
  && \frac{1}{2} P {\delta}^{i j}
  {\partial}_{i}{B_1} {\partial}_{j}{B_1} a^{4} - \frac{1}{2} P
  \phi_1^{2} a^{4} - 2 \phi_1 {\delta}^{i j} \frac{1}{\kappa}
  {a^{\prime}} {\partial}_{i j}{B_1} a + \frac{3}{2} {\delta}^{i j}
  \frac{1}{\kappa} {\partial}_{i}{B_1} {\partial}_{j}{B_1}
  {a^{\prime}}^{2} - \frac{9}{2} \frac{1}{\kappa} \phi_1^{2}
  {a^{\prime}} ^{2} \nonumber\\
  &&
  + \frac{1}{2} {\delta}^{i j} {\delta}^{k l}
  \frac{1}{\kappa} {\partial}_{i k}{B_1} {\partial}_{j l}{B_1} a^{2} -\frac{1}{2} {\delta}^{i j} {\delta}^{k l} \frac{1}{\kappa}
  {\partial}_{i j}{B_1} {\partial}_{k l}{B_1} a^{2} \Big)
\end{eqnarray}}

After integrating by-parts, and dropping off boundary terms, we get,
{\small
\begin{eqnarray}
\label{AP1}
  ^{(2)}\mathcal{S}^{NC} &=& \int d^4x\Big(\frac{1}{2} P_X {\delta}^{i
    j} {\partial}_{i}{B_1} {\partial}_{j}{B_1}
  {\varphi_0^{\prime}}^{2} a^{2} - P_X \phi_1^{2}
  {\varphi_0^{\prime}}^{2} a^{2} + P_X \phi_1 {\varphi_0^{\prime}}
  {\varphi_1^\prime} a^{2} - \frac{1}{2} P_X
  {\varphi_1^\prime}^2 a^{2} + \nonumber\\
  &&P_X {\delta}^{i j} {\varphi_0^{\prime}} {\partial}_{i}{B_1}
  {\partial}_{j}{\varphi_1} a^{2} + \frac{1}{2} P_X {\delta}^{i j}
  {\partial}_{i}{\varphi_1} {\partial}_{j}{\varphi_1} a^{2} +
  \frac{1}{2} P_{XX} \phi_1^{2} {\varphi_0^{\prime}}^{4} - P_{XX}
  \phi_1 {\varphi_1^\prime} {\varphi_0^{\prime}}^{3} +
  \nonumber\\
  && \frac{1}{2} P_{XX} {\varphi_0^{\prime}}^{2} {\varphi_1^\prime}^{2} + \frac{1}{2} P_{\varphi\varphi} \varphi_1^{2} a^{4} + P_{X\varphi}
  \phi_1 {\varphi_0^{\prime}}^{2} a^{2} \varphi_1 - P_{X\varphi}
  {\varphi_0^{\prime}} {\varphi_1^\prime} a^{2} \varphi_1 + P_\varphi
  \phi_1 a^{4} \varphi_1 +  \nonumber\\
  &&\frac{1}{2} P {\delta}^{i j}
  {\partial}_{i}{B_1} {\partial}_{j}{B_1} a^{4} - \frac{1}{2} P \phi_1^{2} a^{4} - 2 \phi_1 {\delta}^{i j}
  \frac{1}{\kappa} {a^{\prime}} {\partial}_{i j}{B_1} a + \frac{3}{2}
  {\delta}^{i j} \frac{1}{\kappa} {\partial}_{i}{B_1}
  {\partial}_{j}{B_1} {a^{\prime}}^{2} - \frac{9}{2} \frac{1}{\kappa}
  \phi_1^{2} {a^{\prime}} ^{2} \Big) \nonumber \\
\end{eqnarray}}

Varying action (\ref{AP1}) with respect to $\varphi_1$, we obtain
first order equation of motion of the scalar field same as
(\ref{P3}). Similarly, varying action with respect to $\phi_1$ and
$B_1$ gives same equations as (\ref{P1}) and (\ref{P2}) respectively,
i.e.,
\begin{eqnarray}
  \left\{\frac{\delta S_2}{\delta \varphi_1}\right\}_{\phi_1, B_1} &\equiv& 
1^{st}~\mbox{order Equation of motion of the scalar field} \nonumber \\
  \left\{\frac{\delta S_2}{\delta \phi_1}\right\}_{\varphi_1, B_1} &\equiv& 
1^{st}~\mbox{order Hamiltonian constraint} \nonumber \\
  \left\{\frac{\delta S_2}{\delta B_1}\right\}_{\varphi_1, \phi_1} &\equiv& 
1^{st}~\mbox{order Momentum constraint} \nonumber
\end{eqnarray}

Similarly, we can expand (\ref{Eq.AA}) upto fourth order by expanding the
field variables ($\varphi_2, \phi_2, B_2$) and vary the action with
respect to second order perturbed field variables to obtain second
order equations. Fourth order action containing only $\varphi_2$ terms
are, 

{\scriptsize
\begin{eqnarray}
\label{AP20}
^{(4)}\mathcal{S}^{NC}_{\varphi_2} &=&\int d^4x \Big(\frac{1}{4} P_X \phi_2
\varphi_0^{\prime} \varphi_2^{\prime} a^{2} + \frac{1}{4} P_X
{\delta}^{i j} \varphi_0^{\prime} \varphi_2^{\prime}
{\partial}_{i}{B_1} {\partial}_{j}{B_1} a^{2} - \frac{3}{4} P_X
\varphi_0^{\prime} \varphi_2^{\prime} \phi_1^{2} a^{2} + \frac{1}{2}
P_X \phi_1 \varphi_1^{\prime} \varphi_2^{\prime} a^{2} - \frac{1}{8}
P_X {\varphi_2^{\prime}}^{2} a^{2} +\nonumber\\
&&
 \frac{1}{4} P_X {\delta}^{i j}
\varphi_0^{\prime} {\partial}_{i}{B_2} {\partial}_{j}{\varphi_2} a^{2}
- \frac{1}{2} P_X \phi_1 {\delta}^{i j} \varphi_0^{\prime}
{\partial}_{i}{B_1} {\partial}_{j}{\varphi_2} a^{2} + \frac{1}{2} P_X
{\delta}^{i j} \varphi_1^{\prime} {\partial}_{i}{B_1}
{\partial}_{j}{\varphi_2} a^{2} + \frac{1}{2} P_X {\delta}^{i j}
\varphi_2^{\prime} {\partial}_{i}{B_1} {\partial}_{j}{\varphi_1} a^{2}
+ \nonumber\\
&&
\frac{1}{8} P_X {\delta}^{i j} {\partial}_{i}{\varphi_2}
{\partial}_{j}{\varphi_2} a^{2} + 
\frac{3}{2} P_{XX}
\varphi_2^{\prime} \phi_1^{2} {\varphi_0^{\prime}} ^{3} - 2 P_{XX}
\phi_1 \varphi_1^{\prime} \varphi_2^{\prime} {\varphi_0^{\prime}}^{2}
+ \frac{1}{2} P_{XX} \phi_1 {\delta}^{i j} {\partial}_{i}{B_1}
{\partial}_{j}{\varphi_2} {\varphi_0^{\prime}}^{3} + \nonumber\\
&&\frac{1}{2}
P_{XX} \phi_1 {\delta}^{i j} {\partial}_{i}{\varphi_1}
{\partial}_{j}{\varphi_2} {\varphi_0^{\prime}}^{2} -
 \frac{1}{4}
P_{XX} \phi_2 \varphi_2^{\prime} {\varphi_0^{\prime}} ^{3} -
\frac{1}{4} P_{XX} {\delta}^{i j} \varphi_2^{\prime}
{\partial}_{i}{B_1} {\partial}_{j}{B_1} {\varphi_0^{\prime}} ^{3} +
\frac{3}{4} P_{XX} \varphi_0^{\prime} \varphi_2^{\prime}
{\varphi_1^{\prime}} ^{2} -  \nonumber\\
&&
\frac{1}{2} P_{XX} {\delta}^{i j}
\varphi_1^{\prime} {\partial}_{i}{B_1} {\partial}_{j}{\varphi_2}
{\varphi_0^{\prime}} ^{2} -\frac{1}{2} P_{XX} {\delta}^{i j}
\varphi_0^{\prime} \varphi_1^{\prime} {\partial}_{i}{\varphi_1}
{\partial}_{j}{\varphi_2} + \frac{1}{8} P_{XX} {\varphi_0^{\prime}}
^{2} {\varphi_2^{\prime} }^{2} - \frac{1}{2} P_{XX} {\delta}^{i j}
\varphi_2^{\prime} {\partial}_{i}{B_1} {\partial}_{j}{\varphi_1}
{\varphi_0^{\prime} }^{2} - \nonumber \\
&&
\frac{1}{4} P_{XX} {\delta}^{i j}
\varphi_0^{\prime} \varphi_2^{\prime} {\partial}_{i}{\varphi_1}
{\partial}_{j}{\varphi_1} +\frac{1}{8} P_{\varphi \varphi}
\varphi_2^{2} a^{4} + \frac{1}{4} P_{X\varphi} \phi_2
{\varphi_0^{\prime} }^{2} a^{2} \varphi_2 + \frac{1}{4} P_{X\varphi}
{\delta}^{i j} {\partial}_{i}{B_1} {\partial}_{j}{B_1}
{\varphi_0^{\prime} }^{2} a^{2} \varphi_2 -  \nonumber\\
&&
\frac{1}{2} P_{X\varphi}
\phi_1^{2} {\varphi_0^{\prime} }^{2} a^{2} \varphi_2 + \frac{1}{2}
P_{X\varphi} \phi_1 \varphi_0^{\prime} \varphi_1^{\prime} a^{2}
\varphi_2 +\frac{1}{2} P_{X\varphi} \phi_1 \varphi_0^{\prime}
\varphi_2^{\prime} a^{2} \varphi_1 - \frac{1}{4} P_{X\varphi}
\varphi_0^{\prime} \varphi_2^{\prime} a^{2} \varphi_2 - \frac{1}{4}
P_{X\varphi} {\varphi_1^{\prime} }^{2} a^{2} \varphi_2 - \nonumber\\
&&
\frac{1}{2}
P_{X\varphi} \varphi_1^{\prime} \varphi_2^{\prime} a^{2} \varphi_1 +
\frac{1}{2} P_{X\varphi} {\delta}^{i j} \varphi_0^{\prime}
{\partial}_{i}{B_1} {\partial}_{j}{\varphi_1} a^{2} \varphi_2 +\frac{1}{2} P_{X\varphi} {\delta}^{i j} \varphi_0^{\prime}
{\partial}_{i}{B_1} {\partial}_{j}{\varphi_2} a^{2} \varphi_1 +
\frac{1}{4} P_{X\varphi} {\delta}^{i j} {\partial}_{i}{\varphi_1}
{\partial}_{j}{\varphi_1} a^{2} \varphi_2 +  \nonumber\\
&&
\frac{1}{2} P_{X\varphi}
{\delta}^{i j} {\partial}_{i}{\varphi_1} {\partial}_{j}{\varphi_2}
a^{2} \varphi_1 - \frac{1}{4} P_{XXX} \varphi_2^{\prime} \phi_1^{2}
{\varphi_0^{\prime} }^{5} a^{-2} +\frac{1}{2} P_{XXX} \phi_1
\varphi_1^{\prime} \varphi_2^{\prime} {\varphi_0^{\prime} }^{4} a^{-2}
- \frac{1}{4} P_{XXX} \varphi_2^{\prime} {\varphi_0^{\prime} }^{3}
{\varphi_1^{\prime} }^{2} a^{-2} +  \nonumber\\
&&
\frac{1}{4} P_{\varphi \varphi
	\varphi} \varphi_1^{2} a^{4} \varphi_2 +\frac{1}{4} P_{XX\varphi}
\phi_1^{2} {\varphi_0^{\prime} }^{4} \varphi_2 - \frac{1}{2}
P_{XX\varphi} \phi_1 \varphi_1^{\prime} {\varphi_0^{\prime} }^{3}
\varphi_2 -\frac{1}{2} P_{XX\varphi} \phi_1 \varphi_2^{\prime}
{\varphi_0^{\prime} }^{3} \varphi_1 + \frac{1}{4} P_{XX\varphi}
{\varphi_0^{\prime}}^{2} {\varphi_1^{\prime} }^{2} \varphi_2 +
 \nonumber\\
&&
\frac{1}{2} P_{XX\varphi} \varphi_1^{\prime} \varphi_2^{\prime}
{\varphi_0^{\prime} }^{2} \varphi_1 + \frac{1}{2} P_{X\varphi \varphi}
\phi_1 {\varphi_0^{\prime} }^{2} a^{2} \varphi_1 \varphi_2 -
\frac{1}{2} P_{X\varphi \varphi} \varphi_0^{\prime} \varphi_1^{\prime}
a^{2} \varphi_1 \varphi_2 -\frac{1}{4} P_{X\varphi \varphi}
\varphi_0^{\prime} \varphi_2^{\prime} \varphi_1^{2} a^{2} +
 \nonumber \\
&&\frac{1}{2} P_X \phi_1 {\delta}^{i j} {\partial}_{i}{\varphi_1}
{\partial}_{j}{\varphi_2} a^{2} + \frac{1}{2} P_{\varphi \varphi}
\phi_1 a^{4} \varphi_1 \varphi_2 + \frac{1}{4} P_\varphi \phi_2 a^{4}
\varphi_2 + \frac{1}{4} P_\varphi {\delta}^{i j} {\partial}_{i}{B_1}
{\partial}_{j}{B_1} a^{4} \varphi_2 - \frac{1}{4} P_\varphi \phi_1^{2}
a^{4} \varphi_2 \Big)
\end{eqnarray}}

Varying the action with respect to $\varphi_2$ leads to the same
equation of motion of $\varphi_2$ (\ref{eos2}). Similarly second order
equations of $\phi_2$ and $B_2$ can be obtained by varying fourth
order action with respect to $\phi_2$ and $B_2$, respectively and these lead to same equations (\ref{G002}) and
(\ref{G0i2}), respectively. This mechanism can be extended up to any
order and we can generalize that, \emph{equations obtained from both
  the approaches are identical and there are no ambiguities as
  discussed in Ref.} \cite{Appignani2010a}.

Another way of seeing constraints is, action (\ref{AP1}) or
(\ref{AP20}) contain no time derivative of $\phi_1, B_1, \phi_2$ and
$B_2$, i.e., Lapse function and Shift vector algebraically enter in
the action. Hence, variation with respect to $\phi$ and $B$ always
lead to constraint equations. So, we can use (\ref{P1}) and (\ref{P2})
constraint equations to eliminate $\phi_1$ and $B_1$ from the action
and use background equations (\ref{Back1}) and (\ref{Back2}) to obtain
a second order single variable action in terms of
$\varphi_1$. Further, writing the action in terms of Mukhanov-Sasaki
variable `$v$',
\begin{equation}\label{1stnc}
  ^{(2)}\mathcal{S}^{NC} = \frac{1}{2} \int d^4x \Big\{ {v^\prime}^2 -
  c_s^2~ \delta^{i j}~\partial_{i}{v}\partial_{j}{v} + \frac{z^{\prime
      \prime}}{z} v^2 \Big\}
\end{equation}
where $v, z$ and $c_s$ are defined in equation (\ref{vzc}). We can
vary the action (\ref{1stnc}) with respect to $v$ to obtain equation
of motion of $v$, which is identical to the equation (\ref{S1}).
Hence \emph{at first order, order-by-order Einstein's equation
  approach and reduced action approach lead to identical result.}

Similar procedure may be followed to obtain reduced second order
single variable equation of motion. Fourth order action does not
contain terms that have time derivatives of $\phi_2$ and $B_2$. Hence,
as in the first order, one should be able to substitute $\phi_2$ and
$B_2$ in the fourth order action to obtain a reduced action in terms
of $\varphi_2$. Malik et al showed, for canonical scalar field under
slow roll approximation, that the single variable equation from both
approaches are same \cite{Malik2008}. Similarly, since in the case of
non-canonical scalar field, equations of Lapse function $\phi_2$ and
Shift vector $\partial_i B_2$ are (1) identical for both the
approaches and (2) are constraint equations, reduced single variable
form of the equation of motion should also be identical.  In Appendix
(\ref{App:Appendix B}), we give the reduced single variable action as
well as equation of motion in terms of `$\varphi_2$' for non-canonical
scalar field in Power law limit.  Hence, \emph{both approaches give
  the identical results up to second order.}

At the first order, equations of motion are linear in first order
variables. At higher orders, only the highest order perturbed variables
appear linearly, where the lowest order perturbed variables contribute
non-linearly to equations of motion. For example, in second order,
equations are linear in second order variables $\varphi_2, \phi_2$ and
$B_2$ but are quadratic in first order variables $\varphi_1, \phi_1$
and $B_1$. Hence, as pointed out in \cite{Appignani2010a}, it does
appear that obtaining equations of field variables $\phi, B$ and
$\varphi$ at higher order from the two approaches may not be identical
and thus the reduced form of the single variable equations of motion
of the two different approaches may differ. However, \emph{instead of
  the non-linear form of the perturbed action, reduced single variable
  equations of motion at second order obtained from both the
  approaches are identical, hence we can generalize that at every
  order, in the case of non-canonical scalar field, both approaches
  lead to identical result.}  This leads to the following question:
\emph{ Why Appignani et al \cite{Appignani2010a} obtained different
  equations of motion from two approaches?}

In the simplified model proposed in Ref. \cite{Appignani2010a},
authors have assumed that the homogeneous universe is filled with
matter fluctuations with no Lapse function $\phi$ and Shift vector
$\partial_{i}{B}$. They have shown that in this simplified model
stress tensor, energy density and pressure are not identical for both
the approaches. Note that, since there are no metric fluctuations,
left hand sides of the perturbed Einstein's equations are zero. This
leads to five equations ($T^{0}_{0} = 0, T^{0}_{i} = 0$ and equation
of motion of scalar field) for a single perturbed variable
$\delta\varphi$, which lead to the inconsistency of the simplified
model, hence the ambiguities. Another way of looking into this is the
following: in the action approach, one can obtain the Hamiltonian
(momentum) constraint of the system by varying the action with respect
to $\phi$ ($B$). Since in this simplified model, both are not present,
this leads to inconsistent results.

This leads to another important question which we address in the rest
of the chapter is:
\emph{For what theories of gravity and matter field, the two approaches
  lead to identical single variable equation of motion?} 

To answer the question, let us look at the procedure of conventional
gauge invariant cosmological perturbation theory, which is based on
two things, first, to obtain gauge invariant variables and second, to
obtain a single variable action/equation of motion in terms of gauge
invariant variables. Gauge invariant variables are model independent
(if the background metric is unchanged), i.e., these are same for
canonical, non-canonical or Galilean models so that we can always
remove two variables out of five by using gauge conditions and define
suitable gauge invariant variables. At each order, we start from five
perturbed variables ($\phi, B, \psi, E$ and $\varphi$). The gauge
choice helps to remove two variables. Carefully choosing a gauge (in
our case, $E=0$ and $\psi=0$) at any order could fix the gauge issue
and reduced variables will coincide with gauge invariant
variables\cite{Malik2009}. So all equations in terms of those
variables also become gauge invariant.

Obtaining a single variable action/equation of motion depends solely
on gauge fixing (the procedure discussed in the above paragraph) and
two constraint equations which differ from model to model. If Lapse
function $N$ ($\phi$ in perturbed case) and Shift vector $N_i$
($\partial_i B$ in perturbed case) remain constraints for any models,
i.e., those functions algebraically enter into the action then
equations of motion of Lapse function and Shift vector contain no time
derivatives of them, we can always eliminate them from action/equation
of motion to get a single variable action/equation. This helps to
reduce the degrees of freedom to one and we can write the
action/equation of motion in a single variable form. However, if
$\phi_1$ or $B_1$ or both become dynamical i.e., if the action
contains terms containing time derivatives of Lapse function and/or
Shift vector such that equations contain double time derivatives of
those variables then it is not possible to substitute those variables
in the action or in the equation of motion of the scalar field and the
method fails. We refer the constrained nature of Lapse function and
Shift vector as `Constraint consistency' condition. If it is satisfied
then the whole method of gauge invariant cosmological perturbation
theory will work. In the next section, we test the `constraint
consistency' condition for several models that are used in the
literature.

In fact, the whole exercise may be done in terms of Lapse $N$, Shift
$N_{i}$ and scalar field $\varphi$ without applying any perturbation
theory. From Hamiltonian theory of General relativity or from
Einstein's equation we obtain constraint equations, i.e., Hamiltonian
and Momentum constraints which are functions of $(N, N_{i}, \gamma_{i
  j}, \varphi, \varphi^\prime, \partial_{i}\varphi)$ in which Lapse
function and Shift vector are constraints. If out of four constraint
equations (1 Hamiltonian equation or 0-0 Einstein's equation and 3
Momentum constraint equation or 0-i Einstein's equation), we can solve
and extract four quantity, one Lapse function and three component
Shift vector then we can substitute those back in the action or in
equation of motion of the scalar field. Unfortunately, GR equations
are so highly non-linear that analytically solving constraint
equations for Lapse function and Shift vector and obtaining a single
variable action or equation is very difficult. Perturbation theory
helps to simplify those equations so that we can invert those
equations in terms of Lapse function and Shift vector and obtain a
simplified solutions of constraint functions.

\section{Specific models}\label{sec3}

In this section, first we start with well known models of inflation
within the framework of general relativity and then move to modified
gravity models. To check for constraint consistency we follow action
formulation, write down second order action in terms of perturbed
variables and identify terms that contain time derivatives of Lapse
function and/or Shift vector.

\subsection{Minimally coupled Galilean field}

We start with the model with derivatives of metric in the action, 

\begin{equation}
 \mathcal{S}^G_m = \int d^4 x \sqrt{-g}~ G(X, \varphi) \Box \varphi 
\end{equation}
which has been proposed by Kobayashi et
al \cite{Kobayashi2010,Kobayashi2011} where `$\Box$' is defined by
equation (\ref{boxx}). Since `$\Box$' contains time derivatives of
metric as well as matter, it is not obvious whether the action can be
expressed in a single variable form. After partial integration, the
second order matter action becomes as follows:

{\scriptsize
\begin{eqnarray}
  ^{(2)}\mathcal{S}^G_{m} &=& \int d^4x \Big\{ 5 \bm{{G_X \phi_1
      \phi_1^\prime} {\varphi_0^\prime }^{3}} + \frac{15}{2} G_X
  \varphi_0^{\prime \prime} \phi_1^{2} {\varphi_0^\prime }^{2} -
  \frac{15}{2} G_X a^\prime \phi_1^{2} {\varphi_0^\prime }^{3} a^{-1}
  - \bm{3 G_X \phi_1^\prime \varphi_1^\prime {\varphi_0^\prime }^{2}}
  - 6 G_X \phi_1 \varphi_0^\prime {\varphi_1^\prime} \varphi_0^{\prime
    \prime} + \nonumber \\
    &&
    9 G_X \phi_1 \varphi_1^\prime a^\prime {\varphi_0^\prime
  }^{2} a^{-1} - G_X {\delta}^{i j} {\partial}_{i}{B_1} {\partial}_{0
    j}{B_1} {\varphi_0^\prime }^{3} - \frac{3}{2} G_X {\delta}^{i j}
  {\partial}_{i}{B_1} {\partial}_{j}{B_1} \varphi_0^{\prime \prime}
  {\varphi_0^\prime }^{2} + \frac{3}{2} G_X {\delta}^{i j} a^\prime
  {\partial}_{i}{B_1} {\partial}_{j}{B_1} {\varphi_0^\prime }^{3}
  a^{-1} - \nonumber \\
  &&
  3 G_X \phi_1 \varphi_1^{\prime \prime} {\varphi_0^\prime
  }^{2} + G_X \varphi_0^{\prime \prime} {\varphi_1^\prime }^{2} + 2
  G_X \varphi_0^\prime \varphi_1^\prime \varphi_1^{\prime \prime} - 3
  G_X \varphi_0^\prime a^\prime {\varphi_1^\prime }^{2} a^{-1} - 
  2 G_X
  {\delta}^{i j} \varphi_0^\prime {\partial}_{i}{B_1}
  {\partial}_{j}{\varphi_1} \varphi_0^{\prime \prime} -\nonumber \\
  &&
  G_X
  {\delta}^{i j} {\partial}_{i}{\varphi_1} {\partial}_{0 j}{B_1}
  {\varphi_0^\prime }^{2} - 2 G_X {\delta}^{i j} {\partial}_{i}{B_1}
  {\partial}_{0 j}{\varphi_1} {\varphi_0^\prime }^{2} + 3 G_X
  {\delta}^{i j} a^\prime {\partial}_{i}{B_1}
  {\partial}_{j}{\varphi_1} {\varphi_0^\prime }^{2} a^{-1} - 
  2 G_X
  {\delta}^{i j} \varphi_0^\prime {\partial}_{i}{\varphi_1}
  {\partial}_{0 j}{\varphi_1} +\nonumber \\
  &&
  G_X {\delta}^{i j} \varphi_0^\prime
  a^\prime {\partial}_{i}{\varphi_1} {\partial}_{j}{\varphi_1} a^{-1}
  + G_X {\delta}^{i j} {\partial}_{i}{B_1} {\partial}_{j}{\phi_1}
  {\varphi_0^\prime }^{3} + G_X {\delta}^{i j} {\partial}_{i}{\phi_1}
  {\partial}_{j}{\varphi_1} {\varphi_0^\prime }^{2} - \bm{G_{XX}
    \phi_1 \phi_1^\prime {\varphi_0^\prime }^{5} a^{(-2)}} -\nonumber \\
    &&
    5 G_{XX}
  \varphi_0^{\prime \prime} \phi_1^{2} {\varphi_0^\prime }^{4}
  a^{(-2)} + 5 G_{XX} a^\prime \phi_1^{2} {\varphi_0^\prime }^{5}
  a^{(-3)} + 7 G_{XX} \phi_1 \varphi_1^\prime \varphi_0^{\prime
    \prime} {\varphi_0^\prime }^{3} a^{(-2)} - 8 G_{XX} \phi_1
  \varphi_1^\prime a^\prime {\varphi_0^\prime }^{4} a^{(-3)} +\nonumber\\
  &&
   G_{XX}
  \phi_1 \varphi_1^{\prime \prime} {\varphi_0^\prime }^{4} a^{(-2)} +
  \frac{1}{2} G_{XX} {\delta}^{i j} {\partial}_{i}{B_1}
  {\partial}_{j}{B_1} \varphi_0^{\prime \prime} {\varphi_0^\prime
  }^{4} a^{(-2)} - \frac{1}{2} G_{XX} {\delta}^{i j} a^\prime
  {\partial}_{i}{B_1} {\partial}_{j}{B_1} {\varphi_0^\prime }^{5}
  a^{(-3)} + \bm{G_{XX} \phi_1^\prime \varphi_1^\prime
    {\varphi_0^\prime }^{4} a^{(-2)}}\nonumber\\
    &&
     - \frac{5}{2} G_{XX}
  \varphi_0^{\prime \prime} {\varphi_0^\prime }^{2} {\varphi_1^\prime
  }^{2} a^{(-2)} + \frac{7}{2} G_{XX} a^\prime {\varphi_0^\prime }^{3}
  {\varphi_1^\prime }^{2} a^{(-3)} - G_{XX} \varphi_1^\prime
  \varphi_1^{\prime \prime} {\varphi_0^\prime }^{3} a^{(-2)} + G_{XX}
  {\delta}^{i j} {\partial}_{i}{B_1} {\partial}_{j}{\varphi_1}
  \varphi_0^{\prime \prime} {\varphi_0^\prime }^{3} a^{(-2)} -\nonumber \\
  &&
  G_{XX}
  {\delta}^{i j} a^\prime {\partial}_{i}{B_1}
  {\partial}_{j}{\varphi_1} {\varphi_0^\prime }^{4} a^{(-3)} +
  \frac{1}{2} G_{XX} {\delta}^{i j} {\partial}_{i}{\varphi_1}
  {\partial}_{j}{\varphi_1} \varphi_0^{\prime \prime}
  {\varphi_0^\prime }^{2} a^{(-2)} - \frac{1}{2} G_{XX} {\delta}^{i j}
  a^\prime {\partial}_{i}{\varphi_1} {\partial}_{j}{\varphi_1}
  {\varphi_0^\prime }^{3} a^{(-3)} -\nonumber\\
  &&
   \bm{G_{XY} \phi_1^\prime
    {\varphi_0^\prime }^{3} \varphi_1} - 3 G_{XY} \phi_1
  \varphi_0^{\prime \prime} {\varphi_0^\prime }^{2} \varphi_1 + 3
  G_{XY} \phi_1 a^\prime {\varphi_0^\prime }^{3} a^{-1} \varphi_1 + 2
  G_{XY} \varphi_0^\prime \varphi_1^\prime \varphi_0^{\prime \prime}
  \varphi_1 - 3 G_{XY} \varphi_1^\prime a^\prime {\varphi_0^\prime
  }^{2} a^{-1} \varphi_1 + \nonumber\\
  &&
  G_{XY} \varphi_1^{\prime \prime}
  {\varphi_0^\prime }^{2} \varphi_1 + \frac{1}{2} G_{XXX}
  \varphi_0^{\prime \prime} \phi_1^{2} {\varphi_0^\prime }^{6}
  a^{(-4)} - \frac{1}{2} G_{XXX} a^\prime \phi_1^{2} {\varphi_0^\prime
  }^{7} a^{(-5)} - G_{XXX} \phi_1 \varphi_1^\prime \varphi_0^{\prime
    \prime} {\varphi_0^\prime }^{5} a^{(-4)} + \nonumber\\
    &&
    G_{XXX} \phi_1
  \varphi_1^\prime a^\prime {\varphi_0^\prime }^{6} a^{(-5)} +
  \frac{1}{2} G_{XXX} \varphi_0^{\prime \prime} {\varphi_0^\prime
  }^{4} {\varphi_1^\prime }^{2} a^{(-4)} - \frac{1}{2} G_{XXX}
  a^\prime {\varphi_0^\prime }^{5} {\varphi_1^\prime }^{2} a^{(-5)} +
  \frac{1}{2} G_{XYY} \varphi_0^{\prime \prime} {\varphi_0^\prime
  }^{2} \varphi_1^{2} -\nonumber \\
  &&
  \frac{1}{2} G_{XYY} a^\prime {\varphi_0^\prime
  }^{3} \varphi_1^{2} a^{-1} + G_{XXY} \phi_1 \varphi_0^{\prime
    \prime} {\varphi_0^\prime }^{4} a^{(-2)} \varphi_1 - G_{XXY}
  \phi_1 a^\prime {\varphi_0^\prime }^{5} a^{(-3)} \varphi_1 - G_{XXY}
  \varphi_1^\prime \varphi_0^{\prime \prime} {\varphi_0^\prime }^{3}
  a^{(-2)} \varphi_1 + \nonumber\\
  &&
  G_{XXY} \varphi_1^\prime a^\prime
  {\varphi_0^\prime }^{4} a^{(-3)} \varphi_1 + \frac{1}{4} G_Y
  {\delta}^{i j} {\partial}_{i}{B_1} {\partial}_{j}{B_1}
  {\varphi_0^\prime }^{2} a^{2} - \frac{3}{4} G_Y \phi_1^{2}
  {\varphi_0^\prime }^{2} a^{2} + G_Y \phi_1 \varphi_0^\prime
  \varphi_1^\prime a^{2} - \frac{1}{2} G_Y {\varphi_1^\prime }^{2}
  a^{2}
  + \nonumber\\
  &&
  G_Y {\delta}^{i j} \varphi_0^\prime {\partial}_{i}{B_1}
  {\partial}_{j}{\varphi_1} a^{2} + \frac{1}{2} G_Y {\delta}^{i j}
  {\partial}_{i}{\varphi_1} {\partial}_{j}{\varphi_1} a^{2} +
  \frac{3}{2} G_{XY} \phi_1^{2} {\varphi_0^\prime }^{4} - \frac{5}{2}
  G_{XY} \phi_1 \varphi_1^\prime {\varphi_0^\prime }^{3} - \frac{1}{4}
  G_{XY} {\delta}^{i j} {\partial}_{i}{B_1} {\partial}_{j}{B_1}
  {\varphi_0^\prime }^{4} + \nonumber\\
  &&
  \frac{5}{4} G_{XY} {\varphi_0^\prime }^{2}
  {\varphi_1^\prime }^{2} - \frac{1}{2} G_{XY} {\delta}^{i j}
  {\partial}_{i}{B_1} {\partial}_{j}{\varphi_1} {\varphi_0^\prime
  }^{3} - \frac{1}{4} G_{XY} {\delta}^{i j} {\partial}_{i}{\varphi_1}
  {\partial}_{j}{\varphi_1} {\varphi_0^\prime }^{2} + \frac{1}{2}
  G_{YY} \phi_1 {\varphi_0^\prime }^{2} a^{2} \varphi_1 - G_{YY}
  \varphi_0^\prime \varphi_1^\prime a^{2} \varphi_1 \nonumber\\
  &&
  - \frac{1}{4}
  G_{XXY} \phi_1^{2} {\varphi_0^\prime }^{6} a^{(-2)} + \frac{1}{2}
  G_{XXY} \phi_1 \varphi_1^\prime {\varphi_0^\prime }^{5} a^{(-2)} -
  \frac{1}{4} G_{XXY} {\varphi_0^\prime }^{4} {\varphi_1^\prime }^{2}
  a^{(-2)} - \frac{1}{4} G_{YYY} {\varphi_0^\prime }^{2} \varphi_1^{2}
  a^{2} -\nonumber\\
  &&
   \frac{1}{2} G_{XYY} \phi_1 {\varphi_0^\prime }^{4} \varphi_1
  + \frac{1}{2} G_{XYY} \varphi_1^\prime {\varphi_0^\prime }^{3}
  \varphi_1 a^{2} \Big\}
\end{eqnarray}
  }
where $G_X \equiv \partial_{X}{G}, G_Y
  \equiv \partial_{\varphi}{G}$ for background and so on. The
  derivatives of the constraints ($\phi_1$) appear linearly in the
  above reduced action (those terms are highlighted in the above
  expression). However, by performing partial integration one can
  rewrite these terms as terms proportional to $\phi_1$ e.g., first
  terms in the action can be written as $ -
  \frac{5}{2} \partial_{0}\{G_X {\varphi_0^\prime}^3\} \phi_1^2$. So,
  \textit{although the action contains time derivative of Lapse
    function but it is reducible to action with no time derivative of
    Lapse or Shift, hence, variation of these terms do not lead
    constraint inconsistencies.}

\subsection{\texorpdfstring{$({\varphi}_{;\lambda} {\varphi}^{; \lambda} \{\Box
  \varphi\}^2)$}{TEXT} model}
Let us consider the following model where the matter action is given
by

\begin{equation}
\mathcal{S}_m = \int d^4x \sqrt{-g} {\varphi}_{;\lambda} {\varphi}^{;\lambda} 
\{\Box \varphi\}^2
\end{equation}
and is minimally coupled to gravity\footnote{Note that, our motivation
  is only about the consistency of the method discussed in the first
  section for different models, not to check its physical
  observational viability or any other problems such as Higher order
  Ostrogradsky's ghost}. Expanding the matter action to second order,
we get,

{\scriptsize
\begin{eqnarray}
\label{box}
^{(2)}\mathcal{S}_m &=& \int d^4x \{- \frac{35}{2} \phi_1^{2}
{\varphi_0^{\prime}} ^{2} {\varphi_0^{\prime \prime} }^{2} a^{-2} - 70
a^\prime \varphi_0^{\prime \prime} \phi_1^{2} {\varphi_0^{\prime}
}^{3} a^{-3} - 14 \phi_1 \phi_1^{\prime} \varphi_0^{\prime \prime}
{\varphi_0^{\prime} }^{3} a^{-2} - 10 \phi_1 {\delta}^{i j}
\varphi_0^{\prime \prime} {\partial}_{i j}{B_1} {\varphi_0^{\prime}
}^{3} a^{-2} - \nonumber\\
&&
70 \phi_1^{2} {\varphi_0^{\prime} }^{4} {a^\prime }^{2}
a^{-4} - 28 \phi_1 \phi_1^{\prime} a^\prime {\varphi_0^{\prime} }^{4}
a^{-3} - 20 \phi_1 {\delta}^{i j} a^\prime {\partial}_{i j}{B_1}
{\varphi_0^{\prime} }^{4} a^{-3} + 10 \phi_1 \varphi_0^{\prime \prime}
\varphi_1^{\prime \prime} {\varphi_0^{\prime} }^{2} a^{-2} - 6 \phi_1
{\delta}^{i j} \varphi_0^{\prime \prime} {\partial}_{j i}{\varphi_1}
{\varphi_0^{\prime} }^{2} a^{-2} \nonumber\\
&&
+ 20 \phi_1 a^\prime
\varphi_1^{\prime \prime} {\varphi_0^{\prime} }^{3} a^{-3} + 60 \phi_1
\varphi_1^{\prime} a^\prime \varphi_0^{\prime \prime}
{\varphi_0^{\prime} }^{2} a^{-3} - 12 \phi_1 {\delta}^{i j} a^\prime
{\partial}_{j i}{\varphi_1} {\varphi_0^{\prime} }^{3} a^{-3} + 80
\phi_1 \varphi_1^{\prime} {\varphi_0^{\prime} }^{3} {a^\prime }^{2}
a^{-4} + 10 \phi_1 \varphi_0^{\prime} \varphi_1^{\prime}
{\varphi_0^{\prime \prime} }^{2} a^{-2}\nonumber\\
&&
 + 6 \phi_1^{\prime}
\varphi_1^{\prime} \varphi_0^{\prime \prime} {\varphi_0^{\prime} }^{2}
a^{-2} + 6 {\delta}^{i j} \varphi_1^{\prime} \varphi_0^{\prime \prime}
{\partial}_{i j}{B_1} {\varphi_0^{\prime} }^{2} a^{-2} + 16
\phi_1^{\prime} \varphi_1^{\prime} a^\prime {\varphi_0^{\prime} }^{3}
a^{-3} + 16 {\delta}^{i j} \varphi_1^{\prime} a^\prime {\partial}_{i
  j}{B_1} {\varphi_0^{\prime} }^{3} a^{-3} - 4 \varphi_0^{\prime}
\varphi_1^{\prime} \varphi_0^{\prime \prime} \varphi_1^{\prime \prime}
a^{-2}
\nonumber\\
&&
+ 4 {\delta}^{i j} \varphi_0^{\prime} \varphi_1^{\prime}
\varphi_0^{\prime \prime} {\partial}_{j i}{\varphi_1} a^{-2} - 12
\varphi_1^{\prime} a^\prime \varphi_1^{\prime \prime}
{\varphi_0^{\prime} }^{2} a^{-3} - 12 \varphi_0^{\prime} a^\prime
\varphi_0^{\prime \prime} {\varphi_1^{\prime} }^{2} a^{-3} + 12
{\delta}^{i j} \varphi_1^{\prime} a^\prime {\partial}_{j i}{\varphi_1}
{\varphi_0^{\prime} }^{2} a^{-3} - 24 {\varphi_0^{\prime} }^{2}
{\varphi_1^{\prime} }^{2} {a^\prime }^{2} a^{-4} \nonumber\\
&&
+ \frac{5}{2}
{\delta}^{i j} {\partial}_{i}{B_1} {\partial}_{j}{B_1}
{\varphi_0^{\prime} }^{2} {\varphi_0^{\prime \prime} }^{2} a^{-2} + 10
{\delta}^{i j} a^\prime {\partial}_{i}{B_1} {\partial}_{j}{B_1}
\varphi_0^{\prime \prime} {\varphi_0^{\prime} }^{3} a^{-3} + 10
{\delta}^{i j} {\partial}_{i}{B_1} {\partial}_{j}{B_1}
{\varphi_0^{\prime} }^{4} {a^\prime }^{2} a^{-4} - {\varphi_1^{\prime}
}^{2} {\varphi_0^{\prime \prime} }^{2} a^{-2} + \nonumber\\
&&
2 {\delta}^{i j}
\varphi_0^{\prime} {\partial}_{j}{B_1} {\partial}_{i}{\varphi_1}
{\varphi_0^{\prime \prime} }^{2} a^{-2} + 8 {\delta}^{i j} a^\prime
{\partial}_{j}{B_1} {\partial}_{i}{\varphi_1} \varphi_0^{\prime
  \prime} {\varphi_0^{\prime} }^{2} a^{-3} + 8 {\delta}^{i j}
{\partial}_{j}{B_1} {\partial}_{i}{\varphi_1} {\varphi_0^{\prime}
}^{3} {a^\prime }^{2} a^{-4} + {\delta}^{i j}
{\partial}_{i}{\varphi_1} {\partial}_{j}{\varphi_1} {\varphi_0^{\prime
    \prime} }^{2} a^{-2} + \nonumber\\
    &&
    4 {\delta}^{i j} \varphi_0^{\prime}
a^\prime {\partial}_{i}{\varphi_1} {\partial}_{j}{\varphi_1}
\varphi_0^{\prime \prime} a^{-3} + 4 {\delta}^{i j}
{\partial}_{i}{\varphi_1} {\partial}_{j}{\varphi_1}
{\varphi_0^{\prime} }^{2} {a^\prime }^{2} a^{-4} + 4 {\delta}^{i j}
{\partial}_{i}{B_1} \varphi_0^{\prime \prime} {\partial}_{j
  0}{\varphi_1} {\varphi_0^{\prime} }^{2} a^{-2} + 2 \phi_1^{\prime}
\varphi_1^{\prime \prime} {\varphi_0^{\prime} }^{3} a^{-2} +\nonumber\\
&&
 4
{\delta}^{i j} a^\prime {\partial}_{i}{B_1} {\partial}_{j}{\varphi_1}
\varphi_0^{\prime \prime} {\varphi_0^{\prime} }^{2} a^{-3} + 2
{\delta}^{i j} {\partial}_{i}{\varphi_1} \varphi_0^{\prime \prime}
{\partial}_{0 j}{B_1} {\varphi_0^{\prime} }^{2} a^{-2} + 2 {\delta}^{i
  j} \varphi_1^{\prime \prime} {\partial}_{i j}{B_1}
{\varphi_0^{\prime} }^{3} a^{-2}
+ 8 {\delta}^{i j} a^\prime {\partial}_{i}{B_1} {\partial}_{j
  0}{\varphi_1} {\varphi_0^{\prime} }^{3} a^{-3} - \nonumber\\
  &&
  2 {\delta}^{i j}
\phi_1^{\prime} {\partial}_{j i}{\varphi_1} {\varphi_0^{\prime} }^{3}
a^{-2} - 2 {\delta}^{i j} {\delta}^{k l} {\partial}_{i j}{B_1}
{\partial}_{l k}{\varphi_1} {\varphi_0^{\prime} }^{3} a^{-2} + 2
{\delta}^{i j} {\partial}_{i}{\phi_1} {\partial}_{j}{\varphi_1}
\varphi_0^{\prime \prime} {\varphi_0^{\prime} }^{2} a^{-2} + 8
{\delta}^{i j} {\partial}_{i}{B_1} {\partial}_{j}{\varphi_1}
{\varphi_0^{\prime} }^{3} {a^\prime }^{2} a^{-4} + \nonumber\\
&&
4 {\delta}^{i j}
a^\prime {\partial}_{i}{\varphi_1} {\partial}_{0 j}{B_1}
{\varphi_0^{\prime} }^{3} a^{-3} + 4 {\delta}^{i j} a^\prime
{\partial}_{i}{\phi_1} {\partial}_{j}{\varphi_1} {\varphi_0^{\prime}
}^{3} a^{-3} + 2 {\delta}^{i j} {\partial}_{i}{B_1} \varphi_0^{\prime
  \prime} {\partial}_{0 j}{B_1} {\varphi_0^{\prime} }^{3} a^{-2} - 2
{\delta}^{i j} {\partial}_{i}{B_1} {\partial}_{j}{\phi_1}
\varphi_0^{\prime \prime} {\varphi_0^{\prime} }^{3} a^{-2} - \nonumber\\
&&
\bm{
  {\phi_1^\prime}^2 {\varphi_0^{\prime} }^{4} a^{-2}} + 4 {\delta}^{i
  j} a^\prime {\partial}_{i}{B_1} {\partial}_{0 j}{B_1}
{\varphi_0^{\prime} }^{4} a^{-3} - 4 {\delta}^{i j} a^\prime
{\partial}_{i}{B_1} {\partial}_{j}{\phi_1} {\varphi_0^{\prime} }^{4}
a^{-3} - 2 {\delta}^{i j} \phi_1^{\prime} {\partial}_{i j}{B_1}
{\varphi_0^{\prime} }^{4} a^{-2} - \nonumber\\
&&
{\delta}^{i j} {\delta}^{k l}
{\partial}_{i j}{B_1} {\partial}_{k l}{B_1} {\varphi_0^{\prime} }^{4}
a^{-2} - 
{\varphi_0^{\prime} }^{2} {\varphi_1^{\prime \prime} }^{2}
a^{-2} + 2 {\delta}^{i j} \varphi_1^{\prime \prime} {\partial}_{j
  i}{\varphi_1} {\varphi_0^{\prime} }^{2} a^{-2} - {\delta}^{i j}
{\delta}^{k l} {\partial}_{j i}{\varphi_1} {\partial}_{l k}{\varphi_1}
{\varphi_0^{\prime} }^{2} a^{-2} \}
\end{eqnarray}}

Following points are worth-noting regarding the above action: (i)
Unlike Galilean scalar fields, the above action contains square of the
derivative of constraints (${\phi^\prime}^2$) [the term is highlighted
in the above expression], (ii) In case of Galilean, it was possible to
rewrite it as boundary term, in this case it is not possible. This
implies that using reduced action approach, we cannot obtain a single
dynamical equation. Hence as discussed in section (2), the constraint
consistency is not satisfied.

\subsection{\texorpdfstring{$(\varphi_{;\lambda} \varphi^{;\lambda}~ \varphi_{; \mu
    \nu} \varphi^{; \mu \nu})$}{TEXT} model}
As like the previous subsection, the following action also contains
higher time derivatives of constraints. Expanding the matter action
\begin{equation}
  \mathcal{S}_m = \int d^4x~\sqrt{-g}~ {\varphi}_{;\lambda} 
{\varphi}^{;\lambda}~ \varphi_{; \mu \nu} \varphi^{; \mu \nu}
\end{equation}
to second order, we get,

{\scriptsize
\begin{eqnarray}
\label{boxd}
^{(2)}\mathcal{S}_m &=&\int d^4x \{ - \frac{35}{2} \phi_1^{2}
{\varphi_0^{\prime} }^{2} {\varphi_0^{\prime \prime} }^{2} a^{-2} + 35
a^\prime \varphi_0^{\prime \prime} \phi_1^{2} {\varphi_0^{\prime}
}^{3} a^{-3} - 14 \phi_1 \phi_1^{\prime} \varphi_0^{\prime \prime}
{\varphi_0^{\prime} }^{3} a^{-2} - 70 \phi_1^{2} {\varphi_0^{\prime}
}^{4} {a^\prime }^{2} a^{-4} + 14 \phi_1 \phi_1^{\prime} a^\prime
{\varphi_0^{\prime} }^{4} a^{-3} - \nonumber\\
&& 10 \phi_1 {\delta}^{i j} a^\prime
{\partial}_{i j}{B_1} {\varphi_0^{\prime} }^{4} a^{-3} + 10 \phi_1
\varphi_0^{\prime \prime} \varphi_1^{\prime \prime}
{\varphi_0^{\prime} }^{2} a^{-2} - 10 \phi_1 a^\prime
\varphi_1^{\prime \prime} {\varphi_0^{\prime} }^{3} a^{-3} - 30 \phi_1
\varphi_1^{\prime} a^\prime \varphi_0^{\prime \prime}
{\varphi_0^{\prime} }^{2} a^{-3} - 6 \phi_1 {\delta}^{i j} a^\prime
{\partial}_{j i}{\varphi_1} {\varphi_0^{\prime} }^{3} a^{-3} \nonumber\\
&& + 80\phi_1 \varphi_1^{\prime} {\varphi_0^{\prime} }^{3} {a^\prime }^{2}
a^{-4} + 10 \phi_1 \varphi_0^{\prime} \varphi_1^{\prime}
{\varphi_0^{\prime \prime} }^{2} a^{-2} + 6 \phi_1^{\prime}
\varphi_1^{\prime} \varphi_0^{\prime \prime} {\varphi_0^{\prime} }^{2}
a^{-2} - 8 \phi_1^{\prime} \varphi_1^{\prime} a^\prime
{\varphi_0^{\prime} }^{3} a^{-3} + 8 {\delta}^{i j} \varphi_1^{\prime}
a^\prime {\partial}_{i j}{B_1} {\varphi_0^{\prime} }^{3} a^{-3} - \nonumber\\
&& 4\varphi_0^{\prime} \varphi_1^{\prime} \varphi_0^{\prime \prime}
\varphi_1^{\prime \prime} a^{-2} + 6 \varphi_1^{\prime} a^\prime
\varphi_1^{\prime \prime} {\varphi_0^{\prime} }^{2} a^{-3} + 6
\varphi_0^{\prime} a^\prime \varphi_0^{\prime \prime}
{\varphi_1^{\prime} }^{2} a^{-3} + 6 {\delta}^{i j} \varphi_1^{\prime}
a^\prime {\partial}_{j i}{\varphi_1} {\varphi_0^{\prime} }^{2} a^{-3}
- 24 {\varphi_0^{\prime} }^{2} {\varphi_1^{\prime} }^{2} {a^\prime
}^{2} a^{-4} + \nonumber\\
&&\frac{5}{2} {\delta}^{i j} {\partial}_{i}{B_1}
{\partial}_{j}{B_1} {\varphi_0^{\prime} }^{2} {\varphi_0^{\prime
    \prime} }^{2} a^{-2} - 5 {\delta}^{i j} a^\prime
{\partial}_{i}{B_1} {\partial}_{j}{B_1} \varphi_0^{\prime \prime}
{\varphi_0^{\prime} }^{3} a^{-3} + 10 {\delta}^{i j}
{\partial}_{i}{B_1} {\partial}_{j}{B_1} {\varphi_0^{\prime} }^{4}
{a^\prime }^{2} a^{-4} - {\varphi_1^{\prime} }^{2} {\varphi_0^{\prime
    \prime} }^{2} a^{-2} + \nonumber\\
    &&
    2 {\delta}^{i j} \varphi_0^{\prime}
{\partial}_{j}{B_1} {\partial}_{i}{\varphi_1} {\varphi_0^{\prime
    \prime} }^{2} a^{-2} - 4 {\delta}^{i j} a^\prime
{\partial}_{j}{B_1} {\partial}_{i}{\varphi_1} \varphi_0^{\prime
  \prime} {\varphi_0^{\prime} }^{2} a^{-3} + 8 {\delta}^{i j}
{\partial}_{j}{B_1} {\partial}_{i}{\varphi_1} {\varphi_0^{\prime}
}^{3} {a^\prime }^{2} a^{-4} + {\delta}^{i j}
{\partial}_{i}{\varphi_1} {\partial}_{j}{\varphi_1} {\varphi_0^{\prime
    \prime} }^{2} a^{-2} - \nonumber\\
    &&
    2 {\delta}^{i j} \varphi_0^{\prime}
a^\prime {\partial}_{i}{\varphi_1} {\partial}_{j}{\varphi_1}
\varphi_0^{\prime \prime} a^{-3} + 6 {\delta}^{i j}
{\partial}_{i}{\varphi_1} {\partial}_{j}{\varphi_1}
{\varphi_0^{\prime} }^{2} {a^\prime }^{2} a^{-4} + 4 {\delta}^{i j}
{\partial}_{i}{B_1} \varphi_0^{\prime \prime} {\partial}_{j
  0}{\varphi_1} {\varphi_0^{\prime} }^{2} a^{-2} + 2 \phi_1^{\prime}
\varphi_1^{\prime \prime} {\varphi_0^{\prime} }^{3} a^{-2} - \nonumber\\
&&
2{\delta}^{i j} a^\prime {\partial}_{i}{B_1} {\partial}_{j}{\varphi_1}
\varphi_0^{\prime \prime} {\varphi_0^{\prime} }^{2} a^{-3} + 2
{\delta}^{i j} {\partial}_{i}{\varphi_1} \varphi_0^{\prime \prime}
{\partial}_{0 j}{B_1} {\varphi_0^{\prime} }^{2} a^{-2} - 4 {\delta}^{i
  j} a^\prime {\partial}_{i}{B_1} {\partial}_{j 0}{\varphi_1}
{\varphi_0^{\prime} }^{3} a^{-3} - 4 {\delta}^{i j}
{\partial}_{i}{\phi_1} {\partial}_{j 0}{\varphi_1} {\varphi_0^{\prime}
}^{3} a^{-2} - \nonumber\\
&&
2 {\delta}^{i j} {\delta}^{k l} {\partial}_{i k}{B_1}
{\partial}_{l j}{\varphi_1} {\varphi_0^{\prime} }^{3} a^{-2} + 2
{\delta}^{i j} {\partial}_{i}{\phi_1} {\partial}_{j}{\varphi_1}
\varphi_0^{\prime \prime} {\varphi_0^{\prime} }^{2} a^{-2} + 8
{\delta}^{i j} {\partial}_{i}{B_1} {\partial}_{j}{\varphi_1}
{\varphi_0^{\prime} }^{3} {a^\prime }^{2} a^{-4} - 2 {\delta}^{i j}
a^\prime {\partial}_{i}{\varphi_1} {\partial}_{0 j}{B_1}
{\varphi_0^{\prime} }^{3} a^{-3} + \nonumber\\
&&
2 {\delta}^{i j} a^\prime
{\partial}_{i}{\phi_1} {\partial}_{j}{\varphi_1} {\varphi_0^{\prime}
}^{3} a^{-3} + 2 {\delta}^{i j} {\partial}_{i}{B_1} \varphi_0^{\prime
  \prime} {\partial}_{0 j}{B_1} {\varphi_0^{\prime} }^{3} a^{-2} - 2
{\delta}^{i j} {\partial}_{i}{B_1} {\partial}_{j}{\phi_1}
\varphi_0^{\prime \prime} {\varphi_0^{\prime} }^{3} a^{-2} -
\bm{{\phi_1^\prime}^2 {\varphi_0^{\prime} }^{4} a^{-2}} - \nonumber\\
&&
2
{\delta}^{i j} a^\prime {\partial}_{i}{B_1} {\partial}_{0 j}{B_1}
{\varphi_0^{\prime} }^{4} a^{-3} + 2 {\delta}^{i j} a^\prime
{\partial}_{i}{B_1} {\partial}_{j}{\phi_1} {\varphi_0^{\prime} }^{4}
a^{-3} + 2 {\delta}^{i j} {\partial}_{i}{\phi_1}
{\partial}_{j}{\phi_1} {\varphi_0^{\prime} }^{4} a^{-2} - {\delta}^{i
  j} {\delta}^{k l} {\partial}_{i k}{B_1} {\partial}_{j l}{B_1}
{\varphi_0^{\prime} }^{4} a^{-2} - \nonumber\\
&&
{\varphi_0^{\prime} }^{2}
{\varphi_1^{\prime \prime} }^{2} a^{-2} + 2 {\delta}^{i j}
{\partial}_{i 0}{\varphi_1} {\partial}_{j 0}{\varphi_1}
{\varphi_0^{\prime} }^{2} a^{-2} - {\delta}^{i j} {\delta}^{k l}
{\partial}_{k i}{\varphi_1} {\partial}_{l j}{\varphi_1}
{\varphi_0^\prime}^{2} a^{-2} - 4 {\delta}^{i j} a^\prime
{\partial}_{i}{\varphi_1} {\partial}_{j 0}{\varphi_1}
{\varphi_0^{\prime} }^{2} a^{-3} \}
\end{eqnarray}}

It is important to note that the second order action contains
${\phi^\prime}^2$ which cannot be absorbed as a boundary term (the
highlighted term in the above expression). Hence the constraint
condition is not satisfied leading to the fact that the reduced action
does not lead to single variable equation of motion.

From the above analysis, we can generalize and apply this to any
higher derivative scalar theory models like $ \{\Box \varphi\}^3,~
\Box \Box \varphi,~ \varphi^{; \alpha}_{;\beta}
\varphi^{;\beta}_{;\gamma} \varphi^{;\gamma}_{;\alpha}, ~\varphi_{;
  \alpha \beta \gamma \delta} \varphi^{\alpha \beta \gamma \delta}$
and that obtaining a single variable equation of motion or action is
not possible for these kind of models.

\subsection{\texorpdfstring{$f(R)$}{TEXT} model}\label{fr}
Until now, we have considered different forms of scalar field action
without modifying gravity. In this subsection, we consider the
simplest modification i.e. $R + \alpha R^2$ while we consider the
matter to be canonical scalar field. Since $R$ and matter part of the
action does not have any inconsistency, we expand $R^2$ up to second
order in perturbed variables to get,

{\scriptsize
\begin{equation}
\begin{split}
 & 36 {\delta}^{i j} {\delta}^{k l} {\partial}_{i j}{B_1} {\partial}_{k
    l}{B_1} {a^{\prime}}^{2} a^{-6} + 12 {\delta}^{i j} {\delta}^{k l}
  a^{\prime} {\partial}_{i j}{B_1} {\partial}_{0 k l}{B_1} a^{-5} + 72
  {\delta}^{i j} \phi_1^{\prime} {\partial}_{i j}{B_1}
  {a^{\prime}}^{2} a^{-6} + \\
  &
  288 \phi_1 {\delta}^{i j} a^{\prime}
  a^{\prime \prime} {\partial}_{i j}{B_1} a^{-6} + 12 {\delta}^{i j}
  {\delta}^{k l} a^{\prime} {\partial}_{i j}{B_1} {\partial}_{k
    l}{\phi_1} a^{-5} + 12 {\delta}^{i j} {\delta}^{k l} a^{\prime}
  {\partial}_{k l}{B_1} {\partial}_{0 i j}{B_1} a^{-5} + 4 {\delta}^{i
    j} {\delta}^{k l} {\partial}_{0 i j}{B_1} {\partial}_{0 k l}{B_1}
  a^{-4} + \\
  &
  24 {\delta}^{i j} \phi_1^{\prime} a^{\prime} {\partial}_{0
    i j}{B_1} a^{-5} + 96 \phi_1 {\delta}^{i j} a^{\prime \prime}
  {\partial}_{0 i j}{B_1} a^{-5} + 4 {\delta}^{i j} {\delta}^{k l}
  {\partial}_{k l}{\phi_1} {\partial}_{0 i j}{B_1} a^{-4} + \bm{36
    {\phi_1^\prime}^2 {a^{\prime} }^{2} a^{-6}} + \\
    &
    432 \phi_1
  \phi_1^{\prime} a^{\prime} a^{\prime \prime} a^{-6} + 24 {\delta}^{i
    j} \phi_1^{\prime} a^{\prime} {\partial}_{i j}{\phi_1} a^{-5} +
  432 \phi_1^{2} {a^{\prime \prime} }^{2} a^{-6} + 96 \phi_1
  {\delta}^{i j} a^{\prime \prime} {\partial}_{i j}{\phi_1} a^{-5} +\\
  &
  12 {\delta}^{i j} {\delta}^{k l} a^{\prime} {\partial}_{k l}{B_1}
  {\partial}_{i j}{\phi_1} a^{-5} + 4 {\delta}^{i j} {\delta}^{k l}
  {\partial}_{i j}{\phi_1} {\partial}_{0 k l}{B_1} a^{-4} + 4
  {\delta}^{i j} {\delta}^{k l} {\partial}_{i j}{\phi_1} {\partial}_{k
    l}{\phi_1} a^{-4} + 24 {\delta}^{i j} \phi_1^{\prime} a^{\prime
    \prime} {\partial}_{i j}{B_1} a^{-5} - \\
    &
    72 {\delta}^{i j}
  a^{\prime} {\partial}_{i}{B_1} a^{\prime \prime} {\partial}_{0
    j}{B_1} a^{-6} + 12 {\delta}^{i j} {\delta}^{k l} a^{\prime
    \prime} {\partial}_{i j}{B_1} {\partial}_{k l}{B_1} a^{-5} + 72
  {\delta}^{i j} a^{\prime} {\partial}_{i}{B_1} {\partial}_{j}{\phi_1}
  a^{\prime \prime} a^{-6} - 72 {\delta}^{i j} {\partial}_{i}{B_1}
  {\partial}_{j}{B_1} {a^{\prime \prime} }^{2} a^{-6} + \\
  &
  24 {\delta}^{i
    j} {\partial}_{i}{\phi_1} {\partial}_{j}{\phi_1} a^{\prime \prime}
  a^{-5} - 12 {\delta}^{i j} {\delta}^{k l} a^{\prime \prime}
  {\partial}_{i k}{B_1} {\partial}_{j l}{B_1} a^{-5}
  \end{split}
\end{equation}}

Following points are interesting to note from the above expression:
(i) $R^2$ term contains time derivative of $\phi$, that cannot be
absorbed as a boundary term. This implies that the above action cannot
lead to the constraint equation. (ii) By doing a conformal
transformation $g_{\mu \nu} \rightarrow \tilde{g_{\mu \nu}} = \Omega^2
g_{\mu \nu} $, the term containing the time derivative of $\phi$ can
be absorbed as a matter field and hence the constraint equation
recovered. (iii) The constraint consistency allows us to identify that
the $f(R)$ gravity models, without conformal transformation lead to
inconsistent dynamics.

\subsection{\texorpdfstring{$[\varphi_{;\lambda} \varphi^{;\lambda}( \{\Box
  \varphi\}^2 - \varphi_{; \mu \nu} \varphi^{; \mu \nu} )]$}{TEXT} model}
As we have shown above, certain higher derivative models do not
satisfy `constraint consistency'. We have also shown which terms in
the second order action spoil the `constraint consistency'. However,
it is interesting to note the terms that contain time derivative of
Lapse functions are identical. Let us consider the following scalar
field action

\begin{equation}
  S = \int d^4x \sqrt{-g}~[~\varphi_{;\lambda} \varphi^{;\lambda}~\left( \{\Box \varphi\}^2 - \varphi_{; \mu \nu} \varphi^{; \mu \nu} \right)~]
\end{equation}
The second order action is given by,

{\scriptsize
\begin{equation}
\begin{split}
  ^{(2)}\mathcal{S}_m =& \int d^4x \{ - 105 a^{\prime}
  \varphi_0^{\prime \prime} \phi_1^{2} {\varphi_0^{\prime}}^{3} a^{-3}
  - 10 \phi_1 {\delta}^{i j} \varphi_0^{\prime \prime} {\partial}_{i
    j}{B_1} {\varphi_0^{\prime} }^{3} a^{-2} - 42 \phi_1
  \phi_1^{\prime} a^{\prime} {\varphi_0^{\prime} }^{4} a^{-3} - 10
  \phi_1 {\delta}^{i j} a^{\prime} {\partial}_{i j}{B_1}
  {\varphi_0^{\prime} }^{4} a^{-3} - \\
  &
  6 \phi_1 {\delta}^{i j}
  \varphi_0^{\prime \prime} {\partial}_{j i}{\varphi_1}
  {\varphi_0^{\prime} }^{2} a^{-2} + 30 \phi_1 a^{\prime}
  \varphi_1^{\prime \prime} {\varphi_0^{\prime} }^{3} a^{-3} + 90
  \phi_1 {\varphi_1^\prime} a^{\prime} \varphi_0^{\prime \prime}
  {\varphi_0^{\prime} }^{2} a^{-3} - 6 \phi_1 {\delta}^{i j}
  a^{\prime} {\partial}_{j i}{\varphi_1} {\varphi_0^{\prime} }^{3}
  a^{-3} + \\
  &
  6 {\delta}^{i j} {\varphi_1^\prime} \varphi_0^{\prime
    \prime} {\partial}_{i j}{B_1} {\varphi_0^{\prime} }^{2} a^{-2} +
  24 \phi_1^{\prime} {\varphi_1^\prime} a^{\prime} {\varphi_0^{\prime}
  }^{3} a^{-3} + 8 {\delta}^{i j} {\varphi_1^\prime} a^{\prime}
  {\partial}_{i j}{B_1} {\varphi_0^{\prime} }^{3} a^{-3} + 4
  {\delta}^{i j} \varphi_0^{\prime} {\varphi_1^\prime}
  \varphi_0^{\prime \prime} {\partial}_{j i}{\varphi_1} a^{-2} - \\
  &
  18
  {\varphi_1^\prime} a^{\prime} \varphi_1^{\prime \prime}
  {\varphi_0^{\prime} }^{2} a^{-3} - 18 \varphi_0^{\prime} a^{\prime}
  \varphi_0^{\prime \prime} {\varphi_1^{\prime} }^{2} a^{-3} + 6
  {\delta}^{i j} {\varphi_1^\prime} a^{\prime} {\partial}_{j
    i}{\varphi_1} {\varphi_0^{\prime} }^{2} a^{-3} + 15 {\delta}^{i j}
  a^{\prime} {\partial}_{i}{B_1} {\partial}_{j}{B_1} \varphi_0^{\prime
    \prime} {\varphi_0^{\prime} }^{3} a^{-3} + \\
    &
    12 {\delta}^{i j}
  a^{\prime} {\partial}_{j}{B_1} {\partial}_{i}{\varphi_1}
  \varphi_0^{\prime \prime} {\varphi_0^{\prime} }^{2} a^{-3} + 6
  {\delta}^{i j} \varphi_0^{\prime} a^{\prime}
  {\partial}_{i}{\varphi_1} {\partial}_{j}{\varphi_1}
  \varphi_0^{\prime \prime} a^{-3} - 2 {\delta}^{i j}
  {\partial}_{i}{\varphi_1} {\partial}_{j}{\varphi_1}
  {\varphi_0^{\prime} }^{2} {a^{\prime} }^{2} a^{-4}%
  + 6 {\delta}^{i j} a^{\prime} {\partial}_{i}{B_1}
  {\partial}_{j}{\varphi_1} \varphi_0^{\prime \prime}
  {\varphi_0^{\prime} }^{2} a^{-3} + \\
  &
  2 {\delta}^{i j}
  \varphi_1^{\prime \prime} {\partial}_{i j}{B_1} {\varphi_0^{\prime}
  }^{3} a^{-2} + 12 {\delta}^{i j} a^{\prime} {\partial}_{i}{B_1}
  {\partial}_{j 0}{\varphi_1} {\varphi_0^{\prime} }^{3} a^{-3} - 2
  {\delta}^{i j} \phi_1^{\prime} {\partial}_{j i}{\varphi_1}
  {\varphi_0^{\prime} }^{3} a^{-2} - 2 {\delta}^{i j} {\delta}^{k l}
  {\partial}_{i j}{B_1} {\partial}_{l k}{\varphi_1}
  {\varphi_0^{\prime} }^{3} a^{-2} + \\
  &
  6 {\delta}^{i j} a^{\prime}
  {\partial}_{i}{\varphi_1} {\partial}_{0 j}{B_1} {\varphi_0^{\prime}
  }^{3} a^{-3} + 2 {\delta}^{i j} a^{\prime} {\partial}_{i}{\phi_1}
  {\partial}_{j}{\varphi_1} {\varphi_0^{\prime} }^{3} a^{-3} + 6
  {\delta}^{i j} a^{\prime} {\partial}_{i}{B_1} {\partial}_{0 j}{B_1}
  {\varphi_0^{\prime} }^{4} a^{-3} - 6 {\delta}^{i j} a^{\prime}
  {\partial}_{i}{B_1} {\partial}_{j}{\phi_1} {\varphi_0^{\prime} }^{4}
  a^{-3} - \\
  &
  2 {\delta}^{i j} \phi_1^{\prime} {\partial}_{i j}{B_1}
  {\varphi_0^{\prime} }^{4} a^{-2} - {\delta}^{i j} {\delta}^{k l}
  {\partial}_{i j}{B_1} {\partial}_{k l}{B_1} {\varphi_0^{\prime}
  }^{4} a^{{-2}} + 2 {\delta}^{i j} \varphi_1^{\prime \prime}
  {\partial}_{j i}{\varphi_1} {\varphi_0^{\prime} }^{2} a^{{-2}} -
  {\delta}^{i j} {\delta}^{k l} {\partial}_{j i}{\varphi_1}
  {\partial}_{l k}{\varphi_1} {\varphi_0^{\prime} }^{2} a^{-2} + \\
  &
  4
  {\delta}^{i j} {\partial}_{i}{\phi_1} {\partial}_{j 0}{\varphi_1}
  {\varphi_0^{\prime} }^{3} a^{-2} + 2 {\delta}^{i j} {\delta}^{k l}
  {\partial}_{i k}{B_1} {\partial}_{l j}{\varphi_1}
  {\varphi_0^{\prime} }^{3} a^{-2} - 2 {\delta}^{i j}
  {\partial}_{i}{\phi_1} {\partial}_{j}{\phi_1} {\varphi_0^{\prime}
  }^{4} a^{-2} + {\delta}^{i j} {\delta}^{k l} {\partial}_{i k}{B_1}
  {\partial}_{j l}{B_1} {\varphi_0^{\prime} }^{4} a^{-2} - \\
  &
  2
  {\delta}^{i j} {\partial}_{i 0}{\varphi_1} {\partial}_{j
    0}{\varphi_1} {\varphi_0^{\prime} }^{2} a^{-2} + {\delta}^{i j}
  {\delta}^{k l} {\partial}_{k i}{\varphi_1} {\partial}_{l
    j}{\varphi_1} {\varphi_0^{\prime} }^{2} a^{-2} + 4 {\delta}^{i j}
  a^{\prime} {\partial}_{i}{\varphi_1} {\partial}_{j 0}{\varphi_1}
  {\varphi_0^{\prime} }^{2} a^{-3}\}
  \end{split}
\end{equation}}

It is interesting to note the following points: (i) the action does
not contain terms having time derivative of Lapse function/Shift
vector. Hence the resultant equation leads to constraint
equation. Similarly, $[\varphi_{;\lambda} \varphi^{;\lambda}(\{\Box
\varphi\}^3 - 3 \Box \varphi ~\varphi_{; \mu \nu}\varphi^{; \mu \nu} +
2 \varphi^{; \alpha}_{;\beta} \varphi^{;\beta}_{;\gamma}
\varphi^{;\gamma}_{;\alpha})]$ model do not have dynamical
Lapse/Shift. (ii) From the above, one may be tempted to relate the
constraint consistency with Ostrogradsky's instabilities. To go about
checking this, let us look at the zeroth order action for the matter
field, i.e.,

\begin{equation}
  ^{(0)}\mathcal{S}_m = \int d^4x \{- 6 a^{\prime} \varphi_0^{\prime
    \prime} {\varphi_0^{\prime} }^{3} a^{(-3)}\}
\end{equation}
which, after integration by-parts, can be re-written as,

\begin{equation}\label{oszero}
^{(0)}\mathcal{S}_m = \int d^4x \{ \frac{3}{2} a^{\prime \prime}a^{-3} {\varphi_0^{\prime} }^{4} - \frac{9}{2} {a^\prime}^2 a^{-4} {\varphi_0^{\prime} }^{4}\}
\end{equation}
Hence the equations of motion will contain $a^{\prime \prime \prime}$
and $\phi_0^{\prime\prime\prime}$. This immediately signals
Ostrogradsky's instability.

This leads us to the important conclusion: \textit{While all
  `constraint' inconsistent models have higher order Ostrogradsky's
  instabilities but the reverse is not true. One can have models with
  constraint Lapse function and Shift vector, though it may have
  Ostrogradsky's instabilities.}

\subsection{Constraint consistent models without instabilities}
In the previous subsection, we showed that identifying the terms in
(\ref{box}) and (\ref{boxd}) that contains higher derivatives of Lapse
function, we can remove these terms by combination of these terms two
terms. However, we noticed that such actions suffer from
Ostrogradsky's instabilities. The term that leads to Ostrogradsky's
instability is $\frac{3}{2} a^{\prime \prime} a^{-3}
{\varphi_{0}^{\prime}}^4$ at zeroth order action (\ref{oszero}). In
order to cancel such term, without involving any higher derivatives of
$\phi$, one needs to introduce non-minimal coupling of the kinetic
term of the scalar field with Ricci scalar. It is easy to show
that the term $-\frac{1}{4} (\varphi_{; \alpha} \varphi^{; \alpha})^2
R$ exactly cancels this and the resulting background action becomes,

\begin{equation}
^{(0)}\mathcal{S}_m = - \frac{9}{2} \int d^4x ~ {a^\prime}^2 a^{-4} {\varphi_0^{\prime} }^{4}
\end{equation}

Similarly, $[\varphi_{; \lambda} \varphi^{; \lambda}(\{\Box
\varphi\}^3 - 3 \Box \varphi ~\varphi_{; \mu \nu}\varphi^{; \mu \nu} +
2 \varphi^{; \alpha}_{;\beta} \varphi^{;\beta}_{;\gamma}
\varphi^{;\gamma}_{;\alpha} - 6 G^{\mu \nu} \varphi^{;\alpha}_{;\mu}
\varphi_{; \alpha} \varphi_{;\nu})]$ does not have any constraint
inconsistencies as well as instabilities.

The following points are worth noting regarding this:
\begin{itemize}
\item[i.] We have arrived at the action $[\varphi_{;\lambda}
  \varphi^{;\lambda}(\{\Box \varphi\}^2 - \varphi_{; \mu
    \nu}\varphi^{; \mu \nu} -\frac{1}{4} (\varphi_{; \alpha}
  \varphi^{; \alpha})^2 R]$ by using the condition that the action
  does not have time derivatives of Lapse function and Shift vector
  and later using the additional condition that Ostrogradsky's
  instability does not arise.
\item[ii.] In a different manner, Nicolis et al \cite{Nicolis2008} and
  Deffayet et al \cite{Deffayet2009} have come up with the same
  action, only with condition of removing Ostrogradsky's
  instability. Here we have verified that those models will result in
  consistent dynamics with `constraint consistency' and lead to single
  variable action as well as equation of motion.
\item[iii.]  Similarly, we find that the third order derivative model
  is $[\varphi_{; \lambda} \varphi^{; \lambda}(\{\Box \varphi\}^3 - 3
  \Box \varphi ~\varphi_{; \mu \nu}\varphi^{; \mu \nu} + 2 \varphi^{;
    \alpha}_{;\beta} \varphi^{;\beta}_{;\gamma}
  \varphi^{;\gamma}_{;\alpha} - 6 G^{\mu \nu} \varphi^{;\alpha}_{;\mu}
  \varphi_{; \alpha} \varphi_{;\nu})]$ which, again has been derived
  in \cite{Nicolis2008} and \cite{Deffayet2009}. This model is also
  constraint consistent and free of Ostrogradsky's instability.

\end{itemize}

\section{Conclusions}
In this chapter, we have revisited the two approaches in cosmological
perturbation theory --- order-by-order approach of the Einstein's
equation and reduced action formalism. Equivalence of the two
approaches were not clear since Gravity equations are highly
non-linear. In this chapter, we have established that equations arising
from order-by-order approach of the Einstein's equations and those
from the action formalism are equivalent for canonical as well as
non-canonical scalar fields up to second order. These results may
easily be extended to any model in Gravity models at any order.

To compare both the approaches, we have identified a `Constraint
consistency condition' where the constrained nature of Lapse function
and Shift vector are studied \cite{Nandi:2015tha}. We have shown that, in order to obtain a
reduced equation for both of the approaches, `Constraint consistency'
relation has to be satisfied, i.e., those variables should appear in
the action algebraically, and no non-reducible partial time
derivatives of Lapse function and Shift vector should be present. In
other words, equations of motion of Lapse function and Shift vector
should not contain second order partial time derivatives. We then
investigated the higher order derivative theories of gravity and found
that models which satisfy the constraint conditions can be applied to
the conventional perturbation theory where we express all equations in
a simplified form with a single variable (Curvature perturbation
$\mathcal{R}$ or Mukhanov-Sasaki variable) equation of motion. One
common problem with higher order derivative theories is that, they may
have Higher derivative equations of motion which can give rise to
Ostrogradsky's instabilities. We showed that all the models which do
not satisfy `Constraint consistency' conditions suffer form
Ostrogradsky's instabilities. However, we also find that, there exist
some models which satisfy constraint conditions though they have
instabilities, i.e., the action can be reduced in a single variable
form but the single variable equation of motion will contain higher
order time derivatives. The method we have proposed here is fast and
efficient and would be useful for inflationary model building \cite{Nandi:2015tha}.

We have also constructed some higher derivative models in such a way
that those models should satisfy constraint consistency condition and
can be free from higher order time derivatives. Those models have
already been derived in the literature in a different manner where the
approach of constructing models are different. This ensures that,
conventional perturbation theory can be used in higher derivative
models which are free from Ostrogradsky's instabilities to obtain a
gauge invariant single variable equation of motion using any
approaches. We summarize the main results,
\begin{enumerate}
\item The two approaches are completely equivalent.
\item Not all models with Lagrangian density $\mathcal{L} = \int
  \sqrt{-g}\left\{R + F(\varphi, \partial\varphi, \partial^2
    \varphi)\right\}$ or $\mathcal{L} = \int \sqrt{-g}
  F(R, \partial\varphi, \partial^2 \varphi)$ can be reduced in a
  single variable form.
\item To obtain a single variable form, `Constraint consistency'
  condition has to be satisfied.
\item Models with constraint inconsistency show Ostrogradsky's
  instability but the reverse is not true.
\end{enumerate}

The analysis here may have implications for models of quantum
cosmology with scalar fields \cite{Bojowald2008}. The quantum
corrections to the matter and gravity can be modelled as effective
stress-tensor \cite{Sotiriou:2008ya}. The effective classical equation
must also satisfy the constraint consistency. This might help to
constraint the quantum cosmology models \cite{Sotiriou:2008ya}.




\pagestyle{fancy}
\chapter{Hamiltonian formalism of cosmological perturbation theory}\label{Chap.3}
  
\section{Introduction}\label{Sec.Intro.Chap3}
As discussed in section (\ref{Sec.DynamicsInflation}), linear order cosmological perturbation
theory~\cite{Bardeen:1980kt,Langlois1994,Mukhanov:1990me,Sasaki1986,Lidsey:1995np,Garriga1999,Kodama01011984}
has been highly successful at describing the CMB
anisotropies~\cite{1994ApJ...420..445F,Ade:2015xua,Komatsu:2008hk}. It also helps to describe the seed
Gaussian density perturbations during inflation and the formation of
large scale
structures~\cite{STAROBINSKY1982175,BardeenSteinhardt1983}. While
linear order perturbations are fairly understood, there are several
open issues in applying theory beyond linear order both early and late
time
universe~\cite{Salopek1990,STAROBINSKY1982175,Sasaki:1995aw,Lyth:1998xn,Lidsey:1995np,Armendariz-Picon1999,Garriga1999}. In
the last decade, the possibility of observing primordial
non-Gaussianity in CMB and potentially ruling out
inflationary
models \cite{Lidsey:1995np,Lyth:1998xn,Lyth:2007qh,Mazumdar:2010sa}
has led to a lot of interest in higher order perturbations.

As mentioned in section (\ref{Sec.Motivation}), currently, there are two formalism in the literature to study gauge
invariant cosmological perturbations --- 
Hamiltonian~\cite{Langlois1994} and Lagrangian
formulation~\cite{Bardeen:1980kt,Mukhanov:1990me,Bruni:1996im,Acquaviva:2002ud,Nakamura:2003wk,PhysRevD.69.104011,Bartolo:2001cw,Rigopoulos:2002mc,Bernardeau:2002jy,Bernardeau:2002jf,Malik:2003mv,
	Bartolo:2003bz, Finelli:2003bp,Bartolo:2004if,Enqvist:2004bk,Vernizzi:2004nc,Tomita:2005et,
	Lyth:2005du, Seery:2005gb, Malik:2005cy,Seery:2006vu,Lifshitz:1963ps,Lukash:1980iv}. In the Lagrangian
formulation, one needs to either perturb Einstein's equations or vary
the perturbed Lagrangian to obtain perturbed equations of motion. In
the Hamiltonian formulation, gauge-invariant first order perturbed
equations are obtained in terms of field variables and their conjugate
momenta.

In the context of early universe, evaluation of $n$-point correlation
functions of the effective (scalar) field requires the quantum
Hamiltonian of this effective field. For matter fields containing
first derivative in time, it is straightforward to obtain the
Hamiltonian by performing Legendre transformation. Equations of motion
of such fields contain upto second derivative of the field variables
which can be linearized as two independent coupled first order
differential equations (Hamilton's equations) --- one corresponding to
the time evolution of the field and other corresponding to the time
evolution of the momentum. This indicates that the phase space is
two-dimensional.

However, for higher derivative field theories, the Hamiltonian
structure and the associated degrees of freedom is not
straightforward. For any higher (more than one) derivative theories,
the equations of motion have upto two times the highest order
derivative of the field. For example, fields with second order time
derivatives, the equation motion contain upto fourth order derivatives
of the field. So, if we linearize the equation, we obtain four
independent coupled first order differential equations which
indirectly imply that, the phase-space is four dimensional and can be
mapped to Hamiltonian of two fields. However, the mapped Hamiltonian
has unbounded negative energy leading to Ostrogradsky's
instability~\cite{Ostro,Woodard:2015zca}. This implies that extra
degrees of freedom (named as ghost) for a higher derivative Lagrangian
causes the instability and hence, in general, quantizing the
Hamiltonian is not possible.

On the contrary, Galilean scalar
field~\cite{Kobayashi2010,Kobayashi2011,Deffayet2009} is a special
higher derivative field which leads to second derivative equations of
motion, implying that the the phase space contains one independent
variable and one corresponding momentum, although, multiple variable as
well as momenta may appear in the Hamiltonian. Also the absence of
extra degrees of freedom leads to the fact that, Hamiltonian of
Galilean field is bounded can be quantized.

The main aim of the chapter is to write the effective Hamiltonian of the
generalized (Galilean) scalar field coupled to gravity at all orders
in perturbations. Hamiltonian approach to the cosmological
perturbations has not been extensively studied in the
literature. Langlois~\cite{Langlois1994} showed that, equations of
motion of canonical scalar field can be obtained in a gauge-invariant
single variable form at first order perturbation. However, Langlois'
approach can not be extended to include higher order due to fact that
the approach requires construction of gauge-invariant conjugate
momentum. Another aim of this chapter is to extend Langlois' analysis to
higher orders. 

It is necessary to extend Langlois' method to higher
order for the following reasons. First, to calculate higher order
correlation functions, currently several approximations are employed
to convert effective Lagrangian to
Hamiltonian~\cite{Huang:2006eha}. Our aim is to provide a simple, yet
robust procedure to calculate Hamiltonian for arbitrary field(s) at
all orders in cosmological perturbations. Second, as mentioned
earlier, we do not have a procedure to perform Hamiltonian analysis
for the Galilean fields. The procedure we adopt here can be extended
to Galilean fields; we explicitly show this in this chapter. Deffayet et
al~\cite{Deffayet:2015qwa} gave a mechanism to deal with Galilean
theory in the context of General relativity. Third, the procedure can
be used to include quantum gravitational
corrections\cite{Barrau:2013ula,Ashtekar:2013xka,Bojowald:2012xy}.

In this chapter, we find a consistent perturbed Hamiltonian
formulation. We use Deffayet's approach~\cite{Deffayet:2015qwa} to
obtain the generalized Hamiltonian of a Galilean theory and along with
canonical scalar field Hamiltonian, we perturb both fields to obtain
all equations of motion as well as interaction Hamiltonian and we
compare with conventional Lagrangian formulation. We find that both
lead to identical results and hence, our Hamiltonian approach leads to
consistent results in a straightforward and efficient way.

As discussed in section (\ref{SubSec.Gauge}) and section (\ref{sec2}), in section (\ref{gauge}), we re-introduce the generic scalar model in the early Universe and briefly discuss gauge fixing and the corresponding gauge invariant equations of motion. In section (\ref{simple model}), we 
take a simple model that highlights the key issues that need to be addressed about the gauge issue in Hamiltonian formulation and also discuss how the same can be addressed. In this simple model, we compare and contrast Lagrangian and Hamiltonian formulation. We calculate third and fourth order perturbed Hamiltonian and replace momenta with the time derivatives of the variables and show that it is consistent with conventional Lagrangian formulation. In section (\ref{EoMLC}), we discuss canonical scalar field in flat-slicing gauge to obtain interaction Hamiltonian in a new and
simple way. In Appendix (\ref{PLeCSF}), we calculate equations of motion of perturbed and unperturbed variables for Canonical scalar field in flat-slicing gauge. We also explicitly obtain the third order interaction Hamiltonian of Canonical scalar field model in phase-space. In order to show that our proposed method works in any gauges, we obtain equations of motion of all variables of Canonical scalar field in uniform density gauge in Appendix (\ref{udgcsf}). In section
(\ref{galileanfield}) and Appendix (\ref{galmod}), we extend the analysis to a very specific Galilean field and evaluate all equations of motion of
perturbed-unperturbed variables. We show that, at every order unlike
higher order generalized Lagrangian, Galilean field model does not
provide extra degrees freedom and behave same as any general first
order derivative Lagrangian model. We also calculate the third and
fourth order perturbed Hamiltonian. In Appendix (\ref{galcan}), we
consider a Galilean model with a canonical scalar field part and
express the full Hamiltonian as well as zeroth and second order
perturbed Hamiltonian and express zeroth and first order perturbed
equations.

\section{Basic models and Gauge choices}\label{gauge}
Action for a generic scalar field $(\varphi)$ minimally coupled to gravity is
\begin{equation}
\label{GravityAction}
\mathcal{S} = \int d^4 x \sqrt{-g} \left[\frac{1}{2 \kappa} R + \mathcal{L}_m(\varphi, \partial\varphi, \partial\partial\varphi) \right],
\end{equation}
where $R$ is the Ricci scalar and the matter Lagrangian, $\mathcal{L}_m$ is of the form.
\begin{equation}\label{boxx}
\mathcal{L}_m = P(X, \varphi) + G(X, \varphi) \Box \varphi,~~~ ~X \equiv 
\frac{1}{2} g^{\mu \nu} \partial_{\mu}{\varphi} \partial_{\mu}{\varphi}, ~~~ 
\Box \equiv -\frac{1}{\sqrt{-g}}\partial_{\mu}\left(\sqrt{-g} g^{\mu \nu} \partial_{\nu}\right).
\end{equation}

It is important to note that the Lagrangian contains
second order derivatives, however, the equation of motion will be of
second order and these are referred as
Galilean\cite{Kobayashi2010,Kobayashi2011,Deffayet2009}. Varying the
action (\ref{GravityAction}) with respect to metric gives Einstein's
equation
\begin{equation}
\label{EinsteinEquation}
R_{\mu \nu} - \frac{1}{2} g_{\mu \nu}~ R = \kappa~ T_{\mu \nu}, 
\end{equation}
where the stress tensor $T_{\mu \nu}$ is
\begin{equation}
\label{EMTensor}
\begin{split}
T_{\mu \nu} &= g_{\mu \nu} \left( P + G_X g^{\alpha
	\beta} \partial_{\alpha}{X} \partial_{\beta}{\varphi} + G_\varphi
g^{\alpha
	\beta} \partial_{\alpha}{\varphi} \partial_{\beta}{\varphi}\right) \\
& - \left(P_X + 2 G_\varphi + G_X \Box
\varphi\right) \partial_{\mu}{\varphi} \partial_{\nu}{\varphi} - 2
G_X \partial_{\mu}{X} \partial_{\nu}{\varphi}.
\end{split}
\end{equation}

Varying the action (\ref{GravityAction}) with respect to the scalar
field `$\varphi$' leads to the following equation of motion
\begin{equation}
\label{EG}
\begin{split}
&\left(2 G_\varphi - 2 X G_{X \varphi} + P_X\right) \Box \varphi - \left(P_{XX} +
2 G_{X \varphi}\right)\partial_{\mu}{\varphi} \partial^{\mu}{\varphi} -
2X
\left(G_{\varphi} + P_{X \varphi} \right) + P_\varphi\\
& - G_X \left( \varphi_{, \mu \nu} \varphi_{,}^{\mu \nu} - \left(\Box
\varphi\right)^2 + R_{\mu
	\nu}\partial^{\mu}{\varphi} \partial^{\nu}{\varphi}\right) - G_{XX}
\left( \partial_{\mu}X \partial^{\mu}X +
\partial_{\mu}{\varphi} \partial^{\mu}X\,\Box \varphi\right) = 0,
\end{split}
\end{equation}
which can also be obtained by using the conservation of
Energy-Momentum tensor, $\nabla_{\mu}{T^{\mu \nu}} = 0$. Setting $G(X,
\varphi) = 0$ corresponds to non-canonical scalar field. Further,
fixing $ P = - X - V(\varphi)$, where $V(\varphi)$ is the potential,
corresponds to canonical scalar field.

The four-dimensional line element in the ADM form is given by,
\begin{eqnarray}
ds^2 &=& g_{\mu \nu} dx^{\mu} dx^{\nu} \nonumber \\
\label{line}
&=& -(N^2 - N_{i} N^{i} ) d\eta^2 + 2 N_{i} dx^{i} d\eta + \gamma_{i j} dx^{i} dx^{j},
\end{eqnarray}
where $N(x^{\mu})$ and $N_i(x^{\mu})$ are Lapse function and Shift
vector respectively, $\gamma_{i j}$ is the 3-D space metric.  Action
(\ref{GravityAction}) for the line element (\ref{line}) takes the
form,
\begin{equation}
\label{AA}
\mathcal{S} = \int d^4x \sqrt{-g} \Big\{ \frac{1}{2 \kappa}
\left(^{(3)}R + K_{i j} K^{i j} - K^2\right) + \mathcal{L}_m \Big\}
\end{equation}
where $K_{i j}$ is extrinsic curvature tensor and is defined by
\begin{eqnarray}
&& K_{i j} \equiv \frac{1}{2N} \Big[ \partial_{0}{\gamma_{i j}} - N_{i|j} - N_{j | i} \Big] \nonumber \\
&& K \equiv \gamma^{i j} K_{i j} \nonumber
\end{eqnarray}

As mentioned in section (\ref{SubSec.Gauge}) and section (\ref{sec2}), perturbatively expanding the metric only in terms of scalar perturbations and the scalar field  about the
flat FRW spacetime in conformal coordinate, we get,

\begin{eqnarray}
\label{PertVar1}
&& g_{0 0} = - a(\eta)^2(1 + 2 \epsilon \phi_1 + \epsilon^2 \phi_2 + ...) \\
\label{PertVar2}
&& g_{0 i} \equiv N_{i} = a(\eta)^2 (\epsilon \partial_{i}{B_1} + \frac{1}{2} \epsilon^2 \partial_{i}{B_2} + ...) \\
\label{PertVar3}
&& g_{i j} =a(\eta)^2 \left((1 - 2 \epsilon \psi_1 - \epsilon^2 \psi_2 -...)\delta^{i j} +  2 \epsilon E_{1 i j} + \epsilon^2 E_{2 i j} + ...\right)\\
\label{PertVar4}
&&\varphi = \varphi_0(\eta) + \epsilon \varphi_1 + \frac{1}{2} \epsilon^2 \varphi_2 + ...
\end{eqnarray}
where $\epsilon$ denotes the order of the perturbation.
To determine the dynamics at every order, we need five scalar functions ($\phi, B,
\psi, E$ and $\varphi$) at each order. Since there are two free gauge
choices, one can fix two of the five scalar functions. In this chapter,
we derive all equations by choosing a specific gauge --- flat-slicing
gauge, i.e., $\psi = 0, E = 0$ --- at all orders:
\begin{eqnarray}
\label{00pmetric}
&& g_{0 0} =- a(\eta)^2(1 + 2 \epsilon \phi_1 + \epsilon^2 \phi_2 + ...) \\
\label{0ipmetric}
&& g_{0 i} \equiv N_{i} = a(\eta)^2 (\epsilon \partial_{i}{B_1} + \frac{1}{2} \epsilon^2 \partial_{i}{B_2} + ...) \\
\label{3metric}
&& g_{i j} =a(\eta)^2 \delta_{i j}\\
\label{field}
&&\varphi = \varphi_0(\eta) + \epsilon \varphi_1 + \frac{1}{2} \epsilon^2 \varphi_2 + ...
\end{eqnarray}

It can be shown that, perturbed equations in flat-slicing
gauge coincide with gauge-invariant equations of motion (in generic
gauge, $\varphi_1$ coincides with $\varphi_1 +
\frac{\varphi_0{}^\prime}{H} {}\psi_1 \equiv
\frac{\varphi_0{}^\prime}{H} \mathcal{R}$ which is a
gauge-invariant quantity, $\mathcal{R}$ is called curvature
perturbation). Similarly, one can choose another suitable gauge with
no coordinate artifacts to obtain gauge-invariant equations of
motion \cite{Malik2009}. Such gauges are Newtonian-conformal gauge
$(B = 0,\, E = 0)$, constant density gauge $(E = 0, \, \delta\varphi
= 0)$, etc. 

One immediate question that needs to be addressed in the Hamiltonian 
formulation is the following: for a given gauge choice, if a particular 
set of variables are set to zero, whether the corresponding conjugate 
momenta also vanish? In other words, in the flat-slicing gauge $\delta g_{ij} = 0$, 
does this mean the corresponding canonical conjugate momentum $\delta \pi^{i
	j}$ vanish? In order to go about understanding this, in the next section, we take a simple 
model of two variables ($x$ and $y$) where one of the variables is perturbed, 
while the other variable is not perturbed and study the Hamiltonian formulation of this model. The above flow-chart provides a bird's eye view of the new Hamiltonian approach developed in this thesis.

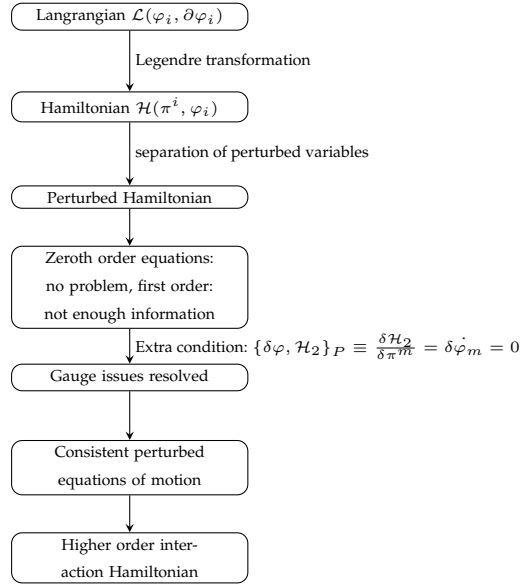
\begin{figure}
{\tiny
	\begin{center}
		\begin{tikzpicture}[node distance=1.2cm]
		
		\node (Lag)[st]  {Langrangian $\mathcal{L}(\varphi_i, \partial\varphi_i)$};
		\node (Ham) [st, below of=Lag] {Hamiltonian $\mathcal{H}(\pi^i, \varphi_i)$};
		\node (pert) [st, below of=Ham]{Perturbed Hamiltonian};
		\node (Zeroth) [st, below of=pert]{Zeroth order equations: no problem, first order: not enough information};
		\node (Gauge) [st, below of=Zeroth] {Gauge issues resolved};
		\node (Cons) [st, below of=Gauge]{Consistent perturbed equations of motion};
		\node (IntHam) [st, below of=Cons]{Higher order interaction Hamiltonian};

		\draw [arrow] (Lag) -- node[anchor=west]{Legendre transformation}(Ham);
		\draw [arrow] (Ham) -- node[anchor=west]{separation of perturbed variables}(pert);
		\draw [arrow] (pert) -- (Zeroth);
		\draw [arrow] (Zeroth) -- node[anchor=west]{Extra condition: $\{\delta\varphi, \mathcal{H}_2\}_P \equiv \frac{\delta\mathcal{H}_2}{\delta\pi^m} = \dot{\delta\varphi_m}= 0$} (Gauge);
		\draw [arrow] (Gauge) -- (Cons);
		\draw [arrow] (Cons) -- (IntHam);
		
		\end{tikzpicture}
	\end{center}
}
\caption{Flow-chart for the proposed Hamiltonian formulation}
\end{figure}
\section{Simple model: Warm up}\label{simple model}

As discussed above, in cosmological perturbation theory, by fixing a `gauge', 
we assume some field variables to be unperturbed where some variables are
perturbed. To go about understanding the procedure in the Hamiltonian
formulation, we consider a simple classical model that consists of
both perturbed and unperturbed variables. We also show that the
Hamilton's equations of unperturbed as well as perturbed variables are
identical to Euler-Lagrange equations of motion. The Lagrangian of the
simple model is
\begin{equation}
\label{fulllagrangian}
\mathcal{L} = \frac{1}{2 y}\, \left((\partial_t x){}^2 + (\partial_ty){}^2\right) - \frac{1}{4}\left(x{}^{4} +  y{}^{4}\right).
\end{equation}
The corresponding momenta are
\begin{equation}
\label{momenta}
\pi_x\, = \,\frac{\partial_tx}{y} ,~~\pi_y\,=\,\frac{\partial_ty}{y},
\end{equation}
and the corresponding Hamiltonian (\ref{fulllagrangian}) is given by
\begin{equation}
\label{fullhamiltonian}
\mathcal{H} = \frac{1}{2} y \left(\pi_x{}^2 + \pi_y{}^2\right) + \frac{1}{4} \left(x^4 + y^4\right).
\end{equation}

\subsection{Perturbed Lagrangian}
As mentioned earlier, we consider $x$ to be unperturbed and 
$y$ to be perturbed, and separate into background and perturbed parts, i.e.,
\begin{equation}
\label{xy}
x = x_0, ~~~y = y_0 + \epsilon\, y_1.
\end{equation}
where $\epsilon$ is the order of perturbation. In this chapter, we mainly
focus on first order perturbation, however, the analysis can be
extended to any higher order perturbations. Using (\ref{xy}), we
separate the Lagrangian (\ref{fulllagrangian}) into a background part
and perturbed parts, and write it as
\begin{equation}
\mathcal{L} = \mathcal{L}_0 + \epsilon \mathcal{L}_1 + \epsilon^2 \mathcal{L}_2 + \epsilon^3 \mathcal{L}_3 + \epsilon^4 \mathcal{L}_4 + ...
\end{equation}
where
\begin{eqnarray}
\label{PerturbedLagrangian}
\label{zerothL}
&&\mathcal{L}_0 = \frac{1}{2}\, (\partial_t{x_0}\, ){}^{2} y_0{}^{-1} + \frac{1}{2}\, (\partial_t{y_0}\, ){}^{2} y_0{}^{-1} - \frac{1}{4}\, \left(x_0{}^{4} + y_0{}^{4}\right)  \\
&& \mathcal{L}_1 = \partial_t{x_0}\,  \partial_t{y_1}\,  y_0{}^{-1} - \frac{1}{2}\, (\partial_t{x_0}\, ){}^{2} y_0{}^{(-2)} y_1 - \frac{1}{2}\, (\partial_t{y_0}\, ){}^{2} y_0{}^{(-2)} y_1 - y	_0{}^{3} y_1  \\
\label{secondL}
&&\mathcal{L}_2 = \frac{1}{2}\, (\partial_t{y_1}\, ){}^{2} y_0{}^{-1} - \partial_t{y_0}\,  \partial_t{y_1}\,  y_0{}^{(-2)} y_1 + \frac{1}{2}\, (\partial_t{x_0}\, ){}^{2} y_0{}^{(-3)} y_1{}^{2} + \nonumber \\
&&~~~~~~\frac{1}{2}\, (\partial_t{y_0}\, ){}^{2} y_0{}^{(-3)} y_1{}^{2} - \frac{3}{2}\, y_0{}^{2} y_1{}^{2}  \\
\label{thirdL}
&&\mathcal{L}_3 = - \frac{1}{2}\, (\partial_t{y_1}\, ){}^{2} y_0{}^{(-2)} y_1 + \partial_t{y_0}\,  \partial_t{y_1}\,  y_0{}^{(-3)} y_1{}^{2} - \frac{1}{2}\, (\partial_t{x_0}\, ){}^{2} y_0{}^{(-4)} y_1{}^{3} -\nonumber \\
&&~~~~~~\frac{1}{2}\, (\partial_t{y_0}\, ){}^{2} y_0{}^{(-4)} y_1{}^{3} - y_1{}^{3} y_0  \\
\label{fourthL}
&&\mathcal{L}_4 = \frac{1}{2}\, (\partial_t{y_1}\, ){}^{2} y_0{}^{(-3)} y_1{}^{2} - \partial_t{y_0}\,  \partial_t{y_1}\,  y_0{}^{(-4)} y_1{}^{3} + \frac{1}{2}\, (\partial_t{x_0}\, ){}^{2} y_0{}^{(-5)} y_1{}^{4} + \nonumber \\
&&~~~~~~\frac{1}{2}\, (\partial_t{y_0}\, ){}^{2} y_0{}^{(-5)} y_1{}^{4} - \frac{1}{4}\, y_1{}^{4}.
\end{eqnarray}
Euler-Lagrange equations of motion of $x$ and $y$ for the Lagrangian (\ref{fulllagrangian}) are
\begin{eqnarray}
&& \partial_t\left(\frac{\partial_tx}{y}\right) + x^3 = 0 \\
\label{eqy}
&& \partial_t\left(\frac{\partial_t{y}}{y}\right) = - \frac{1}{2y^2} \left((\partial_t{x}){}^2 + (\partial_t{y}){}^2\right) - y^3
\end{eqnarray}
We perform order-by-order perturbation of the above equation using
(\ref{xy}) which can also be obtained by varying the perturbed
Lagrangian. Zeroth order equations of $x$, i.e., $x_0$ and $y$, i.e.,
$x_0$ are given by

\begin{eqnarray}
\label{zerothx}
&& \partial_t\left(\frac{\partial_tx_0}{y_0}\right) + x_0^3 = 0 \\
\label{zerothy}
&& \partial_t\left(\frac{\partial_t{y_0}}{y_0}\right) = - \frac{1}{2y_0{}^2} \left((\partial_t{x_0}){}^2 + (\partial_t{y_0}){}^2\right) - y_0^3.
\end{eqnarray}
One can either perturb the equation (\ref{eqy}) or vary the second
order Lagrangian (\ref{secondL}) with respect to $y_1$ to obtain first
order perturbed equation of motion of $y$ and is given by
\begin{equation}
\label{firstoy1}
\partial_t\left(\frac{\partial_ty_1}{y_0} - \frac{\partial_ty_0}{y_0{}^2} y_1\right) = \frac{1}{y_0{}^3}\left((\partial_t{x_0}){}^2 + (\partial_t{y_0}){}^2\right) y_1 - \frac{\partial_t y_0}{y_0{}^2}\partial_ty_1 - 3y_0{}^2 y_1.
\end{equation}
In the next subsection, we explicitly write down the perturbed
equations using Hamiltonian (\ref{fullhamiltonian}).

\subsection{Perturbed Hamiltonian}
It is important to note that even though $x$ is not perturbed, $\pi_x$
contains both perturbed and unperturbed parts.  Using (\ref{xy}), the
following relations can easily be established\footnote{It is apparent from
  equations (\ref{momenta}) that only first order perturbation of $y$
  can produce any higher order perturbed momenta of $x$ and $y$ and it
  is given by
\begin{eqnarray}
\label{pmomx}
&&\pi_x = \pi_{x0} + \epsilon \pi_{x1} + \epsilon^2 \pi_{x2} ... \nonumber \\
\label{pmomy}
&&\pi_y = \pi_{y0} + \epsilon \pi_{y1} +  \epsilon^2 \pi_{y2} ...\nonumber
\end{eqnarray}
where
\begin{eqnarray}
\label{pmomentumx}
&&\pi_{x0} = \frac{d_tx_0}{y_0} ,~~ \pi_{x1} = - \frac{y_1}{y_0} \pi_{x0}, ~~ \pi_{x2} = - \frac{y_1}{y_0} \pi_{x1}, ~~ \pi_{x3} = -\frac{y_1}{y_0} \pi_{x2} ~~~...\nonumber \\
\label{pmomentumy}
&&\pi_{y0} = \frac{d_ty_0}{y_0}, ~~ \pi_{y1} = \frac{d_ty_1}{y_0} - \frac{y_1}{y_0} \pi_{y0}, ~~ \pi_{y2} = - \frac{y_1}{y_0} \pi_{y1}, ~~ \pi_{y3} = - \frac{y_1}{y_0} \pi_{y2} ~~ ...\nonumber
\end{eqnarray}
Since we are not interested in higher order perturbation theory, we neglect higher order momenta and consider only first order momenta in calculating correlation functions.}
\begin{eqnarray}
\label{pmomx}
&&\pi_x = \pi_{x0} + \epsilon {}\pi_{x1}\\
\label{pmomy}
&&\pi_y = {}\pi_{y0} + \epsilon {}\pi_{y1}
\end{eqnarray}
where
\begin{eqnarray}
\label{pmomentumx}
&&{}\pi_{x0} = \frac{\partial_tx_0}{y_0} ,~~ {}\pi_{x1} = - \frac{y_1}{y_0} {}\pi_{x0}\\
\label{pmomentumy}
&&{}\pi_{y0} = \frac{\partial_ty_0}{y_0}, ~~ {}\pi_{y1} = \frac{\partial_ty_1}{y_0} - \frac{y_1}{y_0} {}\pi_{y0}
\end{eqnarray}

Using (\ref{xy}), (\ref{pmomx}) and (\ref{pmomy}), Hamiltonian of the
system (\ref{fullhamiltonian}) can be written as
\begin{equation}
  \mathcal{H} = \mathcal{H}_0 + \epsilon \mathcal{H}_1 + \epsilon^2 \mathcal{H}_2 + \epsilon^3 \mathcal{H}_3 + \epsilon^4 \mathcal{H}_4 + ...
\end{equation}
where
\begin{eqnarray}
\label{PH0}
&&\mathcal{H}_0 = \frac{1}{2}\, y_0 \left({}\pi_{x0}{}^{2} +  {}\pi_{y0}{}^{2}\right) + \frac{1}{4}\, \left( x_0{}^{4} + y_0{}^{4}\right) \\
\label{PH1}
&&\mathcal{H}_1 = \frac{1}{2}\, {}\pi_{x0}{}^{2} y_1 + {}\pi_{x0} {}\pi_{x1} y_0 + \frac{1}{2}\, {}\pi_{y0}{}^{2} y_1 + {}\pi_{y0} {}\pi_{y1} y_0 + y_0{}^{3} y_1 \\
\label{PH2}
&&\mathcal{H}_2 = y_1 \left( {}\pi_{x0} {}\pi_{x1} + {}\pi_{y0} {}\pi_{y1}\right)  + \frac{1}{2}\, y_0 \left({}\pi_{x1}{}^{2} + {}\pi_{y1}{}^2\right)   + \frac{3}{2}\, y_0{}^{2} y_1{}^{2} \\
\label{PH3}
&& \mathcal{H}_3 =  \frac{1}{2}\, y_1\left({}\pi_{x1}{}^{2}  + {}\pi_{y1}{}^{2} \right)  + y_1{}^{3} y_0 \\
\label{PH4}
&& \mathcal{H}_4 = \frac{1}{4} y_1{}^4
\end{eqnarray}

Using (\ref{PH0}), we obtain zeroth order Hamilton's equations
\begin{eqnarray}
\label{zerothHamil1}
&&\partial_tx_0 = y_0 {}\pi_{x0}, ~~~\partial_t{}\pi_{x0} = - x_0{}^3 \\
\label{zerothHamil2}
&&\partial_tx_0 = y_0 {}\pi_{y0}, ~~~\partial_t{}\pi_{y0} = - \frac{1}{2} \left(({}\pi_{x0}{}^2 + {}\pi_{y0}{}^2\right) - y_0{}^3
\end{eqnarray}

Using (\ref{PH2}), first order Hamilton's equations  are

\begin{eqnarray}
\label{firstHamil1}
&& \frac{\partial\mathcal{H}_2}{\partial {}\pi_{x1}} = 0~~~\Rightarrow {}\pi_{x0} y_1 + {}\pi_{x1} y_0 = 0\\
\label{firstHamil2}
&& \partial_ty_1 = \frac{\partial\mathcal{H}_2}{\partial{}\pi_{y1}} = {}\pi_{y0} y_1 + {}\pi_{y1} y_0 \\
\label{firstHamil3}
&& \partial_t{}\pi_{y1} = - \left( {}\pi_{x0} {}\pi_{x1} + {}\pi_{y0} {}\pi_{y1} \right) - 3 y_0{}^2 y_1
\end{eqnarray}

Equation (\ref{firstHamil1}) gives explicit expression for ${}\pi_{x1}$
and leads to identical expression as in (\ref{pmomentumx}). It can
easily be verified that, zeroth order equations (\ref{zerothHamil1})
and (\ref{zerothHamil2}) are identical to equations (\ref{zerothx})
and (\ref{zerothy}), respectively, where (\ref{firstHamil1}),
(\ref{firstHamil2}) and (\ref{firstHamil3}) lead to the equivalent
equation of motion of $y_1$ (\ref{firstoy1}).

To compare the the above expression with that from the Lagrangian
formulation, we rewrite the above expressions using (\ref{pmomentumx})
and (\ref{pmomentumy}). We get
\begin{eqnarray}
&&\mathcal{H}_0 = \frac{1}{2}\, (\partial_t{x_0}\, ){}^{2} y_0{}^{-1} + \frac{1}{2}\, (\partial_t{y_0}\, ){}^{2} y_0{}^{-1} + \frac{1}{4}\, x_0{}^{4} + \frac{1}{4}\, y_0{}^{4} \\
&&\mathcal{H}_1 = - \frac{1}{2}\, (\partial_t{x_0}\, ){}^{2} y_0{}^{(-2)} y_1 - \frac{1}{2}\, (\partial_t{y_0}\, ){}^{2} y_0{}^{(-2)} y_1 + \partial_t{y_0}\,  \partial_t{y_1}\,  y_0{}^{-1} +\nonumber \\
&&~~~~~~~~ y_0{}^{3} y_1 \\
&&\mathcal{H}_2 = - \frac{1}{2}\, (\partial_t{x_0}\, ){}^{2} y_0{}^{(-3)} y_1{}^{2} - \frac{1}{2}\, (\partial_t{y_0}\, ){}^{2} y_0{}^{(-3)} y_1{}^{2} + \frac{1}{2}\, (\partial_t{y_1}\, ){}^{2} y_0{}^{-1} + \nonumber \\
&&~~~~~~~~\frac{3}{2}\, y_0{}^{2} y_1{}^{2} \\
&&\mathcal{H}_3 = \frac{1}{2}\, (\partial_t{x_0}\, ){}^{2} y_0{}^{(-4)} y_1{}^{3} + \frac{1}{2}\, (\partial_t{y_0}\, ){}^{2} y_0{}^{(-4)} y_1{}^{3} - \partial_t{y_0}\,  \partial_t{y_1}\,  y_0{}^{(-3)} y_1{}^{2} +\nonumber \\
&&~~~~~~~~ \frac{1}{2}\, (\partial_t{y_1}\, ){}^{2} y_0{}^{(-2)} y_1 + y_1{}^{3} y_0 \\
&&\mathcal{H}_4 = \frac{1}{4} y_1{}^4
\end{eqnarray}

It is important to note that, only the third order perturbed
Hamiltonian is negative of the third order Lagrangian, i.e.,
\begin{equation}
\mathcal{H}_3 = - \mathcal{L}_3.
\end{equation}

Explicit forms of Interaction Hamiltonians can be obtained using
perturbed parts of the Lagrangian \cite{Huang:2006eha} and using
(\ref{secondL}), (\ref{thirdL}) and (\ref{fourthL}), it can be
verified that, both approaches lead to the identical results. In this
approach, to obtain $n^{th}$ order interaction Hamiltonian, the
Lagrangian is expanded up to $n^{th}$ order perturbation and by
varying the Lagrangian, the momentum corresponding to the perturbed
quantity is obtained as a non-linear combination of time derivative
of the field. Using perturbation techniques, this relation is inverted
and the time derivative of the field is written in terms of non-linear
combination of corresponding momentum. Perturbed Hamiltonian
corresponding to the perturbed parts of the Lagrangian is obtained by
using the conventional definition of Hamiltonian as $\mathcal{H} =
\pi\,\dot{\varphi} - \mathcal{L}$ and $\dot{\varphi}$ is replaced by
the above relation. Once the perturbed Hamiltonian is obtained in
terms of field variable and conjugate momentum, to calculate
correlation functions, the momentum in the Hamiltonian is replaced in
terms of time derivative of the field and in this time, the linear
relation of momentum and time derivative of the field is used. The
above procedure is rather cumbersome and involves series of
approximations. Since we start with the general Hamiltonian, our
approach is very straightforward and efficient.

As mentioned earlier, while we focus on first order perturbations, our
approach can easily be extended to any higher order perturbation,
e.g., to obtain second order equations of motion of field variables,
we have to consider up to second order field perturbation and its
corresponding momentum up to second order and calculate the fourth
order perturbed Hamiltonian. Since, we already
have obtained zeroth and first order equations, second order perturbed
field equations are obtained by varying fourth order perturbed
Hamiltonian with respect to second order perturbed variables and their
corresponding momenta.

\section{Canonical scalar field}\label{EoMLC}
In order to show the advantages of the Hamiltonian formulation, we
first focus on canonical scalar field. The action (\ref{AA}) for canonical
scalar field in the ADM formulation is
\begin{eqnarray}
\label{ActionCanonical}
&& \mathcal{S}_C = \int d^4 x \Big[N  \frac{\gamma^{\frac{1}{2}}}{2 \kappa}
\left(^{(3)}R + K_{i j} K^{i j} - K^2 \right) + \frac{1}{2}\, N{}^{-1} \gamma{}^{\frac{1}{2}} ({\partial}_{0}{\varphi}\, ){}^{2} -{N}^{i} {\partial}_{0}{\varphi}\,  {\partial}_{i}{\varphi}\,  N{}^{-1} \gamma{}^{\frac{1}{2}} \nonumber \\
&&~~~~~~~~~~  - \frac{1}{2}\, N {\gamma}^{i j} {\partial}_{i}{\varphi}\,  {\partial}_{j}{\varphi}\,  \gamma{}^{\frac{1}{2}} + \frac{1}{2}\, {N}^{i} {N}^{j} {\partial}_{i}{\varphi}\,  {\partial}_{j}{\varphi}\,  N{}^{-1} \gamma{}^{\frac{1}{2}} - N V \gamma{}^{\frac{1}{2}}\Big].
\end{eqnarray}

To obtain the equations of motion of all variables, one can simply use
the Einstein's equation or one can directly vary the action with
respect to field variables. 0-0 and 0-i components of the Einstein's
equations represent the equations of motion of $g_{0 0}$ and $g_{0
  i}$, which is $N$ and $N^{i}$ respectively. Hence, the above two
equations are identical to Hamiltonian and Momentum constraints
respectively and i-j component of the Einstein's equations and
conservation of Energy-Momentum tensor lead to equation of motion of
3-metric and matter field, respectively. In the rest of this
section, we will use the definitions (\ref{00pmetric}),
(\ref{0ipmetric}), (\ref{3metric}) and (\ref{field}) in the the above
equations to obtain perturbed equations of motion of
gravitational field variables at all order. In Appendix (\ref{PLeCSF}),
  we derive zeroth and first order perturbed field equations of
  canonical scalar field and in the following subsection, we apply the
  procedure discussed in section (\ref{simple model}) to canonical scalar field
  model to obtain consistent equations of motion as well as
  interaction Hamiltonian using Hamiltonian formulation.

\subsection{Hamitonian formulation}\label{EoMHC}
Conjugate momenta of all field variables $\gamma_{i j},\, \varphi,\,
N$ and $N^{i}$ are defined as
\begin{equation}
\pi^{i j} \equiv \frac{\delta L}{ \delta \dot{\gamma_{i j}}}, ~~\pi_\varphi \equiv \frac{\delta L}{ \delta \dot{\varphi}},~~~\pi_N \equiv \frac{\delta L}{ \delta \dot{N}}, ~~~\pi_{i}  \equiv \frac{\delta L}{ \delta \dot{N^{i}}}.
\end{equation}
Using the action (\ref{ActionCanonical}), conjugate momenta are given by
\begin{eqnarray}
\label{pimnC}
&& {\pi}^{i j} = \frac{1}{2}\, \kappa{}^{-1} \gamma{}^{\frac{1}{2}} ({\gamma}^{i j} {\gamma}^{k l} - {\gamma}^{i k} {\gamma}^{j l}) {K}_{k l} \\
\label{pipC}
&& \pi_\varphi  = N{}^{-1} \gamma{}^{\frac{1}{2}} \varphi^\prime\,  - {N}^{i} {\partial}_{i}{\varphi}\,  N{}^{-1} \gamma{}^{\frac{1}{2}} \\
\label{pinpini}
&& \Phi^N \equiv \pi_N = 0, ~~~\Phi^{N^{i}}_i \equiv \pi_{i} = 0.
\end{eqnarray}

From equation (\ref{pinpini}), it is apparent that, for canonical
scalar field, Lapse function $N$ and shift vector $N^{i}$ are
constraints and behave like Lagrange multipliers and it gives
4-primary constraint relations. Inverse relations of equations
(\ref{pimnC}) and (\ref{pipC}) are
\begin{eqnarray}
&& \gamma_{m n}^\prime = {\gamma}_{n k} {{N}^{k}{}_{|m}}\,  + {\gamma}_{m k} {{N}^{k}{}_{|n}}\,   - 2\, N {K}_{m n}, ~~ {K}_{i j} \equiv \kappa \gamma{}^{-\frac{1}{2}} ({\gamma}_{i j} {\gamma}_{k l} - 2\, {\gamma}_{i k} {\gamma}_{j l}) {\pi}^{k l} \\
&& \varphi^\prime = N \pi_\varphi \gamma{}^{-\frac{1}{2}} + {N}^{i} {\partial}_{i}{\varphi}
\end{eqnarray}

Using the above definitions of canonical momenta and
(\ref{ActionCanonical}), we get the Hamiltonian density of the system
\begin{eqnarray}
\label{HamiltonianCanonical}
&& \mathcal{H}_C = \pi^{i j} \gamma_{i j}^\prime + \pi_\varphi \varphi^\prime - L \nonumber\\
&&~~~~={\gamma}_{j k} {\partial}_{i}{{N}^{k}}\,  {\pi}^{i j} + {N}^{k} {\partial}_{k}{{\gamma}_{i j}}\,  {\pi}^{i j} + {\gamma}_{i k} {\partial}_{j}{{N}^{k}}\,  {\pi}^{i j} - N {\gamma}_{i j} {\gamma}_{k l} \kappa {\pi}^{k l} {\pi}^{i j} \gamma{}^{-\frac{1}{2}} + \nonumber \\
&&~~~~~~2\, N {\gamma}_{i k}\, {\gamma}_{j l}\, \kappa {\pi}^{k l} {\pi}^{i j} \gamma{}^{-\frac{1}{2}} + \frac{1}{2}\, N \pi_\varphi{}^{2} \gamma{}^{-\frac{1}{2}} + {N}^{i} \pi_\varphi {\partial}_{i}{\varphi}\,  - \frac{1}{2}\, N ^{(3)}R \gamma{}^{\frac{1}{2}} \kappa{}^{-1} +\nonumber \\
&&~~~~~~ \frac{1}{2}\, N {\gamma}^{i j} {\partial}_{i}{\varphi}\,  {\partial}_{j}{\varphi}\,  \gamma{}^{\frac{1}{2}} + N V \gamma{}^{\frac{1}{2}}.
\end{eqnarray}

Since the action (\ref{ActionCanonical}) has
diffeomorphism-invariance, Hamiltonian (\ref{HamiltonianCanonical})
vanishes identically, i.e., $$\mathcal{H}_C = 0.$$ Evolution of primary
constraints vanishes weakly and gives rise to four secondary
constraints: one Hamiltonian constraint due to $\mathcal{H}_N \equiv
\frac{d\Phi^N}{dt} =\{\pi_N, \mathcal{H}_C\}\equiv -\frac{\delta \mathcal{H}_C}{\delta N}
\approx 0$ and three Momentum constraints due to $\mathcal{H}_i \equiv
\frac{d\Phi^{N^{i}}_i}{dt} = \{\pi_i, \mathcal{H}_C\} \equiv - \frac{\delta
  \mathcal{H}_C}{\delta N^{i}} \approx 0$ .
\begin{eqnarray}
\label{HamConsCan}
&& \mathcal{H}_N \equiv  N \kappa \gamma{}^{-\frac{1}{2}} (2 {\gamma}_{i k}\, {\gamma}_{j l} - {\gamma}_{i j} {\gamma}_{k l}) \, \kappa {\pi}^{k l} {\pi}^{i j}  + \frac{1}{2}\, N \pi_\varphi{}^{2} \gamma{}^{-\frac{1}{2}} \nonumber \\
&& ~~~~~~ - \frac{1}{2}\, N ^{(3)}R \gamma{}^{\frac{1}{2}} \kappa{}^{-1} + \frac{1}{2}\, N {\gamma}^{i j} {\partial}_{i}{\varphi}\,  {\partial}_{j}{\varphi}\,  \gamma{}^{\frac{1}{2}} + N V \gamma{}^{\frac{1}{2}} \approx 0 \\
\label{MomConsCan}
&&\mathcal{H}_i \equiv  - 2 \partial_{k}{\gamma_{i j} \pi^{j k}} + \pi^{j k} \partial_{i}{\gamma_{j k}} + \pi_\varphi {\partial}_{i}{\varphi} \approx 0 .
\end{eqnarray}

Hamiltonian density can be written in terms Hamiltonian and Momentum
constraint as
\begin{equation}
\mathcal{H}_C = N \mathcal{H}_N + N^{i} \mathcal{H}_i \approx 0.
\end{equation}

\subsubsection{Zeroth order Hamilton's equations}
Using $\gamma_{i j} = a^2 \delta_{i j}$ and all background quantities
being independent of spatial coordinates, Hamiltonian density
(\ref{HamiltonianCanonical}) becomes
\begin{equation}
\label{HCB1}
\mathcal{H}_0^C =  2\, N_0\, a\, \kappa \,({\delta}_{i k} {\delta}_{j l}  - {\delta}_{i j} {\delta}_{k l}) \pi_0{}^{i j} {{}\pi_0{}}^{k l} + \frac{1}{2}\, N_0 {}\pi_{\varphi0}{}^{2} a{}^{(-3)} + N_0 V a{}^{3} \approx 0.
\end{equation}
At zeroth order, conjugate momentum of $a$, $\pi_a \equiv \frac{\delta
  L}{\delta \dot{a}}$ is directly related with ${}\pi_0{}^{i j}$ by the
simple relation ${}\pi_0{}^{i j} = \frac{1}{6 a} \pi_a \delta^{i j},~\pi_a
= 2 a \delta_{i j} {}\pi_0{}^{i j}$. The zeroth order Hamiltonian
(\ref{HCB1}), in terms of $\pi_a$ takes the simple form

\begin{equation}
\label{HCBF}
\mathcal{H}_0^C = N_0 \left[- \frac{1}{12}\,  \kappa \pi_a{}^{2} a{}^{-1} + \frac{1}{2}\,  {}\pi_{\varphi0}{}^{2} a{}^{(-3)} +  V a{}^{3}\right] \equiv N_0\, ^{(0)}\mathcal{H}_N.
\end{equation}

The terms inside the bracket in the right hand side is the Hamiltonian
constraint. At zeroth order it is independent of $N_0$. So, as we have
mentioned earlier, $N_0$ cannot be determined uniquely and we can
choose it arbitrarily. In this chapter, we use comoving coordinate, $N_0
= a$.

Varying the Hamiltonian (\ref{HCBF}) with respect to the momenta, we get
\begin{eqnarray}
&& a^\prime = - 1/6 N_0 \kappa \pi_a a^{-1} \\
&&\Rightarrow \pi_a = - 6 a a^\prime N_0^{-1} \kappa^{-1}~~~\Rightarrow {}\pi_0{}^{i j} = -N_0{}^{-1} \kappa{}^{-1} a^{\prime}\,  {\delta}^{i j} \\
&&\varphi_0^\prime = N_0 {}\pi_{\varphi0} a{}^{(-3)}
\end{eqnarray}

Hamiltonian constraint in conformal or comoving coordinate in terms of
field derivatives is
\begin{eqnarray}
\label{HConsCan0}
&&\mathcal{H}_{N0} \equiv - \frac{1}{12}\, \kappa \pi_a{}^{2} a{}^{-1} + \frac{1}{2}\, {}\pi_	{\varphi0}{}^{2} a{}^{(-3)} + V a{}^{3} = 0\\
\label{friedmann00}
&&\Rightarrow - 3\, \kappa{}^{-1} H^2 + \frac{1}{2}\,  \varphi_0^{\prime}\, {}^{2} + V\, a^2 = 0, \quad \mbox{where} \quad H \equiv \frac{a^\prime}{a}
\end{eqnarray}

Variation of the Hamiltonian with respect to the field variables and
relating with the time derivatives of the momenta lead to the
dynamical equation of motion of the field variables. Hence, equation
of motion of $a$ in comoving coordinate in terns of field derivatives
becomes
\begin{eqnarray}
\label{EoaHC0}
&&\pi_a^\prime\,  + \frac{\delta \mathcal{H}^C_0}{\delta a} = 0 \nonumber \\
&&\Rightarrow \pi_a^\prime\,  + \frac{1}{12} \, N_0 \kappa \pi_a{}^{2} a{}^{(-2)} - \frac{3}{2}\, N_0 {}\pi_{\varphi0}{}^{2} a{}^{(-4)} + 3\, N_0 V a{}^{2} = 0 \\
\label{eomah}
&&\Rightarrow 3\, \kappa{}^{-1} H^2 - 6\, \frac{a^{\prime \prime}}{a}\,  \kappa{}^{-1} - \frac{3}{2}\, \varphi_0^{\prime}\, {}^{2} a + 3\, V a{}^{2} = 0.
\end{eqnarray}

Similarly, the equation of motion of $\varphi_0$ takes the form
\begin{eqnarray}
\label{EovarphiHC0}
&&~~~~{}\pi_{\varphi0}\,{}^\prime  + N_0 V_\varphi\,  a{}^{3} = 0\\
\label{eomphih}
&& \Rightarrow \varphi_0^{\prime \prime} + 2 H\, \varphi_0^{\prime}+ V_\varphi\,  a{}^{2} = 0.
\end{eqnarray}
The three equations (\ref{friedmann00}), (\ref{eomah}) and
(\ref{eomphih}) are, as expected, identical to the equations
(\ref{B00EE}), (\ref{BijEE}) and (\ref{BEoMS}) respectively.

\subsubsection{First order perturbed Hamilton's equation}

As mentioned earlier, we consider flat-slicing gauge, hence there is
no perturbation in the 3-metric, i.e., $\delta g_{i j} = 0$. As we
pointed out in the simple model, while $x$ is treated as unperturbed,
$\pi_x$ is non-zero. In the flat-slicing gauge, there is no
perturbation in the 3-metric, however, canonical conjugate momentum
corresponding to 3-metric will have non-zero perturbed
contributions. This becomes transparent if we perturb (\ref{pimnC}),
i.e., $$\delta {\pi}^{m n} = \frac{1}{2}\, \kappa{}^{-1}
\gamma^{\frac{1}{2}} ({\gamma}^{m n} {\gamma}^{k l} - {\gamma}^{m k}
{\gamma}^{n l}) \delta {K}_{k l}.$$ Hence, perturbed part of $K_{i
  j}$, i.e., $\delta K_{i j}$ is not zero and it contributes to the
perturbed part of $\pi^{i j}$, i.e., $\delta \pi^{i j}$.

We can separate unperturbed and perturbed parts of field variables and
their corresponding momenta as
\begin{eqnarray}
\label{perturbfm1}
&& N = N_0 + \epsilon N_1, ~~~N^{i} = \epsilon N_1^i,~~~\varphi = \varphi_0 + \epsilon \varphi_1 \\
\label{perturbfm2}
&& \pi^{i j} = {}\pi_0{}^{i j} + \epsilon \pi_1^{i j}, ~~~\pi_\varphi = {}\pi_{\varphi0} + \epsilon \pi_{\varphi1}
\end{eqnarray}

Comparing (\ref{00pmetric}) and (\ref{0ipmetric}) with $N_1$ and $N_1^{i}$, we obtain 
\begin{equation}
\label{ncomp}
N_1 = a \,\phi_1,~~~~~~N_1^{i} = \delta^{i j}\, \partial_{i j}B_1
\end{equation}

The second order Hamiltonian density can be obtained by substituting
(\ref{perturbfm1}) and (\ref{perturbfm2}) in
(\ref{HamiltonianCanonical})
\begin{eqnarray}
\label{PHamSecCan}
&&\mathcal{H}^C_2 = {\delta}_{i j} {\partial}_{k}{{N_1}^{j}}\,  {\pi_1}^{i k} a{}^{2} + {\delta}_{i j} {\partial}_{k}{{N_1}^{j}}\,  {\pi_1}^{i k} a{}^{2} - N_0 {\delta}_{i j} {\delta}_{k l} \kappa {\pi_1}^{i j} {\pi_1}^{k l} a -  \nonumber \\
&&~~~~~~~~2\, N_1 {\delta}_{i j} {\delta}_{k l} \kappa \pi_0{}^{i j} {\pi_1}^{k l} a +2\, N_0 {\delta}_{i j} {\delta}_{k l} \kappa {\pi_1}^{i k} {\pi_1}^{j l} a + 4\, N_1 {\delta}_{i j} {\delta}_{k l} \kappa {{}\pi_0{}}^{i k} {\pi_1}^{j l} a +  \nonumber \\
&&~~~~~~~~\frac{1}{2}\, N_0 \pi_{\varphi1}{}^{2} a{}^{(-3)} + N_1 {}\pi_{\varphi0} \pi_{\varphi1} a{}^{(-3)} +{N_1}^{i} {}\pi_{\varphi0} {\partial}_{i}{\varphi_1}\,  + \nonumber \\
&&~~~~~~~~\frac{1}{2}\, N_0 {\delta}^{i j} {\partial}_{i}{\varphi_1}\,  {\partial}_{j}{\varphi_1}\,  a + N_1 V_\varphi\,  a{}^{3} \varphi_1 + \frac{1}{2}\, N_0 V_{\varphi \varphi}\,  \varphi_1{}^{2} a{}^{3}.
\end{eqnarray}

Since there is no perturbation in the 3-metric $\gamma_{i j}$,
variation with respect to $\pi_1^{i j}$ gives rise to a separate
constraint equation from which we can extract the derived value of
$\pi_1^{i j}$.
\begin{eqnarray}
&& ~~~\frac{\delta \mathcal{H}_2^C}{\delta \pi_1^{i j}} = 0 \\
&& \Rightarrow {\delta}_{n j} {\partial}_{m}{{N_1}^{j}}\,  a{}^{2} + {\delta}_{m j} {\partial}_{n}{{N_1}^{j}}\,  a{}^{2} - 2\, N_0 {\delta}_{m n} {\delta}_{k l} \kappa {\pi_1}^{k l} a + 4\, N_0 {\delta}_{m k} {\delta}_{n l} \kappa {\pi_1}^{k l} a - \nonumber \\
&& ~~~2\, N_1 {\delta}_{m n} {\delta}_{k l} \kappa {{}\pi_0{}}^{k l} a + 4\, N_1 {\delta}_{m k} {\delta}_{n l} \kappa {{}\pi_0{}}^{k l} a = 0.
\end{eqnarray}

Multiplying above expression with $(\delta^{m n} \delta^{i j} -
\delta^{m i} \delta^{n j})$ gives
\begin{eqnarray}
\label{pi1ij}
&&\pi_1^{i j} = \frac{1}{2}\, N_0{}^{-1} \kappa{}^{-1} a {\delta}^{i j} {\partial}_{k}{{N_1}^{k}}\,  - \frac{1}{4}\, N_0{}^{-1} \kappa{}^{-1} a {\delta}^{k i} {\partial}_{k}{{N_1}^{j}}\,  - \frac{1}{4}\, N_0{}^{-1} \kappa{}^{-1} a {\delta}^{k j} {\partial}_{k}{{N_1}^{i}}\,  \nonumber \\
&&~~~~~~~~- N_0{}^{-1} N_1 \pi_0{}^{i j}.
\end{eqnarray}

The relation of time derivative of perturbed matter field $\varphi_1$
and conjugate momentum of $\varphi_1$, $(\pi_{\varphi1})$ are
\begin{eqnarray}
&&~~~\varphi_1^\prime = \frac{\delta \mathcal{H}_2^C}{\delta \pi_{\varphi1}} \nonumber \\
&&\Rightarrow \varphi_1^\prime =  N_0 \pi_{\varphi1} a{}^{(-3)} +  N_1 {}\pi_{\varphi0} a{}^{(-3)}.
\end{eqnarray}

In the conformal coordinate, the perturbed Hamiltonian constraint is
obtained by varying the second order perturbed Hamiltonian density
(\ref{PHamSecCan}) with respect to perturbed Lapse function
$N_1$. Using above relations for the momenta with the time derivatives of
the field variables and (\ref{ncomp}), the perturbed Hamiltonian
constraint becomes
\begin{eqnarray}
&& ~~~~~~~\frac{\delta \mathcal{H}_2^C}{\delta N_1} = 0 \\
&& \Rightarrow 2\, {\delta}^{i j} H\,  {\partial}_{i j}{B_1}\,  \kappa{}^{-1} + 6\, \phi_1 \kappa{}^{-1} H^2 + \varphi_0^{\prime}\,  \varphi_1^\prime\,   -  \nonumber \\
\label{HConsH1}
&&~~~~~~~~\phi_1 \varphi_0^{\prime}\, {}^{2}  +V_\varphi\,  a{}^{2} \varphi_1 = 0.
\end{eqnarray}

Similarly, Momentum constraint is given by
\begin{eqnarray}
&& ~~~M_{i} \equiv \frac{\delta \mathcal{H}_2^C}{\delta N_1^{i}} = 0 \\
\label{MConsH1}
&&\Rightarrow \varphi_0^{\prime}\,  {\partial}_{i}{\varphi_1}\,   - 2\, H\,  {\partial}_{i}{\phi_1}\,  \kappa{}^{-1}  = 0.
\end{eqnarray}
and the equation of motion of the perturbed scalar field $\varphi_1$ becomes
\begin{eqnarray}
  && ~~~{\pi_{\varphi1}{}^\prime} + \frac{\delta \mathcal{H}_2^C}{\delta \varphi_1} = 0 \\
  && \Rightarrow  \varphi_1^{\prime \prime}\, + 2H\, \varphi_1^{\prime}\,    - {\phi_1^\prime}\,  \varphi_0^{\prime}\,  + 2\, V_\varphi \,  \phi_1 a{}^{2} - {\delta}^{i j} \varphi_0^{\prime}\,  {\partial}_{i j}{B_1}\,    \nonumber \\
\label{EConsH1}
&&~~~~~~~~- {\delta}^{i j} {\partial}_{i j}{\varphi_1}\, + V_{\varphi \varphi}\,  a{}^{2} \varphi_1 = 0.
\end{eqnarray}

Equation (\ref{HConsH1}), (\ref{MConsH1}) and (\ref{EConsH1}) are
identical to the equations (\ref{100EE}), (\ref{10iEE}) and
(\ref{1EoMS}), respectively.

This is a very important result. Unlike Lagrangian formalism, choosing
a gauge in Hamiltonian formalism is not trivial.  But using the above
simple mechanism, it is possible to construct the perturbed
Hamiltonian and its equations of motion and can now be treated in the
same manner as Lagrangian formalism. It can, even, be extended to any
order of perturbation, e.g., to obtain second order equations of
motion of field variables, we have to extend the Hamiltonian at fourth
order perturbation in terms of second order field variables and the
second order momenta and vary the Hamiltonian with respect to second
order variables and its conjugate momenta. At second order, in
flat-slicing gauge, we will again obtain a constraint equation
$\frac{\partial \mathcal{H}_4}{\partial \pi_2^{i j}} = 0$ that will give the
expression of $\pi_2^{i j}$. 

Our proposed mechanism works for any other arbitrary
  gauge also. In Appendix (\ref{udgcsf}), we obtain all consistent
  perturbed and unperturbed Hamilton's equations for Canonical scalar
  field in uniform density gauge. This mechanism can be applied to
  any generalized first order derivative theory like non-canonical
  scalar field.

\subsubsection{Single variable-Momentum Hamiltonian}
Since the second order Hamiltonian holds the dynamics of first order
perturbed variables, we can obtain a single variable-single
momentum effective Hamiltonian of the background and constraint 
equations, and replacing background momenta in
terms of time derivatives of the fields. Substituting $\pi_1^{i j}$
using (\ref{pi1ij}) and ${}\pi_0{}^{i j}$ and ${}\pi_{\varphi0}$ in the
second order Hamiltonian and using (\ref{ncomp}), we get

\begin{eqnarray}
  && \mathcal{H}_2^C = \frac{1}{2}\, {\delta}^{i j} {\delta}^{k l} {\partial}_{i j}{B_1}\,  {\partial}_{k l}{B_1}\,  \kappa{}^{-1} a{}^{2} - \frac{1}{2} {\delta}^{i j} {\delta}^{k l} {\partial}_{i k}{B_1}\,  {\partial}_{j l}{B_1}\,  \kappa{}^{-1} a{}^{2} + 2\, {\delta}^{i j} H\,  {\partial}_{i j}{B_1}\,  \phi_1 \kappa{}^{-1} a^2 \nonumber \\
  &&~~~~~~+3\, \kappa{}^{-1} H^{2} \phi_1{}^{2} a^2 + \frac{1}{2}\, \pi_{\varphi1}{}^{2} a{}^{(-2)} + \pi_{\varphi1} \varphi_0^{\prime}\,  \phi_1 + {\delta}^{i j} \varphi_0^{\prime}\,  {\partial}_{i}{B_1}\,  {\partial}_{j}{\varphi_1}\,  a{}^{2} \nonumber \\
  &&~~~~~~+ \frac{1}{2}\, {\delta}^{i j} {\partial}_{i}{\varphi_1}\,  {\partial}_{j}{\varphi_1}\,  a{}^{2} + V_{\varphi}\,  \phi_1 a{}^{4} \varphi_1 + \frac{1}{2}\, V_{\varphi \varphi}\,  \varphi_1{}^{2} a{}^{4}
\end{eqnarray}

First and second terms in the right hand side lead to boundary
term. Moreover, performing integration by-parts to the seventh term
and substituting \[\phi_1 = \frac{\kappa}{2 H}
\varphi_0^\prime\, \varphi_1\] in the Hamiltonian, we get
\begin{eqnarray}
&&\mathcal{H}_2^C =\frac{3}{4}\, \kappa \varphi_0^{\prime}\, {}^{2} \varphi_1{}^{2} a{}^{2} + \frac{1}{2}\, \pi_{\varphi1}{}^{2} a{}^{(-2)} + \frac{\kappa}{2 H}\, \pi_{\varphi1}  \varphi_0^{\prime}\, {}^{2}  \varphi_1  + \frac{1}{2}\, {\delta}^{i j} {\partial}_{i}{\varphi_1}\,  {\partial}_{j}{\varphi_1}\,  a{}^{2}  \nonumber \\
&&~~~~~~+\frac{\kappa}{2 H}\,  V_\varphi\,  \varphi_0^{\prime}\,   \varphi_1{}^{2} a{}^{4} + \frac{1}{2}\, V_{\varphi \varphi}\,  \varphi_1{}^{2} a{}^{4}
\end{eqnarray}

This is the single variable-single momentum Hamiltonian density
$(\varphi_1, ~\pi_{\varphi1})$.  Further, this can be expressed in
terms of Mukhanov-Sasaki variable-Momentum\footnote{ Mukhanov-Sasaki variable is also a gauge invariant quantity related to curvature perturbation. At first order, it is given by $$u_1 = \frac{a \varphi_0^\prime}{H} \mathcal{R}_1.$$} form. In flat-slicing gauge,
Mukhanov-Sasaki variable $u_1 = a \varphi_1$, hence, 
\[\pi_{\varphi1} = a \pi_{u1}.\]

Hence the Hamiltonian in terms of Mukhanov-Sasaki variable and its conjugate momenta takes the form\footnote{Lagrangian can be written as
\begin{equation}
\mathcal{L} = 	\pi_{\varphi} \varphi^\prime - \mathcal{H}, \nonumber
\end{equation}

Changing of variables $\varphi \rightarrow \frac{u}{a}, \pi_\varphi \rightarrow a \pi_u$ leaves the Lagrangian unchanged. Hence,
\begin{eqnarray}
\mathcal{L} &=& a \pi_u \left( \frac{u^\prime}{a} - \frac{a^\prime}{a^2} u  \right) - \mathcal{H} \nonumber \\
&=& \pi_u u^\prime - \frac{a^\prime}{a} u \pi_u - \mathcal{H} \nonumber
\end{eqnarray}

Hence the new Hamiltonian in terms of $u$ and $\pi_u$ takes the form
\begin{equation}
\mathcal{H}^u = \mathcal{H}(u) + \frac{a^\prime}{a} u \pi_u \nonumber
\end{equation}

}

\begin{eqnarray}
\mathcal{H}_2^u &=& \mathcal{H}^C_2(u) + \frac{a^\prime}{a}\, \pi_{u1} u_1 \nonumber \\ 
&=&\frac{3}{4}\, \kappa u_1{}^{2} \varphi_0^{\prime}\, {}^{2} + \frac{1}{2}\, \pi_{u1}{}^{2} + \frac{1}{2}\, \pi_{u1} u_1 \kappa \varphi_0^{\prime}\, {}^{2} H{}^{-1} + \frac{1}{2}\, {\delta}^{i j} {\partial}_{i}{u_1}\,  {\partial}_{j}{u_1}\,  \nonumber \\
&&+ \frac{1}{2}\, \kappa V_\varphi\,  \varphi_0^{\prime}\,  u_1{}^{2} H^{-1} a{}^{2} + \frac{1}{2}\, V_{\varphi\varphi}\,  u_1{}^{2} a{}^{2} + H\, \pi_{u1} u_1.
\end{eqnarray}

One can verify that the equation of motion of $u_1$ becomes $$u_1^{\prime \prime} - \nabla^2 u_1 - \frac{z^{\prime\prime}}{z} u_1 = 0, \quad \mbox{where} \quad z \equiv \frac{a \varphi_0^\prime}{H}.$$

\subsubsection{Interaction Hamiltonian for calculating higher order
  correlations} 
  
Expanding Hamiltonian
  (\ref{HamiltonianCanonical}) to third order, we obtain third order
  Interaction Hamiltonian whose explicit form in phase space is given
  in Appendix (\ref{ThirdOrderInteractionCanonical}). Replacing the
  momenta in terms of terms of time derivatives of the fields,
  interaction Hamiltonian for canonical scalar field becomes
{
\begin{eqnarray}
&& \mathcal{H}_3^C \, = - \frac{1}{2}\, {\delta}^{i j} {\delta}^{k l} {\partial}_{i j}{B_1}\,  {\partial}_{k l}{B_1}\,  \phi_1 \kappa{}^{-1} a{}^{2} - 2\, {\delta}^{i j} H\,  {\partial}_{i j}{B_1}\,  \kappa{}^{-1} \phi_1{}^{2} a^2 - 
3\, \kappa{}^{-1} H^{2} \phi_1{}^{3} a^2 + \nonumber \\
&&\frac{1}{2}\, {\delta}^{i j} {\delta}^{k l} {\partial}_{i k}{B_1}\,  {\partial}_{j l}{B_1}\,  \phi_1 \kappa{}^{-1} a{}^{2} +  \frac{1}{2}\, \phi_1 {\varphi_1^\prime}\, {}^{2} a{}^{2} - {\varphi_0^\prime}\,  {\varphi_1^\prime}\,  \phi_1{}^{2} a{}^{2} + \frac{1}{2}\, {\varphi_0^\prime}\, {}^{2} \phi_1{}^{3} a{}^{2} + \nonumber \\
&&{\delta}^{i j} {\varphi_1^\prime}\,  {\partial}_{i}{B_1}\,  {\partial}_{j}{\varphi_1}\,  a{}^{2} - {\delta}^{i j} {\varphi_0^\prime}\,  {\partial}_{i}{B_1}\,  {\partial}_{j}{\varphi_1}\,  \phi_1 a{}^{2} + \frac{1}{2}\, {\delta}^{i j} {\partial}_{i}{\varphi_1}\,  {\partial}_{j}{\varphi_1}\,  \phi_1 a{}^{2} + \frac{1}{2}\, V_{\varphi\varphi}\,  \phi_1 \varphi_1{}^{2} a{}^{4} + \nonumber \\
&&\frac{1}{6}\, V_{\varphi\varphi\varphi}\,  \varphi_1{}^{3} a{}^{4}.
\end{eqnarray}}

It can be verified that $^{(3)}\mathcal{H}_C = - ^{(3)}\mathcal{L}_C$. Similarly, fourth order Interaction Hamiltonian for canonical scalar field takes the form
{
\begin{equation}
\mathcal{H}_4^C = \frac{1}{6}\, \phi_1 V_{\varphi \varphi \varphi}\,  \varphi_1{}^{3} a{}^{4} + \frac{1}{24}\,  V_{\varphi \varphi \varphi \varphi}\,  \varphi_1{}^{4} a{}^{4}
\end{equation}}
which is independent of kinetic part (time derivatives of fields) of
the field. This can be verified by looking at the Hamiltonian
(\ref{HamiltonianCanonical}). Terms containing momenta only contribute
up to third order Hamiltonian in flat-slicing gauge since $\gamma_{i j}$ is unperturbed in flat-slicing gauge.
Hence, fourth or higher order perturbed Hamiltonian is independent of
the kinetic part of the fields. Furthermore, the higher order interaction Hamiltonian can be expressed as single variable form by using constrained equations (\ref{HConsH1}) and (\ref{MConsH1}) \cite{Mizuno:2009mv, Arroja:2013dya, Izumi:2011di}.

\section{Galilean single scalar field model}\label{galileanfield}
In the action (\ref{boxx}), $G(X, \varphi) \neq 0$ leads to Galilean
field where action contains second derivative terms of the field
variables. In this chapter, to simplify our calculations, we take
a specific and simple form of the Galilean model with $P(X, \varphi) =
- V(\varphi)$ and $G(X, \varphi) = -2 X$, i.e.,
\begin{equation}
\label{action}
\mathcal{S}_G = \int d^4 x \sqrt{-g} \left[\frac{1}{2 \kappa} R -  g^{\mu \nu} \partial_{\mu}{\varphi} \partial_{\nu}{\varphi} \Box \varphi  - V(\varphi) \right] \equiv \int \mathcal{L}_G~d^4 x.
\end{equation}

Zeroth and first order perturbed Euler-Lagrange equations of motion are provided in Appendix (\ref{galmod}).

\subsection{Hamiltonian formulation of the Galilean scalar field}

Hamiltonian formulation of Higher derivative fields is not unique and there
exist several ways~\cite{Ostro,Andrzejewski:2010kz,Deffayet:2015qwa,Nesterenko,Chen:2012au,Woodard:2015zca}
to rewrite the Hamiltonian as there are infinite ways to absorb the higher
derivative terms. For our case, one easy way is to let 
$S \equiv \Box \varphi$ and re-write the action (\ref{action}) as
\begin{equation}
\label{full action galilean}
\mathcal{S}_G = \int d^4 x \sqrt{-g} \left[\frac{1}{2 \kappa} R - \frac{1}{2} g^{\mu \nu} \partial_{\mu}{\varphi} \partial_{\nu}{\varphi} S  - V(\varphi) \right] + \int d^4 x ~\lambda \left(S - \Box \varphi\right).
\end{equation}
where $\lambda$ is the Langrange multiplier whose variation leads to 
$S = \Box \varphi$. Since $\Box \varphi$ appears linearly in the 
action, we can rewrite the action in terms of
first order derivatives of the fields by performing integration
by-parts as
\begin{eqnarray}
&& \mathcal{S}_G = \int d^4 x \sqrt{-g} \left[\frac{1}{2 \kappa} R - \frac{1}{2} g^{\mu \nu} \partial_{\mu}{\varphi} \partial_{\nu}{\varphi} \, S  - V(\varphi) \right] + \int d^4 x ~\lambda \left(S - \Box \varphi\right) \nonumber\\
&& ~~= ... + \int d^4x \left[\lambda S - \lambda~ g^{\mu \nu} \left(\partial_{\mu \nu}{\varphi} - \Gamma^{\alpha}_{\mu \nu} ~\partial_{\alpha}{\varphi} \right) \right] \nonumber \\
&& ~~= ... + \int d^4x \left[\lambda S + \lambda~ g^{\mu \nu}~\Gamma^{\alpha}_{\mu \nu} ~\partial_{\alpha}{\varphi} + g^{\mu \nu} \partial_{\mu}{\varphi} \partial_{\nu}\lambda + \lambda \, \partial_{\nu}{g^{\mu \nu}} \partial_{\mu}{\varphi}\right] + \mbox{Boundary term} \nonumber \\
\label{full action}
&& ~~= \int d^4 x \sqrt{-g} \left[\frac{1}{2 \kappa} R - \frac{1}{2} g^{\mu \nu} \partial_{\mu}{\varphi} \partial_{\nu}{\varphi} \, S  - V(\varphi) \right] + \int d^4x [\, \lambda S + \lambda~ g^{\mu \nu}~\Gamma^{\alpha}_{\mu \nu} ~\partial_{\alpha}{\varphi} + \nonumber \\
&& ~~~~~~~~g^{\mu \nu} \partial_{\mu}{\varphi} \partial_{\nu}\lambda + \lambda \, \partial_{\nu}{g^{\mu \nu}} \partial_{\mu}{\varphi}\,]. 
\end{eqnarray}
Although the action, now, contains extra variables $\lambda $ and $S$,
action (\ref{action}) and (\ref{full action}) lead to equivalent
equations of motion. Since, the action (\ref{full action}) contains no
double derivative terms, we can construct the Hamiltonian of the
system.

Using the line element (\ref{line}), the action (\ref{full
  action}) can be decomposed and rewritten as

{\small
\begin{eqnarray}
\label{full lagrangian}
 &&\mathcal{S}_G \equiv \int \mathcal{L}_G\,d^4x = \int d^4 x \,(S \lambda - N V \gamma{}^{\frac{1}{2}} + {\gamma}^{i j} {\partial}_{i}{\lambda}\,  {\partial}_{j}{\varphi}\,  + \lambda {\partial}_{i}{{\gamma}^{i j}}\,  {\partial}_{j}{\varphi}\,  - {\lambda^\prime}\,  {\varphi^\prime}\,  N{}^{(-2)} + \frac{1}{2}\, N ^{(3)}R \gamma{}^{\frac{1}{2}} \kappa{}^{-1} + \nonumber \\
 &&~~~~{N}^{i} {\lambda^\prime}\,  {\partial}_{i}{\varphi}\,  N{}^{(-2)} + {N}^{i} {\varphi^\prime}\,  {\partial}_{i}{\lambda}\,  N{}^{(-2)} + S N{}^{-1} \gamma{}^{\frac{1}{2}} {\varphi^\prime}\, {}^{2} + \lambda {N^\prime}\,  {\varphi^\prime}\,  N{}^{(-3)} - {N}^{i} {N}^{j} {\partial}_{i}{\lambda}\,  {\partial}_{j}{\varphi}\,  N{}^{(-2)} - \nonumber \\
 &&~~~~{N}^{i} \lambda {N^\prime}\,  {\partial}_{i}{\varphi}\,  N{}^{(-3)} - {N}^{i} \lambda {\varphi^\prime}\,  {\partial}_{i}{N}\,  N{}^{(-3)} + {\gamma}^{i j} {\gamma}^{k l} \lambda {\partial}_{i}{{\gamma}_{j k}}\,  {\partial}_{l}{\varphi}\,  - \frac{1}{2}\, {\gamma}^{i j} {\gamma}^{k l} \lambda {\partial}_{i}{{\gamma}_{k l}}\,  {\partial}_{j}{\varphi}\,  + \nonumber \\
 &&~~~~\frac{1}{2}\, {\gamma}^{i j} \lambda {{\gamma}_{i j}}{}^\prime\,  {\varphi}^\prime\,  N{}^{(-2)} - {\gamma}^{i j} \lambda {\partial}_{i}{N}\,  {\partial}_{j}{\varphi}\,  N{}^{-1} - N S {\gamma}^{i j} {\partial}_{i}{\varphi}\,  {\partial}_{j}{\varphi}\,  \gamma{}^{\frac{1}{2}} + {N}^{i} {N}^{j} \lambda {\partial}_{i}{N}\,  {\partial}_{j}{\varphi}\,  N{}^{(-3)}%
 - \nonumber \\
 &&~~~~2\, {N}^{i} S {\varphi^\prime}\,  {\partial}_{i}{\varphi}\,  N{}^{-1} \gamma{}^{\frac{1}{2}} - \frac{1}{2}\, {N}^{i} {\gamma}^{j k} \lambda {{\gamma}_{j k}}{}^\prime\,  {\partial}_{i}{\varphi}\,  N{}^{(-2)} - \frac{1}{2}\, {N}^{i} {\gamma}^{j k} \lambda {\varphi^\prime}\,  {\partial}_{i}{{\gamma}_{j k}}\,  N{}^{(-2)} - \nonumber \\
 &&~~~~\frac{1}{2}\, {K}^{i j} {K}^{k l} N {\gamma}_{i j} {\gamma}_{k l} \gamma{}^{\frac{1}{2}} \kappa{}^{-1} + \frac{1}{2}\, {K}^{i j} {K}^{k l} N {\gamma}_{i k} {\gamma}_{j l} \gamma{}^{\frac{1}{2}} \kappa{}^{-1} + {N}^{i} {N}^{j} S {\partial}_{i}{\varphi}\,  {\partial}_{j}{\varphi}\,  N{}^{-1} \gamma{}^{\frac{1}{2}} +\nonumber \\
 &&~~~~ \frac{1}{2}\, {N}^{i} {N}^{j} {\gamma}^{k l} \lambda {\partial}_{i}{{\gamma}_{k l}}\,  {\partial}_{j}{\varphi}\,  N{}^{(-2)}).
\end{eqnarray}}

Momenta corresponding to the variables are defined as
\begin{eqnarray}
\label{dmomenta}
&&\pi^{i j} = \frac{\delta \mathcal{S}_G}{\delta \gamma_{i j}{}^\prime} ~~~~~~~~ \pi_N = \frac{\delta \mathcal{S}_G}{\delta N^\prime} ~~~~~~~~ \pi_{i} = \frac{\delta \mathcal{S}_G}{\delta N^{i}{}^\prime} \nonumber \\
&&\pi_\varphi = \frac{\delta \mathcal{S}_G}{\delta \varphi^\prime}~~~~~~~~\pi_\lambda = \frac{\delta \mathcal{S}_G}{\delta \lambda^\prime}~~~~~~~~ \pi_S = \frac{\delta \mathcal{S}_G}{\delta S^\prime} \nonumber ~~
\end{eqnarray}

Using the action (\ref{full lagrangian}), we obtain the following relations:
\begin{eqnarray}
\label{momentag}
&& \pi^{i j} = \frac{1}{2}\, \kappa{}^{-1} \gamma{}^{\frac{1}{2}} ({\gamma}^{m n} {\gamma}^{k l} - {\gamma}^{m k} {\gamma}^{n l}) {K}_{k l} + \frac{1}{2}\, {\gamma}^{m n} \lambda {\varphi^\prime}\,  N{}^{(-2)} - \frac{1}{2}\, {N}^{i} {\partial}_{i}{\varphi}\,  \lambda N{}^{(-2)} {\gamma}^{m n} ~~~~~~~~ \\
\label{momentaN}
&& \pi_\varphi =  - {\lambda^\prime}\,  N{}^{(-2)} + {N}^{i} {\partial}_{i}{\lambda}\,  N{}^{(-2)} + 2\, S N{}^{-1} \gamma{}^{\frac{1}{2}} {\varphi^\prime}\,  + \lambda {N^\prime}\,  N{}^{(-3)} - {N}^{i} \lambda {\partial}_{i}{N}\,  N{}^{(-3)}\nonumber \\
&& ~~~~~~~~+ \frac{1}{2}\, {\gamma}^{i j} \lambda {{\gamma}_{i j}{}^\prime}\,  N{}^{(-2)} - 2\, {N}^{i} S {\partial}_{i}{\varphi}\,  N{}^{-1} \gamma{}^{\frac{1}{2}} - \frac{1}{2}\, {N}^{i} {\gamma}^{j k} \lambda {\partial}_{i}{{\gamma}_{j k}}\,  N{}^{(-2)} \\
\label{momentaL}
&& \pi_\lambda = - {\varphi^\prime}\,  N{}^{(-2)} + {N}^{i} {\partial}_{i}{\varphi}\,  N{}^{(-2)} \\
&& \pi_N = {\varphi^\prime}\,  \lambda N{}^{(-3)} - {N}^{i} \lambda {\partial}_{i}{\varphi}\,  N{}^{(-3)} \nonumber \\
\label{momentaNN}
&& ~~~~~~~~ = - \lambda N{}^{-1} \pi_\lambda \\
&& \pi_{i} = 0 \\
&& \pi_S = 0.
\end{eqnarray}

Equations (\ref{momentag}), (\ref{momentaN}) and (\ref{momentaL}) are
invertible and $\partial_{0}{\gamma_{i j}}$, $\partial_{0}{\lambda}$
and $\partial_{0}{\varphi}$ can be written in terms of $\pi^{i j}$,
$\pi_\varphi$ and $\pi_\lambda$ as

\begin{eqnarray}
\label{g0}
&& \gamma_{m n}{}^\prime = {\gamma}_{n k} {N}^{k}{}_{|m}\,  + {\gamma}_{m k} {{N}^{k}{}_{|n}}\,  - 2\, N {K}_{m n} \\
\label{k0}
&& K_{i j} = \kappa \gamma{}^{-1/2} ({\gamma}_{i j} {\gamma}_{m n} - 2\, {\gamma}_{i m} {\gamma}_{j n}) {\pi}^{m n} + \frac{1}{2}\, {\gamma}_{i j} \lambda \pi_\lambda \\
\label{v0}
&& {\varphi}{}^\prime = N^{i} \partial_{i}{\varphi} - N^2 \pi_\lambda \\
&& {\lambda}{}^\prime = N^2 ( - \pi_\varphi + 2 S N^{-1} \gamma^{1/2} {\varphi^\prime} - 2 N^{i} S \partial_{i}{\varphi} N^{-1} \gamma^{1/2} 
 + N^{i} \partial_{i}{\lambda} N^{(-2)} + \lambda {N^\prime} N^{(-3)} \nonumber \\
 \label{l0}
 &&~~~~~~~~- N^{i} \lambda \partial_{i}{N} N^{(-3)} + \frac{1}{2} \gamma^{i j} \lambda {\gamma_{i j}^\prime} N^{(-2)} - \frac{1}{2} N^{i} \gamma^{j k} \lambda \partial_{i}{\gamma_{j k}} N^{(-2)}).
\end{eqnarray}

 Hence the Hamiltonian density is given by

\begin{equation}
 \mathcal{H}_G = \pi^{i j} {\gamma_{i j}^\prime} + \pi_\varphi {\varphi^\prime} + \pi_{\lambda} {\lambda^\prime} + \pi_N {N^\prime} - \mathcal{L}_G.
\end{equation}

Using the action (\ref{full lagrangian}) and (\ref{g0}), (\ref{k0}), (\ref{v0}) (\ref{l0}) and (\ref{momentaNN}), the Hamiltonian density becomes
{\small
\begin{eqnarray}
\label{fullH}
&&\mathcal{H}_G =  - S \lambda + N V \gamma{}^{\frac{1}{2}} + {N}^{i} \pi_\lambda {\partial}_{i}{\lambda}\,  + {N}^{i} \pi_\varphi {\partial}_{i}{\varphi}\,  - \pi_\lambda \pi_\varphi N{}^{2} + \pi_\lambda \lambda {\partial}_{i}{{N}^{i}}\,  + 2 {\gamma}_{i j} {\partial}_{k}{{N}^{i}}\,  {\pi}^{ j k} +\nonumber \\
&&~~~~ {\gamma}^{i j} \lambda {\partial}_{i}{N}\,  {\partial}_{j}{\varphi}\,  N{}^{-1} - {\gamma}^{i j} {\partial}_{i}{\lambda}\,  {\partial}_{j}{\varphi}\,  - \lambda {\partial}_{i}{{\gamma}^{i j}}\,  {\partial}_{j}{\varphi}\,  - \frac{1}{2}\, N ^{(3)}R \gamma{}^{\frac{1}{2}} \kappa{}^{-1} - S N{}^{3} \pi_\lambda{}^{2} \gamma{}^{\frac{1}{2}} - \nonumber \\
&&~~~~\frac{3}{4}\, N \kappa \pi_\lambda{}^{2} \gamma{}^{-\frac{1}{2}} \lambda{}^{2} - {N}^{i} \pi_\lambda \lambda {\partial}_{i}{N}\,  N{}^{-1} +  {N}^{i} {\partial}_{i}{{\gamma}_{l m}}\,  {\pi}^{l m} -   N^{i} {\partial}_{l}{{\gamma}_{i m}}\,  {\pi}^{l m} + \nonumber \\
&&~~~~ {N}^{i} {\partial}_{m}{{\gamma}_{i l}}\,  {\pi}^{l m} - {\gamma}^{i j} {\gamma}^{k l} \lambda {\partial}_{i}{{\gamma}_{j k}}\,  {\partial}_{l}{\varphi}\,  + \frac{1}{2}\, {\gamma}^{i j} {\gamma}^{k l} \lambda {\partial}_{i}{{\gamma}_{k l}}\,  {\partial}_{j}{\varphi}\,  + N S {\gamma}^{i j} {\partial}_{i}{\varphi}\,  {\partial}_{j}{\varphi}\,  \gamma{}^{\frac{1}{2}} - \nonumber \\
&&~~~~    N \pi_\lambda {\gamma}_{i j} \kappa \lambda {\pi}^{i j} \gamma{}^{-\frac{1}{2}}+ 2\, N {\gamma}_{i j} {\gamma}_{k l} \kappa {\pi}^{i k} {\pi}^{j l} \gamma{}^{-\frac{1}{2}} - N {\gamma}^{i j} {\gamma}^{k l} \kappa {\pi}^{i j} {\pi}^{k l} \gamma{}^{-\frac{1}{2}}.
\end{eqnarray}}

Since $\pi_N$ does not appear in the Hamiltonian, complete dynamics of
the fields are obtained from the Dirac Hamiltonian
\begin{equation}
\label{DHamiltonian}
\mathcal{H}_D = \mathcal{H}_G + \xi \left( \pi_N + \frac{\lambda}{N} \pi_\lambda \right).
\end{equation}

The above Dirac-Hamiltonian can be used to obtain the perturbed
equations of motion.

\subsubsection{Background equations} \label{GalBack}
At zeroth order, field variables are defined as
\begin{eqnarray}
&&N = N_0, ~~~~ N^{i} = 0,~~~~ \gamma_{i j} = a^2 \delta_{i j}, ~~~~S= S_0,~~~~\lambda=\lambda_0,~~~~\varphi = \varphi_0 \nonumber\\
&&\pi^{i j} = {}\pi_0{}^{i j}, ~~~~ \pi_N = \pi_{N0},~~~~\pi_\varphi= {}\pi_{\varphi0},~~~~\pi_\lambda = {}\pi_{\lambda0}.
\end{eqnarray}
Using the above relations, the Dirac-Hamiltonian (\ref{DHamiltonian})
takes the form

\begin{eqnarray}
\label{BackHD}
&&\mathcal{H}_{D0} = - S_0 \lambda_0 + N_0 V_0 a{}^{3} - {}\pi_{\lambda0} {}\pi_{\varphi0} N_0{}^{2} - S_0 N_0{}^{3} {}\pi_{\lambda0}{}^{2} a{}^{3} - \frac{3}{4}\, N_0 \kappa {}\pi_{\lambda0}{}^{2} \lambda_0{}^{2} a{}^{(-3)} - \nonumber\\
&&~~~~~~~~\frac{1}{2}\, N_0 \pi_a {}\pi_{\lambda0} \kappa \lambda_0 a{}^{(-2)} - \frac{1}{12}\, N_0 \kappa \pi_a{}^{2} a{}^{-1} + \xi_0 \left( \pi_{N0}  + {}\pi_{\lambda0} \lambda_0 N_0{}^{-1} \right).
\end{eqnarray}

Since the momentum corresponding to $S$ does not appear in the
Hamiltonian (\ref{BackHD}), variation of the Hamiltonian with
respect to $S$ leads to the first secondary constraint and it is given
by
\begin{equation}
\label{lamb0}
\lambda_0 = -N_0{}^{3} {}\pi_{\lambda0}{}^{2} a{}^{3}
\end{equation}
and varying the Dirac-Hamiltonian with respect to $\xi_0$ recovers the primary constraint
\begin{equation}
\label{pin0constraint}
\pi_{N0} = - \lambda_0 {}\pi_{\lambda0} N_0^{-1}.
\end{equation}

The Hamilton's equations which relate time variation of the field
variables with momenta, are given by
\begin{eqnarray}
\label{del0phi0}
&& {\varphi_0{}^\prime} = - {}\pi_{\lambda0} N_0^2 \\
\label{del0a0}
&& a^{\prime}\,  =  - \frac{1}{6}\, N_0 \kappa a{}^{-1} \pi_a - \frac{1}{2}\, N_0 {}\pi_{\lambda0} \kappa \lambda_0 a{}^{(-2)} \\
\label{del0n0}
&& {N_0{}^\prime} = \xi_0 \\
\label{del0lam0}
&& {\lambda_0{}^\prime} = - {}\pi_{\varphi0} N_0{}^{2} - 2\, S_0 N_0{}^{3} {}\pi_{\lambda0} a{}^{3} - \frac{3}{2}\, N_0 \kappa {}\pi_{\lambda0} \lambda_0{}^{2} a{}^{(-3)} - \nonumber \\
&&~~~~~~~~\frac{1}{2}\, N_0 \pi_a \kappa \lambda_0 a{}^{(-2)} + \lambda_0 N_0{}^{-1} \xi_0 ~~~~~~~~~~
\end{eqnarray}
which relate the momenta and time derivatives of the fields. The above
relations are invertible and all momenta can be written in terms of
the time derivatives of the field variables. Hamilton's equations
corresponding to the time variation of momenta are given by:
\begin{eqnarray}
\label{pia}
&& {\pi_a{}^\prime} = - 3\, N_0 V_0 a{}^{2} + 3\, S_0 N_0{}^{3} {}\pi_{\lambda0}{}^{2} a{}^{2} - \frac{9}{4}\, N_0 \kappa {}\pi_{\lambda0}{}^{2} \lambda_0{}^{2} a{}^{(-4)} - N_0 \pi_a {}\pi_{\lambda0} \kappa \lambda_0 a{}^{(-3)}\nonumber \\
&&~~~~~~~~ - \frac{1}{12}\, N_0 \kappa \pi_a{}^{2} a{}^{(-2)} \\
\label{pil}
&& {{}\pi_{\lambda0}{}^\prime} =   S_0 - 3\, {}\pi_{\lambda0} a^{\prime}\,  a{}^{-1} - {}\pi_{\lambda0} {\partial}_{0}{N_0}\,  N_0{}^{-1} \\
\label{pin}
&& {\pi_{N0}{}^\prime} = - V_0 a{}^{3} + 2\, {}\pi_{\lambda0} {}\pi_{\varphi0} N_0 + 3\, S_0 N_0{}^{2} {}\pi_{\lambda0}{}^{2} a{}^{3} + \frac{3}{4}\, \kappa {}\pi_{\lambda0}{}^{2} \lambda_0{}^{2} a{}^{(-3)} \nonumber \\
&&~~~~~~~~+ \frac{1}{2}\, \pi_a {}\pi_{\lambda0} \kappa \lambda_0 a{}^{(-2)} + \frac{1}{12}\, \kappa \pi_a{}^{2} a{}^{-1} + {}\pi_{\lambda0} \lambda_0 N_0{}^{(-2)} \xi_0 \\
\label{pip}
&& {{}\pi_{\varphi0}{}^\prime} = - N_0 V_{\varphi} a^3.
\end{eqnarray}
which, by using other background Hamilton's equations, lead to the
identical dynamical equations of motion of the field variables
obtained from action formulation.  Using background Hamilton's
equations in conformal time coordinate, equation (\ref{pil}) becomes
\begin{equation}
\label{s0}
S_0 = - 2H\, \varphi_0^{\prime}\,  a{}^{(-2)} - \varphi_0^{\prime \prime}\,  a{}^{(-2)}.
\end{equation}

Variation of action (\ref{full action galilean}) or (\ref{full
  action}) with respect to $\lambda$ lead to $S = \Box \varphi$. At
zeroth order, $\Box \varphi = - 2H\, \varphi_0^{\prime}\, 
a{}^{(-2)} - \varphi_0^{\prime \prime}\, a{}^{(-2)}$ implying that the
dynamical equation (\ref{s0}) obtained using Hamiltonian formulation
is consistent. Similarly, equation (\ref{pin}) leads to the zeroth
order Hamiltonian constraint of the Galilean scalar field model. In
conformal time, it is given by
\begin{equation}
\label{hc0g}
V_0 a{}^{2} - 6\, H\,  a{}^{(-2)} \varphi_0^{\prime}\, {}^{3} - 3\,  \kappa{}^{-1} H^2 = 0
\end{equation}

Equations (\ref{pia}) and (\ref{pip}) lead to the equation of motion
of $a$ and $\varphi_0$ respectively and are given by
\begin{eqnarray}
\label{eoma0g}
&&6\, H\,  \varphi_0^{\prime}\, {}^{3} a{}^{(-2)} - 6\, \varphi_0^{\prime \prime}\,  \varphi_0^{\prime}\, {}^{2} a{}^{-2} + 3\, \kappa{}^{-1} H^2 - 6\, a^{\prime \prime}\, a{}^{-1} \kappa{}^{-1} + 3\, V_0 a{}^{2}= 0 ~~~~~~~~\\
\label{eomphig}
&& \frac{1}{2}\,\varphi_0^{\prime}{}^2\,a^{\prime \prime}\,a^{-1}H{}^{-1} -  \frac{H}{2}\,\varphi_0^{\prime}{}^2 + \,\varphi_0^{\prime \prime}\varphi_0^\prime\,  - \frac{1}{12}\, V_{\varphi}\,   H{}^{-1} a{}^{4} = 0.
\end{eqnarray}

Equations (\ref{hc0g}), (\ref{eoma0g}) and (\ref{eomphig}) are
identical to the Lagrangian equations (\ref{hc0g1}), (\ref{eomag1}) and (\ref{eomphig1}),
respectively. Hence, at zeroth order, Hamiltonian formulation is
consistent with Lagrangian formulation.

\paragraph{\underline{Counting scalar degrees of freedom at zeroth order:}}
As one can see in (\ref{BackHD}), background phase space contains 10
variable ($a,\, N, \, \varphi_0, \, \lambda_0, \,
S_0$ and corresponding momenta). There are two primary
constrained equations:
\begin{eqnarray}
\label{pc01}
&& \Phi^1_p \equiv \pi_{S0} = 0 \\
\label{pc02}
&& \Phi_p^2 \equiv \pi_{N0} + \lambda_0 {}\pi_{\lambda0} N_0^{-1} = 0
\end{eqnarray}

Conservation of primary constraints gives rise to secondary constraints:
\begin{eqnarray}
\label{sc01}
\Phi_s^1 \equiv \{\Phi_p^1, \mathcal{H}_{D0}\} \approx 0 && 
\Rightarrow \lambda_0 + N_0{}^{3} {}\pi_{\lambda0}{}^{2} a{}^{3} \approx 0 \\
\label{sc02}
\Phi_s^2 \equiv \{\Phi_p^2, \mathcal{H}_{D0}\} \approx 0 
&& \Rightarrow - V_0 a^3 + {}\pi_{\lambda0} {}\pi_{\varphi0} N_0 + \frac{3}{4 a^3} \kappa {}\pi_{\lambda0}{}^2 \lambda_0{}^2  + 
\frac{1}{12 \, a} \kappa \pi_a^2 \nonumber \\
&&~~~~+ \frac{\kappa}{2 \, a^2}  N_0 \pi_a {}\pi_{\lambda0} \lambda_0 = 0 
\end{eqnarray}

Equation (\ref{sc02}) leads to zeroth order Hamiltonian constraint and
is equivalent to equation (\ref{hc0l}). Further, conservation of
secondary constraint (\ref{sc01}) leads to tertiary constraint
\begin{eqnarray}
\label{tc0}
\Phi_t \equiv \{\Phi_s^1, \mathcal{H}_{D0}\} \approx 0 
&& \Rightarrow {}\pi_{\varphi0} - \kappa N_0^2 {}\pi_{\lambda0}{}^2 a - 3 \kappa N_0^2 {}\pi_{\lambda0}{}^3 \lambda_0 = 0
\end{eqnarray}
and generates quaternary constraint
\begin{eqnarray}
&&~~~~\Phi_q \equiv \{\Phi_t, \mathcal{H}_{D0}\} \approx 0 \\
&&\Rightarrow S_0 \left(- 2 \kappa N_0^2 {}\pi_{\lambda0} a + 15 \kappa N_0^5{}\pi_{\lambda0}{}^5 \pi_a a \right) - N_0 V_{\varphi} - \frac{5}{2} \kappa^2 N_0^3 {}\pi_{\lambda0}{}^3 \lambda_0 a^{-2}   \nonumber \\
&&~~~~ + \frac{5}{6} \kappa^2 N_0^3 {}\pi_{\lambda0}{}^2 \pi_a  a^{-1}+ 18 \kappa^2 N_0^6 {}\pi_{\lambda0}{}^6 \lambda_0  + 6 \kappa^2 N_0^6 {}\pi_{\lambda0}{}^6 \pi_a a = 0
\end{eqnarray}

Out of the 6 constrained equations, equations (\ref{pc02}) and
(\ref{sc02}) are first class and rest are second class
constraints. Hence, in coordinate space, number of degrees of freedom
is \[\frac{1}{2} \times (10 - 2 \times 2 - 4) = 1\] which is same as
canonical scalar field model.

\subsubsection{First order perturbed equations}
\label{galfirst}
The first order perturbation of the field variables and the momenta
are defined as
\begin{eqnarray}
\label{fog}
&&N = N_0 + \epsilon N_1, ~~~~ N^{i} = \epsilon N_1^{i},~~~~ \gamma_{i j} = a^2 \delta_{i j}, ~~~~S= S_0 + \epsilon S_1,\nonumber\\
&&\lambda=\lambda_0 + \epsilon \lambda_1, \varphi = \varphi_0 + \epsilon \varphi_1 ~~~~\pi^{i j} = {}\pi_0{}^{i j} + \epsilon \pi_1^{i j}, ~~~~ \pi_N = \pi_{N0} + \epsilon \pi_{N1},\nonumber\\
&&\pi_\varphi= {}\pi_{\varphi0} + \epsilon \pi_{\varphi1},~~~~\pi_\lambda = {}\pi_{\lambda0} + \epsilon \pi_{\lambda 1}.
\end{eqnarray}

First order perturbed Hamiltonian equations are obtained by varying
the second order perturbed Hamiltonian. Using above definitions of
perturbations in the Hamiltonian (\ref{DHamiltonian}), the second
order perturbed Hamiltonian becomes 
\begin{eqnarray}
\label{HG2}
&&\mathcal{H}_{D2} = - S_1 \lambda_1 + N_1 V_\varphi  a{}^{3} \varphi_1 + \frac{1}{2}\, N_0 V_{\varphi \varphi}  \varphi_1{}^{2} a{}^{3} + {N_1}^{i} {}\pi_{\lambda0} {\partial}_{i}{\lambda_1}\,  + {N_1}^{i} {}\pi_{\varphi0} {\partial}_{i}{\varphi_1}\,   -  \nonumber \\
&&2\, N_0 N_1 {}\pi_{\lambda0} \pi_{\varphi1} - 2\, N_0 N_1 \pi_{\lambda 1} {}\pi_{\varphi0} - \pi_{\lambda 1} \pi_{\varphi1} N_0{}^{2} + \pi_{\lambda 1} \lambda_0 {\partial}_{i}{{N_1}^{i}}\,  + {}\pi_{\lambda0} \lambda_1 {\partial}_{i}{{N_1}^{i}}\,   - \nonumber \\
&& {\delta}^{i j} {\partial}_{i}{\lambda_1}\,  {\partial}_{j}{\varphi_1}\,  a{}^{(-2)} - S_0 N_0{}^{3} \pi_{\lambda 1}{}^{2} a{}^{3} - 6\, N_1 {}\pi_{\lambda0} \pi_{\lambda 1} S_0 N_0{}^{2} a{}^{3} - 3\, N_0 S_0 N_1{}^{2} {}\pi_{\lambda0}{}^{2} a{}^{3} -\nonumber \\
&&2\, {}\pi_{\lambda0} \pi_{\lambda 1} S_1 N_0{}^{3} a{}^{3} - 3\, N_1 S_1 N_0{}^{2} {}\pi_{\lambda0}{}^{2} a{}^{3}%
 - \frac{3}{4}\, N_0 \kappa {}\pi_{\lambda0}{}^{2} \lambda_1{}^{2} a{}^{(-3)}  - \nonumber \\
&& \frac{3}{4}\, N_0 \kappa \pi_{\lambda 1}{}^{2} \lambda_0{}^{2} a{}^{(-3)} -\frac{3}{2}\, N_1 \kappa \lambda_0 \lambda_1 {}\pi_{\lambda0}{}^{2} a{}^{(-3)} - \frac{3}{2}\, N_1 {}\pi_{\lambda0} \pi_{\lambda 1} \kappa \lambda_0{}^{2} a{}^{(-3)}  + \nonumber \\
&&{\delta}^{i j} \lambda_0 {\partial}_{i}{N_1}\,  {\partial}_{j}{\varphi_1}\,  N_0{}^{-1} a{}^{(-2)} + N_0 S_0 {\delta}^{i j} {\partial}_{i}{\varphi_1}\,  {\partial}_{j}{\varphi_1}\,  a - N_0 \pi_{\lambda 1} {\delta}_{i j} \kappa \lambda_0 {\pi_1}^{i j} a{}^{-1} - \nonumber \\
&&N_1 {}\pi_{\lambda0} {\delta}_{i j} \kappa \lambda_0 {\pi_1}^{i j} a{}^{-1} - N_1 \pi_{\lambda 1} {\delta}_{i j} \kappa \lambda_0 \pi_0{}^{i j} a{}^{-1} - N_0 {}\pi_{\lambda0} {\delta}_{i j} \kappa \lambda_1 {\pi_1}^{i j} a{}^{-1} - \nonumber \\
&&N_1 {}\pi_{\lambda0} {\delta}_{i j} \kappa \lambda_1 \pi_0{}^{i j} a{}^{-1} - N_0 {\delta}_{i j} {\delta}_{k l} \kappa {\pi_1}^{i j} {\pi_1}^{k l} a - 2\, N_1 {\delta}_{i j} {\delta}_{k l} \kappa \pi_0{}^{i j} {\pi_1}^{k l} a  + \nonumber \\
&&4\, N_1 {\delta}_{i j} {\delta}_{k l} \kappa {{}\pi_0{}}^{i k} {\pi_1}^{j l} a %
 + \pi_{N1} \xi_1 - N_1 {}\pi_{\lambda0} \lambda_0 N_0{}^{(-2)} \xi_1 + {}\pi_{\lambda0} \lambda_0 N_0{}^{(-3)} N_1{}^{2} \xi_0  -\nonumber \\
&& N_1 \pi_{\lambda 1} \lambda_0 N_0{}^{(-2)} \xi_0 + {}\pi_{\lambda0} \lambda_1 N_0{}^{-1} \xi_1 -N_1 {}\pi_{\lambda0} \lambda_1 N_0{}^{(-2)} \xi_0 +\pi_{\lambda 1} \lambda_1 N_0{}^{-1} \xi_0\nonumber \\
&&- {}\pi_{\lambda0} {}\pi_{\varphi0} N_1{}^{2}+ 2 {\delta}_{i j} {\partial}_{k}{{N_1}^{i}}\,  {\pi_1}^{j k} a{}^{2}- 3\, N_0 {}\pi_{\lambda0} \pi_{\lambda 1} \kappa \lambda_0 \lambda_1 a{}^{(-3)}+ \pi_{\lambda 1} \lambda_0 N_0{}^{-1} \xi_1\nonumber \\
&&+ 2\, N_0 {\delta}_{i j} {\delta}_{k l} \kappa {\pi_1}^{i k} {\pi_1}^{j l} a- N_0 \pi_{\lambda 1} {\delta}_{i j} \kappa \lambda_1 \pi_0{}^{i j} a{}^{-1} - {N_1}^{i} {}\pi_{\lambda0} \lambda_0 {\partial}_{i}{N_1}\,  N_0{}^{-1}.
\end{eqnarray}

Equations of motion corresponding to the perturbed Hamiltonian (\ref{HG2}) is expressed in Appendix (\ref{firstGaleq}).

Variation of the perturbed Hamiltonian density (\ref{HG2}) with
respect to $S_1$ lead to the secondary constraint
\begin{equation}
\label{lam1}
\lambda_1 =  - 2\, \pi_{\lambda 1} {}\pi_{\lambda0} N_0{}^{3} a{}^{3} - 3\, N_1 N_0{}^{2} {}\pi_{\lambda0}{}^{2} a{}^{3}.
\end{equation}

Variation of (\ref{HG2}) with respect to $\xi_1$ leads to the equation
\begin{equation}
\label{fcons1}
\pi_{N1} =  - \lambda_0 \pi_{\lambda 1} N_0{}^{-1} - \lambda_1 {}\pi_{\lambda0} N_0{}^{-1} + \lambda_0 {}\pi_{\lambda0} N_0{}^{(-2)} N_1
\end{equation}
which constrains $\pi_{N1}$ with $\pi_{\lambda 1}$. Since there is no
perturbation in the 3-metric, variation with respect to $\pi_1^{i j}$,
as expected, is equal to zero and contracting with $(\delta^{m n} \delta^{i j} -
\delta^{m i} \delta^{n j})$, we get
{\small
\begin{eqnarray}
\label{fcons2}
&& {\pi_1}^{i j} = \frac{1}{2}\, N_0{}^{-1} a \kappa{}^{-1} {\delta}^{i j} {\partial}_{k}{{N_1}^{k}}\,  - \frac{1}{2}\, a{}^{(-2)} {\delta}^{i j} \pi_{\lambda 1} \lambda_0 - \frac{1}{2}\, a{}^{(-2)} N_1 N_0{}^{-1} {\delta}^{i j} {}\pi_{\lambda0} \lambda_0 \nonumber \\
&&- \frac{1}{2}\, a{}^{(-2)} {\delta}^{i j} {}\pi_{\lambda0} \lambda_1 - \frac{1}{4}\, N_0{}^{-1} \kappa{}^{-1} a {\delta}^{k j} {\partial}_{k}{{N_1}^{i}}\,  - \frac{1}{4}\, N_0{}^{-1} \kappa{}^{-1} a {\delta}^{k i} {\partial}_{k}{{N_1}^{j}}\,  - N_0{}^{-1} N_1 \pi_0{}^{i j}~~~~~~~
\end{eqnarray}}

At first order perturbation, Hamilton's equations corresponding to the
time variation of field variables are given by {\small
\begin{eqnarray}
\label{del0lam1}
&& {\lambda_1{}^\prime}\,  = - 2\, N_0 N_1 {}\pi_{\varphi0} - \pi_{\varphi1} N_0{}^{2} + \lambda_0 {\partial}_{i}{{N_1}^{i}}\,  - 2\, \pi_{\lambda 1} S_0 N_0{}^{3} a{}^{3} - 6\, N_1 {}\pi_{\lambda0} S_0 N_0{}^{2} a{}^{3} - \nonumber \\
&&~~~~~~~~ 2\, {}\pi_{\lambda0} S_1 N_0{}^{3} a{}^{3} - 3\, N_0 {}\pi_{\lambda0} \kappa \lambda_0 \lambda_1 a{}^{(-3)} - \frac{3}{2}\, N_0 \pi_{\lambda 1} \kappa \lambda_0{}^{2} a{}^{(-3)} - \frac{3}{2}\, N_1 {}\pi_{\lambda0} \kappa \lambda_0{}^{2} a{}^{(-3)} \nonumber \\
&&~~~~~~~~ - N_0 {\delta}_{i j} \kappa \lambda_0 {\pi_1}^{i j} a{}^{-1} - N_1 {\delta}_{i j} \kappa \lambda_0 \pi_0{}^{i j} a{}^{-1} - N_0 {\delta}_{i j} \kappa \lambda_1 \pi_0{}^{i j} a{}^{-1} + \nonumber \\
&&~~~~~~~~\lambda_0 N_0{}^{-1} \xi_1 - N_1 \lambda_0 N_0{}^{(-2)} \xi_0 + \lambda_1 N_0{}^{-1} \xi_0 \\
&& \varphi_1^{\prime}\,  = - 2\, N_0 N_1 {}\pi_{\lambda0} - \pi_{\lambda 1} N_0{}^{2} \\
&& {N_1^\prime} = \xi_1
\end{eqnarray}}

Hamilton's equation corresponding to the time variation of the momentum
$\pi_{\lambda 1}$ is given by {\small
\begin{eqnarray}
&& {\pi_{\lambda 1}{}^\prime} + \frac{\delta  \mathcal{H}_{D2}}{\delta \lambda_1} = 0 \\
&& \Rightarrow {\pi_{\lambda 1}{}^\prime}\,  - S_1 + {\delta}^{i j} {\partial}_{i j}{\varphi_1}\,  a{}^{(-2)} - \frac{3}{2}\, N_0 \kappa {}\pi_{\lambda0}{}^{2} \lambda_1 a{}^{(-3)} - 3\, N_0 {}\pi_{\lambda0} \pi_{\lambda 1} \kappa \lambda_0 a{}^{(-3)} -  \nonumber \\
&&~~~~~~~~\, N_0 {}\pi_{\lambda0} {\delta}_{i j} \kappa {\pi_1}^{i j} a{}^{-1}- \frac{3}{2}\, N_1 \kappa \lambda_0 {}\pi_{\lambda0}{}^{2} a{}^{(-3)} -  N_0 \pi_{\lambda 1} {\delta}_{i j} \kappa \pi_0{}^{i j} a{}^{-1}   \nonumber \\
&&~~~~~~~~-  N_1 {}\pi_{\lambda0} {\delta}_{i j} \kappa \pi_0{}^{i j} a{}^{-1}  +{}\pi_{\lambda0} N_0{}^{-1} \xi_1 - N_1 {}\pi_{\lambda0} N_0{}^{(-2)} \xi_0 + \pi_{\lambda 1} N_0{}^{-1} \xi_0 = 0~~~~~~~~~~
\end{eqnarray}}
which, using other first order equations and (\ref{ncomp}), becomes
{\small
\begin{eqnarray}
&&- S_1 - \varphi_1^{\prime \prime}\,  a{}^{(-2)} + {\delta}^{i j} {\partial}_{i j}{\varphi_1}\,  a{}^{(-2)} + {\phi_1^\prime}\,  \varphi_0^{\prime}\,  a{}^{(-2)} - 2\, \varphi_1^{\prime}\,  a^{\prime}\,  a{}^{(-3)} + \nonumber \\
&&~~~~~~~~2\, \varphi_0^{\prime \prime}\,  \phi_1 a{}^{(-2)} + {\delta}^{i j} \varphi_0^{\prime}\,  {\partial}_{i j}{B_1}\,  a{}^{(-2)} + 4\, \varphi_0^{\prime}\,  a^{\prime}\,  \phi_1 a{}^{(-3)} = 0 \nonumber \\
&&\Rightarrow S_1 = - \varphi_1^{\prime \prime}\,  a{}^{(-2)} + {\delta}^{i j} {\partial}_{i j}{\varphi_1}\,  a{}^{(-2)} + {\phi_1^\prime}\,  \varphi_0^{\prime}\,  a{}^{(-2)} - 2\, \varphi_1^{\prime}\,  H\,  a{}^{(-2)}\nonumber \\
&&~~~~~~~~ + 2\, \varphi_0^{\prime \prime}\,  \phi_1 a{}^{(-2)} + {\delta}^{i j} {\partial}_{i j}{B_1}\,  \varphi_0^{\prime}\,  a{}^{(-2)} + 4\, \varphi_0^{\prime}\,  H\,  \phi_1 a{}^{(-2)}.
\end{eqnarray}}

Right hand side of the above equation is the explicit form of first
order perturbed $\Box \varphi$. Hence, the first equation obtained
from perturbed Hamiltonian is consistent.

First order perturbed Hamiltonian constraint is obtained by the time
variation of $\pi_{N1}$ and is given by {\small
\begin{eqnarray}
&&~~~~{\pi_{N1}{}^\prime} + \frac{\delta  \mathcal{H}_{D2}}{\delta N_1} = 0 \\
&& \Rightarrow {\pi_{N1}{}^\prime}\,  + V_\varphi  a{}^{3} \varphi_1 - 2\, N_1 {}\pi_{\lambda0} {}\pi_{\varphi0} - 2\, N_0 {}\pi_{\lambda0} \pi_{\varphi1} - 2\, N_0 \pi_{\lambda 1} {}\pi_{\varphi0} - 6\, {}\pi_{\lambda0} \pi_{\lambda 1} S_0 N_0{}^{2} a{}^{3} -   \nonumber \\
&&~~~~6\, N_0 N_1 S_0 {}\pi_{\lambda0}{}^{2} a{}^{3} -3\, S_1 N_0{}^{2} {}\pi_{\lambda0}{}^{2} a{}^{3} -\frac{3}{2}\, \kappa \lambda_0 \lambda_1 {}\pi_{\lambda0}{}^{2} a{}^{(-3)} - \frac{3}{2}\, {}\pi_{\lambda0} \pi_{\lambda 1} \kappa \lambda_0{}^{2} a{}^{(-3)}  \nonumber \\
&&~~~~+ {}\pi_{\lambda0} \lambda_0 {\partial}_{i}{{N_1}^{i}}\,  N_0{}^{-1} - {\delta}^{i j} \lambda_0 {\partial}_{i j}{\varphi_1}\,  N_0{}^{-1} a{}^{(-2)} - {}\pi_{\lambda0} {\delta}_{i j} \kappa \lambda_0 {\pi_1}^{i j} a{}^{-1} -  \nonumber \\
&&~~~~\pi_{\lambda 1} {\delta}_{i j} \kappa \lambda_0 \pi_0{}^{i j} a{}^{-1} - {}\pi_{\lambda0} {\delta}_{i j} \kappa \lambda_1 \pi_0{}^{i j} a{}^{-1} -2\, {\delta}_{i j} {\delta}_{k l} \kappa \pi_0{}^{i j} {\pi_1}^{k l} a + 4\, {\delta}_{i j} {\delta}_{k l} \kappa {{}\pi_0{}}^{i k} {\pi_1}^{j l} a  \nonumber \\
&&~~~~ - {}\pi_{\lambda0} \lambda_0 N_0{}^{(-2)} \xi_1 + 2\, N_1 {}\pi_{\lambda0} \lambda_0 N_0{}^{(-3)} \xi_0%
 -\pi_{\lambda 1} \lambda_0 N_0{}^{(-2)} \xi_0 -{}\pi_{\lambda0} \lambda_1 N_0{}^{(-2)} \xi_0 = 0~~~~~~~~~~
\end{eqnarray}}

In flat-slicing gauge, it becomes
\begin{eqnarray}
\label{hc1gh}
&&\mathcal{H}_{N1} \equiv 24\, H\,  \phi_1 \varphi_0^{\prime}\, {}^{3} a{}^{(-2)} - 18\, \varphi_1^{\prime}\,  H\,  \varphi_0^{\prime}\, {}^{2} a{}^{(-2)} + V_{\varphi}\,  a{}^{2} \varphi_1 + 2\, {\delta}^{i j} {\partial}_{i j}{B_1}\,  \varphi_0^{\prime}\, {}^{3} a{}^{-2} + \nonumber \\
&&2\, {\delta}^{i j} {\partial}_{i j}{\varphi_1}\,  \varphi_0^{\prime}\, {}^{2} a{}^{-2} + 2\, {\delta}^{i j} H\,  {\partial}_{i j}{B_1}\,  \kappa{}^{-1} + 6\, \phi_1 \kappa{}^{-1} H^2 = 0.
\end{eqnarray}

Since there is no momentum $\pi^1_i$ corresponding to $N_1^{i}$
appeared in the second order Hamiltonian (\ref{HG2}), variation of the
Hamiltonian with respect to $N_1^{i}$ leads to Momentum constraint of
Galilean field
\begin{equation}
\label{mcg1}
M_{i1} \equiv {}\pi_{\varphi0} {\partial}_{i}{\varphi_1}\,  - \lambda_0 {\partial}_{i}{\pi_{\lambda 1}}\,  - 2\, {\delta}_{i j} a{}^{2} {\partial}_{k}{{\pi_1}^{j k}}\,  - {}\pi_{\lambda0} \lambda_0 {\partial}_{i}{N_1}\,  N_0{}^{-1} = 0,
\end{equation}
which, in flat-slicing gauge, becomes
\begin{equation}
\label{mc1gh}
M_{i1} \equiv - 6\, H\,  {\partial}_{i}{\varphi_1}\,  \varphi_0^{\prime}\, {}^{2}  - 2\, {\partial}_{i}{\phi_1}\,  \varphi_0^{\prime}\, {}^{3} + 2\, {\partial}_{i}{\varphi_1{}^\prime}\,  \varphi_0^{\prime}\, {}^{2} - 2\, H\,  {\partial}_{i}{\phi_1}\,  \kappa{}^{-1} a^2 = 0.
\end{equation}

Similarly, equation of motion of $\varphi_1$ can be obtained from
\begin{eqnarray}
&&~~~~{\pi_{\varphi1}{}^\prime} + \frac{\delta \mathcal{H}_{D2}}{\delta \varphi_1} = 0 \\
&& \Rightarrow {\pi_{\varphi1}{}^\prime}\,  + N_1 V_{\varphi}\,  a{}^{3} + N_0 V_{\varphi \varphi}\,  \varphi_1 a{}^{3} - {\partial}_{i}{{N_1}^{i}}\,  {}\pi_{\varphi0} + {\delta}^{i j} {\partial}_{i j}{\lambda_1}\,  a{}^{(-2)} \nonumber \\
&& ~~~~~~~~- {\delta}^{i j} \lambda_0 {\partial}_{i j}{N_1}\,  N_0{}^{-1} a{}^{(-2)} -2\, N_0 S_0 {\delta}^{i j} {\partial}_{i j}{\varphi_1}\,  a  = 0.
\end{eqnarray}

Using (\ref{ncomp}), the above equation can be written as
\begin{eqnarray}
\label{eomphi1gh}
&& - 18\, \phi_1 \varphi_0^{\prime}\, {}^{2} H^2 + 12\, \varphi_0^{\prime}\,  \varphi_1^{\prime}\,  H^2 + 18\, \phi_1^\prime\,  H\,  \varphi_0^{\prime}\, {}^{2}  \nonumber \\
&&+ 36\, \varphi_0^{\prime}\,  H\,  \varphi_0^{\prime \prime}\,  \phi_1  - 12\, \varphi_0^{\prime}\,  H\,  \varphi_1^{\prime \prime} - 12\, \varphi_1^{\prime}\,  H\,  \varphi_0^{\prime \prime} \nonumber \\
&& - 12\, \varphi_0^{\prime}\,  \varphi_1^{\prime}\,  a^{\prime \prime}\,  a{}^{-1} + 6\, {\delta}^{i j} H\,  {\partial}_{i j}{B_1}\,  \varphi_0^{\prime}\, {}^{2}  + 4\, {\delta}^{i j} \varphi_0^{\prime}\,  \varphi_0^{\prime \prime}\,  {\partial}_{i j}{B_1}\,  + \nonumber \\
&&4\, {\delta}^{i j} \varphi_0^{\prime}\,  H\,  {\partial}_{i j}{\varphi_1}\,  a{}^{-1} + 4\, {\delta}^{i j} \varphi_0^{\prime \prime}\,  {\partial}_{i j}{\varphi_1}\,  + 2\, {\delta}^{i j} {\partial}_{i j}{B_1^\prime}\,  \varphi_0^{\prime}\, {}^{2} + V_{\varphi}\,  \phi_1 a{}^{4} + \nonumber \\
&&V_{\varphi \varphi}\,  a{}^{4} \varphi_1 + 2\, {\delta}^{i j} {\partial}_{i j}{\phi_1}\,  \varphi_0^{\prime}\, {}^{2}+ 18\, a^{\prime \prime}\,  \phi_1 \varphi_0^{\prime}\, {}^{2} a{}^{-1} = 0.
\end{eqnarray}

Equations (\ref{hc1gh}), (\ref{mc1gh}) and (\ref{eomphi1gh}) obtained
from Hamiltonian formulation are identical to (\ref{hc1ga}),
(\ref{mc1ga}) and (\ref{eomphi1ga}), respectively. 

Third and fourth order interaction Hamiltonian are given in Appendix \ref{intHamGal}.

\paragraph{\underline{Counting scalar degrees of freedom at first order:}}
Since we are considering only scalar first order perturbations,
$N_1^i$ contains one scalar variable ($N_1^i = \delta^{i
  j} \partial_{j}{B_1}$) and hence, $\pi_i = 0$ and Momentum
constraint (\ref{mcg1}) lead to two constrained equations. Along with
(\ref{fcons1}) and Hamiltonian constraint (\ref{hc1gh}), we get 4
constrained equations. Similarly, as we have seen in zeroth order, at
first order, we get four second class constraints:
\begin{eqnarray}
&& \Phi_p \equiv \pi_S = 0 \\
&& \Phi_s \equiv \{\pi_S, \mathcal{H}_{D2}\} = \lambda_1 + 2\, \pi_{\lambda 1} {}\pi_{\lambda0} N_0{}^{3} a{}^{3} + 3\, N_1 N_0{}^{2} {}\pi_{\lambda0}{}^{2} a{}^{3} = 0\\
\label{fcons3}
&&\Phi_t \equiv \{\Phi_s, \mathcal{H}_{D2}\} = \pi_{\varphi1} + 2\, N_1 {}\pi_{\varphi0} N_0{}^{-1} - 2\, N_0 {\partial}_{i}{{N_1}^{i}}\,  {}\pi_{\lambda0}{}^{2} a{}^{3} + 2\, N_0 {}\pi_{\lambda0} {\delta}^{i j} {\partial}_{i j}{\varphi_1}\,  a + \nonumber \\
&&~~~~~~~~9\, N_1 \kappa N_0{}^{4} {}\pi_{\lambda0}{}^{5} a{}^{3} +  6\, \pi_{\lambda 1} \kappa N_0{}^{5} {}\pi_{\lambda0}{}^{4} a{}^{3} -3\, N_0 N_1 \pi_a \kappa {}\pi_{\lambda0}{}^{2} a \nonumber \\
&&~~~~~~~~- 2\, \pi_a {}\pi_{\lambda0} \pi_{\lambda 1} \kappa N_0{}^{2} a = 0~~~\\
&& \Phi_q \equiv \{\Phi_t, \mathcal{H}_{D2}\} \approx 0
\end{eqnarray}

We also get 12 more second class constraints: $\delta\gamma_{i j} = 0$
and $\{\delta\gamma_{i j}, ^{(2)}\mathcal{H}_D\}$ which leads to
equation (\ref{fcons2}).  Since, we have fixed the gauge, all
constraints become second class. Our Galilean phase space contains 22
variables. Hence in configuration space, the number of degrees of
freedom is \[\frac{1}{2}\times(22 - 4 - 4- 12)= 1.\]

This procedure can be extended to higher order and at 
any order it can be shown that the degrees of freedom is one. 
So, at any order, Galilean scalar field
produces no extra degrees of freedom due to the higher derivative
terms present in the Lagrangian and behave exactly same as any single
derivative Lagrangian system.

However, if we consider generalized Lagrangian containing second order 
derivative terms, the above analysis can not be extended. In that 
case, unlike Galilean field, the Lapse function and Shift vector 
will not act like constraints and hence, 
$\pi_N + ... \neq 0,~\pi_i + ...\neq 0$ since
$\pi_N$ and $\pi_i$ contain time derivatives of $N$ and
$N^{i}$\cite{Nandi:2015tha}, and dynamical degrees of freedom can be
generated from those. This means that for any higher derivative
gravitational theory, lack of first class constraints lead to extra
degrees of freedom. Similarly, extra degrees of freedom will always be
generated for any generalized second order derivative Lagrangian.

\section{Conclusions and discussions}

In this chapter, using Hamiltonian formulation, we have formulated 
a consistent cosmological perturbation theory at all orders. We
have adopted the following procedures: we choose a particular gauge
that does not lead to any particular gauge artifact \cite{Malik2009}
such that some variables remain unperturbed while others can be
separated as zeroth order part and perturbation part. In order to make
the procedure transparent, we considered a simple model of two
variables where one variable is unperturbed and other variable can be
perturbed. At first order, we confronted the gauge-issue and found
that, even canonical conjugate momentum of unperturbed quantity has
perturbation part that leads a constrained equation at every perturbed
order and by using the equation we can get the exact form of perturbed
momentum. We fixed the gauge-issue and obtained all first order
perturbed equations as well as third and fourth order perturbed
Hamiltonian which is consistent with Lagrangian formulation. The
procedure is simple and robust and can be extended to any order of
perturbation. Table below provides a bird's eye view of the both the formulations and 
advantages of the Hamiltonian formulation that is proposed in this chapter:
\pagebreak

\begin{table}
	\label{Table.Comp}
{\scriptsize\begin{center}
\begin{tabular}{|m{2cm}|m{6cm}|m{6cm}|}
\hline
 & Lagrangian formulation & Hamiltonian formulation \\
\hline
Gauge conditions and {guage-invariant equations} & At any order, choose a gauge {which} does not lead to gauge-artifacts & 
Choose a gauge {with no gauge-artifacts}, however, momentum corresponding  to unperturbed quantity is non-zero leading to consistent equations of motion.\\
\hline
Dynamical variables & {Counting true dynamical degrees of freedom is difficult.} & Using Dirac's procedure, constraints can 
easily be obtained and is easy to determine the degrees of freedom.\\
\hline
Quantization at all orders &  Difficult to quantize constrained system. & Since constraints are obtained systematically and reduced phase space contains only true degrees of freedom, it is straightforward to quantize the theory using Hamiltonian formulation.\\
\hline
Calculating the observables & Requires to invert the expressions at each order and hence non-trivial 
to compute higher-order correlation function. & Once the relation between $\varphi$ and Curvature perturbation\footnotemark[4]  is known, 
calculating the correlation functions from the Hamiltonian is simple and straightforward to obtain.\\
\hline
\end{tabular}
\end{center}}
\caption{Comparison between two approaches}
\end{table}

\footnotetext[4]{It is important to note that, in the case of first order, relation between $\varphi$ and three-curvature is straight forward. However, it is more subtle in the case of higher-order perturbations \cite{Malik2009}.}
We have applied the procedure to canonical scalar field minimally
coupled to gravity and similarly obtained all equations as well as
interaction Hamiltonian using both Lagrangian and Hamiltonian
formulation \cite{Nandi:2015ogk}. Both lead to identical results. We also showed that,
obtaining interaction Hamiltonian by using Hamiltonian formulation is efficient and straightforward. Unlike the Lagrangian 
formulation, we do not need to invert the expressions at each order~\cite{Huang:2006eha}.

We, then obtained a consistent perturbed Hamiltonian formulation for
Galilean scalar fields. Using flat-slicing gauge, we have obtained zeroth and 
first order Hamilton's equations that are consistent with Lagrangian formulation. 
We carefully analyzed the constraints in the system and counted degrees of freedom at 
every order in the system that is consistent with the results of Deffayet \textit{et
  al} \cite{Deffayet:2015qwa} results. It has been shown that general higher derivative
models lead to dynamical equations of Lapse function/shift vector which
increases the number degrees of freedom of the system. But, only in
higher derivative Galilean theory, Lapse function/shift vector remain
constraint, therefore, no extra degrees of freedom flows in the
system. Similar type of problem has been encountered in a different manner in Ref. \cite{Nandi:2015tha}.

To make the Physics transparent, in this chapter, we have  neglected vector and tensor perturbations.  Our 
approach can be applied to higher-order perturbations including vector and tensor perturbations \cite{Nandi:2015ogk}. In the 
presence of tensor or vector or any mixed modes at any higher perturbed order, since modes do not 
decouple, $\pi^{i j}$ cannot decouple and act as the momentum corresponding the overall 3-metric 
$\delta \gamma_{i j}$ which contains mixed modes. Hence, 
\begin{eqnarray}
&& \frac{\delta \mathcal{H}}{\delta \pi^{i j} } = \partial_0 (\delta \gamma_{i j}) \nonumber \\
&& ~~~~~~= \partial_0 (\delta \gamma^S_{i j} + \delta \gamma^V_{i j} + \delta \gamma^T_{i j}), \nonumber 
\end{eqnarray}
where $\gamma^S_{i j}, \gamma^V_{i j}, \gamma^T_{i j}$ are scalar, vector and tensor modes, respectively.

Our approach can be applied to any model of gravity and matter fields to obtain any higher order interaction Hamiltonian without invoking any approximation such as slow roll, etc. It can also been shown that, the mechanism can be applied for any generalized tensor fields and it can even extract any higher order cross-correlation interaction Hamiltonian.

{Our approach can also be used for modified gravity models including $f(R)$ model, other scalar-tensor theories like Gauss-Bonnet inflation, Lovelock gravity,
Hordenski theory. In fact, we can say unequivocally that our approach can be used for any kind of gravity models.
}




\pagestyle{fancy}
\chapter{Hamiltonian formalism for generalized non-canonical scalar fields}\label{chap.4}
\section{Introduction}
As mentioned in sections (\ref{Sec.Motivation}), (\ref{Sec.Intro.Chap2}) and (\ref{Sec.Intro.Chap3}), the inflationary paradigm is highly successful and an attractive way
of resolving some of the puzzles of standard cosmology. During
inflation, the early universe undergoes an accelerated expansion,
stretching quantum fluctuations to super-horizon scales which we
observe today as CMB anisotropy \cite{1994ApJ...420..445F,Ade:2015xua,Komatsu:2008hk}. Since
Einstein's equations are highly non-linear, comparison of the
predictions of inflation with the observations require one to expand
the equations order-by-order. At linear order, predictions of
inflation is consistent with CMB. However, the linear order
observables, like scalar spectral index and tensor-to-scalar ratio,
can not rule out models of inflation; physically measurable
observables corresponding to higher-order quantities like bispectrum
or trispectrum will help to rule out some models of inflation
\cite{Lidsey:1995np,Lyth:1998xn,Lyth:2007qh,Mazumdar:2010sa}.

Since inflation takes place at high-energies, quantum field theory is
the best description of the matter at these energies.  Hence,
evaluation of any physical quantity, like the $n-$point correlation
functions, require us to either promote effective field variables
(using Heisenberg picture) to operators or integrate over all possible
field configurations on all of space–time (path - integral
picture). Since it is unclear what effective field configurations to
integrate over, Heisenberg picture is the preferred approach. In other
words, we obtain the effective Hamiltonian operator and evaluate the
correlation functions of the relevant operator.

As mentioned earlier in sections (\ref{Sec.Motivation}), (\ref{Sec.Intro.Chap2}) and (\ref{Sec.Intro.Chap3}), in the context of cosmological perturbation theory, there are
currently two approaches in the literature to evaluate the effective
Hamiltonian --- Lagrangian fomalism and Hamiltonian formalism. In case
of Lagrangian formalism,  the Lagrangian is expanded upto
  a particular order, i.e., if we are interested in obtaining third
  order interaction Hamiltonian, effective Lagrangian needs to be
  expanded up to third order and constraints are systematically
  removed from the system to obtain the effective perturbed
  Lagrangian. Then, the momentum $\pi$ corresponding to $\varphi$ is
  obtained as a polynomial of $\dot{\varphi}$ and using order-by-order
  approximations, $\dot{\varphi}$ is expressed as a polynomial of
  $\pi$. Next, using Legendre transformation, Hamiltonian is expressed
  in terms of $\pi$ and $\varphi$. In order to express the Hamiltonian
  in terms of $\dot{\varphi}$ and $\varphi$, only the leading order
  relation between $\pi$ and $\dot{\varphi}$ is used
  \cite{Huang:2006eha, Seery:2005wm, Chen:2006nt, Bartolo:2001cw,
    Bartolo:2003bz, Bartolo:2004if}. There are some difficulties with
  the previous method:
\begin{enumerate}
\item In case of cosmological perturbations, $\pi$ and $\dot{\varphi}$
  are perturbed quantities (curvature perturbation), expressing one in
  terms of polynomial of other is an approximation.
\item At the end, to express the Hamiltonian in terms of
  configuration-space variable, we use only the leading order relation
  between $\pi$ and $\dot{\varphi}$, not the polynomial
  relation. Hence, several approximations are employed to convert
  effective Lagrangian to effective Hamiltonian.
\item The above method is also very restrictive and it is 
  difficult to extend the method for a generalized constrained
  system.
\item Also, it is not possible to use this method for higher
  order perturbations or mixed-mode (e.g.,
  scalar-tensor) interaction Hamiltonian.
\end{enumerate}

Langolois \cite{Langlois1994} first introduced a consistent
Hamiltonian formulation of canonical scalar field. However, Langlois'
approach is also difficult to extend to higher order of perturbations
or to any different types of field due to the fact that it requires
construction of gauge-invariant conjugate momentum. In our recent work and in the previous chapter (\ref{Chap.3}) 
\cite{Nandi:2015ogk}, henceforth referred as I, we have introduced a
different Hamiltonian approach that can address and deal with all the
issues in the previous methods and provides an effective and robust
way to obtain interaction Hamiltonian for any model for any order of
perturbations. Table (\ref{Table.Comp}) (in page 96) below provides a bird's eye view of the both the
formulations and advantages of the Hamiltonian formulation that is
proposed in the previous chapter (\ref{Chap.3}) \cite{Nandi:2015ogk}.
  
 In chapter (\ref{Chap.3}) \cite{Nandi:2015ogk}, we applied our new method to canonical and a specific higher derivative (Galilean) scalar field model, and showed
explicitly that the method can efficiently obtain Hamiltonian at all
orders. In the case of certain non-canonical scalar field models, if
$\dot{\varphi}$ can be expressed uniquely in terms of the canonical
conjugate momentum, it is then possible to obtain Hamiltonian and the
results of chapter (\ref{Chap.3}) \cite{Nandi:2015ogk} can be extended. However, for a general non-canonical
scalar field, it is not possible to do the procedure as we do not have
a way to rewrite $\dot{\varphi} $ in terms of the canonical conjugate
momentum and, hence, it is not possible to obtain the Hamiltonian for
general non-canonical scalar field. In this chapter, we explicitly obtain
Hamiltonian for a general non-canonical scalar fields and obtain
interaction Hamiltonian upto fourth order.

This chapter is divided into two parts. In the first part, we provide the
procedure to obtain Hamiltonian for the non-canonical scalar field by
introducing a phase-space variable. Then, by choosing different
models, we explicitly show that the Hamiltonian leads to consistent
equations of motion as well as perturbed interaction Hamiltonian by
implementing our approach. We also find a new definition of speed of
sound in terms of phase-space variables. In the second part, in order
to retrieve generalized equations of motion in configuration-space
from phase-space, we provide a systematic way to invert generalized
non-canonical phase-space variables to configuration-space variables
and vice versa and show that, all equations are consistent. Finally,
we extend the method to generalized higher derivative
  scalar fields. A flow-chart above illustrates the method for
non-canonical scalar fields:

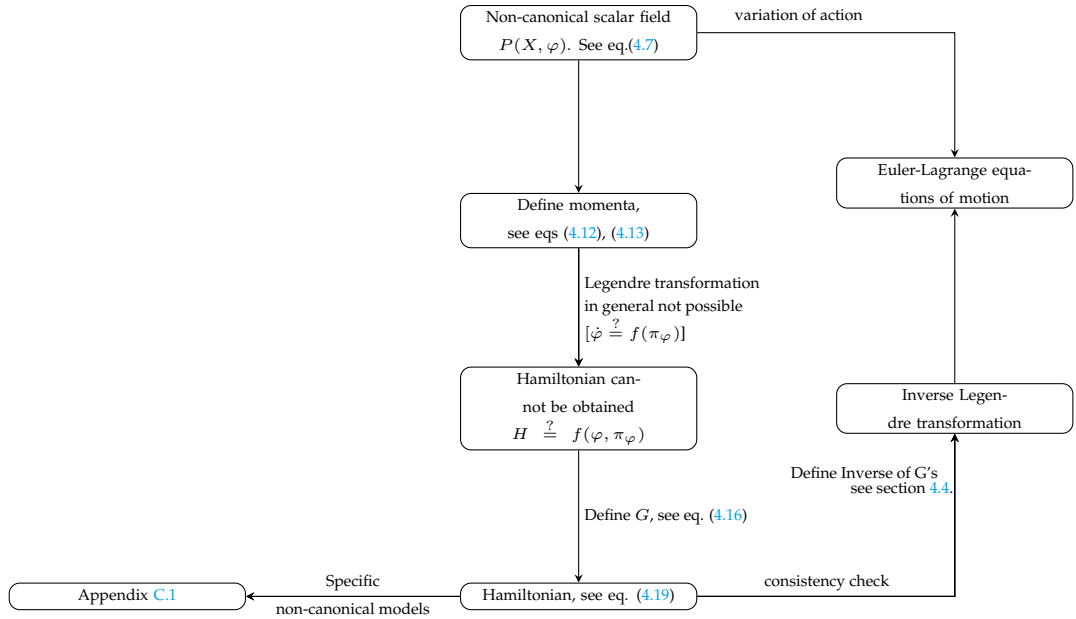
\begin{figure}
{\tiny
\begin{center}
\begin{tikzpicture}[node distance=2.5cm]

  \node (start)[st] {Non-canonical scalar field \\ $P(X,
    \varphi)$. See eq.(\ref{ADM-Non-canonical})}; \node (defmom) [st,
  below of=start] {Define momenta, see eqs (\ref{piijgeneral}),
    (\ref{piphigeneral})}; \node (nonham) [st, below
  of=defmom]{Hamiltonian cannot be obtained\\$H \stackrel{?}{=}
    f(\varphi, \pi_\varphi) $}; \node (ham) [st, below of=nonham
  ]{Hamiltonian, see eq. (\ref{Hamfull})}; \node (Appndix) [st, below
  of=nonham, xshift= -6cm] {Appendix \ref{Inversion}}; \node (Inv)
  [st, below of=defmom, xshift=5cm]{Inverse Legendre transformation};
  \node (EL) [st, below of=ham, yshift=8cm, xshift=5cm]{Euler-Lagrange
    equations of motion};

\draw [arrow] (start) -- (defmom);
\draw [arrow] (defmom) --node[anchor=west, yshift=.3cm]{Legendre transformation} (nonham);
\draw [arrow] (defmom) --node[anchor=west]{in general not possible} (nonham);
\draw [arrow] (defmom) --node[anchor=west, yshift=-.3cm]{$[\dot{\varphi} \stackrel{?}{=} f(\pi_\varphi)$]} (nonham);
\draw [arrow] (nonham) -- node[anchor=west]{Define $G$, see eq. (\ref{G})}(ham);
\draw [arrow] (ham) -- node[anchor=south]{Specific}(Appndix);
\draw [arrow] (ham) -- node[anchor=north]{non-canonical models}(Appndix);
\draw [arrow] (ham) -| node[anchor=south, xshift=-1.7cm]{consistency check} (Inv);
\draw [arrow] (ham) -| node[anchor=north, xshift=-.7cm, yshift=1.6cm]{see section \ref{InversionNC}.} (Inv);
\draw [arrow] (ham) -| node[anchor=south, xshift=-1.2cm, yshift=1.5cm]{Define Inverse of G's} (Inv);
\draw [arrow] (start) -| node[anchor=west, xshift=-3cm, yshift=.25cm]{variation of action} (EL);
\draw [arrow] (Inv) -- (EL);

\end{tikzpicture}
\end{center}
}
\caption{Flow-chart for the proposed Hamiltonian formulation of non-canonical scalar field}
\end{figure}
In the next section, we briefly discuss about non-canonical scalar
fields. We also discuss the gauge choices and corresponding
gauge-invariant variables. In section (\ref{Hamiltonian-nonC}),
Hamiltonian formulation of generalized non-canonical scalar field is
introduced by defining a new phase-space function which provides
consistent equations of motion. In section (\ref{first}), we extend the
results of I to non-canonical scalar field in flat-slicing gauge to
obtain perturbed equations of motion. In section (\ref{InversionNC}), we
provide a partial inversion method between phase-space variables and
configuration-space variables and in section (\ref{InteractionHam}), we
provide the third and fourth order Interaction Hamiltonian for
non-canonical scalar field. In section (\ref{ExGal}), we briefly discuss
the application of our method to generalized Galilean scalar field
model. Finally, in section (\ref{Conclu}), we end with discussions and
conclusions of the results. In Appendix (\ref{Inversion}), functional
form of the new variable is obtained for different scalar field
models. In Appendix (\ref{Langlois-nonC}), we implement Langlois'
approach to non-canonical scalar field model.

\section{Model and gauge choices}\label{BasicModel}

As mentioned in previous chapters, action for non-canonical scalar field minimally coupled
to gravity is
\begin{equation}
\label{GravityAction}
\mathcal{S} = \int d^4 x \sqrt{-g} \left[\frac{1}{2 \kappa} R + \mathcal{L}_m(\varphi, \partial\varphi) \right],
\end{equation}
where $R$ is the Ricci scalar and the matter Lagrangian, $\mathcal{L}_m$ is of the form.
\begin{equation}\label{boxx}
  \mathcal{L}_m = P(X, \varphi)  ,~~~ ~X \equiv 
  \frac{1}{2} g^{\mu \nu} \partial_{\mu}{\varphi} \partial_{\mu}{\varphi}.
\end{equation}

{$P(X, \varphi)$ corresponds to non-standard kinetic term
  and hence the name non-canonical scalar
  field \cite{Armendariz-Picon1999,Garriga1999,Armendariz-Picon2001}. Further,
  fixing $ P = - X - V(\varphi)$, where $V(\varphi)$ is the potential,
  one can retrieve the well-known canonical scalar field model.}
Varying the action (\ref{GravityAction}) with respect to metric gives
Einstein's equation
\begin{equation}
\label{EinsteinEquation}
R_{\mu \nu} - \frac{1}{2} g_{\mu \nu}~ R = \kappa~ T_{\mu \nu}, 
\end{equation}
where the stress tensor $(T_{\mu \nu})$ for non-canonical scalar field is
\begin{equation}
\label{EMTensor}
T_{\mu \nu} = - P_X \partial_{\mu}{\varphi} \partial_{\nu}{\varphi} + g_{\mu \nu}  P.
\end{equation}

Varying the action (\ref{GravityAction}) with respect to the scalar
field `$\varphi$' leads to the following equation of motion
\begin{equation}
\label{EG}
   P_X \Box \varphi - P_{XX} \partial_{\mu}{\varphi} \partial^{\mu}{\varphi} -
  2X
   P_{X \varphi}  + P_\varphi = 0
\end{equation}
which can also be obtained from the conservation of Energy-Momentum
tensor, $\nabla_{\mu}{T^{\mu \nu}} = 0$.

The four-dimensional line element in the ADM form is given by,
\begin{eqnarray}
ds^2 &=& g_{\mu \nu} dx^{\mu} dx^{\nu} \nonumber \\
\label{line}
&=& -(N^2 - N_{i} N^{i} ) d\eta^2 + 2 N_{i} dx^{i} d\eta + \gamma_{i j} dx^{i} dx^{j},
\end{eqnarray}
where $N(x^{\mu})$ and $N_i(x^{\mu})$ are Lapse function and Shift
vector respectively, $\gamma_{i j}$ is the 3-D space metric.  Action
(\ref{GravityAction}) for the line element (\ref{line}) takes the form
\begin{equation}
\label{ADM-Non-canonical}
\mathcal{S}_{NC} = \int d^4x\, N\, \sqrt{\gamma}\, \left[ \frac{1}{2 \kappa}
\left(^{(3)}R + K_{i j} K^{i j} - K^2\right) + P(X, \varphi) \right]
\end{equation}
where $K_{i j}$ is extrinsic curvature tensor and is defined by
\begin{eqnarray}
  && K_{i j} \equiv \frac{1}{2N} \left( \partial_{0}{\gamma_{i j}} - N_{i|j} - N_{j | i} \right), \nonumber \\
  && K \equiv \gamma^{i j} K_{i j}. \nonumber
\end{eqnarray}

As mentioned before in (\ref{SubSec.Gauge}), perturbatively expanding the metric only in terms of scalar perturbations and the scalar field about the flat FRW space-time in
conformal coordinate, we get in equations (\ref{PertVar1}), (\ref{PertVar2}), (\ref{PertVar3}) and (\ref{PertVar4}). In this chapter, we derive all equations by choosing a
specific gauge --- flat-slicing gauge, i.e., $\psi = 0, E = 0$ --- at
all orders:
\begin{eqnarray}
	\label{00pmetric}
	&& g_{0 0} =- a(\eta)^2(1 + 2 \epsilon \phi_1 + \epsilon^2 \phi_2 + ...) \\
	\label{0ipmetric}
	&& g_{0 i} \equiv N_{i} = a(\eta)^2 (\epsilon \partial_{i}{B_1} + \frac{1}{2} \epsilon^2 \partial_{i}{B_2} + ...) \\
	\label{3metric}
	&& g_{i j} =a(\eta)^2 \delta_{i j}\\
	\label{field}
	&&\varphi = \varphi_0(\eta) + \epsilon \varphi_1 + \frac{1}{2} \epsilon^2 \varphi_2 + ...
	\end{eqnarray}

It can be shown that, perturbed equations in flat-slicing gauge
coincide with gauge-invariant equations of motion (in generic gauge,
$\varphi_1$ coincides with $\varphi_1 + \frac{\varphi_0{}^\prime}{H}
{}\psi_1 \equiv \frac{\varphi_0{}^\prime}{H} \mathcal{R}$ which is a
gauge-invariant quantity, $\mathcal{R}$ is called curvature
perturbation \cite{Malik2009}). Similarly, one can choose another
suitable gauge with no coordinate artifacts to obtain gauge-invariant
equations of motion. Such gauges are Newtonian-conformal gauge $(B =
0,\, E = 0)$, constant density gauge $(E = 0, \, \delta\varphi = 0)$,
etc.
  
Before we proceed with the Hamiltonian formulation, it is important to
clarify issues related to the quantization in the cosmological
perturbation theory: While the field variable $\varphi_i$ and metric
variables $\phi_i, B_i, \psi_i$ (where $i$ takes values $1, 2,
\cdots$) are expanded perturbatively, it is important to note that the
operators corresponding to these variables (i. e. $\varphi_1,
\varphi_2, \cdots$) can not be treated as independent operators as
higher orders in perturbation theory do not lead to independent
degrees of freedom. As otherwise, the unperturbed theory should have
infinitely many local degrees of freedom.  In a canonical
quantization, there is one operator and one for its momentum, on which
the quantum Hamiltonian depends\footnote{We thank Martin Bojowald for
  discussion regarding this point.}.

\section{Hamiltonian formulation}\label{Hamiltonian-nonC}

In the previous chapter ({\ref{Chap.3}}) \cite{Nandi:2015ogk}, we provided an efficient way of
obtaining consistent perturbed Hamiltonian for any gravity
models. However, it works only if the form of the action is specified
in such a way that the Legendre transformation $\dot{\varphi}
\rightarrow \pi_\varphi$ is invertible in both ways. For non-canonical
scalar fields, momenta corresponding to the action
(\ref{ADM-Non-canonical}) are
\begin{eqnarray}
\label{piijgeneral}
\pi^{i j} \equiv \frac{\delta \mathcal{S}_{NC}}{\delta {\gamma^\prime_{i j}}}  &=& \frac{\sqrt{\gamma}}{2 \kappa}\,  ({\gamma}^{i j} {\gamma}^{k l} - {\gamma}^{i k} {\gamma}^{j l}) {K}_{k l} \\
 \label{piphigeneral}
 \pi_\varphi \equiv \frac{\delta \mathcal{S}_{NC}}{\delta {\varphi^\prime}} &=& - \sqrt{\gamma}P_X \sqrt{- 2X +  Y}, \quad \mbox{where}\quad Y \equiv \gamma^{i j} \partial_i \varphi \partial_j \varphi.
\end{eqnarray}

As one can see, equation (\ref{piijgeneral}) is invertible and the
inversion relation is given by
\begin{equation}\label{InvPiij}
 {\gamma}^\prime_{m n} = \gamma_{n k} N^k_{|m} + \gamma_{m k} N^k_{|n} - 2 N K_{m n}, ~~~ K_{i j} = \frac{\kappa}{\sqrt{\gamma}} \left(\gamma_{i j} \gamma_{k l} - 2 \gamma_{i k} \gamma_{j l}\right)
\pi^{k l} 
\end{equation}
but equation (\ref{piphigeneral}) is non-invertible for arbitrary
function of $P(X, \varphi)$. However, if $P(X, \varphi)$ is specified,
it may be possible to invert the equation and $X$ can be written in
terms of $\pi_\varphi$. Inversion relations for commonly used
non-canonical models are given in Appendix (\ref{Inversion}). Using
equation (\ref{InvPiij}), we can write the Hamiltonian density as
\begin{eqnarray}
  \mathcal{H}_{NC} &=& \pi^{i j} \gamma_{i j}^\prime + \pi_\varphi \varphi^\prime - \mathcal{L}_{NC} \nonumber \\
  &=& 2 {\gamma}_{i j} {\partial}_{k}{{N}^{j}}\,  {\pi}^{i k} + {N}^{i} {\partial}_{i}{{\gamma}_{j k}}\,  {\pi}^{j k}  - \frac{N \kappa}{\sqrt{\gamma}} \left(\gamma_{i j} {\gamma}_{k l} - 2\gamma_{i k} \gamma_{j l}\right) {\pi}^{i j} {\pi}^{k l}  - \frac{N \sqrt{\gamma}}{2 \kappa} \,{}^{(3)}R  -\nonumber \\
 \label{fullhamiltonian}
 &&  N \sqrt{\gamma}~\tilde{G}(X, \gamma, Y, \varphi) + {N}^{i} \pi_\varphi {\partial}_{i}{\varphi}, \quad\mbox{where}\quad\tilde{G}\equiv \left( P - P_X \left(2X - Y\right)\right).
\end{eqnarray}

Note that, the above expression is still not the Hamiltonian since
$\tilde{G}$ is not a phase-space variable and it is not invertible for
arbitrary form of $P(X, \varphi)$ since equation (\ref{piphigeneral})
is not invertible, in general. Hence, a natural question that arises
is: \emph{How to invert configuration-space variables to phase-space
  variables so that we can obtain generalized consistent Hamiltonian
  for non-canonical scalar field?}
 
In this section, we show that, by defining a new phase-space function,
the above problem can be resolved. The new phase-space quantity is
defined as
\begin{equation}
\label{G}
G(\pi_\varphi, \gamma, Y, \varphi) = \tilde{G}(X, \gamma, Y, \varphi) \equiv P - P_X \left(2X - Y\right).
\end{equation}

Since momenta corresponding to $N$ and $N^i$ vanish, i.e., $\pi_N =
\pi_i = 0$, using the above defined variable, Hamiltonian constraint
can be written as
\begin{equation}
\label{HamitonianConstraint}
\mathcal{H}_N \equiv \{\pi_N, \mathcal{H}_{NC}\} = \frac{\delta \mathcal{H}_{NC}}{\delta N} = - \frac{ \kappa}{\sqrt{\gamma}} \left(\gamma_{i j} {\gamma}_{k l} - 2\gamma_{i k} \gamma_{j l}\right) {\pi}^{i j} {\pi}^{k l}  - \frac{ \sqrt{\gamma}}{2 \kappa} \,{}^{(3)}R  -  \sqrt{\gamma}\,\, G(\pi_\varphi, \gamma, Y, \varphi) = 0,
\end{equation}
and momentum constraint is given by
\begin{equation}
\label{MomentumConstraint}
M_i \equiv \{\pi_i, \mathcal{H}_{NC}\} = \frac{\delta \mathcal{H}_{NC}}{\delta N^i}= -2 \partial\left(\gamma_{i m}\pi^{m n}\right) + \pi^{k l} \partial_{i}\gamma_{k l} + \pi_\varphi\partial_i \varphi = 0.
\end{equation}

Due to diffeomorphic invariance, total Hamiltonian density can be written as
\begin{equation}\label{Hamfull}
\mathcal{H}_{NC} = N \mathcal{H}_N + N^i \mathcal{H}_i = 0. 
\end{equation}

Instead of defining $G$, one can define any other phase-space
variable(s) and express the Hamiltonian in a different form and the
possibilities are infinite. However, as one can see, since
$\tilde{G}(X, \gamma, Y, \varphi)$ automatically appears directly in
the Hamiltonian, this is the simplest and effective way to express the
Hamiltonian for non-canonical scalar field. $G$ not only resolves the
issue of expressing Hamiltonian for non-canonical scalar field and is
also uniquely defined for different non-canonical scalar
fields. Hence, $G$ carries the signature of the non-canonical scalar
fields in phase-space. Explicit forms of $G(\pi_\varphi, \gamma, Y,
\varphi)$ for different types of scalar fields are given in Appendix
(\ref{Inversion}).

\subsection{Zeroth order}\label{zero}
At zeroth order, since $\gamma_{i j} = a^2 \delta_{i j}$ and all
quantities are independent of spatial coordinates, we then get
\begin{equation}
\pi_0^{i j} = \frac{1}{6 a} \pi_a \delta^{i j} 
\end{equation}
and Hamiltonian density at zeroth order becomes
\begin{equation}
\label{zerothH}
\mathcal{H}_{0} = - \frac{N_0 \kappa}{12 a} \pi_a^2 - G \,N_0\, a^3.
\end{equation}

Variation of the Hamiltonian (\ref{zerothH}) with respect to the
momenta leads to two equations and are given by
\begin{eqnarray}
\label{Zerotha'}
a^\prime &=& - \frac{N_0 \kappa}{6 a} \pi_a \\
\label{Zerothphi'}
\varphi_0^\prime &=& - N_0 \,a^3\, G_{ \pi_\varphi}.
\end{eqnarray}

Hamiltonian constraint leads to the equation of motion of $N$ and at
zeroth order, it is given by
\begin{equation}\label{ZerothHam}
\mathcal{H}_{N0}  \equiv - \frac{\kappa}{12 a} \pi_a^2 - G\, a^3 = 0.
\end{equation}

Equations of motion are obtained by varying the Hamiltonian
(\ref{zerothH}) with respect to field variables. Hence, equation of
motion of $a$ is obtained by the relation
\begin{equation}\label{ZerothEoMa}
  \pi_a^\prime = - \frac{\delta \mathcal{H}_{0}}{\delta a} = - \frac{N_0 \kappa}{12 a^2} \pi_a^2 + 3\, G\, N_0\, a^2 + G_{a}\, N_0\, a^3.
\end{equation}

Similarly, equation of motion of $\varphi_0$ can be obtained from
\begin{equation}\label{ZerothEoMphi}
\pi_{\varphi0}^\prime = N_0\, a^3\,G_{\varphi}.
\end{equation}

\subsection{First order}\label{first}

As we have mentioned in the introduction and in chapter (\ref{Chap.3}), there are two ways to obtain
Hamiltonian --- Langlois' approach \cite{Langlois1994}, and the
approach used in chapter (\ref{Chap.3}) \cite{Nandi:2015ogk}. In this chapter, we use both the
approaches and explicitly show that it is possible to obtain a
consistent Hamiltonian for non-canonical scalar fields. In Appendix
(\ref{Langlois-nonC}), we extend Langlois' approach to non-canonical
scalar field and in the rest of the section, we extend I to obtain a
consistent Hamiltonian for non-canonical scalar field.

The field variables and their corresponding momenta can be separated
into unperturbed and perturbed parts as
\begin{eqnarray}
\label{perturbfm1}
&& N = N_0 + \epsilon N_1, ~~~N^{i} = \epsilon N_1^i,~~~\varphi = \varphi_0 + \epsilon \varphi_1 \\
\label{perturbfm2}
&& \pi^{i j} = {}\pi_0{}^{i j} + \epsilon \pi_1^{i j}, ~~~\pi_\varphi = {}\pi_{\varphi0} + \epsilon \pi_{\varphi1}
\end{eqnarray}
and by using Taylor expansion of phase-space variable $G(\pi_\varphi,
\gamma, Y, \varphi)$, the second order perturbed Hamiltonian density
is given by
\begin{eqnarray}
\label{firstH}
&& \mathcal{H}_{2} =  
{\delta}_{i j} {\partial}_{k}{{N_1}^{j}}\,  ({\pi_1}^{i k} + \pi_1^{k i}) a{}^{2}  - N_0  \kappa a\,({\delta}_{i j} {\delta}_{k l} - 2 \delta_{i k} \delta_{j l}){\pi_1}^{i j} {\pi_1}^{k l} - 2\, N_1 \kappa a \,({\delta}_{i j} {\delta}_{k l} - 2 \delta_{i k} \delta_{j l}){\pi_0}^{i j} {\pi_1}^{k l} \nonumber \\
&&~~~~~ - G_\varphi N_1 a{}^{3} \varphi_1 - G_{\pi_\varphi} N_1 \,\pi_{\varphi 1} \,a{}^{3} - \frac{1}{2}\, G_{\varphi\varphi} N_0\, \varphi_1{}^{2} a{}^{3} - \frac{1}{2}\, G_{\pi_\varphi \pi_\varphi} N_0\, \pi_{\varphi 1} {}^{2} a{}^{3} - G_{\varphi \pi_\varphi } N_0 \pi_{\varphi 1} a{}^{3} \varphi_1 \nonumber \\
&&~~~~~- G_Y N_0 {\delta}^{i j} {\partial}_{i}{\varphi_1}\,  {\partial}_{j}{\varphi_1}\,  a + {N_1}^{i} \pi_{0 \varphi} {\partial}_{i}{\varphi_1}.\,
\end{eqnarray}

Note that, as we have pointed out in chapter (\ref{Chap.3}) \cite{Nandi:2015ogk},
perturbed momentum corresponding to an unperturbed variable may arise
due to the presence of other perturbed phase-space variables, thus
$\pi^{ij}_1$ is non-zero and can be obtained by varying the
Hamiltonian (\ref{firstH}) with respect to $\pi^{ij}_1$:
\begin{equation}
  \frac{\delta \mathcal{H}_2}{\delta \pi_1^{i j}} = 0 \Rightarrow \pi^{i j}_1 = \frac{a}{2 N_0 \kappa} \delta^{i j} \partial_k N_1^k - \frac{a}{4 N_0 \kappa} \delta^{k i} \partial_k N_1^j - \frac{a}{4 N_0 \kappa} \delta^{k j} \partial_k N_1^i - \frac{N_1}{N_0} \pi_0^{i j}.
\end{equation}

Varying the perturbed Hamiltonian (\ref{firstH}) with respect to $\pi_{\varphi1}$ leads to the following equation
\begin{eqnarray}
  \varphi_1^\prime &=& - G_{\pi_\varphi} N_1 a^3 - G_{\pi_\varphi \pi_\varphi}N_0\, \pi_{\varphi1} a^3 - G_{\varphi \pi_\varphi} N_0\, \varphi_1 a^3 \\
  \Rightarrow \pi_{\varphi1} &=& - \frac{1}{N_0\, a^3\,G_{\pi_\varphi \pi_\varphi} }\left(\varphi_1^\prime + G_{\pi_\varphi} N_1 a^3 + G_{\varphi \pi_\varphi} N_0\, \varphi_1 a^3 \right).
\end{eqnarray}

Hamiltonian constraint is obtained by varying the Hamiltonian with
respect to Lapse function. Hence, varying (\ref{firstH}) with respect
to $N_1$ leads to first order Hamiltonian constraint and takes the
form
\begin{equation}
\label{FirstHamiltonian}
-2 \delta_{i j} \delta_{k l} \,\pi_0^{i j}\, \pi_1^{k l} + 4 \delta_{i j} \delta_{k l} \,\pi_0^{i k}\, \pi_1^{j l} - G_\varphi a^2 \varphi_1 - G_{\pi_\varphi} \pi_{\varphi1}\, a^2 = 0.\\
\end{equation}

Similarly, by varying Hamiltonian with respect to $N_1^i$ we get the following Momentum constraint,
\begin{equation}
\label{FirstMomentum}
\pi_{\varphi0} \, \partial_i \varphi_1 - 2 a^2\, \delta_{i j}\,\partial_k  \pi_1^{k j}=0.
\end{equation}

Finally, equation of motion of $\varphi_1$ is obtained by varying the Hamiltonian with respect to $\varphi_1$, i.e.,
\begin{equation}
\label{FirstEoMvarphi}
\pi_{\varphi_1}^\prime = a^3 \,G_\varphi N_1  + a^3\,G_{\varphi\varphi}N_0 \varphi_1  + a^3\,G_{\varphi \pi_\varphi} N_0 \pi_{\varphi1} - 2\,a\, G_Y N_0 \nabla^2 \varphi_1 + \pi_0^{i j}\,\partial_iN_1^i.
\end{equation}

Since the perturbed scalar field equation is linear in nature and
follows wave equation, speed of sound is defined as the ratio of
negative of the coefficient of $\nabla^2 \varphi_1$ and
$\varphi_1^{\prime\prime}$ and in phase-space, it takes the form
\begin{equation}
c_s^2 = 2 \,N_0^2\,a^4 \, G_{\pi_\varphi \pi_\varphi}\, G_Y
\end{equation}
which, in conformal coordinate can be expressed as
\begin{equation}\label{sonic-conf}
c_s^2 =  2 \,a^6 \, G_{\pi_\varphi \pi_\varphi}\, G_Y.
\end{equation}

The relation between generalized phase-space derivatives of $G$
($G_\varphi,~ G_Y,~G_{\varphi \pi_\varphi}$ etc.) and
configuration-space derivatives of $P(X, \varphi)$
($P,~P_{\varphi},~P_{\varphi X}$ etc.) is unknown, hence, it is not
possible to invert above Hamilton's equations to Euler-Lagrange
equations and hence, it is not possible to compare both the
formalisms. However, for a particular scalar field, the exact form of
$G$ is known to us (see Appendix \ref{Inversion}), and hence, for
those model it is possible to write down equations of motion in
configuration space and can be verified that Hamiltonian formulation
of non-canonical scalar field is consistent.

\section{Inversion of non-canonical terms}\label{InversionNC}
In the last section, we showed that it is possible to obtain
Hamiltonian for a non-canonical scalar field by defining a new
variable G (see eqs. (\ref{G}) and (\ref{Hamfull})).  In order to
understand the importance of this new function G, we ask the following
question: \emph{Starting from the Hamiltonian (\ref{zerothH}) and
  (\ref{firstH}), can we invert the expressions leading to generalized
  equations in configuration-space?} In this section, we show that,
inversion can be established.

To invert the equations, one needs to invert the coefficients like
$G_\varphi,~ G_Y,~G_{\varphi \pi_\varphi}$ from phase-space to
configuration-space. Since the form of $G$ in configuration space is
known, by carefully looking at the equations, it is apparent that only
the phase-space derivatives of $G$ are needed to invert which, in
general, is not possible.

To begin with, let us take a phase-space function $F \equiv
F(\pi_\varphi, \gamma, Y, \varphi) = \tilde{F}(X, \gamma, Y,
\varphi)$, i.e.,
\begin{eqnarray}
 F &=& F(\pi_\varphi, \gamma, Y, \varphi) \nonumber \\
\Rightarrow dF &=& F_{\pi_\varphi} d\pi_\varphi + F_\gamma d\gamma + F_Y  dY + F_\varphi d\varphi \nonumber
\end{eqnarray}

Note that, tilde is used for configuration-space functions only. The invertibility of Legendre transformation implies that, if $X=X(\pi_\varphi, \gamma, Y, \varphi)$ then $\pi_\varphi = \pi_\varphi (\pi_\varphi, \gamma, Y, \varphi)$, i.e., 

\begin{eqnarray}
&& \pi_\varphi = \pi_\varphi ( X, \gamma, Y, \varphi) \nonumber \\
&& d\pi_\varphi = \frac{\partial \pi_\varphi}{\partial X} dX + \frac{\partial \pi_\varphi}{\partial \gamma}d\gamma + \frac{\partial \pi_\varphi}{\partial Y} dY + \frac{\partial \pi_\varphi}{\partial \varphi} d \varphi \nonumber 
\end{eqnarray}
implying that
\begin{eqnarray}
 d\tilde{F}(X, \gamma, Y, \varphi) &=& F_{\pi_\varphi} \frac{\partial \pi_\varphi}{\partial X} dX + \left( F_\gamma + F_{\pi_\varphi} \frac{\partial \pi_\varphi}{\partial \gamma}\right)  d\gamma + \left( F_Y + F_{\pi_\varphi}\frac{\partial \pi_\varphi}{\partial Y}\right)  dY \nonumber \\
&&+\left( F_\varphi +  F_{\pi_\varphi}\frac{\partial \pi_\varphi}{\partial \varphi}\right) d\varphi.
\end{eqnarray}

Hence, the relations between phase-space variables and
configuration-space variables are

\begin{eqnarray}
 F_{\pi_\varphi} &=& \frac{\tilde{F}_X}{\frac{\partial \pi_\varphi}{\partial X}}, \\
 F_\gamma &=& \tilde{F}_\gamma -  F_{\pi_\varphi} \frac{\partial \pi_\varphi}{\partial \gamma}, \\
 F_Y &=& \tilde{F}_Y -  F_{\pi_\varphi} \frac{\partial \pi_\varphi}{\partial Y},\\
 F_\varphi &=& \tilde{F}_\varphi -  F_{\pi_\varphi} \frac{\partial \pi_\varphi}{\partial \varphi}.
\end{eqnarray}

In our case, for arbitrary non-canonical scalar field, we do not know
the exact form of $G(\pi_\varphi, \gamma, Y, \varphi)$, however, we
know $\tilde{G}(X, \gamma, Y, \varphi) = P - P_X \left(2X -
  Y\right)$. Using equation (\ref{piphigeneral}) and the above
established relations, we get

\begin{eqnarray}
&&G_{\pi_\varphi} =  - \frac{\sqrt{-2X}}{a^3},~~~G_{\pi_\varphi \pi_\varphi} = \frac{1}{a^6 \left(P_X + 2X P_{XX}\right)},\nonumber \\
 &&G_{\pi_\varphi \pi_\varphi \pi_\varphi} = -\frac{\sqrt{-2X} \left(3P_{XX} + 2X P_{XXX}\right)}{a^9 \left(P_X + 2X P_{XX}\right)^3} \nonumber \\
 &&G_\varphi = P_\varphi, ~~~~~G_{\varphi \pi_\varphi} = \frac{\sqrt{-2X}\,P_{X\varphi}}{a^3 \left(P_X + 2X P_{XX}\right)}\nonumber \\
&& G_{\varphi \varphi} = P_{\varphi\varphi} - \frac{2X\,P_{X\varphi}^2}{ \left(P_X + 2X P_{XX}\right)}
\end{eqnarray}

Using these definitions it is possible to invert all phase-space
quantities to the ones in configuration-space. To start with, we first
consider speed of sound. In conformal coordinate, it is given by
equation (\ref{sonic-conf}) and hence using above relations, we get
\begin{equation}
c_s^2 = \frac{P_X}{P_X + 2XP_{XX}}
\end{equation}
which matches with the conventional configuration-space definition of
sound of speed.  Similarly, the zeroth order equations
(\ref{ZeroHamP}), (\ref{ZeroEomaP}) and (\ref{ZeroEoMphiP}), after
inversion, become
\begin{eqnarray}
\label{ZeroHamP}
 && H^2 =- \frac{\kappa}{3}  (P_X {\varphi^{\prime}_0}^{2} + P a^2), ~~~ \mbox{Hubble  Constant:}~ H \equiv \frac{a^\prime}{a}    \\
 \label{ZeroEomaP}
&&- 2 \frac{a^{\prime \prime}}{a} +   H^2 = \kappa P a^2, \\
\label{ZeroEoMphiP}
&& P_X {\varphi_0^{\prime \prime}} - P_{XX} {\varphi_0^{\prime \prime}}
{\varphi_0^{\prime}}^{2} a^{-2} + P_{X\varphi}
{\varphi_0^{\prime}}^{2} + 2 P_X {\varphi_0^{\prime}} H +
P_{XX} H {\varphi_0^{\prime}}^{3} a^{-2} + P_{\varphi}
a^{2} = 0,
\end{eqnarray}
respectively. At first order, $N_1 = a \phi_1$ and $N^i = \delta^{i
  j} \partial_j B_1$, which helps to reduce first order perturbed
Hamiltonian constraint (\ref{FirstHamiltonian}) into
\begin{eqnarray}
&& \frac{H}{\kappa} \nabla^2 B_1 + \frac{3\, H^2}{\kappa} \phi_1 + \frac{G^2_{\pi_\varphi}}{2\,G_{\pi_\varphi \pi_\varphi}} \phi_1 a^2 + \frac{G_{\pi_\varphi}}{2\, G_{\pi_\varphi \pi_\varphi} a^2} \varphi_1^\prime + \frac{G_{\pi_\varphi}\,G_{\varphi \pi_\varphi}}{2\, G_{\pi_\varphi \pi_\varphi}} \varphi_1 a^2 \nonumber \\
&&~~~~~~~~ - \frac{G_\varphi}{2} \varphi_1 a^2 = 0
\end{eqnarray}

which, further, can be inverted back to configuration-space, again by
using above relations as
\begin{eqnarray}
\label{firstHamP}
&&\mathcal{H} \nabla^2{B_1} = \frac{\kappa}{2} \Big[ P_X \phi_1
{\varphi_0^{\prime}}^{2} + 2 P a^2 \phi_1 + 
P_X \varphi_0^{\prime} \varphi_1^{\prime} + P_{XX}
\phi_1 {\varphi_0^{\prime}}^{4} a^{-2} 
- P_{XX} \varphi_1^{\prime} {\varphi_0^{\prime}}^{3} a^{-2}
+ \nonumber \\
&& ~~~~~~~~~~~~~P_{X \varphi} {\varphi_0^{\prime}}^{2} \varphi_1 
+ P_\varphi \varphi_1 a^2 \Big].
\end{eqnarray}

Similarly, first order Momentum constraint becomes
\begin{equation}
\label{fisrtMomP}
\partial_i\phi_1= - \frac{\kappa}{2 \,H}P_X \varphi_0^{\prime} \partial_i\varphi_1
\end{equation}

and equation of motion of scalar field $\varphi_1$, i.e., equation (\ref{FirstEoMvarphi}) takes the form

\begin{eqnarray}
\label{FirstEoMP}
&&- P_X {\varphi_1^{\prime \prime}} a^{2} - P_{XX}
{\phi_1^{\prime}}{\varphi_0^{\prime}}^{3} + P_{XX} {\varphi_1^{\prime
    \prime}} {\varphi_0^{\prime}}^{2} - P_{XX\varphi} \phi_1
{\varphi_0^{\prime}}^{4} 
+ P_{XX\varphi} {\varphi_1^{\prime}}
{\varphi_0^{\prime}}^{3} -P_{\varphi} \phi_1 a^{4}- \nonumber\\
&& P_{\varphi \varphi} a^{4} \varphi_1 + P_X \phi_1 {\varphi_0^{\prime \prime}}
a^{2} + P_X \nabla^2{\varphi_1} a^{2} + P_X {\phi_1^{\prime}}{\varphi_0^{\prime}} a^{2} - 2 P_X {\varphi_1^{\prime}} H
a^2 - 4 P_{XX} \phi_1 {\varphi_0^{\prime
    \prime}}{\varphi_0^{\prime}}^{2} +\nonumber\\
  && 3 P_{XX} {\varphi_0^{\prime}}
{\varphi_1^{\prime}}{\varphi_0^{\prime \prime}} + P_{XX\varphi}
{\varphi_0^{\prime \prime}}{\varphi_0^{\prime}}^{2} \varphi_1 -
P_{X\varphi} {\varphi_0^{\prime}} {\varphi_1^{\prime}} a^{2} -
P_{X\varphi} {\varphi_0^{\prime \prime}} a^{2} \varphi_1 -P_{X\varphi\varphi} {\varphi_0^{\prime}}^{2} a^{2} \varphi_1 + \nonumber\\
&&  2 P_X
\phi_1 {\varphi_0^{\prime}} H a^2 + P_X {\varphi_0^{\prime}}
\nabla^2{B_1} a^{2} + P_{XX} \phi_1 H
{\varphi_0^{\prime}}^{3} - P_{XX} {\varphi_1^{\prime}} H
{\varphi_0^{\prime}}^{2} - P_{XXX} \phi_1 H
{\varphi_0^{\prime}}^{5} a^{-2} +\nonumber \\
&& P_{XXX} \phi_1 {\varphi_0^{\prime
    \prime}}{\varphi_0^{\prime}}^{4} a^{-2} + P_{XXX}
{\varphi_1^{\prime}}H{\varphi_0^{\prime}}^{4} a^{-2} -
P_{XXX}{\varphi_1^{\prime}}{\varphi_0^{\prime
    \prime}}{\varphi_0^{\prime}}^{3} a^{-2} - P_{XX\varphi}
H {\varphi_0^{\prime}}^{3} \varphi_1 - \nonumber\\
&&  2 P_{X\varphi}
{\varphi_0^{\prime}} H \varphi_1 a^2 = 0.
\end{eqnarray} 

Equations (\ref{ZeroHamP}), (\ref{ZeroEomaP}), (\ref{ZeroEoMphiP}),
(\ref{firstHamP}), (\ref{fisrtMomP}) and (\ref{FirstEoMP}) are
consistent with the zeroth and first order perturbed Euler-Lagrange
equations of motion.

\section{Interaction Hamiltonian}\label{InteractionHam}

The higher-order physical observables like Bi-spectrum/Tri-spectrum
are related to higher-order correlation functions; in order to compute
higher-order correlation functions, we need higher-order interaction
Hamiltonian. In this section, we obtain the interaction Hamiltonian of
the non-canonical field. Third order perturbed generalized interaction
Hamiltonian for non-canonical scalar field in terms of phase-space
variables is obtained directly by expanding the Hamiltonian
(\ref{Hamfull}) upto third order of perturbation \cite{Nandi:2015ogk}
and it takes the form
\begin{eqnarray}
\label{ThirdIntHam}
\mathcal{H}_3 &=&  - N_1 {\delta}_{i j} {\delta}_{k l} \kappa {\pi_1}^{i j} {\pi_1}^{k l} a + 2\, N_1 {\delta}_{i j} {\delta}_{k l} \kappa {\pi_1}^{i k} {\pi_1}^{j l} a - \frac{1}{2}\, G_{\varphi\varphi} N_1 \varphi_1{}^{2} a{}^{3} - \frac{1}{2}\, G_{\pi_\varphi \pi_\varphi} N_1 \pi_{\varphi1}{}^{2} a{}^{3} -\nonumber \\
&& G_{\varphi\pi_\varphi} N_1 \pi_{\varphi1} a{}^{3} \varphi_1 - G_Y N_1 {\delta}^{i j} {\partial}_{i}{\varphi_1}\,  {\partial}_{j}{\varphi_1}\,  a - G_{Y \pi_\varphi} N_0 \pi_{\varphi1} {\delta}^{i j} {\partial}_{i}{\varphi_1}\,  {\partial}_{j}{\varphi_1}\,  a -\nonumber \\
&& G_{\varphi Y} N_0 {\delta}^{i j} {\partial}_{i}{\varphi_1}\,  {\partial}_{j}{\varphi_1}\,  \varphi_1 a - \frac{1}{6}\, G_{\pi_\varphi \pi_\varphi \pi_\varphi} N_0 \pi_{\varphi1}{}^{3} a{}^{3} - \frac{1}{6}\, G_{\varphi \varphi \varphi} N_0 \varphi_1{}^{3} a{}^{3} - \nonumber \\
&&\frac{1}{2}\, G_{\varphi \pi_\varphi \pi_\varphi} N_0 \pi_{\varphi1}{}^{2} a{}^{3} \varphi_1 - \frac{1}{2}\, G_{\varphi\varphi\pi_\varphi} N_0 \pi_{\varphi1} \varphi_1{}^{2} a{}^{3} + {N_1}^{i} \pi_{\varphi1} {\partial}_{i}{\varphi_1}
\end{eqnarray}
and similarly, fourth order interaction Hamiltonian takes the form
\begin{eqnarray}
\label{FourthIntHam}
\mathcal{H}_4 &=&  - \frac{1}{2}\, G_{YY} N_0 {\delta}^{i j} {\delta}^{k l} {\partial}_{i}{\varphi_1}\,  {\partial}_{j}{\varphi_1}\,  {\partial}_{k}{\varphi_1}\,  {\partial}_{l}{\varphi_1}\,  a{}^{-1} - G_{\pi_\varphi Y} N_1 \pi_{\varphi1} {\delta}^{i j} {\partial}_{i}{\varphi_1}\,  {\partial}_{j}{\varphi_1}\,  a - \nonumber \\
&& G_{\varphi Y} N_1 {\delta}^{i j} {\partial}_{i}{\varphi_1}\,  {\partial}_{j}{\varphi_1}\,  \varphi_1 a - \frac{1}{6}\, G_{\pi_\varphi \pi_\varphi \pi_\varphi} N_1 \pi_{\varphi1}{}^{3} a{}^{3} - \frac{1}{6}\, G_{\varphi\varphi\varphi} N_1 \varphi_1{}^{3} a{}^{3} - \nonumber \\
&&\frac{1}{2}\, G_{\varphi \pi_\varphi \pi_\varphi} N_1 \pi_{\varphi1}{}^{2} a{}^{3} \varphi_1 - \frac{1}{2}\, G_{\pi_\varphi \pi_\varphi Y} N_0 {\delta}^{i j} {\partial}_{i}{\varphi_1}\,  {\partial}_{j}{\varphi_1}\,  \pi_{\varphi1}{}^{2} a - \frac{1}{2}\, G_{\varphi\varphi \pi_\varphi} N_1 \pi_{\varphi1} \varphi_1{}^{2} a{}^{3} - \nonumber \\
&&\frac{1}{2}\, G_{\varphi\varphi Y} N_0 {\delta}^{i j} {\partial}_{i}{\varphi_1}\,  {\partial}_{j}{\varphi_1}\,  \varphi1{}^{2} a - G_{\varphi \pi_\varphi Y} N_0 \pi_{\varphi1} {\delta}^{i j} {\partial}_{i}{\varphi_1}\,  {\partial}_{j}{\varphi_1}\,  \varphi_1 a - \nonumber \\
&&\frac{1}{24}\, G_{\pi_\varphi \pi_\varphi \pi_\varphi \pi_\varphi} N_0 \pi_{\varphi1}{}^{4} a{}^{3} - \frac{1}{24}\, G_{\varphi\varphi\varphi\varphi} N_0 \varphi_1{}^{4} a{}^{3} - \frac{1}{6}\, G_{\varphi\pi_\varphi\pi_\varphi\pi_\varphi} N_0 \pi_{\varphi1}{}^{3} a{}^{3} \varphi_1 - \nonumber \\
&&\frac{1}{6}\, G_{\varphi\varphi\varphi\pi_\varphi} N_0 \pi_{\varphi1} \varphi_1{}^{3} a{}^{3} - \frac{1}{4}\, G_{\varphi\varphi\pi_\varphi\pi_\varphi} N_0 \pi_{\varphi1}{}^{2} \varphi_1{}^{2} a{}^{3}.
\end{eqnarray}

Again, using inversion formulae mentioned in above section,
phase-space form of interaction Hamiltonian can be written in terms of
configuration-space variables.

\section{Extension to generalized higher-derivative models}\label{ExGal}

As we have shown above, for an arbitrary non-canonical scalar field,
it is possible to define a canonical conjugate momenta and the
Hamiltonian. In this section, we extend our method to generalize higher-derivative models. First, we extend the analysis to 
  G-Inflation \cite{Kobayashi2010, Kobayashi2011} model with
  generalized functions $P(X, \varphi)$ and $K(X, \varphi)$ .
  
Action for G-Inflation scalar field minimally coupled to gravity is
given by

\begin{equation}\label{galaction}
  \mathcal{S}_G = \int d^4 x \sqrt{-g} \left[\frac{1}{2 \kappa} R + P(X, \varphi) + K(X, \varphi) \Box \varphi \right].
\end{equation}

Directly obtaining the Hamiltonian for the above action is difficult
since it contains second order derivatives of the scalar
field. However, using the approach of Deffayet \emph{et al.}
\cite{Deffayet:2015qwa, Nandi:2015ogk}, action (\ref{galaction}) can
be re-written as
\begin{eqnarray}\label{GalileanAction}
\mathcal{S}_G &=& \int d^4 x \sqrt{-g} \left[\frac{1}{2 \kappa} R + P(X, \varphi) + K(X, \varphi)\, S \right] + \int d^4 x \,\lambda\, (S - \Box \varphi) \nonumber \\
&=& \int d^4 x \sqrt{-g} \left[\frac{1}{2 \kappa} R + P(X, \varphi) + K(X, \varphi)\, S \right]+ \int d^4x [\, \lambda S + \lambda~ g^{\mu \nu}~\Gamma^{\alpha}_{\mu \nu} ~\partial_{\alpha}{\varphi} + \nonumber \\
&& ~~~~~~~~g^{\mu \nu} \partial_{\mu}{\varphi} \partial_{\nu}\lambda + \lambda \, \partial_{\nu}{g^{\mu \nu}} \partial_{\mu}{\varphi}\,].
\end{eqnarray}

Linearizing the action costs two extra variables in
configuration-space, thus four extra phase-space variables. We have
discussed the issue in I \cite{Nandi:2015ogk} and proved that those
variables are not dynamic in nature and thus, there are no extra
degrees of freedom.

Since the action (\ref{GalileanAction}) is converted in terms of first
derivatives of fields, it is now possible to define momenta in terms
of time derivative of the fields. However, the action still contains
two generalized configuration-space variables $P(X, \varphi)$ and
$G(X, \varphi)$. Hence, using the above approach for generalized
non-canonical scalar field, a consistent perturbed Hamiltonian
formalism for generalized Galilean scalar field can be established.

 The approach can also be extended to any other
  higher-derivative models like Hordenski scalar field models,
  modified gravity models or an arbitrary higher-derivative
  theory. The above case is the quadratic and cubic parts of the
  Hordenski's scalar field model\cite{Horndeski1974}. In case of general Hordenski's
  scalar field model, action depends on $R_{\mu \nu}$, $\nabla_{\mu
    \nu}\varphi$, $\partial_{\mu}{\varphi}, g_{\mu \nu}$ and
  $\varphi$. The metric-part can be written in terms of extrinsic
  tensor, $K_{i j}$ and 3-Ricci scalar, $^{(3)}R$ with Lapse function,
  $N$ and Shift vector, $N^i$. Since the action contains
  $\nabla_{\mu \nu} \varphi$ instead of only $\Box \varphi$, above
  method for linearizing action will not work. Instead, we have to
  linearize the action by adding $$ S_H + \int d^4x~\lambda^{\mu
    \nu}\left(S_{\mu \nu} - \nabla_{\mu \nu}\varphi\right) $$ for
  general Hordenski's model \cite{Deffayet:2015qwa}. Hordenski's full
  action contains four unknown functions, $G_n(X, \varphi),\,
  n=2\cdots5$, hence, using the approach for non-canonical
  scalar field, we can also deal with Hamiltonian formulation for
  Hordenski's theory.

  By using the same argument and method, it is possible to obtain
  Hamiltonian beyond Hordenski's model, i.e., for any higher-order
  derivative gravity models with arbitrary functions. To deal with
  arbitrary functions, using the above approach for generalized
  non-canonical scalar fields, we can define a new and unique
  phase-space variable(s) and write the corresponding canonical
  Hamiltonian of the system and inversion formulae can be used to
  invert from phase-space variable to configuration-space variable and
  vice-versa. Once the Hamiltonian is obtained, we can use the
  approach in I to get the consistent Hamiltonian formalism of
  cosmological perturbation at any perturbed order for the specific
  model.

  Hamiltonian approach in chapter (\ref{Chap.3}) \cite{Nandi:2015ogk}, is independent of how we construct the
  Hamiltonian and is readily applicable once we successfully write
  down a consistent Hamiltonian for a specific model. Hence, the
  Hamiltonian approach for higher derivative theory is not restricted
  only by the Deffayet's approach \cite{Deffayet:2015qwa}. Recently,
  Langlois and Noui \cite{Langlois:2015cwa, Langlois:2015skt} have
  also provided a simpler way to obtain Hamiltonian for higher
  derivative theory and the Hamiltonian approach for perturbation can
  also be extended to these models. 

\section{Conclusions and discussions}\label{Conclu}

In this chapter, we have explicitly provided the Hamiltonian formulation
of cosmological perturbation theory for generalized non-canonical
scalar fields. The following procedure was adopted: first we provided
the essential information regarding gauge-choices and related
gauge-invariant quantities. Next, we performed Legendre transformation
for the generalized non-canonical scalar fields and showed that, since
($\varphi^\prime \rightarrow \pi_\varphi$) transformation is not
possible, Hamiltonian for generalized non-canonical scalar fields
cannot be obtained by using conventional method.

We introduced a new generalized phase-space variable $G(\pi_\varphi,
\gamma, Y, \varphi)$ that is unique for different non-canonical scalar
fields and obtained Hamiltonian of a non-canonical scalar field. We
showed that, this is the simplest and efficient way to obtain the
Hamiltonian \cite{Nandi:2016pfr}. We extended the approach in (\ref{Chap.3}) \cite{Nandi:2015ogk} to generalized
non-canonical scalar fields in the flat-slicing that doesn't lead to
gauge-artifacts and obtained perturbed Hamilton's equations in terms
of phase-space variables. In parallel, we also extended Langlois'
approach to generalized non-canonical scalar field and showed that
both approaches lead to identical speed of sound.

In order to compare Hamiltonian approach with Lagrangian approach,
Hamilton's equations are to be converted to Euler-Lagrange equation
and in doing so, we provided explicit forms of $G(\pi_\varphi, \gamma,
Y, \varphi)$ for different non-canonical scalar field models and
showed that the Hamiltonian formulation is consistent.

Since we do not know how, in general, phase-space derivatives of
$G(\pi_\varphi, \gamma, Y, \varphi)$ transform to configuration-space
derivatives, hence for an arbitrary field, it is not possible to
directly invert the generalized phase-space Hamilton's equations to
Euler-Lagrange equations. In order to overcome this, we prescribed an
inversion mechanism from generalized phase-space variables to
generalized configuration-space variables (and vice versa) and showed
that all generalized phase-space equations lead to consistent E-L
equations. We also retrieved the conventional form of speed of sound
in configuration-space.

We also obtained the interaction Hamiltonian in terms of phase-space
variables for generalized non-canonical scalar field at third and
fourth order of perturbation for scalar perturbations. These can also
be expressed in terms of $(\varphi^\prime,~ \varphi)$ using the
general inversion formulae. Note that, we considered only the first
order scalar perturbations. Vector or tensor modes can similarly be
implemented by considering $\delta\gamma_{i j} \neq 0$ and decomposing
the metric using vector and tensor modes. Hamiltonian as well as
equations of motions for vector or tensor modes also change
accordingly. For the linear order, three modes decouple and
$\delta\pi^{i j}$ can also be decomposed as ${\delta\pi_{S}^{i j} +
  \delta\pi_{V}^{ij} + \delta\pi_{T}^{i j}}$, so the equations of
motion. However, for higher order of perturbations, modes are highly
coupled to each other, hence similar decomposition is not possible.

Finally, we briefly discussed the Hamiltonian formulation for
generalized higher derivative scalar fields. The method
is not restricted to gravity related models, it can also be applied to
any other models where the Lagrangian is not specified properly.

Throughout the chapter, we carried out the method by assuming that the
field allows Legendre transformation, which, most of the known models
follow. However, if a certain model is specified in such a way that
$\varphi^\prime$ cannot be written in terms of $\pi_\varphi$ or the
mapping is one-to-many, then the current formalism cannot be applied
to obtain a unique form of $G(\pi_\varphi, \gamma, Y, \varphi)$ and
hence, for those kind of models, this approach is not applicable.

\chapter{Primordial magnetogenesis and vector Galileon}\label{Chap.5}

\section{Introduction}\label{Chap5.Intro}

Since the early days of quantum electrodynamics, higher derivative field theories \cite{RevModPhys.20.40,Pais:1950za} have been proposed to improve the divergence structure. However, higher-derivative theories suffer 
from Ostrogradsky instability \cite{Ostro,Woodard:2015zca}. These negative energy states can be traded by negative norm states leading to non-unitary theories \cite{Barth:1983hb,Simon:1990ic,Hawking:2001yt}. 

Recently, it has been realized that it is possible to construct scalar field theories whose action can have higher derivatives, however, the equations of motion are still 
second order. These are referred to as Galilean models and do not suffer from Ostrogradsky instabilities \cite{Horndeski:1974wa,Nicolis2008,Deffayet2009,Deffayet:2015qwa,Nandi:2015ogk}. Scalar Galilean theories have a lot in common with Lovelock theories of gravity \cite{Lovelock:1971yv,Lovelock:1972vz}. Lovelock theories are obtained by imposing three conditions --- gravity must be described by metric and its derivatives, diffeomorphism invariance and equations of motion be quasi-linear. Using these conditions, it can be shown that Einstein's gravity is unique in 4-D. In higher dimensions, $R^2 - 4 R_{ab} R^{ab} + R_{abcd} R^{abcd}$ also lead to quasi-linear equations of motions. Lovelock extensions of Einstein gravity are shown to be free of ghost  and  evade problems of Unitarity \cite{Zumino:1985dp,Padmanabhan:2013xyr}. 

It is natural to ask: \emph{Can we construct a higher derivative Electromagnetic (EM) field action by demanding following three conditions:  theory be described by vector potential $A^{\mu}$ and its derivatives, Gauge invariance is satisfied and equations of motion be second order?} In this chapter, we explicitly construct a higher-derivative EM action satisfying the above conditions. We show that the higher-derivative terms in the action vanishes in the flat space-time and hence, do not have any observable consequences in terrestrial experiments \cite{Deffayet:2013tca}. We provide one concrete observational consequence and show that our model provides an elegant solution to the generation of primordial magnetic field during inflation.

The standard EM action in 4-D space-time, 
\begin{equation}
\label{eq:SEM}
\mathcal{S}_{SEM} = - \int \frac{d^4x}{4} \sqrt{-g} \,F_{\mu \nu} F^{\mu \nu}; F_{\mu \nu} = \partial_{\mu}A_\nu - \partial_{\nu}A_\mu
\end{equation}
is conformally invariant \cite{Birrell:1982ix}. Hence, the equations of motion of the magnetic field in FRW space-time 
are time independent \cite{Grasso:2000wj,Widrow:2002ud,Subramanian:2009fu,Widrow:2011hs}: 

\begin{equation}
\left(\partial^2_\eta - \nabla^2 \right) (a^2 {\bf B}) = 0.
\end{equation}
Thus, to generate magnetic field during inflation, it is necessary to break conformal 
invariance of the EM action. Starting from Turner and Widrow \cite{Turner:1987bw,Dolgov:1993vg,Gasperini:1995dh,Davidson:1996rw,Calzetta:1997ku,Prokopec:2001nc,Giovannini:2000dj,Durrer:2010mq,Byrnes:2011aa,Atmjeet:2013yta,Atmjeet:2014cxa,Sriramkumar:2015yza}, several authors have suggested many ways to break the conformal invariance of the electromagnetic field by introducing  
(i) coupling of the electromagnetic field  with the Ricci/Riemann vectors, 
(ii) non-minimal coupling of the electromagnetic field with scalar/axion/fermionic field and (iii) compactifaction from higher dimensional space-time. Here, we show that the 
higher derivative EM action generates the required seed magnetic fields at small length scales. The advantages of our model compared to earlier approaches is that the new EM action do not have any non-minimal coupling with other scalar/axion/fermionic fields, and do not require new physics.

\section{Modified Electromagnetic action}
We consider the following modification to the standard EM action (\ref{eq:SEM}),

\begin{eqnarray}\label{VecGalAction}
\mathcal{S}_{VG} = \lambda\,\int \,d^4 x\, \sqrt{-g}\, \epsilon^{\alpha \gamma \nu} \epsilon^{\mu \eta \beta}\, \nabla_{\alpha \beta}A_\gamma \, \nabla_{\mu \nu}A_\eta
\end{eqnarray}
where $\nabla$ is covariant derivative, $\lambda$ --- whose dimension is inverse mass square --- is the coupling constant that determines the effect of the higher-derivative terms in the propagation of the EM field. In 4-D space-time, the product of two anti-symmetric epsilon tensors is given by 
\begin{equation} 
\label{eq:epsilontensor}
\epsilon^{\alpha\gamma\nu}\,\epsilon^{\mu \eta\beta} = \left|\begin{array}{ccc}
g^{\alpha\mu}& g^{\alpha\eta} & g^{\alpha\beta}\\
g^{\gamma\mu} & g^{\gamma\eta} & g^{\gamma\beta}\\
g^{\nu\mu} & g^{\nu\eta} & g^{\nu\beta}\\
\end{array}\right|.
\end{equation}
Before we proceed, it is important to understand how the above action behaves in the flat Minkowski space-time: First, the contraction between the first and third indices of the epsilon tensor and the derivative of the vector potential $\epsilon^{\alpha \gamma \nu} \epsilon^{\mu \eta \beta} \nabla_{\alpha\beta} A_{\gamma} \nabla_{\mu \nu} A_{\eta}$ ensures no higher derivative terms in the equations of motion. Second, since the metric is constant, ${\cal S}_{VG}$ is a non-dynamical term (boundary term or vanishes identically) and does not affect the equations of motion. 
This is the no-go theorem by Deffayet et al \cite{Deffayet:2013tca}. 
Third, the covariant derivatives $\nabla$ are replaced by partial derivatives $\partial$. Hence, in flat space-time, contraction between first two indices of the epsilon tensor and derivative of the vector potential $\epsilon^{\alpha \gamma \delta} \partial_{\alpha \delta}A_{\gamma}$ preserves gauge-invariance.

However, in curved space-time,  action (\ref{VecGalAction}) is not a boundary term. The product of epsilon tensors (\ref{eq:epsilontensor}) is a 
function of metric, thus the above action in curved space-time can not be written as a boundary term. Hence, the above action modifies the equations of 
motion of the electromagnetic field. Also, covariant derivatives of the vector potential lead to extra connection terms in the action which are not gauge-invariant. Thus, these additional terms can also lead to higher-derivative terms in the equation of motion.

The fact that the action (\ref{VecGalAction}) is gauge-invariant in flat space-time implies that we need to include non-minimal coupling terms of the electromagnetic potential and its derivatives with the Riemann/Ricci tensors and 
Ricci scalars. However, {\it ab initio} we do not know the non-minimal coupling terms 
and, we need to consider all possible terms. In Appendix (\ref{app:model construction}), we have listed twelve possible non-minimal coupling terms. Thus, the modification to the electromagnetic action is:
\begin{eqnarray}\label{FullAction}
\mathcal{S}_{VEC} = \mathcal{S}_{VG} + \lambda \sum_{i=1}^{12} \mathcal{S}_i.
\end{eqnarray}
where $S_i$'s for $i=1, \cdots, 12$ are given in Appendix (\ref{app:model construction}), $D_i$'s are the twelve unknown coefficients and are dimensionless 
and $\lambda$ is the coupling constant in action (\ref{VecGalAction}). 

Demanding that the above action is gauge-invariant in curved space-time and that 
the equations of motion do not contain higher order terms, the coefficients $D_i$'s can be fixed uniquely in-terms of $D_1$ (See Appendix \ref{app:Gaugefixing}). Thus, the model has one unknown coupling parameter $\lambda$ that can be fixed from observations. 
The modified electromagnetic action is
\begin{eqnarray}\label{FullEM}
\mathcal{S}_{EM} = \int d^4x \sqrt{-g} \left(-\frac{1}{4}\,F_{\mu \nu} F^{\mu \nu} + \,A_\mu J^\mu\right) +\mathcal{S}_{VEC}~~
\end{eqnarray} 
where the first two terms correspond to the standard electromagnetic action. 

This is the first key result of this chapter, regarding which we would like to stress the following points. First, in the flat space-time, $\mathcal{S}_{VEC}$ becomes zero, thus the above reduces to standard electromagnetic action. Second, from the equation of motion of $A_0$ in the FRW background, 
\begin{equation}
\label{eq:FRW}
ds^2 = - N^2 d\eta^2 + a^2 d{\bf x}^2
\end{equation}
where $N(\eta)$ is an arbitrary Lapse function, the scalar potential is given by:
\begin{eqnarray}\label{Permit}
\Phi \equiv - A_0 = \frac{1}{4\pi\left(1 - 4 \,D\,H^2\right)}\,\frac{\rho(\vec{r_0})}{r}
\end{eqnarray}
where $D \equiv \lambda \,(1 + 3\, D_1)$. Thus, the effect of action (\ref{FullAction}) is to change the permitivity to 
$\epsilon \equiv (1 - 4 \,D\,H^2)$ where $H$ is the Hubble constant. The electrostatic potential still goes as inverse of the distance. Permitivity being 
positive provides a condition on the value of $D$. If $D$ is negative all values are allowed, however, if $D$ is positive, $4 D H^2 < 1$. Thus, the modified action does not have any observable consequence in the terrestrial experiments. However, as we will show in the rest of this chapter, the above modified action has important consequence in the early Universe.

\section{Breaking of conformal invariance and consequences}

Having discussed the model and the effect on the Coloumb potential, let us now 
look at the effects in the early Universe. Since the FRW background is conformally flat, the background gravitational field does not produce particles in the case of standard electromagnetic action (\ref{eq:SEM}) \cite{Birrell:1982ix}. However, the modified action (\ref{FullAction}) explicitly breaks conformal invariance. 
The modified action can have significant contribution in the early Universe,  thus leading to production of magnetic fields. Since we are interested in particle production, we consider only (\ref{FullAction}) and drop the standard EM action.

Since action (\ref{FullAction}) is gauge-invariant, we choose 
Coulomb gauge $(A_0 = 0)$ and obtain all the physical quantities. 
In the FRW background (\ref{eq:FRW}), action (\ref{FullAction}) becomes:
\begin{eqnarray}
\label{TheModelFRW}
\mathcal{S}_{VEC} =  D \, \int\,d^4x \Big[-2 \,\frac{a^\prime{}^2}{N^3\, a}\,A_i^\prime{}^2 + 2\,  \frac{a^{\prime\prime}}{N\,a^2}\, \left(\partial_i A_j\right)^2  - 2\,\frac{a^{\prime}\,N^\prime}{N^2\,a^2}\,\left(\partial_i A_j\right)^2\Big] \, .
\end{eqnarray} 
Varying the above action with respect to $A_i$ and setting $N(\eta) = a (\eta)$, 
leads to the following equations of motion:
\begin{eqnarray}\label{EoMASpace}
A_i^{\prime\prime} + 2\frac{J^\prime}{J}\,A_i^\prime  - \frac{\left(aJ\right)^\prime}{\left(aJ\right)^2} \nabla^2 A_i = 0,~~ \mbox{where}~~J \equiv \frac{\mathcal{H}}{a}.
\end{eqnarray}
Fourier decomposing the vector potential $A_i$ \cite{Grasso:2000wj,Widrow:2002ud,Subramanian:2009fu,Widrow:2011hs}, we get 
\begin{eqnarray}\label{fourierd}
\hat{A}_i (\eta, {\bf x}) = \sqrt{4\pi} \int \frac{d^3 {\bf k}}{(2\pi)^{3/2}} \sum_{\Lambda=1}^{2}
\epsilon_{\Lambda i}({\bf k}) \Big[\hat{b}_{\bf k}^\Lambda A_k(\eta)e^{i{\bf k.x}} + \hat{b}_{\bf k}^{\Lambda\dagger} A^*_k(\eta) e^{- i{\bf k.x}}\Big] \, ,
\end{eqnarray}
where $\Lambda$ corresponds to two orthonormal transverse polarizations and $\epsilon_{\Lambda i}$ are the polarization vectors. Substituting (\ref{fourierd}) in (\ref{EoMASpace}), we get 
\begin{eqnarray}\label{EoMAK}
A_k^{\prime\prime} + 2\frac{J^\prime}{J}\,A_k^\prime + k^2 \frac{\left(aJ\right)^\prime}{\left(aJ\right)^2}  A_k = 0.
\end{eqnarray}
By fixing the initial state of the electromagnetic field, we can evaluate the vector potential at a later times. 

To compare with the observations, we need to evaluate the energy density \cite{Durrer:2010mq,Byrnes:2011aa,Atmjeet:2013yta,Atmjeet:2014cxa}. $0-0$ component of the energy momentum tensor $T_{\mu\nu}$ in the FRW background (\ref{eq:FRW}) is 
$$ T_{0 0} = - \frac{N^2}{a^3} \frac{\delta \mathcal{L}}{\delta N} \, .$$
The energy density in conformal coordinates is:
\begin{eqnarray}
\rho \equiv - T^{0}_{0} 
=  - 6\, D \,\frac{\mathcal{H}^2}{a^6}\, \delta^{i j} A_i^\prime A_j^\prime +   4\, D\, \frac{\mathcal{H}^2}{a^6}\, \delta^{i k} \delta^{j l}\, \partial_i A_j \,\partial_k A_l  + 4\,D\,\frac{\mathcal{H}}{a^6}\, \delta^{i j}\, A_i{}^\prime\, \nabla^2 A_j 
\end{eqnarray}
The first term is the energy density of the Electric field $(\rho_E)$. Second and third terms are the energy densities of the magnetic field $(\rho_B)$ and $(\rho_{B.B^\prime})$, respectively:

Using the decomposition (\ref{fourierd}), the electric, magnetic part of the 
perturbation spectrum per lograthmic interval can be written as: 
\begin{eqnarray}\label{SpectraB}
&& \mathcal{P}_{B}(k) \equiv  \frac{\mbox{d}}{\mbox{dln}k} \langle 0|\hat{\rho}_{B^2}|0 \rangle = \frac{16\, D\, \mathcal{H}^2}{\pi} \frac{k^5}{a^6} \left|A_k\right|^2 \\
\label{SpectraE}
&&  \mathcal{P}_{E}(k) \equiv  \frac{\mbox{d}}{\mbox{dln}k} \langle0|\hat{\rho}_{E^2}|0\rangle =  -\frac{24 \,D\, \mathcal{H}^2}{\pi} \frac{k^3}{a^6} \left|A_k^\prime\right|^2 \\
\label{SpectraBB}
&& \mathcal{P}_{_{B.B^\prime}}(k) \equiv  \frac{\mbox{d}}{\mbox{dln}k} \langle0|\hat{\rho}_{_{B.B^\prime}}|0\rangle = -\frac{16\, D\, \mathcal{H}}{\pi} \frac{k^5}{a^6}  A_k^\prime A_k^*~~~~.
\end{eqnarray}
It is important to note the following:  In the standard electromagnetic action, the energy density is always positive and can be written as $\left( B_iB^i + E_i E^i \right)$. However, in our case, it is given by $ D \left( H^2 \, B_iB^i - H^2 \, E_i E^i - H B_i^\prime B_i \right)$. During most part of the evolution of the Universe, electrical 
conductivity is high \cite{1990-Kolb.Turner-Book}, hence, electric fields decay and 
do not contribute to the energy density. This implies that $D > 0$. 

Until now the analysis has been general and can be applied at any stage of the Universe evolution. In the rest of this chapter, we calculate the energy density of the electromagnetic field during inflation. We assume that the inflation is driven a 
scalar field and that the energy density of the electromagnetic do not contribute 
to the accelerated expansion during inflation. In other words, we treat the electromagnetic field as a test field and obtain the power spectrum. 

Let us first consider power-law inflation i. e. $a(t) = a_0 t^p; a(\eta) = a_0 \left(-\eta\right)^{1+\beta}$, where $p > 1; \beta \leq - 2$. Note that $\beta = -2$ corresponds to de Sitter.  Substituting $a(\eta)$ in (\ref{EoMAK}), we have: 
\begin{eqnarray}
\mathcal{A}_k^{\prime\prime} + \left(c_s^2 k^2 - \frac{(2+\beta)(3+\beta)}{\eta^2}\right)\mathcal{A}_k = 0
\end{eqnarray}
where  $\mathcal{A}_k \equiv J(\eta)\,A_k$ and  $c_s \equiv -\frac{1}{1+\beta} >  0$. This is the second key result of this chapter. The electromagnetic perturbations do not propagate at the speed of light. This is not unusual, as the scalar perturbations in Galileon inflation also propagate less than the speed of light \cite{Kobayashi2010,Kobayashi2011,Unnikrishnan2014}, however, the two speeds are not the same.

During power-law $c_s$ is a constant and the solution to the above differential equation is given by:
\begin{eqnarray}\label{GenSol}
\mathcal{A}_k = \sqrt{-\eta}\left[C_1 J_{\beta+\frac{5}{2}}(-c_sk\eta) + C_2 J_{-\beta-\frac{5}{2}}(-c_sk\eta)\right].~~
\end{eqnarray}
Imposing the initial condition in the sub-Hubble scales ($-k\eta \rightarrow \infty$)
that the field is in vacuum state corresponds to  $\mathcal{A}_k \to \frac{1}{\sqrt{2c_s k}}e^{-ic_sk\eta}$. This leads to:
\begin{equation}
C_1 = \sqrt{\frac{\pi}{4}}\frac{e^{i(\beta+1)\frac{\pi}{2}}}{\mbox{cos}\left(\beta\pi\right)},~~ C_2 = \sqrt{\frac{\pi}{4}}\frac{e^{-i\beta\frac{\pi}{2}}}{\mbox{cos}\left(\beta\pi\right)}.
\end{equation}
It is important to note that for $\beta \leq -5/2$, $J_{\beta + 5/2}$ dominates, however,  $J_{- (\beta + 5/2)}$ dominates for $\beta \geq -5/2$.

From (\ref{GenSol}), we can obtain the spectra of the energy-densities (\ref{SpectraB}, \ref{SpectraE}) and (\ref{SpectraBB}) at the crossing of the sound horizon ($c_s k_* = a_* H_* = \frac{1+\beta}{\eta_*}$). The magnetic part of the energy density is (see Appendix \ref{app:otherEner}):
	\begin{eqnarray}\label{PBFinal}
	\mathcal{P}_B &=& \frac{16 D}{\pi\,c_s^{11 + 2\beta}} \mathcal{F}_1(\beta)\,H_*^4\,\left( \frac{k}{k_*}\right)^{10+2\beta} ~~\mbox{for}~~\beta < -\frac{5}{2},~~ \mathcal{F}_1(\beta) = \frac{|C_1|^2}{2^{2\beta+5}\left(\Gamma(\beta+ 7/2)\right)^2} \nonumber \\
	&=& \frac{16 D}{\pi\,c_s^{1-2\beta}} \mathcal{F}_2(\beta)\,H_*^4 \left(\frac{k}{k_*}\right)^{-2\beta} ~~~~\,\mbox{for}~~\beta > -\frac{5}{2},~~\mathcal{F}_2(\beta) = \frac{|C_2|^2}{2^{-2\beta-5}\left(\Gamma(-\beta- 3/2)\right)^2} ~~~~~~~~~~
	\end{eqnarray}

This is the third key result regarding which we would like to stress the following points: First, for $\beta = -5$, the magnetic spectra is scale invariant. However, 
for $\beta = -5$, the electric field energy density diverges. Hence, $\beta = -5$ is ruled out as that will lead to negative energy density (since $D > 0$). Second, for $\beta \simeq -2$, the spectra is highly blue-titled. To go about understanding the consequence of the same, the energy spectra during the slow-roll inflation \cite{Bassett:2005xm} is given by (see Appendix \ref{app:slow-roll}):
\begin{eqnarray}
\label{eq:PBGenerated}
\mathcal{P}_B = \frac{8\,D}{\pi\,c_s^5} \,H_*^4 \left(\frac{k}{k_*}\right)^4 \, ;
&& \qquad c_s = 1 - \epsilon_1
\end{eqnarray}
where $\epsilon_1$ is the first slow-roll parameter \cite{Bassett:2005xm}. It is interesting to note that at the start of inflation $\epsilon_1 \ll 1$ and the speed of the EM perturbations is close to unity. However, 
during inflation, as $\epsilon_1$  increases,
the speed of perturbations decrease, hence, leading to larger value of the energy 
spectrum. The condition for the exit of inflation is $\epsilon_1 = 1$ and, the 
speed of these perturbations vanishes and the above analysis fails at $\epsilon_1 = 1$.  Our model predicts a highly blue-tilted magnetic and electric spectrum \cite{Kahniashvili:2010wm}. Finally, it is important to 
note that the power-spectrum in our model has the same blue-tilt as that of the vacuum polarization power-spectrum in the standard electromagnetic action. However, the power-spectrum evaluated here is due to particle production during inflation and depends on $D$ and $c_s$ \cite{Turner:1987bw,Birrell:1982ix}. To fix these values and compare with observations, we need to evolve 
magnetic fields to the current epoch. 

\section{Post inflationary evolution}

Reheating is expected to convert the energy in inflaton field to radiation \cite{Bassett:2005xm} and Universe for most cosmic history has been good conductor ($\sigma \ll 1$).  Assuming instantaneous reheating, the equation of motion of $A_i$ for large wavelength modes is \cite{Grasso:2000wj,Widrow:2002ud,Subramanian:2009fu,Widrow:2011hs}:
\begin{eqnarray}
\ddot{A_i} + \frac{ \sigma + H\left(1 - 8 D \dot{H} -  4 D H^2\right)}{1 -  4 D H^2}\,\dot{A_i} = 0
\end{eqnarray}
where $J^i  = - g^{i j} \sigma \dot{A_j}$. At late times, using Eq.~(\ref{Permit}), we have $D H^2 \ll 1$. Hence, the above equation reduces to:
\begin{eqnarray}
\ddot{A_i} +  \sigma\,  \dot{A_i} = 0 \quad \Rightarrow \quad A_i = C_1({\bf x}) t^{-\sigma\,t} + C_2({\bf x}) \, ,
\end{eqnarray} 
which is same as standard EM action~(\ref{eq:SEM}). Thus, the vector potential $A_i$ is constant in time implying that the electric field vanishes and magnetic field decays as $a^{-2}$. During Radiation-dominated era, $H \propto a^{-2}$, and the energy density corresponding to ${\cal S}_{VEC}$ decays as $a^{-6}$. However, the energy density of the standard EM action goes as $a^{-4}$. At late times, only EM action (\ref{eq:SEM}) contributes.

\section{Constraints from observations}

To compare whether the generated magnetic field (\ref{eq:PBGenerated}) 
has the right magnitude needed to seed galactic fields, we need to compare $\rho_B$ 
with radiation background energy density $\rho_{\gamma} \propto T^4$. This is because, the magnetic field generated during inflation evolve as $\rho_{B} \propto a^{-4}$ \cite{Grasso:2000wj,Widrow:2002ud,Subramanian:2009fu,Widrow:2011hs,Turner:1987bw,Bassett:2005xm} which is same as $\rho_\gamma$. Hence, the dimensionless quantity $r \equiv \rho_{B}/\rho_\gamma$ remains approximately constant and provides a convenient method to constrain the primordial magnetic field \cite{Turner:1987bw}.  From Eq.~(\ref{eq:PBGenerated}), we get,
\begin{eqnarray}
r \sim \frac{D}{c_s} 10^{-104}\, \lambda_{\text Mpc}^{-4} {\rm eV}^2 \, .
\end{eqnarray}    
Note that $D$ has dimensions of inverse mass square. The field strength required to seed galactic fields with an efficient galactic dynamo translates to $r \sim 10^{-34}$ \cite{Grasso:2000wj,Widrow:2002ud,Subramanian:2009fu,Widrow:2011hs,Turner:1987bw}. 
For length scales of $1 {\text Mpc}$, this translates to $D/c_s \sim 10^{70}$. Using the fact that permitivity has to be positive, from Eq. (\ref{Permit}), we get 
$D \sim 10^{-46}$ eV$^{-2}$. Thus, near the exit of inflation, $c_s \sim 10^{-116}$. 

This is the last key result of this chapter and we would like the stress the following points: First, at the early epoch of inflation $\epsilon_1 \ll 1$, implying that, $c_s \sim 1$, Hence, the energy density of the magnetic fields generated at the early epoch of inflation is tiny and the magnetic fields, present at decoupling and homogeneous on scales larger than the horizon at that time is much less than the current limit of $B \leq 10 \, nG$ \cite{Grasso:2000wj,Widrow:2002ud,Subramanian:2009fu,Widrow:2011hs}. Second, appreciable seed magnetic fields are generated only close to the exit of inflation. Thus, our model naturally generates appreciable magnetic field at Mpc scale as the modes that leave  the horizon close to the exit of inflation re-enter early during radiation epoch and an efficient dynamo mechanism can generate the observed magnetic field. Thus, our model generates appreciable magnetic fields \emph{only} for smaller wavelength modes. This is the key unique feature of our model compared other proposed models for magnetogenesis. 

\section{Conclusions and discussions}

In this chapter, by demanding that the theory be described by vector potential $A^{\mu}$ and its derivatives, Gauge invariance be satisfied, and equations of motion be linear in second derivatives of vector potential, we have constructed a higher derivative electromagnetic action that does not have ghosts and preserve gauge invariance. We have shown that the higher order terms vanish in the flat space-time and hence, consistent with the no-go theorem by Deffayet et al \cite{Deffayet:2013tca}. 

We have shown that the model breaks conformal invariance and hence, can generate magnetic field during inflation, the modes generated propagate less than the speed of light and the speed of propagation depends on the slow-roll parameter (\ref{eq:PBGenerated}). We have explicitly shown that our model generates appreciable magnetic field for small wavelength modes ($\sim$ Mpc) while the model generates tiny magnetic fields for large wavelength modes. This is one of the unique feature of our model compared other models that generate magnetic field during inflation. The energy density of the magnetic field is appreciable only at the end of inflation and hence, our model does not lead to any back-reaction. 

For the inflation to end, $\epsilon_1 = 1$. The model proposed can generate appreciable magnetic field near the exit and, in principle, can provide a dynamical mechanism for the exit of inflation. This precise exit calculation is under investigation.

The magnetic field spectra generated in our model is blue tilted. This should be contrasted with other models where the spectra can be fine tuned \cite{Durrer:2010mq,Byrnes:2011aa,Atmjeet:2013yta,Atmjeet:2014cxa}. Recently, Kahniashvili et al \cite{Kahniashvili:2010wm} have done a detailed analysis to place constraints on the primordial magnetic field 
from the cosmological data including models that have blue tilt. It is interesting to investigate how the mass dispersion $\sigma(M,z)$ behaves and its effect on the structure formation. This is currently under investigation.

\pagestyle{fancy}
\chapter{Conclusions and future outlook}\label{Chap.Conc}

One of the main aim of the thesis is to use Hamiltonian analysis of cosmological perturbations to improve theoretical predictions about the early Universe. As mentioned earlier, there are several formalisms and mathematically they always do not lead to same equations of motion \cite{Appignani2010a}. Therefore, we not only extensively studied these formalisms, but also compared with each other to analyze the inconsistency of each formalisms. We learnt that in order to constrain models of the early Universe, we need to evaluate higher order correlation functions of curvature perturbations and match them with observations. To evaluate higher order correlations, we need to to go beyond linear order perturbations and evaluate interaction Hamiltonian of the system. There exists many approaches to calculate interaction Hamiltonian, however, Hamiltonian framework has never been used to evaluate this. In this thesis, we obtained a consistent Hamiltonian framework for cosmological perturbation theory and showed that, this framework is not only simple, robust and efficient in calculating interaction Hamiltonian but also provides extra information about the true degrees of freedom of the system.

Second aim of the thesis is to extend the standard electromagnetic theory by including higher derivative terms in the action such that the equations of motion are second order and gauge-invariance is preserved. Recently, Deffayet et al. \cite{Deffayet:2013tca} proved a no-go theorem that states that constructing such a Lagrangian in flat Minkowski space-time is not possible. We showed that it is possible to construct such Lagrangian in curved space-time. The U(1) gauge symmetry is preserved, however, conformal symmetry is broken. Hence, it can lead to generation of magnetic fields in early Universe.

This chapter is divided into two parts --- the first part discusses the main conclusions of the thesis, more specifically, chapter-wise conclusions, and the second part looks at the future outlook of the thesis. 

In chapter (\ref{Chap.2}), we, first, focused on the equivalence of two approaches --- order-by-order gravity equations approach and reduced action approach \cite{Nandi:2015tha}. In this chapter, we established that, by choosing all perturbations consistently, both the approaches lead to same equations of motion for canonical and non-canonical scalar fields. We also proved that the equivalence between the two approaches are generally true for any gravity model at any order of perturbations. While comparing both the approaches, we obtained `constraint consistency' condition where the constrained nature of Lapse function and shift vector are studied. We showed that, to obtain the reduced equation(s), perturbed Lapse function and perturbed Shift vector have to act as `constraints' and the equations of motion should not contain time derivatives of Lapse function and Shift vector. In other words, Lapse function and Shift vector should not behave as dynamical variables. We then applied `constraint consistency' to higher derivative theories of gravity. One common problem with higher derivative theories is that they suffer from Ostrogradsky's instabilities, i.e., Hamiltonian is not bounded . We showed that the models with `constraint inconsistency' also suffers from Ostrogradsky's instabilities, however, we also found that, there exist some models where constraint consistency is satisfied but they still suffer from instabilities. In other words, the action can be reduced to a single variable form but the reduced equation contains higher time derivative.

As mentioned earlier, in order to match theory with observations, we need to know the Hamiltonian of the system for the canonical quantization of the reduced field. Till now, the approach has been to start with a Lagrangian and at every order convert it to Hamiltonian. However, a proper Hamiltonian framework has never been used before to obtain this. Langlois \cite{Langlois1994} first provided Hamiltonian formalism for canonical scalar fields at linear order perturbation. However, the approach is very difficult to extend for any generalized field at any order of perturbations. The main problem of Hamiltonian formalism is to obtain gauge-invariance in Hamiltonian framework. In chapter (\ref{Chap.3}), we constructed generalized Hamiltonian formalism for cosmological perturbations by successfully incorporating gauge conditions as constraints. We first applied the mechanism for canonical scalar fields and then  applied for Galilean scalar fields \cite{Nandi:2015ogk}. An interesting property of Galileons is that, the action contains second order time derivative of the field, however, equations of motion are still second order. Using consistent Hamiltonian framework, we showed that,  in case of Galileons, there are no extra degrees of freedom even though the action contains higher order time derivative. We also showed that, our method is model independent, can be applied to any types of fields at any order of perturbations. Finally, we obtained interaction Hamiltonian without any approximations unlike the previous method and showed that, our approach is simple, robust and efficient than the earlier approach.

In chapter (\ref{Chap.3}), we first provided a generic Hamiltonian formulation for arbitrary fields for cosmological perturbations and showed that, it can even be applied for any order of perturbations. However, the approach can be applied only if the Hamiltonian of the system is known. In certain non-canonical scalar fields, it may be possible but, due to non-inversion of Legendre transformation, Hamiltonian of generalized non-canonical scalar fields cannot be obtained. In chapter (\ref{chap.4}), we showed that, we can obtain Hamiltonian of generalized non-canonical scalar fields by introducing a new phase-space variable \cite{Nandi:2016pfr}. We then  applied our Hamiltonian approach discussed in chapter (\ref{Chap.3}) \cite{Nandi:2015ogk} to non-canonical scalar fields and showed that, the equations of motion are consistent. While studying perturbation theory for non-canonical scalar fields, we also discovered that, due to the new phase-space variable, we obtained a new expression of speed of sound in terms of phase-space variables. We also provided the inversion relations between phase-space variables (and its derivatives) and configuration-space variables (and its derivatives) to compare and contrast general Hamilton's equations to Euler-Lagrange equations. We provided third and fourth order interaction Hamiltonian for non-canonical scalar fields without any approximations. Finally, using the technique to write down the Hamiltonian for models where Legendre transformation in non-invertible, we extended our perturbed Hamiltonian approach to generalized higher derivative theories like Horndeski model.

In chapter (\ref{Chap.5}), we focused on a different aspect of early Universe: \emph{primordial magnetogenesis} using higher derivative theories \cite{Nandi:2017ajk}. The origin of $\mu$Gauss strength magnetic fields in the galaxies and clusters of galaxies is one of the long-standing problems in astrophysics and cosmology. These fields could have arisen from the dynamo amplification of seed fields. However, the dynamo mechanism needs the seed magnetic fields. The key question is whether these seed magnetic fields were generated in the early Universe. The origin and detection of these seed magnetic fields is a subject of intense study. Cosmological inflation has been successful in providing a causal mechanism for the generation of the large scale density perturbations and temperature fluctuations in the Cosmic microwave background. However, inflation can not provide the necessary amplitude for the seed magnetic field. One reason is that the 4-dimensional electromagnetic action is conformally invariant. Several mechanisms have been proposed in the literature to break the conformal invariance of the electromagnetic action. All these mechanisms either require coupling of 4-vector potential to scalar field or breaking of gauge invariance or introducing ghosts. In chapter (\ref{Chap.5}), we provided a new way of breaking conformal invariance to the 4-dimensional electromagnetic  action without introducing ghosts or breaking gauge invariance or coupling to any scalar field. In the modified electromagnetic (vector Galileon) model, as in standard electrodynamics, the scalar potential between two static charges scale as inverse of the distance. One of the interesting feature of this model is that the appreciable magnetic fields can be generated at Mpc length scales and tiny magnetic fields are generated in Gpc scales.

\section*{Future outlook}

Based on the conclusions listed above, the thesis seeks further work in an effort to investigate the signature of the early Universe in the context of higher order perturbation theory. Below, we list out some of the important future outlook of the thesis.

In chapters (\ref{Chap.3}) and (\ref{chap.4}), we have developed a generalized Hamiltonian approach for cosmological perturbations. The main application of this is to provide a simple approach to calculate interaction Hamiltonian which is needed to evaluate correlation functions. We also have shown that, our results match with the result obtained in the previous approach only in the case of linear order perturbations. However, if we consider higher order perturbations in the metric as well as matter in the first place, and include that to calculate interaction Hamiltonian, it would give a significant corrections to it. Till now, we do not know how to interpret such higher order perturbations as in the previous method, only linear order perturbations are considered. But, in our proposed approach, that is not the case and hence, it can give us a new effects of higher order perturbations in the correlations functions. Another useful application of our approach is that, it can be applied to any kind of model at any order of perturbations, hence, it can even be applied where the previous approach fails. Although we have focused on early Universe, the Hamiltonian formalism can be applied to the late Universe, especially to study the effects of dark energy perturbations on the structure formation.

As mentioned in chapter (\ref{Chap.5}), appreciable magnetic field is generated at the end of the inflation. Our model can provide a natural way to end the inflation. One important consequence is to see how the coupling of the electromagnetic field with standard model particles can provide new features during reheating.

\clearpage
\thispagestyle{empty}
\phantom{a}
\vfill
\vfill

\begin{appendices}
	\chapter{}\label{app1}
\section{Coefficients of second order equation of motion of
  non-canonical scalar field} \label{App:Appendix A}
  
   {\small
\begin{eqnarray}
 \qquad C_X &= &\varphi_2^{\prime \prime} - 2 \phi_1 \varphi_1^{\prime \prime}
  - \phi_2 \varphi_0^{\prime \prime} - \nabla^2{\varphi_2} - 2
  \phi_1^{\prime} \varphi_1^{\prime} - \phi_2 \varphi_0^{\prime} + 2
  \mathcal{H} \varphi_2^{\prime} + 3\varphi_0^{\prime \prime}
  \phi_1^{2} - 2 \phi_1 \nabla^2{\varphi_1} + 6 \phi_1 \phi_1^{\prime}
  \varphi_0^{\prime} - \nonumber\\
  &&
  4 \mathcal{H} \phi_1 {\varphi_1^{\prime}} - 2
  \mathcal{H} \phi_2 \varphi_0^{\prime} - \varphi_0^{\prime}
  \nabla^2{B_2} - 2 {\varphi_1^{\prime}} \nabla^2{B_1} - 4 {\delta}^{i
    j} {\partial}_{i}{B_1} {\partial}_{j}{\varphi_1^{\prime}} - 2
  {\delta}^{i j} {\partial}_{i}{\phi_1} {\partial}_{j}{\varphi_1} - 2
  {\delta}^{i j} {\partial}_{i}{\varphi_1}
  {\partial}_{j}{B_1^{\prime}} + \nonumber\\
  &&
  6 \mathcal{H} \varphi_0^{\prime}
  \phi_1^{2} + 2 \phi_1 {\varphi_0^{\prime}} \nabla^2{B_1} + 2
  {\delta}^{i j} \varphi_0^{\prime} {\partial}_{i}{B_1}
  {\partial}_{j}{\phi_1} - 2 {\delta}^{i j} {\varphi_0^{\prime}}
  {\partial}_{i}{B_1} {\partial}_{j}{B_1^{\prime}} - 4 \mathcal{H}
  {\delta}^{i j} {\partial}_{i}{B_1} {\partial}_{j}{\varphi_1} -\nonumber\\
  &&
  {\delta}^{i j} {\partial}_{i}{B_1} {\partial}_{j}{B_1}
  {\varphi_0^{\prime \prime}} -
   2 \mathcal{H} {\delta}^{i j}
  {\varphi_0^{\prime}} {\partial}_{i}{B_1} {\partial}_{j}{B_1} \\
  \qquad C_{XX} &=& a^{-2}\phi_2^{\prime} {\varphi_0^{\prime}}^{3} - 3 a^{-2}
  {\varphi_0^{\prime \prime}} {\varphi_1^{\prime}}^{2} - a^{-2}
  {\varphi_2^{\prime \prime}}{\varphi_0^{\prime}}^{2} - 12 a^{-2}
  \phi_1 \phi_1^{\prime} {\varphi_0^{\prime}}^{3} + 8 a^{-2} \phi_1
  {\varphi_1^{\prime \prime}} {\varphi_0^{\prime}}^{2} + \nonumber\\
  &&
  4 a^{-2}
  \phi_2 {\varphi_0^{\prime \prime}} {\varphi_0^{\prime}}^{2} + 8
  a^{-2} {\phi_1^{\prime}}
  {\varphi_1^{\prime}}{\varphi_0^{\prime}}^{2} - 6 a^{-2}
  {\varphi_0^{\prime}}{\varphi_1^{\prime}}{\varphi_1^{\prime \prime}}
  - 3 a^{-2} {\varphi_0^{\prime}}
  {\varphi_2^{\prime}}{\varphi_0^{\prime \prime}} - 21 a^{-2}
  {\varphi_0^{\prime \prime}} \phi_1^{2} {\varphi_0^{\prime}}^{2} - \nonumber\\
  &&
  2
  a^{-2} \phi_1 \nabla^2{B_1}{\varphi_0^{\prime}}^{3} - 2 a^{-2}
  \phi_1 \nabla^2{\varphi_1} {\varphi_0^{\prime}}^{2} + 18 a^{-2}
  \phi_1 {\varphi_0^{\prime}} {\varphi_1^{\prime}} {\varphi_0^{\prime
      \prime}} - a^{-2} \phi_2 \mathcal{H} {\varphi_0^{\prime}}^{3} +
  2 a^{-2} {\varphi_0^{\prime}}{\varphi_1^{\prime}}
  \nabla^2{\varphi_1} + \nonumber\\
  &&
  4 a^{-2} {\delta}^{i j} {\varphi_0}
  {\partial}_{i}{\varphi_1} {\partial}_{j}{\varphi_1^{\prime}} + 2
  a^{-2} {\varphi_1^{\prime}} \nabla^2{B_1} {\varphi_0^{\prime}}^{2} -
  2 a^{-2} {\delta}^{i j} {\partial}_{i}{B_1} {\partial}_{j}{\phi_1}
  {\varphi_0^{\prime}}^{3} + \nonumber\\
  &&
  2 a^{-2} {\delta}^{i j}
  {\partial}_{i}{B_1}
  {\partial}_{j}{B_1^{\prime}}{\varphi_0^{\prime}}^{3} + 4 a^{-2}
  {\delta}^{i j} {\partial}_{i}{B_1}
  {\partial}_{j}{\varphi_1^{\prime}} {\varphi_0^{\prime}}^{2} -
   2
  a^{-2} {\delta}^{i j} {\partial}_{i}{\phi_1}
  {\partial}_{j}{\varphi_1} {\varphi_0^{\prime}}^{2} + a^{-2}
  {\delta}^{i j} {\partial}_{i}{\varphi_1} {\partial}_{j}{\varphi_1}
  {\varphi_0^{\prime \prime}} + \nonumber\\
  &&
  2 a^{-2} {\delta}^{i j}
  {\partial}_{i}{\varphi_1} {\partial}_{j}{B_1^{\prime}}
  {\varphi_0^{\prime}}^{2} + 
  a^{-2} \mathcal{H} \varphi_2^{\prime}
  {\varphi_0^{\prime}}^{2} + 3 a^{-2} \mathcal{H} \phi_1^{2}
  {\varphi_0^{\prime}}^{3} - 
  2 a^{-2} \mathcal{H} \phi_1
  {\varphi_1^{\prime}} {\varphi_0^{\prime}}^{2} + \nonumber\\
  &&
  6 a^{-2} {\delta}^{i
    j} {\varphi_0^{\prime}} {\partial}_{i}{B_1}
  {\partial}_{j}{\varphi_1} {\varphi_0^{\prime \prime}} + 4 a^{-2}
  {\delta}^{i j} {\partial}_{i}{B_1} {\partial}_{j}{B_1}
  {\varphi_0^{\prime \prime}} {\varphi_0^{\prime}}^{2} -
   a^{-2}
  \mathcal{H} {\delta}^{i j} {\partial}_{i}{B_1} {\partial}_{j}{B_1}
  {\varphi_0^{\prime}}^{3} -\nonumber\\
  &&
   2 a^{-2} \mathcal{H} {\delta}^{i j}
  {\partial}_{i}{B_1} {\partial}_{j}{\varphi_1}
  {\varphi_0^{\prime}}^{2} \\
 C_{XXX} &=& 2 a^{-4} \phi_1 {\phi_1^{\prime}} {\varphi_0^{\prime}}^{5}
  - 2 a^{-4} \phi_1 {\varphi_1^{\prime
      \prime}}{\varphi_0^{\prime}}^{4} + a^{-4} \mathcal{H} \phi_2
  {\varphi_0^{\prime}} ^{5} - a^{-4} \phi_2 \varphi_0^{\prime \prime}
  {\varphi_0^{\prime}}^{4} - 2 a^{-4} \phi_1^{\prime}
  \varphi_1^{\prime} {\varphi_0^{\prime}} ^{4} + \nonumber\\
  &&
  2 a^{-4}
  \varphi_1^{\prime} \varphi_1^{\prime \prime}
  {\varphi_0^{\prime}}^{3} - a^{-4} \mathcal{H}\varphi_2^{\prime}
  {\varphi_0^{\prime}}^{4} + a^{-4} \varphi_2^{\prime}
  \varphi_0^{\prime \prime} {\varphi_0^{\prime}}^{3} - 8 a^{-4}
  \mathcal{H} \phi_1^{2} {\varphi_0^{\prime}} ^{5} - 5 a^{-4}
  \mathcal{H} {\varphi_0^{\prime}}^{3} {\varphi_1^{\prime}}^{2} + \nonumber\\
  &&
  11
  a^{-4} \varphi_0^{\prime \prime} \phi_1^{2} {\varphi_0^{\prime}}
  ^{4} + 6 a^{-4} \varphi_0^{\prime \prime} {\varphi_0^{\prime}}^{2}
  {\varphi_1^{\prime}} ^{2} + 12 a^{-4} \mathcal{H} \phi_1
  \varphi_1^{\prime} {\varphi_0^{\prime}}^{4} - 16 a^{-4} \phi_1
  \varphi_1^{\prime} \varphi_0^{\prime \prime}
  {\varphi_0^{\prime}}^{3} + \nonumber\\
  &&
  a^{-4} \mathcal{H} {\delta}^{i j}
  {\partial}_{i}{B_1} {\partial}_{j}{B_1} {\varphi_0^{\prime}}^{5} +
  2
  a^{-4} \mathcal{H} {\delta}^{i j} {\partial}_{i}{B_1}
  {\partial}_{j}{\varphi_1} {\varphi_0^{\prime}} ^{4} + a^{-4}
  \mathcal{H} {\delta}^{i j} {\partial}_{i}{\varphi_1}
  {\partial}_{j}{\varphi_1} {\varphi_0^{\prime}}^{3} - \nonumber\\
  &&
  a^{-4}
  {\delta}^{i j} {\partial}_{i}{B_1} {\partial}_{j}{B_1}
  \varphi_0^{\prime \prime} {\varphi_0^{\prime}} ^{4}
  - 2 a^{-4}
  {\delta}^{i j} {\partial}_{i}{B_1} {\partial}_{j}{\varphi_1}
  \varphi_0^{\prime \prime} {\varphi_0^{\prime}}^{3} -
  a^{-4}
  {\delta}^{i j} {\partial}_{i}{\varphi_1} {\partial}_{j}{\varphi_1}
  \varphi_0^{\prime \prime} {\varphi_0^{\prime} }^{2}\\
 C_{XXXX} &=& a^{-6} \mathcal{H} \phi_1^{2} {\varphi_0^{\prime}}^{7} +
  a^{-6} \mathcal{H} {\varphi_0^{\prime}}^{5} {\varphi_1^{\prime}}^{2}
  - a^{-6} {\varphi_0^{\prime \prime}} \phi_1^{2}
  {\varphi_0^{\prime}}^{6} - a^{-6} {\varphi_0^{\prime \prime}}
  {\varphi_0^{\prime}}^{4} {\varphi_1^{\prime}}^{2} - \nonumber\\
  &&
  2
  a^{-6}\mathcal{H} \phi_1 \varphi_1^{\prime} {\varphi_0^{\prime}}^{6}
  + 2 a^{-6} \phi_1 \varphi_1^{\prime} {\varphi_0^{\prime \prime}}
  {\varphi_0^{\prime}}^{5}\\
  C_{XXX \varphi} &=& a^{-4} \phi_1^{2} {\varphi_0^{\prime}}^{6} +
  a^{-4} {\varphi_0^{\prime}} ^{4} {\varphi_1^{\prime}}^{2} - 2 a^{-4}
  \phi_1 {\varphi_1^{\prime}} {\varphi_0^{\prime}} ^{5} + 2 a^{-4}
  \mathcal{H} \phi_1 {\varphi_0^{\prime}}^{5} \varphi_1 - 2 a^{-4}
  \phi_1 \varphi_0^{\prime \prime} {\varphi_0^{\prime}}^{4} \varphi_1
  - \nonumber\\
  &&
  2 a^{-4} \mathcal{H} {\varphi_1^{\prime}} {\varphi_0^{\prime}}
  ^{4} \varphi_1 + 2 a^{-4} {\varphi_1^{\prime}} \varphi_0^{\prime
    \prime} {\varphi_0^{\prime}}^{3} \varphi_1\\
  C_{XX\varphi} &=& a^{-2} \phi_2 {\varphi_0^{\prime}}^{4} - a^{-2}
  \varphi_2^{\prime \prime} {\varphi_0^{\prime}} ^{3} - 5 a^{-2}
  \phi_1^{2} {\varphi_0^{\prime}} ^{4} - 4 a^{-2} {\varphi_0^{\prime}}
  ^{2} {\varphi_1^{\prime}}^{2} + 8 a^{-2} \phi_1 \varphi_1^{\prime}
  {\varphi_0^{\prime}} ^{3} + \nonumber\\
  &&
  2 a^{-2} {\phi_1^{\prime}}
  {\varphi_0^{\prime}}^{3} \varphi_1 - a^{-2} \varphi_0^{\prime
    \prime} {\varphi_0^{\prime}} ^{2} \varphi_2 - 2 a^{-2}
  \varphi_1^{\prime \prime} {\varphi_0^{\prime}} ^{2} \varphi_1 + 8
  a^{-2} \phi_1 \varphi_0^{\prime \prime} {\varphi_0^{\prime}} ^{2}
  \varphi_1 + \nonumber\\
  &&
  a^{-2} {\delta}^{i j} {\partial}_{i}{B_1}
  {\partial}_{j}{B_1} {\varphi_0^{\prime}} ^{4} + 2 a^{-2} {\delta}^{i
    j} {\partial}_{i}{B_1} {\partial}_{j}{\varphi_1}
  {\varphi_0^{\prime}} ^{3} + a^{-2} {\delta}^{i j}
  {\partial}_{i}{\varphi_1} {\partial}_{j}{\varphi_1}
  {\varphi_0^{\prime}} ^{2} - 6 a^{-2} \varphi_0^{\prime}
  \varphi_1^{\prime} \varphi_0^{\prime \prime} \varphi_1 +\nonumber\\
  &&
   a^{-2}
  \mathcal{H} {\varphi_0^{\prime}} ^{3} \varphi_2 - 2 a^{-2}
  \mathcal{H} \phi_1 {\varphi_0^{\prime}} ^{3} \varphi_1 + 2 a^{-2}
  \mathcal{H} \varphi_1^{\prime} {\varphi_0^{\prime}} ^{2} \varphi_1\\
  C_{XX \varphi \varphi} &=& 2 a^{-2} \phi_1 {\varphi_0^{\prime}}^{4}
  \varphi_1 - 2 a^{-2} {\varphi_1^{\prime}} {\varphi_0^{\prime}}^{3}
  \varphi_1 - a^{-2} {\varphi_0^{\prime \prime}}
  {\varphi_0^{\prime}}^{2} \varphi_1^{2} + a^{-2} \mathcal{H}
  {\varphi_0^{\prime}}^{3} \varphi_1^{2} \\
  C_{X\varphi} &=& {\varphi_1^{\prime}} ^{2} + \varphi_0^{\prime}
  \varphi_2^{\prime} + \varphi_0^{\prime \prime} \varphi_2 + 2
  \varphi_1^{\prime \prime} \varphi_1 + \phi_1^{2}
  {\varphi_0^{\prime}}^{2} - 2 \phi_1 \varphi_0^{\prime}
  \varphi_1^{\prime} -
  2 \phi_1 \varphi_0^{\prime \prime} \varphi_1 -\nonumber\\
  &&
  {\delta}^{i j} {\partial}_{i}{\varphi_1} {\partial}_{j}{\varphi_1} -
  2 {\delta}^{i j} {\partial}_{i j}{\varphi_1} \varphi_1 - 2
  {\phi_1^{\prime}} \varphi_0^{\prime} \varphi_1 + 
  2 \mathcal{H}
  \varphi_0^{\prime} \varphi_2 + 4 \mathcal{H} \varphi_1^{\prime}
  \varphi_1 -\nonumber\\
  &&
   4 \mathcal{H} \phi_1 \varphi_0^{\prime} \varphi_1 - 2
  {\delta}^{i j} \varphi_0^{\prime} {\partial}_{i}{B_1}
  {\partial}_{j}{\varphi_1} - 2 {\delta}^{i j} \varphi_0^{\prime}
  {\partial}_{i j}{B_1} \varphi_1 \\
  C_{X\varphi\varphi} &=& {\varphi_0^{\prime \prime}} \varphi_1^{2} +
  {\varphi_0^{\prime}}^{2} \varphi_2 + 2 \varphi_0^{\prime}
  {\varphi_1^{\prime}} \varphi_1 + 2 \mathcal{H} {\varphi_0^{\prime}}
  \varphi_1^{2} \\
  C_{X\varphi\varphi\varphi} &=& {\varphi_0^{\prime}}^{2} \varphi_1^{2}\\
  C_{\varphi} &=& a^{2} \phi_2 - a^{2} \phi_1^{2} + a^{2} {\delta}^{i j}
  {\partial}_{i}{B_1} {\partial}_{j}{B_1} \\
  C_{\varphi \varphi} &=& a^{2} \varphi_2 + 2 a^{2} \phi_1 \varphi_1 \\
  C_{\varphi\varphi\varphi} &=& a^{2} \varphi_1^{2}
\end{eqnarray}}

\section{Second order single variable equation of motion for
  non-canonical scalar field for Power-law
  inflation} \label{App:Appendix B} Fourth order action for
non-canonical scalar field, after partial integration is,

{\scriptsize
\begin{equation}\label{nonc}
\begin{split}
  ^{(4)}\mathcal{S} =& \int d^4x \Big( 3 \phi_1 \phi_2 {\delta}^{i j}
  {a^\prime} {\partial}_{i j}{B_1} \kappa^{-1} a + \frac{1}{8} P_{X}
  \phi_2^{2} {\varphi_0^\prime}^{2} a^{2} + \frac{1}{4} P \phi_2^{2}
  a^{4} + \frac{1}{4} P_{X} \phi_2 {\delta}^{i j} {\partial}_{i}{B_1}
  {\partial}_{j}{B_1} {\varphi_0^\prime}^{2} a^{2} + 
  \frac{1}{2} P
  \phi_2 {\delta}^{i j} {\partial}_{i}{B_1} {\partial}_{j}{B_1} a^{4}
  - \\
  &
  \frac{3}{2} P_{X} \phi_2 \phi_1^{2} {\varphi_0^\prime}^{2} a^{2} -
  3 P \phi_2 \phi_1^{2} a^{4} - \frac{1}{4} \phi_2 {\delta}^{i j}
  {\delta}^{k l} {\partial}_{i k}{B_1} {\partial}_{j l}{B_1}
  \kappa^{-1} a^{2} + \frac{1}{4} \phi_2 {\delta}^{i j} {\delta}^{k l}
  {\partial}_{i j}{B_1} {\partial}_{k l}{B_1} \kappa^{-1} a^{2} -
  \frac{3}{2} P_{X} \phi_1 \phi_2 {\varphi_0^\prime}
  {\varphi_1^\prime} a^{2} + \\
  &
  \frac{1}{4} P_{X} \phi_2
  {\varphi_0^\prime} {\varphi_2^\prime} a^{2} + \frac{1}{4} P_{X}
  {\delta}^{i j} {\varphi_0^\prime} {\varphi_2^\prime}
  {\partial}_{i}{B_1} {\partial}_{j}{B_1} a^{2} - \frac{3}{4} P_{X}
  {\varphi_0^\prime} {\varphi_2^\prime} \phi_1^{2} a^{2} + \frac{1}{4}
  P_{X} \phi_2 {\varphi_1^\prime}^{2} a^{2} + \frac{1}{2} P_{X} \phi_1
  {\varphi_1^\prime} {\varphi_2^\prime} a^{2} - \frac{1}{8} P_{X}
  {\varphi_2^\prime}^{2} a^{2} - \\
  &
  \frac{1}{2} P_{X} \phi_2 {\delta}^{i
    j} {\varphi_0^\prime} {\partial}_{i}{B_1}
  {\partial}_{j}{\varphi_1} a^{2} - \frac{1}{2} P_{X} \phi_1
  {\delta}^{i j} {\varphi_0^\prime} {\partial}_{i}{B_1}
  {\partial}_{j}{\varphi_2} a^{2} + \frac{1}{2} P_{X} {\delta}^{i j}
  {\varphi_1^\prime} {\partial}_{i}{B_1} {\partial}_{j}{\varphi_2}
  a^{2}
  + \frac{1}{2} P_{X} {\delta}^{i j} {\varphi_2^\prime}
  {\partial}_{i}{B_1} {\partial}_{j}{\varphi_1} a^{2} + \\
  &
  \frac{1}{8}
  P_{X} {\delta}^{i j} {\partial}_{i}{\varphi_2}
  {\partial}_{j}{\varphi_2} a^{2} - \frac{9}{4} P_{XX} \phi_2
  \phi_1^{2} {\varphi_0^\prime}^{4} + 3 P_{XX} \phi_1 \phi_2
  {\varphi_1^\prime} {\varphi_0^\prime}^{3} + \frac{3}{2} P_{XX}
  {\varphi_2^\prime} \phi_1^{2} {\varphi_0^\prime}^{3} -
  2 P_{XX}
  \phi_1 {\varphi_1^\prime} {\varphi_2^\prime} {\varphi_0^\prime}^{2}
  + \\
  &
  \frac{1}{2} P_{XX} \phi_1 {\delta}^{i j} {\partial}_{i}{B_1}
  {\partial}_{j}{\varphi_2} {\varphi_0^\prime} ^{3} + \frac{1}{2}
  P_{XX} \phi_1 {\delta}^{i j} {\partial}_{i}{\varphi_1}
  {\partial}_{j}{\varphi_2} {\varphi_0^\prime}^{2} + \frac{1}{8}
  P_{XX} \phi_2^{2} {\varphi_0^\prime}^{4} + \frac{1}{4} P_{XX} \phi_2
  {\delta}^{i j} {\partial}_{i}{B_1} {\partial}_{j}{B_1}
  {\varphi_0^\prime}^{4} - \frac{1}{4} P_{XX} \phi_2
  {\varphi_2^\prime} {\varphi_0^\prime}^{3} - \\
  &
  P_{XX} \phi_2 {\varphi_0^\prime}^{2} {\varphi_1^\prime}^{2} + \frac{1}{2} P_{XX}
  \phi_2 {\delta}^{i j} {\partial}_{i}{B_1} {\partial}_{j}{\varphi_1}
  {\varphi_0^\prime}^{3} + \frac{1}{4} P_{XX} \phi_2 {\delta}^{i j}
  {\partial}_{i}{\varphi_1} {\partial}_{j}{\varphi_1}
  {\varphi_0^\prime}^{2} - \frac{1}{4} P_{XX} {\delta}^{i j}
  {\varphi_2^\prime} {\partial}_{i}{B_1} {\partial}_{j}{B_1}
  {\varphi_0^\prime} ^{3} + \\
  &
  \frac{3}{4} P_{XX} {\varphi_0^\prime}
  {\varphi_2^\prime} {\varphi_1^\prime}^{2} - \frac{1}{2} P_{XX}
  {\delta}^{i j} {\varphi_1^\prime} {\partial}_{i}{B_1}
  {\partial}_{j}{\varphi_2} {\varphi_0^\prime}^{2} - \frac{1}{2}
  P_{XX} {\delta}^{i j} {\varphi_0^\prime} {\varphi_1^\prime}
  {\partial}_{i}{\varphi_1} {\partial}_{j}{\varphi_2} + \frac{1}{8}
  P_{XX} {\varphi_0^\prime} ^{2} {\varphi_2^\prime}^{2} - \frac{1}{2}
  P_{XX} {\delta}^{i j} {\varphi_2^\prime} {\partial}_{i}{B_1}
  {\partial}_{j}{\varphi_1} {\varphi_0^\prime}^{2}\\
  &
  - \frac{1}{4} P_{XX} {\delta}^{i j} {\varphi_0^\prime}
  {\varphi_2^\prime} {\partial}_{i}{\varphi_1}
  {\partial}_{j}{\varphi_1} + \frac{1}{8} P_{YY} \varphi_2^{2} a^{4} +
  \frac{1}{4} P_{XY} \phi_2 {\varphi_0^\prime} ^{2} a^{2} \varphi_2 +
  \frac{1}{4} P_{XY} {\delta}^{i j} {\partial}_{i}{B_1}
  {\partial}_{j}{B_1} {\varphi_0^\prime} ^{2} a^{2} \varphi_2 -
  \frac{1}{2} P_{XY} \phi_1^{2} {\varphi_0^\prime}^{2} a^{2} \varphi_2  \\
  &
  - P_{XY} \phi_1 \phi_2 {\varphi_0^\prime} ^{2} a^{2} \varphi_1 +
  \frac{1}{2} P_{XY} \phi_1 {\varphi_0^\prime} {\varphi_1^\prime}
  a^{2} \varphi_2 + \frac{1}{2} P_{XY} \phi_2 {\varphi_0^\prime}
  {\varphi_1^\prime} a^{2} \varphi_1 + \frac{1}{2} P_{XY} \phi_1
  {\varphi_0^\prime} {\varphi_2^\prime} a^{2} \varphi_1 - \frac{1}{4}
  P_{XY} {\varphi_0^\prime} {\varphi_2^\prime} a^{2} \varphi_2 -\\
  &
  \frac{1}{4} P_{XY} {\varphi_1^\prime}^{2} a^{2} \varphi_2 -
  \frac{1}{2} P_{XY} {\varphi_1^\prime} {\varphi_2^\prime} a^{2}
  \varphi_1 + \frac{1}{2} P_{XY} {\delta}^{i j} {\varphi_0^\prime}
  {\partial}_{i}{B_1} {\partial}_{j}{\varphi_1} a^{2} \varphi_2 +
  \frac{1}{2} P_{XY} {\delta}^{i j} {\varphi_0^\prime}
  {\partial}_{i}{B_1} {\partial}_{j}{\varphi_2} a^{2} \varphi_1 +
  \frac{1}{4} P_{XY} {\delta}^{i j} {\partial}_{i}{\varphi_1}
  {\partial}_{j}{\varphi_1} a^{2} \varphi_2 + \\
  &
  \frac{1}{2} P_{XY}
  {\delta}^{i j} {\partial}_{i}{\varphi_1} {\partial}_{j}{\varphi_2}
  a^{2} \varphi_1 + \frac{1}{4} P_{XXX} \phi_2 \phi_1^{2}
  {\varphi_0^\prime}^{6} a^{(-2)} - \frac{1}{4} P_{XXX}
  {\varphi_2^\prime} \phi_1^{2} {\varphi_0^\prime} ^{5} a^{(-2)} -
  \frac{1}{2} P_{XXX} \phi_1 \phi_2 {\varphi_1^\prime}
  {\varphi_0^\prime}^{5} a^{(-2)} + \\
  &
  \frac{1}{2} P_{XXX} \phi_1
  {\varphi_1^\prime} {\varphi_2^\prime} {\varphi_0^\prime} ^{4}
  a^{(-2)}
  + \frac{1}{4} P_{XXX} \phi_2 {\varphi_0^\prime}^{4}
  {\varphi_1^\prime}^{2} a^{(-2)} - \frac{1}{4} P_{XXX}
  {\varphi_2^\prime} {\varphi_0^\prime}^{3} {\varphi_1^\prime}^{2}
  a^{(-2)} + \frac{1}{4} P_{YYY} \varphi_1^{2} a^{4} \varphi_2 +
  \frac{1}{4} P_{XXY} \phi_1^{2} {\varphi_0^\prime}^{4} \varphi_2 +\\
  &
  \frac{1}{2} P_{XXY} \phi_1 \phi_2 {\varphi_0^\prime}^{4} \varphi_1 -
  \frac{1}{2} P_{XXY} \phi_1 {\varphi_1^\prime} {\varphi_0^\prime}
  ^{3} \varphi_2 - \frac{1}{2} P_{XXY} \phi_1 {\varphi_2^\prime}
  {\varphi_0^\prime}^{3} \varphi_1 - \frac{1}{2} P_{XXY} \phi_2
  {\varphi_1^\prime} {\varphi_0^\prime}^{3} \varphi_1 + \frac{1}{4}
  P_{XXY} {\varphi_0^\prime} ^{2} {\varphi_1^\prime}^{2} \varphi_2 +\\
  &
  \frac{1}{2} P_{XXY} {\varphi_1^\prime} {\varphi_2^\prime}
  {\varphi_0^\prime}^{2} \varphi_1 + \frac{1}{2} P_{XYY} \phi_1
  {\varphi_0^\prime}^{2} a^{2} \varphi_1 \varphi_2 + \frac{1}{4}
  P_{XYY} \phi_2 {\varphi_0^\prime}^{2} \varphi_1^{2} a^{2} -
  \frac{1}{2} P_{XYY} {\varphi_0^\prime} {\varphi_1^\prime} a^{2}
  \varphi_1 \varphi_2 - \frac{1}{4} P_{XYY} {\varphi_0^\prime}
  {\varphi_2^\prime} \varphi_1^{2} a^{2} + \\
  &
  \frac{1}{2} P_{X} \phi_1
  {\delta}^{i j} {\partial}_{i}{\varphi_1} {\partial}_{j}{\varphi_2}
  a^{2} + \frac{1}{2} P_{YY} \phi_1 a^{4} \varphi_1 \varphi_2 +
  \frac{1}{4} P_{X} \phi_2 {\delta}^{i j} {\partial}_{i}{\varphi_1}
  {\partial}_{j}{\varphi_1} a^{2} + \frac{1}{4} P_{Y} \phi_2 a^{4}
  \varphi_2 + \frac{1}{4} P_{YY} \phi_2 \varphi_1^{2} a^{4} +\\
  &
  \frac{1}{4} P_{Y} {\delta}^{i j} {\partial}_{i}{B_1}
  {\partial}_{j}{B_1} a^{4} \varphi_2%
  - \frac{1}{4} P_{Y} \phi_1^{2} a^{4} \varphi_2 - \frac{1}{2} P_{Y}
  \phi_1 \phi_2 a^{4} \varphi_1 \Big)
  \end{split}
\end{equation}
}

\noindent We can eliminate $\phi_2,~ \phi_1$ and $B_1$ using
constraint equations. While this is possible for canonical scalar
field, it is highly non-trivial for non-canonical scalar fields. Hence
we consider Power law inflation to reduce the action in a single
variable form.

For Power-law, we have

\begin{eqnarray}
&&a = a_0 (-\eta)^\beta \\
&&\Rightarrow \mathcal{H} \equiv \frac{a^\prime}{a}= - \frac{\beta}{(-\eta)}\\
&&\mbox{and}\quad\frac{a^{\prime \prime}}{a} = - \frac{\beta (1 -\beta)}{(-\eta)^2} \\
&&\varphi_0 = \varphi_{c} (-\eta)^\alpha \\
&&\Rightarrow{\varphi_0}^\prime = \varphi_{c} \alpha (-\eta)^\alpha \\
  &&\mathcal{H}^2 = -\frac{\kappa}{3} (P_X {{\varphi_0}^{\prime 2}} + P_0 a^2) \\
  &&- 2 \frac{a^{\prime \prime}}{a} + \mathcal{H}^2 = \kappa P_0 a^2
\end{eqnarray}

For Power-law inflation, $P(X, \varphi) = \frac{C}{\varphi^2} ~(X^2 +
X)$. Solving for $C, \varphi_{c}$ and $\alpha$ using above equations,
\begin{eqnarray}
  &&\alpha = 1 + \beta \\
  &&C = \frac{- 6 \beta (1 - \beta)}{\kappa (1 + \beta)^2} \\
  &&\varphi_{0c} = \sqrt{\frac{2}{3}} \frac{a_0}{1 + \beta} \sqrt{\frac{1- 2 \beta}{1- \beta}}
\end{eqnarray}
Using first order Einstein's equations,
\begin{eqnarray}
  &&\phi_1 = \sqrt{\frac{3}{2}} \sqrt{\frac{1- \beta}{1-2\beta}} \frac{1 + \beta}{a_0(-\eta)^{1 + \beta}} \varphi_1 \\
  &&\phi_1 = F_3 \varphi_1 \\
  &&\nabla^2 B_1 = \frac{3}{a_0} \sqrt{\frac{3}{2}} \sqrt{\frac{1- \beta}{1-2\beta}} (1-3\beta) \Big\{ \frac{1 + \beta}{(-\eta)^{2 + \beta}} \varphi_1 + \frac{1}{(-\eta)^{ 1 + \beta}} \varphi_1^\prime \Big\}\\
  &&\nabla^2 B_1 = F_1 \varphi_1 + F_2 \varphi_1^\prime
\end{eqnarray}
where,
\begin{eqnarray}
  &&F_1 = \frac{3}{a_0} \sqrt{\frac{3}{2}} \sqrt{\frac{1- \beta}{1-2\beta}} 
(1-3\beta)\frac{1 + \beta}{(-\eta)^{2 + \beta}} \\
  &&F_2 = \frac{3}{a_0} \sqrt{\frac{3}{2}} \sqrt{\frac{1- \beta}{1-2\beta}} 
(1-3\beta)\frac{1}{(-\eta)^{ 1 + \beta}} \\
  &&F_3 = \frac{1}{a_0}\sqrt{\frac{3}{2}} \sqrt{\frac{1- \beta}{1-2\beta}} 
\frac{1 + \beta}{(-\eta)^{1 + \beta}}
\end{eqnarray}

Similarly using second order Einstein's equations, 
\begin{eqnarray}
  &&\partial_{i}{\phi_2} = C_1 \partial_{i}{\varphi_2} + C_2 \varphi_1 
\partial_{i}{\varphi_1} + C_3 \delta^{j k} \partial_{j}{\varphi_1} 
\partial_{i k}{B_1} + C_4 \delta^{j k} \partial_{j}{B_1} \partial_{i k}{B_1}+ 
C_5 \varphi_1^\prime \partial_{i}{\varphi_1} ~~~~~~~~\\
  &&\phi_2 = C_1 \varphi_2 + \frac{1}{2} C_2 \varphi_1^2 - \delta^{i j} 
\partial_{i}{B_1} \partial_{j}{B_1} + Q \\
  &&Q = {\partial^{-1 i}} \{C_3 \delta^{j k} \partial_{j}{\varphi_1} 
\partial_{i k}{B_1} + C_5 \varphi_1^\prime \partial_{i}{\varphi_1}\} 
\end{eqnarray}
where
\begin{eqnarray}
  &&C_1 = \sqrt{\frac{3}{2}} \sqrt{\frac{1- \beta}{1-2\beta}} 
\frac{1 + \beta}{a_0~(-\eta)^{1 + \beta}}\\
  &&C_2 = - \frac{3}{2 a_0^2} \frac{(1+\beta) 
(1-\beta)(3 - 14\beta +7\beta^2)}{\ (1-2\beta)}\frac{1}{(-\eta)^{2+ 2\beta}}\\
  &&C_3 = \sqrt{\frac{3}{2}}\sqrt{\frac{1-\beta}{1-2\beta}}
\frac{(1+ \beta)}{a_0 \beta} \frac{1}{(-\eta)^{\beta}}\\
  &&C_4 = -2\\
  &&C_5 = - \frac{9}{2 a_0^2} \frac{(1-\beta)^2 (1-3\beta)}{\beta(1-2\beta)}
\frac{1}{(-\eta)^{1+2\beta}}
\end{eqnarray}
Multiplying the fourth order action (\ref{nonc}) by $\frac{8}{3} (1- 2\beta)^2$,
\begin{equation}
\begin{split}
  &\int d^4x \Big[ \Big(A_1 \varphi_2^{\prime 2} +
  A_2 \partial^{i}{\varphi_2} \partial_{i}{\varphi_2} + A_3 \varphi_2
  \varphi_2^\prime + A_4 \varphi_2^2 \Big) + \partial^{i}{\varphi_2}
  ~\Big(B_1 \varphi_1 \partial_{i}{\varphi_1} +
   B_2
  \varphi_1 \partial_{i}{B_1}+ \\
  &
  B_3\varphi_1^\prime \partial_{i}{B_1} +
  B_4 \varphi_1^\prime \partial_{i}{\varphi_1} \Big) +
  \varphi_2^\prime~ \Big(D_1 \varphi_1^2 + D_2 \varphi_1^\prime
  \varphi_1 + D_3 Q + D_4 {\varphi_1^\prime}^2 +
  D_5 \partial^{i}{B_1} \partial_{i}{\varphi_1} +\\
  &
  D_6 \partial^{i}{\varphi_1}\partial_{i}{\varphi_1}\Big) + \varphi_2
  ~ \Big(E_1 \varphi_1^2 + E_2 (\nabla^2 B_1)^2 + E_3 \partial^{i
    j}{B_1} \partial_{i j}{B_1} + E_4 \varphi_1 \varphi_1^\prime + E_5
  {\varphi_1^\prime}^2 + E_6 Q \\
  &
  +
  E_7 \partial^{i}{B_1} \partial_{i}{\varphi_1} +
  E_8 \partial^{i}{\varphi_1} \partial_{i}{\varphi_1} \Big) \Big]
  \end{split}
\end{equation}
where,
\begin{eqnarray}
  &&A_1 = -  \frac{3\beta (1 -\beta)(1-2\beta)(1-3\beta)}{\kappa (-\eta)^2} \\
  &&A_2 = - \frac{\beta(1-\beta)(1+\beta)(1-2\beta)}{\kappa (-\eta)^2} \\
  &&A_3 = -  \frac{2\beta(1-\beta)(1+\beta)(1-2\beta)(1-11\beta)}{\kappa (-\eta)^3} \\
  &&A_4 =  \frac{3\beta(1-\beta)(1+\beta)^2 (1-2\beta)(1+3\beta)}{\kappa (-\eta)^4}\\
  &&B_1 = -\frac{36\beta(1-\beta)(1+\beta)(1-2\beta)(1-3\beta)}{\sqrt{6} 
a_0 ~\kappa (-\eta)^{3+\beta}} \sqrt{\frac{1-\beta}{1-2\beta}} \\
  &&B_2 = \frac{4\beta(1-\beta)(1+\beta)(1-2\beta)(1-11\beta)}{\kappa 
(-\eta)^{3}}\\
  &&B_3 = \frac{12\beta(1-\beta)(1-2\beta)(1-3\beta)}{\kappa (-\eta)^2} \\
  &&B_4 = -\frac{48\beta(1-2\beta)^2 (1-\beta)}{\sqrt{6}~a_0~\kappa 
(-\eta)^{2+\beta} } \sqrt{\frac{1-\beta}{1-2\beta}}\\
  &&D_1 = \frac{9 \sqrt{(1-\beta)(1-2\beta)} (1+\beta)(1-\beta)
(3-5\beta + 5 \beta^2 -83 \beta^3)}{\sqrt{6}~a_0~\kappa (-\eta)^{4+\beta}}\\
  &&D_2 = \frac{36\beta\sqrt{(1-\beta)(1-2\beta)}(1-\beta)(1+\beta)
(7-17\beta)}{\sqrt{6}~a_0~\kappa (-\eta)^{3+\beta}}\\
  &&D_3 = - \frac{-12~a_0~\beta\sqrt{(1-\beta)(1-2\beta)}(1-2\beta)
(1-3\beta)}{\sqrt{6} \kappa (-\eta)^{2-\beta}}\\
  &&D_4 = \frac{72\beta\sqrt{(1-\beta)(1-2\beta)} 
(1-\beta)(1-2\beta)}{\sqrt{6}~a_0~\kappa (-\eta)^{2+\beta}}\\
  &&D_5 = \frac{12\beta(1-\beta)(1-2\beta)(1-3\beta)}{\kappa (-\eta)^2}\\
  &&D_6 = -\frac{24 \beta \sqrt{(1-2\beta)(1-\beta)}(1-2\beta)(1-\beta)
 }{\sqrt{6}~a_0~\kappa (-\eta)^{2+\beta}}\\
  &&E_1 = \frac{9\sqrt{(1-\beta)(1-2\beta)}(1-\beta)(1+\beta)^2(3-19\beta 
- 3\beta^2 -77\beta^3)}{\sqrt{6}~a_0~\kappa (-\eta)^{5+\beta}}\\
  &&E_2 = \frac{2~a_0~\sqrt{(1-\beta)(1-2\beta)}
(1+\beta)(1-2\beta)}{\sqrt{6}~\kappa (-\eta)^{1-\beta}}\\
  &&E_3 = -\frac{2~a_0~\sqrt{(1-\beta)(1-2\beta)}
(1+\beta)(1-2\beta)}{\sqrt{6}~\kappa (-\eta)^{1-\beta}}\\
  &&E_4 = \frac{72\beta\sqrt{(1-\beta)(1-2\beta)}(1-\beta)^2 
(1+\beta)(3+11\beta)}{\sqrt{6}~a_0~\kappa (-\eta)^{4+\beta}}\\
  &&E_5 = \frac{18\beta\sqrt{(1-\beta)(1-2\beta)}(1-\beta) 
(1+\beta)(7-17\beta)}{\sqrt{6}~a_0~\kappa (-\eta)^{3+\beta}}\\
  &&E_6 = -\frac{12~a_0~\beta\sqrt{(1-\beta)(1-2\beta)}(1-2\beta) 
(1+\beta)(1-3\beta)}{\sqrt{6}~\kappa (-\eta)^{3+\beta}}\\
  &&E_7 = \frac{4\beta(1-\beta)(1-2\beta)(1+\beta) 
(1-11\beta)}{\kappa (-\eta)^{3}}\\
  &&E_8 = - \frac{18\beta\sqrt{(1-\beta)(1-2\beta)}(1-\beta) 
(1+\beta)(1-3\beta)}{\sqrt{6}~a_0~\kappa (-\eta)^{3+\beta}}\\
\end{eqnarray}

After taking partial derivative,
\begin{equation}
\begin{split}
  & \int d^4x \Big[ \Big(A_1 \varphi_2^{\prime 2} +
  A_2 \partial^{i}{\varphi_2} \partial_{i}{\varphi_2} + A_5
  \varphi_2^2 \Big) + \varphi_2 ~ \Big(G_1 \varphi_1^2 + E_2 (\nabla^2
  B_1)^2 + E_3 \partial^{i j}{B_1} \partial_{i j}{B_1} + \\
  &
  G_2 \varphi_1
  \varphi_1^\prime + G_3 {\varphi_1^\prime}^2 + G_4 \varphi_1
  \varphi_1^{\prime \prime} + G_5 \varphi_1^\prime \varphi_1^{\prime
    \prime} + G_6 \varphi \nabla^2 \varphi_1 + G_7 \varphi^\prime
  \nabla^2 \varphi_1 +\\
  &
  G_8 \partial^{i}{\varphi_1}\partial_{i}{\varphi_1} +
  G_9 \partial^{i}{\varphi_1^\prime}\partial_{i}{\varphi_1} +
  G_{10} \partial^{i}{\varphi_1}\partial_{i}{B_1} +
  G_{11}\partial^{i}{\varphi_1^\prime}\partial_{i}{B_1} + G_{12} Q +
  G_{13} Q^\prime \Big) \Big]
  \end{split}
\end{equation}
where,
\begin{eqnarray}
  &&G_1 = E_1 - B_2 F_1 - D_1^\prime \\
  &&G_2 = E_4 - (B_2 F_2 + B_3 F_1) - (D_2^\prime + 2 D_1)\\
  &&G_3 = E_5 - B_3 F_2 - (D_2 + D_4^\prime)\\
  &&G_4 = - D_2 \\
  &&G_5 = -2D_4\\
  &&G_6 = -B_1\\
  &&G_7 = -B_4\\
  &&G_8 = E_8 - B_1 - (D_6^\prime - D_5 F_3)\\
  &&G_9 = - B_4 - 2 D_6\\
  &&G_{10} = E_7 - B_2 - (D_5^\prime - 2 \mathcal{H} D_5)\\
  &&G_{11} = -B_3 - D_5\\
  &&G_{12} = E_6 - D_3^\prime\\
  &&G_{13} = - D_3
\end{eqnarray}
so the equation of motion of scalar field for power law K-Inflation is,
\begin{equation}
\begin{split}
  &2 A_1 \varphi_2^{\prime \prime} + 2 A_1^\prime \varphi_2^{\prime} +
  A_2 \nabla^2{\varphi_2} + A_5 \varphi_2^2 = G_1 \varphi_1^2 + E_2
  (\nabla^2 B_1)^2 +
  E_3 \partial^{i j}{B_1} \partial_{i j}{B_1} + \\
  &
  G_2
  \varphi_1 \varphi_1^\prime +
  G_3 {\varphi_1^\prime}^2
  + G_4
  \varphi_1 \varphi_1^{\prime \prime} + G_5 \varphi_1^\prime
  \varphi_1^{\prime \prime} + G_6 \varphi \nabla^2 \varphi_1 + G_7
  \varphi^\prime \nabla^2 \varphi_1 +  \\
  &
  G_8 \partial^{i}{\varphi_1}\partial_{i}{\varphi_1} +
  G_9 \partial^{i}{\varphi_1^\prime}\partial_{i}{\varphi_1} +
  G_{10} \partial^{i}{\varphi_1}\partial_{i}{B_1} +
  G_{11}\partial^{i}{\varphi_1^\prime}\partial_{i}{B_1} + G_{12} Q +
  G_{13} Q^\prime
  \end{split}
\end{equation}

$a_0$ dependency in the action as well as in the equation of motion
can be resolved by rescaling $\varphi_1 \rightarrow a_0 \varphi_1,
\varphi_2 \rightarrow a_0 \varphi_2$.  
	\chapter{}\label{app2}

\section{Perturbed equations of motion of Canonical scalar field in flat-slicing gauge}\label{PLeCSF}
 
\subsection{Background equations}

0-0 component of the Einstein's equation or the Hamiltonian constraint
in conformal coordinate is given by
\begin{equation}
\label{B00EE}
 {H}^2 \equiv a^{\prime}{}^{2} a^{-2} = \frac{\kappa}{3} \left[\frac{1}{2}\,  \varphi_0^{\prime}\, {}^{2} + V a^2 \right].
\end{equation}

Trace of i-j component of the Einstein's equation gives the equation
of motion of $a$, which can also be obtained by varying the zeroth
order action with respect to $a$ and is given by
\begin{equation}
\label{BijEE}
3\, \kappa{}^{-1} H^2 - 6\, \frac{a^{\prime \prime}}{a}\,  \kappa{}^{-1} - \frac{3}{2}\, \varphi_0^{\prime}\, {}^{2}  + 3\, V a{}^{2} = 0
\end{equation}
and the equation of motion of scalar field at zeroth order takes the
form
\begin{equation}
\label{BEoMS}
\varphi_0^{\prime \prime} + 2 {H}\, \varphi_0^{\prime}+ V_\varphi\,  a{}^{2} = 0.
\end{equation}

\subsection{First order perturbed equations}
Using the perturbed metric and perturbed scalar field defined in
(\ref{00pmetric}), (\ref{0ipmetric}), (\ref{3metric}) and
(\ref{field}), perturbed Hamiltonian constraint or the first order
perturbed 0-0 component of the Einstein's equation becomes
\begin{equation}
\label{100EE}
2\, {\delta}^{i j} H\,  {\partial}_{i j}{B_1}\,  \kappa{}^{-1} + 6\, \phi_1 \kappa{}^{-1} H^2 + \varphi_0^{\prime}\,  \varphi_1^{\prime} - \phi_1 \varphi_0^{\prime}\, {}^{2}  + V_\varphi \,  a{}^{2} \varphi_1 = 0.
\end{equation}

Similarly, perturbed Momentum constraint or perturbed 0-i component of
the Einstein's equation is
\begin{eqnarray}
\label{10iEE}
&& ~~\frac{2 H}{\kappa} \partial_{i}{\phi_1} = \varphi_0^\prime ~ \partial_{i}{\varphi_1} \\
&& \Rightarrow \phi_1 = \frac{\kappa}{2 {H}} \varphi_0^\prime ~\varphi_1
\end{eqnarray}
and equation of motion of the scalar field $\varphi_1$, using
(\ref{BEoMS}) is given by
\begin{equation}
\label{1EoMS}
\varphi_1^{\prime \prime}\, + 2\,H \varphi_1^\prime  - \phi_1^\prime\,  \varphi_0^{\prime}\,  + 2\, V_{\varphi}\,  \phi_1 a{}^{2} - {\delta}^{i j} \varphi_0^{\prime}\,  {\partial}_{i j}{B_1}\,  - {\delta}^{i j} {\partial}_{i j}{\varphi_1}\,  + V_{\varphi \varphi}\,  a{}^{2} \varphi_1 = 0.
\end{equation}

\subsection{Third order interaction Hamiltonian of Canonical scalar field  in terms of phase-space variables in flat-slicing gauge}\label{ThirdOrderInteractionCanonical}

Third or higher order perturbed Hamiltonian in terms of first order
perturbed variables is needed to calculate the interaction Hamiltonian
which helps to calculate higher order correlation functions. The third
order Hamiltonian is obtained by expanding the Hamiltonian (\ref{HamiltonianCanonical}) up to third order and is given by

\begin{eqnarray}
&& \mathcal{H}_3^C\, (\pi, \varphi) = - \frac{1}{2}\, N_1 {\partial}_{i}{{N_1}^{i}}\,  {\partial}_{j}{{N_1}^{j}}\,  N_0{}^{(-2)} \kappa{}^{-1} a{}^{3} + 2\, {\delta}_{i j} {\partial}_{k}{{N_1}^{k}}\,  \pi_0{}^{i j} N_0{}^{(-2)} N_1{}^{2} a{}^{2} - \nonumber \\
&&~~~~~~~~~~2\, {\delta}_{i j} {\delta}_{k l} \kappa \pi_0{}^{i j} {{}\pi_0{}}^{k l} N_0{}^{(-2)} N_1{}^{3} a + \frac{1}{4}\, N_1 {\delta}_{i j} {\delta}^{l k} {\partial}_{k}{{N_1}^{i}}\,  {\partial}_{l}{{N_1}^{j}}\,  N_0{}^{(-2)} \kappa{}^{-1} a{}^{3} + \nonumber \\
&&~~~~~~~~~~\frac{1}{4}\, N_1 {\partial}_{i}{{N_1}^{j}}\,  {\partial}_{j}{{N_1}^{i}}\,  N_0{}^{(-2)} \kappa{}^{-1} a{}^{3} + \frac{1}{2}\, {\delta}_{i j} {\partial}_{k}{{N_1}^{i}}\,  {{}\pi_0{}}^{k j} N_0{}^{(-2)} N_1{}^{2} a{}^{2} + \nonumber \\
&&~~~~~~~~~~\frac{1}{2}\, {\delta}_{i j} {\partial}_{k}{{N_1}^{i}}\,  {{}\pi_0{}}^{j k} N_0{}^{(-2)} N_1{}^{2} a{}^{2} - 2\, {\delta}_{i j} {\partial}_{k}{{N_1}^{k}}\,  \pi_0{}^{i j} N_0{}^{(-2)} N_1{}^{2} a{}^{2} + \nonumber \\
&&~~~~~~~~~~\frac{1}{2}\, {\delta}_{i j} {\partial}_{k}{{N_1}^{j}}\,  {{}\pi_0{}}^{k i} N_0{}^{(-2)} N_1{}^{2} a{}^{2} + \frac{1}{2}\, {\delta}_{i j} {\partial}_{k}{{N_1}^{j}}\,  {{}\pi_0{}}^{i k} N_0{}^{(-2)} N_1{}^{2} a{}^{2} + \nonumber \\
&&~~~~~~~~~~  2\, {\delta}_{i j} {\delta}_{k l} \kappa {{}\pi_0{}}^{i k} {{}\pi_0{}}^{j l} N_0{}^{(-2)} N_1{}^{3} a + \frac{1}{2}\, N_1 \pi_{\varphi1}{}^{2} a{}^{(-3)} + {N_1}^{i} \pi_{\varphi1} {\partial}_{i}{\varphi_1}\,  + \nonumber \\
&&~~~~~~~~~~{\delta}_{i j} {\delta}_{k l} \kappa \pi_0{}^{i j} {{}\pi_0{}}^{l k} N_0{}^{(-2)} N_1{}^{3} a + \frac{1}{2}\, N_1 {\delta}^{i j} {\partial}_{i}{\varphi_1}\,  {\partial}_{j}{\varphi_1}\,  a + \frac{1}{2}\, N_1 V_{\varphi \varphi}\,  \varphi_1{}^{2} a{}^{3} + \nonumber \\
&&~~~~~~~~~~\frac{1}{6}\, N_0 V_{\varphi \varphi \varphi}\,  \varphi_1{}^{3} a{}^{3}
\end{eqnarray}

\section{Hamiltonian formulation of Canonical scalar field in uniform density gauge}\label{udgcsf}
In uniform-density gauge, $E = \delta\varphi = 0$ and $\gamma_{i j} = a^2 (1 - 2 \epsilon {}\psi_1) \delta_{i j}, \gamma^{i j} = a^2 (1 + 2 \epsilon {}\psi_1 + 4 \epsilon^2 {}\psi_1^2) \delta^{i j}, \sqrt{\gamma} = a^6 ( 1 - 6 \epsilon {}\psi_1 + 12 \epsilon^2 {}\psi_1^2)$. The second order perturbed Hamiltonian is obtained by expanding the  Hamiltonian (\ref{HamiltonianCanonical}) up to second order by using above definitions and is given by

\begin{eqnarray}
\label{hamuniden}
&& \mathcal{H}_2^C = {\delta}_{i j} {\partial}_{k}{{N_1}^{j}}\,  {\pi_1}^{i k} a{}^{2} - 2\, {\delta}_{i j} {\partial}_{k}{{N_1}^{j}}\,  {{}\pi_0{}}^{i k} a{}^{2} {}\psi_1  -  2\, {N_1}^{i} {\delta}_{j k} {\partial}_{i}{{}\psi_1}\,  {{}\pi_0{}}^{j k} a{}^{2} +  \nonumber \\
&&{\delta}_{i j} {\partial}_{k}{{N_1}^{j}}\,  {\pi_1}^{i k} a{}^{2} -2\, {\delta}_{i j} {\partial}_{k}{{N_1}^{j}}\,  {{}\pi_0{}}^{i k} a{}^{2} {}\psi_1 - N_0 {\delta}_{i j} {\delta}_{k l} \kappa {\pi_1}^{i j} {\pi_1}^{k l} a - \nonumber \\
&&2\, N_1 {\delta}_{i j} {\delta}_{k l} \kappa \pi_0{}^{i j} {\pi_1}^{k l} a + 2\, N_0 {\delta}_{i j} {\delta}_{k l} \kappa \pi_0{}^{i j} {\pi_1}^{k l} {}\psi_1 a + N_1 {\delta}_{i j} {\delta}_{k l} \kappa \pi_0{}^{i j} {{}\pi_0{}}^{k l} {}\psi_1 a + \nonumber \\
&& \frac{1}{2}\, N_0 {\delta}_{i j} {\delta}_{k l} \kappa \pi_0{}^{i j} {{}\pi_0{}}^{k l} {}\psi_1{}^{2} a + 2\, N_0 {\delta}_{i j} {\delta}_{k l} \kappa {\pi_1}^{i k} {\pi_1}^{j l} a + 4\, N_1 {\delta}_{i j} {\delta}_{k l} \kappa {{}\pi_0{}}^{i k} {\pi_1}^{j l} a   + \nonumber \\
&&N_1 {}\pi_{\varphi0} \pi_{\varphi1} a{}^{(-3)} + 3\, N_0 {}\pi_{\varphi0} \pi_{\varphi1} a{}^{(-3)} {}\psi_1%
 + \frac{3}{2}\, N_1 {}\pi_{\varphi0}{}^{2} a{}^{(-3)} {}\psi_1 + \frac{15}{4}\, N_0 {}\pi_{\varphi0}{}^{2} {}\psi_1{}^{2} a{}^{(-3)}  - \nonumber \\
&&2\, N_0 {\delta}^{i j} {\partial}_{i j}{{}\psi_1}\,  \kappa{}^{-1} {}\psi_1 a - 2\, N_1 {\delta}^{i j} {\partial}_{i j}{{}\psi_1}\,  \kappa{}^{-1} a - 3\, N_1 V_0 a{}^{3} {}\psi_1 + \frac{3}{2}\, N_0 V_0 {}\psi_1{}^{2} a{}^{3}\nonumber \\
&&- 2\, N_1 {\delta}_{i j} {\delta}_{k l} \kappa {{}\pi_0{}}^{i k} {{}\pi_0{}}^{j l} {}\psi_1 a - N_0 {\delta}_{i j} {\delta}_{k l} \kappa {{}\pi_0{}}^{i k} {{}\pi_0{}}^{j l} {}\psi_1{}^{2} a + \frac{1}{2}\, N_0 \pi_{\varphi1}{}^{2} a{}^{(-3)}\nonumber \\
&&-4\, N_0 {\delta}_{i j} {\delta}_{k l} \kappa {{}\pi_0{}}^{i k} {\pi_1}^{j l} {}\psi_1 a- 3\, N_0 {\delta}^{i j} {\partial}_{i}{{}\psi_1}\,  {\partial}_{j}{{}\psi_1}\,  \kappa{}^{-1} a.
\end{eqnarray}

Since, $\varphi$ is unperturbed, variation of (\ref{hamuniden}) with respect to $\pi_{\varphi1}$ vanishes. 
\begin{eqnarray}
&&~~~\frac{\delta \mathcal{H}_2^C}{\delta \pi_{\varphi1}} = 0 \nonumber \\
&&\Rightarrow \pi_{\varphi1} = - \frac{N_1}{N_0} {}\pi_{\varphi0} - 3 {}\pi_{\varphi0} {}\psi_1
\end{eqnarray}

Explicit expression of $\pi_1^{i j}$ is obtained by varying the above Hamiltonian with respect to $\pi_1^{i j}$.
\begin{eqnarray}
&&~~~\partial_{0}\gamma_{i j} = \frac{\delta \mathcal{H}_2^C}{\delta \pi_1^{i j}} \nonumber \\
&&\Rightarrow \pi_1^{i j} = \kappa{}^{-1} (\frac{1}{2}\, N_0{}^{-1} a {\delta}^{i j} {\partial}_{k}{{N_1}^{k}}\,  + 2\, N_0{}^{-1} {\delta}^{i j} {\partial}_{0}{a}\,  {}\psi_1 + N_0{}^{-1} a {\delta}^{i j} {\partial}_{0}{\psi_1}\,  - \frac{1}{2}\, N_0{}^{-1} a {\delta}^{k j} {\partial}_{k}{{N_1}^{i}}\, )\nonumber \\
&&~~~~~~~~~~~ - N_0{}^{-1} N_1 \pi_0{}^{i j} + \pi_0{}^{i j} {}\psi_1
\end{eqnarray}

Using the above definitions and varying the Hamiltonian (\ref{hamuniden}) with respect to $N_1$, we get the Hamiltonian Constraint, which, in conformal coordinate becomes

\begin{eqnarray}
&&~~~\frac{\delta \mathcal{H}_2^C}{\delta N_1} = 0 \nonumber \\
&&\Rightarrow - 2\, {\delta}_{i j} {\delta}_{k l} \kappa \pi_0{}^{i j} {\pi_1}^{k l} a + {\delta}_{i j} {\delta}_{k l} \kappa \pi_0{}^{i j} {{}\pi_0{}}^{k l} {}\psi_1 a + 4\, {\delta}_{i j} {\delta}_{k l} \kappa {{}\pi_0{}}^{i k} {\pi_1}^{j l} a - 2\, {\delta}_{i j} {\delta}_{k l} \kappa {{}\pi_0{}}^{i k} {{}\pi_0{}}^{j l} {}\psi_1 a + \nonumber \\
&&~~~~~~~~~~{}\pi_{\varphi0} \pi_{\varphi1} a{}^{(-3)} + \frac{3}{2}\, {}\pi_{\varphi0}{}^{2} a{}^{(-3)} {}\psi_1 - 2\, {\delta}^{i j} {\partial}_{i j}{{}\psi_1}\,  \kappa{}^{-1} a - 3\, V_0 a{}^{3} {}\psi_1 = 0 \nonumber \\
&&\Rightarrow - 3 H \psi_1^\prime +  \delta^{i j} \partial_{i j}{}\psi_1 -  H \delta^{i j} \partial_{i j}B_1 -  \kappa \phi_1 V_0 a^2 = 0.
\end{eqnarray}

Similarly, varying the Hamiltonian with respect to $N_1^i$ leads to Momentum constraint and in conformal coordinate, it becomes

\begin{eqnarray}
&&~~~ \frac{\delta \mathcal{H}_2^C}{\delta N_1^{i}} = 0 \nonumber \\
&&\Rightarrow - 2 \delta_{i j} \partial_i \pi_1^{j k} + 4 \delta_{i j} \partial_k \psi_1 {}\pi_0{}^{i j} - 2 \delta_{j k} {}\pi_0{}^{j k} \partial_i \psi_1 = 0 \nonumber \\
&& \Rightarrow \partial_{i}{\psi_1^\prime} + H \partial_i \phi_1 = 0.
\end{eqnarray}

Finally, the equation of motion of $\psi_1$ is given by
\begin{eqnarray}
&&~~~\delta^{i j} \left(\partial_{0}{\pi_1^{i j}} + \frac{\delta \mathcal{H}_2^C}{\delta \gamma_{i j}}\right) = 0 \nonumber \\
&&\Rightarrow \delta^{i j}\left(\partial_{0}{\pi_1^{i j}} + \frac{\delta \mathcal{H}_2^C}{\delta \psi_1} \frac{\partial \psi}{\partial \gamma_{i j}}\right) = 0 \nonumber \\
&& \Rightarrow 2\, {\delta}^{i j} H\,  {\partial}_{i j}{B_1} + {\delta}^{i j} {\partial}_{i j}{B_1^\prime}\,   + 6H\, {\psi^\prime} + 3\, \psi^{\prime \prime}\,   + 3\kappa
\, V_0 \phi_1 a{}^{2} +  \nonumber \\
&&~~~~3H\, {\phi_1^\prime} - {\delta}^{i j} {\partial}_{i j}{\psi}\,   + {\delta}^{i j} {\partial}_{i j}{\phi_1}\,   = 0.
\end{eqnarray}

It can be verified that above equations are consistent with Lagrangian equations of motion.

\section{Perturbed Lagrangian equations of motion of Galilean scalar field in flat-slicing gauge}\label{galmod}

\subsection{Background equations}
In flat-slicing gauge, using (\ref{perturbfm1}) and
(\ref{perturbfm2}), (\ref{3metric}) and (\ref{field}), we get the
zeroth order Lagrangian density of the Galilean field (\ref{action})
\begin{equation}
\label{lb}
 \mathcal{L}_{G0} = - 3\, N_0{}^{-1} \kappa{}^{-1} a^{\prime}\, {}^{2} a - 2 a^{\prime}\,  N_0{}^{(-3)} \varphi_0^{\prime}\, {}^{3} a{}^{2} - N_0 V_0 a{}^{3}.
\end{equation}

Varying the Lagrangian (\ref{lb}) with respect to the
zeroth order Lapse function $N_0$, we get Hamiltonian constraint at
zeroth order and it is given by
\begin{equation}
\label{hc0l}
 3 N_0^{-2} \kappa^{-1} {a^\prime}{}^2 a + 6 N_0^{-4} \varphi_0^\prime \,{}^3 a^2 a^\prime - V_0 a^3 = 0.
\end{equation}

In conformal coordinate, the above equation takes the form
\begin{equation}
\label{hc0g1}
V_0 a{}^{2} - 6\, H a{}^{-2} \varphi_0^{\prime}\, {}^{3} - 3\,  \kappa{}^{-1} H^{2} = 0.
\end{equation}

Similarly, equation of motion of $a$ in conformal coordinate is given by
\begin{equation}
\label{eomag1}
6\, H\,  \varphi_0^{\prime}\, {}^{3} a{}^{(-2)} - 6\, \varphi_0^{\prime \prime}\,  \varphi_0^{\prime}\, {}^{2} a{}^{-2} + 3\, \kappa{}^{-1} H^2 - 6\, \frac{a^{\prime \prime}}{a}\,  \kappa{}^{-1} + 3\, V_0 a{}^{2}= 0
\end{equation}
and equation of motion of $\varphi_0$ in conformal coordinate is given by
\begin{equation}
\label{eomphig1}
 \frac{a^{\prime \prime}}{2\, a\, H}\,\varphi_0^{\prime}{}^2\,   - \frac{1}{2}\,H \varphi_0^{\prime}{}^2 + \varphi_0^{\prime \prime} \varphi_0^\prime\,  - \frac{1}{12\, H}\, V_{\varphi}\,   a{}^{4} = 0.
\end{equation}

\subsection{First order perturbation}\label{firstGaleq}

In conformal coordinate, it can be shown that, the equation of motion
of $N_1$ or the first order Hamiltonian constraint is
\begin{eqnarray}
\label{hc1ga}
&&24\, H\,  \phi_1 \varphi_0^{\prime}\, {}^{3} a{}^{(-2)} - 18\, \varphi_1^{\prime}\,  H\,  \varphi_0^{\prime}\, {}^{2} a{}^{(-2)} + V_{\varphi}\,  a{}^{2} \varphi_1 + 2\, {\delta}^{i j} {\partial}_{i j}{B_1}\,  \varphi_0^{\prime}\, {}^{3} a{}^{-2} + \nonumber \\
&&2\, {\delta}^{i j} {\partial}_{i j}{\varphi_1}\,  \varphi_0^{\prime}\, {}^{2} a{}^{-2} + 2\, {\delta}^{i j} H\,  {\partial}_{i j}{B_1}\,  \kappa{}^{-1} + 6\, \phi_1 \kappa{}^{-1} H = 0.
\end{eqnarray}

Similarly, equation of motion of $N_1^{i}$ or the first order
perturbed Momentum constraint is the following
\begin{equation}
\label{mc1ga}
- 6\, H\,  {\partial}_{i}{\varphi_1}\,  \varphi_0^{\prime}\, {}^{2}  - 2\, {\partial}_{i}{\phi_1}\,  \varphi_0^{\prime}\, {}^{3} + 2\, {\partial}_{i}{\varphi_1^\prime}\,  \varphi_0^{\prime}\, {}^{2} - 2\, H\,  {\partial}_{i}{\phi_1}\,  \kappa{}^{-1} a^2 = 0
\end{equation}
and equation of motion of $\varphi_1$ is given by
\begin{eqnarray}
\label{eomphi1ga}
&& - 18\, \phi_1 \varphi_0^{\prime}\, {}^{2} H^2 + 12\, \varphi_0^{\prime}\,  \varphi_1^{\prime}\,  H^2 + 18\, {\phi_1}^\prime\,  H\,  \varphi_0^{\prime}\, {}^{2}  \nonumber \\
&&+ 36\, \varphi_0^{\prime}\,  H\,  \varphi_0^{\prime \prime}\,  \phi_1  - 12\, \varphi_0^{\prime}\,  H\,  \varphi_1^{\prime \prime} - 12\, \varphi_1^{\prime}\,  H\,  \varphi_0^{\prime \prime} \nonumber \\
&& - 12\, \varphi_0^{\prime}\,  \varphi_1^{\prime}\,  a^{\prime \prime}\,  a{}^{-1} + 6\, {\delta}^{i j} H\,  {\partial}_{i j}{B_1}\,  \varphi_0^{\prime}\, {}^{2}  + 4\, {\delta}^{i j} \varphi_0^{\prime}\,  \varphi_0^{\prime \prime}\,  {\partial}_{i j}{B_1}\,  + \nonumber \\
&&4\, {\delta}^{i j} \varphi_0^{\prime}\,  H\,  {\partial}_{i j}{\varphi_1} + 4\, {\delta}^{i j} \varphi_0^{\prime \prime}\,  {\partial}_{i j}{\varphi_1}\,  + 2\, {\delta}^{i j} {\partial}_{i j}{B_1^\prime}\,  \varphi_0^{\prime}\, {}^{2} + V_{\varphi}\,  \phi_1 a{}^{4} + \nonumber \\
&&V_{\varphi \varphi}\,  a{}^{4} \varphi_1 + 2\, {\delta}^{i j} {\partial}_{i j}{\phi_1}\,  \varphi_0^{\prime}\, {}^{2}+ 18\, a^{\prime \prime}\,  \phi_1 \varphi_0^{\prime}\, {}^{2} a{}^{-1} = 0.
\end{eqnarray}

\section{Interaction Hamiltonian for higher order correlations of Galilean scalar field}\label{intHamGal}
Third order Interaction Hamiltonian of Galilean scalar field model
(\ref{action}), which is needed to compute Bi-spectrum, can be
obtained by substituting (\ref{fog}) in the Hamiltonian (\ref{fullH})
and extract the third order perturbed part as

\begin{eqnarray}
&& \mathcal{H}_{3} = \frac{1}{2}\, N_1 V_{\varphi \varphi} \varphi_1{}^{2} a{}^{3} + {N_1}^{i} \pi_{\lambda 1} {\partial}_{i}{\lambda_1}\,  + {N_1}^{i} \pi_{\varphi1} {\partial}_{i}{\varphi_1}\,  - {}\pi_{\lambda0} \pi_{\varphi1} N_1{}^{2} - \pi_{\lambda 1} {}\pi_{\varphi0} N_1{}^{2}   \nonumber \\
&&- 2\, N_0 N_1 \pi_{\lambda 1} \pi_{\varphi1} + \pi_{\lambda 1} \lambda_1 {\partial}_{i}{{N_1}^{i}}\,  -3\, N_1 S_0 N_0{}^{2} \pi_{\lambda 1}{}^{2} a{}^{3} - 6\, N_0 {}\pi_{\lambda0} \pi_{\lambda 1} S_0 N_1{}^{2} a{}^{3}  \nonumber \\
&&- S_0 N_1{}^{3} {}\pi_{\lambda0}{}^{2} a{}^{3} - S_1 N_0{}^{3} \pi_{\lambda 1}{}^{2} a{}^{3} -6\, N_1 {}\pi_{\lambda0} \pi_{\lambda 1} S_1 N_0{}^{2} a{}^{3} -3\, N_0 S_1 N_1{}^{2} {}\pi_{\lambda0}{}^{2} a{}^{3} \nonumber \\
&& - \frac{3}{2}\, N_0 {}\pi_{\lambda0} \pi_{\lambda 1} \kappa \lambda_1{}^{2} a{}^{(-3)} - \frac{3}{2}\, N_0 \kappa \lambda_0 \lambda_1 \pi_{\lambda 1}{}^{2} a{}^{(-3)} -\frac{3}{4}\, N_1 \kappa {}\pi_{\lambda0}{}^{2} \lambda_1{}^{2} a{}^{(-3)}  
 - \nonumber \\
&&{N_1}^{i} \pi_{\lambda 1} \lambda_0 {\partial}_{i}{N_1}\,  N_0{}^{-1} - {N_1}^{i} {}\pi_{\lambda0} \lambda_1 {\partial}_{i}{N_1}\,  N_0{}^{-1} - N_1 {\delta}^{i j} \lambda_0 {\partial}_{i}{N_1}\,  {\partial}_{j}{\varphi_1}\,  N_0{}^{(-2)} a{}^{(-2)} + \nonumber \\
&&~~~~~{\delta}^{i j} \lambda_1 {\partial}_{i}{N_1}\,  {\partial}_{j}{\varphi_1}\,  N_0{}^{-1} a{}^{(-2)} + N_1 S_0 {\delta}^{i j} {\partial}_{i}{\varphi_1}\,  {\partial}_{j}{\varphi_1}\,  a + N_0 S_1 {\delta}^{i j} {\partial}_{i}{\varphi_1}\,  {\partial}_{j}{\varphi_1}\,  a - \nonumber \\
&&~~~~~N_1 \pi_{\lambda 1} {\delta}_{i j} \kappa \lambda_0 {\pi_1}^{i j} a{}^{-1} - N_0 \pi_{\lambda 1} {\delta}_{i j} \kappa \lambda_1 {\pi_1}^{i j} a{}^{-1} - N_1 {}\pi_{\lambda0} {\delta}_{i j} \kappa \lambda_1 {\pi_1}^{i j} a{}^{-1}  - \nonumber \\
&&~~~~~N_1 {\delta}_{i j} {\delta}_{k l} \kappa {\pi_1}^{i j} {\pi_1}^{k l} a + 2\, N_1 {\delta}_{i j} {\delta}_{k l} \kappa {\pi_1}^{i k} {\pi_1}^{j l} a- \frac{3}{4}\, N_1 \kappa \pi_{\lambda 1}{}^{2} \lambda_0{}^{2} a{}^{(-3)} + \nonumber \\
&&N_1 {N_1}^{i} {}\pi_{\lambda0} \lambda_0 {\partial}_{i}{N_1}\,  N_0{}^{(-2)}- N_1 \pi_{\lambda 1} {\delta}_{i j} \kappa \lambda_1 \pi_0{}^{i j} a{}^{-1}- 3\, N_1 {}\pi_{\lambda0} \pi_{\lambda 1} \kappa \lambda_0 \lambda_1 a{}^{(-3)}.
\end{eqnarray}

First order and zeroth order Hamiltonian relations along with
(\ref{ncomp}) can be used to express the third order Hamiltonian in
terms of a single field and its derivatives so that we can compute the
correlation function.

Similarly, fourth Order interaction Hamiltonian, which helps to
compute Tri-spectrum, is given by
\begin{eqnarray}
&&\mathcal{H}_4 = - \pi_{\lambda 1} \pi_{\varphi1} N_1{}^{2} - 3\, N_0 S_0 N_1{}^{2} \pi_{\lambda 1}{}^{2} a{}^{3} - 2\, {}\pi_{\lambda0} \pi_{\lambda 1} S_0 N_1{}^{3} a{}^{3}   - \nonumber \\
&&~~~~S_1 N_1{}^{3} {}\pi_{\lambda0}{}^{2} a{}^{3} - \frac{3}{4}\, N_0 \kappa \pi_{\lambda 1}{}^{2} \lambda_1{}^{2} a{}^{(-3)} - \frac{3}{2}\, N_1 {}\pi_{\lambda0} \pi_{\lambda 1} \kappa \lambda_1{}^{2} a{}^{(-3)}  -  \nonumber \\
&&~~~~{N_1}^{i} {}\pi_{\lambda0} \lambda_0 {\partial}_{i}{N_1}\,  N_0{}^{(-3)} N_1{}^{2} +N_1 {N_1}^{i} \pi_{\lambda 1} \lambda_0 {\partial}_{i}{N_1}\,  N_0{}^{(-2)}  - \nonumber \\
&&~~~~{N_1}^{i} \pi_{\lambda 1} \lambda_1 {\partial}_{i}{N_1}\,  N_0{}^{-1} + {\delta}^{i j} \lambda_0 {\partial}_{i}{N_1}\,  {\partial}_{j}{\varphi_1}\,  N_0{}^{(-3)} N_1{}^{2} a{}^{(-2)}  +  \nonumber \\
&&~~~~N_1 S_1 {\delta}^{i j} {\partial}_{i}{\varphi_1}\,  {\partial}_{j}{\varphi_1}\,  a - N_1 \pi_{\lambda 1} {\delta}_{i j} \kappa \lambda_1 {\pi_1}^{i j} a{}^{-1}- 6\, N_0 {}\pi_{\lambda0} \pi_{\lambda 1} S_1 N_1{}^{2} a{}^{3}\nonumber \\
&&~~~~ -3\, N_1 S_1 N_0{}^{2} \pi_{\lambda 1}{}^{2} a{}^{3}+ N_1 {N_1}^{i} {}\pi_{\lambda0} \lambda_1 {\partial}_{i}{N_1}\,  N_0{}^{(-2)}- \frac{3}{2}\, N_1 \kappa \lambda_0 \lambda_1 \pi_{\lambda 1}{}^{2} a{}^{(-3)}\nonumber \\
&&~~~~- N_1 {\delta}^{i j} \lambda_1 {\partial}_{i}{N_1}\,  {\partial}_{j}{\varphi_1}\,  N_0{}^{(-2)} a{}^{(-2)}.
\end{eqnarray}

\section{Galilean and Canonical scalar field}\label{galcan}
Now we proceed to the action where both canonical part is present in a
Galilean fields model. The action is given by

\begin{eqnarray}
&&\mathcal{S} = \int d^4 x \sqrt{-g} \left[\frac{1}{2 \kappa} R - \frac{1}{2} \beta\, \partial_{\mu}{\varphi} \partial_{\nu}{\varphi} - \alpha\, g^{\mu \nu} \partial_{\mu}{\varphi} \partial_{\nu}{\varphi} \Box \varphi  - V(\varphi)\right] \nonumber \\
&&~~= \int d^4 x \sqrt{-g} \left[\frac{1}{2 \kappa} R - \frac{1}{2} \beta\, \partial_{\mu}{\varphi} \partial_{\nu}{\varphi} - \alpha\, g^{\mu \nu} \partial_{\mu}{\varphi} \partial_{\nu}{\varphi} S  - V(\varphi)\right] + \int d^4x \lambda \left( S - \Box \varphi \right).~~~~~~~~ \\
&&~~= \int \mathcal{L} \,d^4x 
\end{eqnarray}

Expanding the above action using ADM decomposition, we get {\small
\begin{eqnarray}
&& \mathcal{L} = S \lambda - N V \gamma{}^{\frac{1}{2}} + {\gamma}^{i j} {\partial}_{i}{\lambda}\,  {\partial}_{j}{\varphi}\,  + \lambda {\partial}_{i}{{\gamma}^{i j}}\,  {\partial}_{j}{\varphi}\,  - {\lambda^\prime}\,  {\varphi^\prime}\,  N{}^{(-2)} + \frac{1}{2}\, N ^{(3)}R \gamma{}^{\frac{1}{2}} \kappa{}^{-1} + {N}^{i} {\lambda^\prime}\,  {\partial}_{i}{\varphi}\,  N{}^{(-2)} + \nonumber \\
&&~~~~{N}^{i} {\varphi^\prime}\,  {\partial}_{i}{\lambda}\,  N{}^{(-2)} + \frac{1}{2}\, \beta N{}^{-1} \gamma{}^{\frac{1}{2}} {\varphi^\prime}\, {}^{2} + \lambda {N^\prime}\,  {\varphi^\prime}\,  N{}^{(-3)} - {N}^{i} {N}^{j} {\partial}_{i}{\lambda}\,  {\partial}_{j}{\varphi}\,  N{}^{(-2)} - {N}^{i} \lambda {N^\prime}\,  {\partial}_{i}{\varphi}\,  N{}^{(-3)} - \nonumber \\
&&~~~~{N}^{i} \lambda {\varphi^\prime}\,  {\partial}_{i}{N}\,  N{}^{(-3)} + S \alpha N{}^{-1} \gamma{}^{\frac{1}{2}} {\varphi^\prime}\, {}^{2} + {\gamma}^{i j} {\gamma}^{k l} \lambda {\partial}_{i}{{\gamma}_{j k}}\,  {\partial}_{l}{\varphi}\,  - \frac{1}{2}\, {\gamma}^{i j} {\gamma}^{k l} \lambda {\partial}_{i}{{\gamma}_{k l}}\,  {\partial}_{j}{\varphi}\,  + \nonumber \\
&&~~~~\frac{1}{2}\, {\gamma}^{i j} \lambda {{\gamma}_{i j}}{}^\prime\,  {\varphi^\prime}\,  N{}^{(-2)} - {\gamma}^{i j} \lambda {\partial}_{i}{N}\,  {\partial}_{j}{\varphi}\,  N{}^{-1} - \frac{1}{2}\, N \beta {\gamma}^{i j} {\partial}_{i}{\varphi}\,  {\partial}_{j}{\varphi}\,  \gamma{}^{\frac{1}{2}}%
 + {N}^{i} {N}^{j} \lambda {\partial}_{i}{N}\,  {\partial}_{j}{\varphi}\,  N{}^{(-3)} - \nonumber \\
&&~~~~{N}^{i} \beta {\varphi^\prime}\,  {\partial}_{i}{\varphi}\,  N{}^{-1} \gamma{}^{\frac{1}{2}} - \frac{1}{2}\, {N}^{i} {\gamma}^{j k} \lambda {{\gamma}_{j k}}{}^\prime\,  {\partial}_{i}{\varphi}\,  N{}^{(-2)} - \frac{1}{2}\, {N}^{i} {\gamma}^{j k} \lambda {\varphi^\prime}\,  {\partial}_{i}{{\gamma}_{j k}}\,  N{}^{(-2)} - \nonumber \\
&&~~~~\frac{1}{2}\, {K}^{i j} {K}^{k l} N {\gamma}_{i j} {\gamma}_{k l} \gamma{}^{\frac{1}{2}} \kappa{}^{-1} + \frac{1}{2}\, {K}^{i j} {K}^{k l} N {\gamma}_{i k} {\gamma}_{j l} \gamma{}^{\frac{1}{2}} \kappa{}^{-1} - N S \alpha {\gamma}^{i j} {\partial}_{i}{\varphi}\,  {\partial}_{j}{\varphi}\,  \gamma{}^{\frac{1}{2}} + \nonumber \\
&&~~~~\frac{1}{2}\, {N}^{i} {N}^{j} \beta {\partial}_{i}{\varphi}\,  {\partial}_{j}{\varphi}\,  N{}^{-1} \gamma{}^{\frac{1}{2}} + \frac{1}{2}\, {N}^{i} {N}^{j} {\gamma}^{k l} \lambda {\partial}_{i}{{\gamma}_{k l}}\,  {\partial}_{j}{\varphi}\,  N{}^{(-2)} - 2\, {N}^{i} S \alpha {\partial}_{0}{\varphi}\,  {\partial}_{i}{\varphi}\,  N{}^{-1} \gamma{}^{\frac{1}{2}} + \nonumber \\
&&~~~~{N}^{i} {N}^{j} S \alpha {\partial}_{i}{\varphi}\,  {\partial}_{j}{\varphi}\,  N{}^{-1} \gamma{}^{\frac{1}{2}}
\end{eqnarray}}

Definition of all momenta are same as we derived in the Galilean case except $\pi_\varphi$ and it is given by
\begin{eqnarray}
&& \pi_\varphi = - {\lambda^\prime}\,  N{}^{(-2)} + {N}^{i} {\partial}_{i}{\lambda}\,  N{}^{(-2)} + \beta N{}^{-1} \gamma{}^{\frac{1}{2}} {\varphi^\prime}\,  + \lambda {N^\prime}\,  N{}^{(-3)} - {N}^{i} \lambda {\partial}_{i}{N}\,  N{}^{(-3)} + \nonumber \\
&&~~~~ 2\, S \alpha N{}^{-1} \gamma{}^{\frac{1}{2}} {\varphi^\prime}\,  + \frac{1}{2}\, {\gamma}^{i j} \lambda {{\gamma}_{i j}}{}^\prime\,  N{}^{(-2)} - {N}^{i} \beta {\partial}_{i}{\varphi}\,  N{}^{-1} \gamma{}^{\frac{1}{2}} - \frac{1}{2}\, {N}^{i} {\gamma}^{j k} \lambda {\partial}_{i}{{\gamma}_{j k}}\,  N{}^{(-2)} - \nonumber \\
&&~~~~2\, {N}^{i} S \alpha {\partial}_{i}{\varphi}\,  N{}^{-1} \gamma{}^{\frac{1}{2}}.
\end{eqnarray}

Then the Dirac-Hamiltonian becomes,
\begin{eqnarray}
&&  \mathcal{H}_D = - S \lambda + N V \gamma{}^{\frac{1}{2}} + {N}^{i} \pi_\lambda {\partial}_{i}{\lambda}\,  + {N}^{i} \pi_\varphi {\partial}_{i}{\varphi}\,  - \pi_\lambda \pi_\varphi N{}^{2} + \pi_\lambda \lambda {\partial}_{i}{{N}^{i}}\,  +  2 {\gamma}_{i j} {\partial}_{k}{{N}^{i}}\,  {\pi}^{j k} -  \nonumber \\
&&~~~~~~ {\gamma}^{i j} {\partial}_{i}{\lambda}\,  {\partial}_{j}{\varphi}\,  - \lambda {\partial}_{i}{{\gamma}^{i j}}\,  {\partial}_{j}{\varphi}\,  - \frac{1}{2}\, N ^{(3)}R \gamma{}^{\frac{1}{2}} \kappa{}^{-1} - \frac{1}{2}\, \beta N{}^{3} \pi_\lambda{}^{2} \gamma{}^{\frac{1}{2}} - \frac{3}{4}\, N \kappa \pi_\lambda{}^{2} \gamma{}^{-\frac{1}{2}} \lambda{}^{2} -\nonumber \\
&&~~~~~~ {N}^{i} \pi_\lambda \lambda {\partial}_{i}{N}\,  N{}^{-1} +  {N}^{i}  {\partial}_{i}{{\gamma}_{l m}}\,  {\pi}^{l m} -  {N}^{i}  {\partial}_{l}{{\gamma}_{i m}}\,  {\pi}^{l m} + {N}^{i}  {\partial}_{m}{{\gamma}_{i l}}\,  {\pi}^{l m}  - S \alpha N{}^{3} \pi_\lambda{}^{2} \gamma{}^{\frac{1}{2}} -  \nonumber \\
&&~~~~~~{\gamma}^{i j} {\gamma}^{k l} \lambda {\partial}_{i}{{\gamma}_{j k}}\,  {\partial}_{l}{\varphi}\,  +\frac{1}{2}\, {\gamma}^{i j} {\gamma}^{k l} \lambda {\partial}_{i}{{\gamma}_{k l}}\,  {\partial}_{j}{\varphi}\,  + {\gamma}^{i j} \lambda {\partial}_{i}{N}\,  {\partial}_{j}{\varphi}\,  N{}^{-1} + \frac{1}{2}\, N \beta {\gamma}^{i j} {\partial}_{i}{\varphi}\,  {\partial}_{j}{\varphi}\,  \gamma{}^{\frac{1}{2}} -  \nonumber \\
&&~~~~~~ N \pi_\lambda {\gamma}_{i j} \kappa \lambda {\pi}^{i j} \gamma{}^{-\frac{1}{2}} + N S \alpha {\gamma}^{i j} {\partial}_{i}{\varphi}\,  {\partial}_{j}{\varphi}\,  \gamma{}^{\frac{1}{2}} -N {\gamma}_{i j} {\gamma}_{k l} \kappa {\pi}^{i j} {\pi}^{k l} \gamma{}^{-\frac{1}{2})} + 2\, N {\gamma}_{i j} {\gamma}_{k l} \kappa {\pi}^{i k} {\pi}^{j l} \gamma{}^{-\frac{1}{2}}  \nonumber \\
&&~~~~~~ + \xi (\pi_N + \lambda N{}^{-1} \pi_\lambda)
\end{eqnarray}

\subsection{Zeroth order}
Zeroth order Hamiltonian, in terms of $\pi_a$, takes the form
\begin{eqnarray}
&& \mathcal{H}_{D0} = - S_0 \lambda_0 + N_0 V a{}^{3} - {}\pi_{\lambda0} {}\pi_{\varphi0} N_0{}^{2} - \frac{1}{2}\, \beta N_0{}^{3} {}\pi_{\lambda0}{}^{2} a{}^{3} - \frac{3}{4}\, N_0 \kappa {}\pi_{\lambda0}{}^{2} \lambda_0{}^{2} a{}^{(-3)} \nonumber \\
&& - S_0 \alpha N_0{}^{3} {}\pi_{\lambda0}{}^{2} a{}^{3}- \frac{1}{2}\, N_0 \pi_a {}\pi_{\lambda0} \kappa \lambda_0 a{}^{(-2)} - \frac{1}{12}\, N_0 \kappa \pi_a{}^{2} a{}^{-1} + \xi_0 \left( \pi_{N0} + {}\pi_{\lambda0} \lambda_0 N_0{}^{-1} \right).~~~~~~
\end{eqnarray}

Zeroth order Hamiltonian constraint or the equation of motion of $N_0$
is given by
\begin{equation}
V a{}^{2} + \frac{1}{2}\, \beta  \varphi_0^{\prime}\, {}^{2} - 3\,  \kappa{}^{-1} H^2 - 6\, \alpha H\,  a{}^{(-2)} \varphi_0^{\prime}\, {}^{3} = 0.
\end{equation}

Similarly, equations of motion of $a$ and $\varphi_0$ are given by
 \begin{eqnarray}
&& 6\, \alpha H\,  \varphi_0^{\prime}\, {}^{3} a{}^{(-2)} - 6\, \alpha \varphi_0^{\prime \prime}\,  \varphi_0^{\prime}\, {}^{2} a{}^{-2} + 3\, \kappa{}^{-1} H^2 - \nonumber \\
&&~~~~~~~~6\, a^{\prime \prime}\,a^{-1}  \kappa{}^{-1} + 3\, V a{}^{2} - \frac{3}{2}\, \beta \varphi_0^{\prime}\, {}^{2}  = 0, \\ 
&& V_\varphi\,  a{}^{2}- 6\, \alpha a^{\prime \prime}\,  \varphi_0^{\prime}\, {}^{2} a{}^{(-3)} + 6\, \alpha \varphi_0^{\prime}\, {}^{2} H^2 a{}^{(-2)} - 12\, \alpha \varphi_0^{\prime}\,  H\,  \varphi_0^{\prime \prime}\,  a{}^{(-2)} \nonumber \\
&&~~~~+ 2\, \beta \varphi_0^{\prime}\,  H + \beta \varphi_0^{\prime \prime}\, = 0.
 \end{eqnarray}

\subsection{First order}
first order Hamilton's equations are obtained from second order
perturbed Hamiltonian and it is given by
\begin{eqnarray}
&& \mathcal{H}_{D2} = - S_1 \lambda_1 + N_1 V_\varphi\,  a{}^{3} \varphi_1 + \frac{1}{2}\, N_0 V_{\varphi \varphi}\,  \varphi_1{}^{2} a{}^{3} + {N_1}^{i} {}\pi_{\lambda0} {\partial}_{i}{\lambda_1}\,  + {N_1}^{i} {}\pi_{\varphi0} {\partial}_{i}{\varphi_1}\,   - \nonumber \\
&&~~~~2\, N_0 N_1 {}\pi_{\lambda0} \pi_{\varphi1} - 2\, N_0 N_1 \pi_{\lambda 1} {}\pi_{\varphi0} - \pi_{\lambda 1} \pi_{\varphi1} N_0{}^{2} + \pi_{\lambda 1} \lambda_0 {\partial}_{i}{{N_1}^{i}}\,   -\nonumber \\
&&~~~~  {\delta}^{i j} {\partial}_{i}{\lambda_1}\,  {\partial}_{j}{\varphi_1}\,  a{}^{(-2)} - \frac{1}{2}\, \beta N_0{}^{3} \pi_{\lambda 1}{}^{2} a{}^{3} - 3\, N_1 {}\pi_{\lambda0} \pi_{\lambda 1} \beta N_0{}^{2} a{}^{3} - \frac{3}{2}\, N_0 \beta N_1{}^{2} {}\pi_{\lambda0}{}^{2} a{}^{3} -\nonumber \\
&&~~~~ \frac{3}{4}\, N_0 \kappa {}\pi_{\lambda0}{}^{2} \lambda_1{}^{2} a{}^{(-3)} - 3\, N_0 {}\pi_{\lambda0} \pi_{\lambda 1} \kappa \lambda_0 \lambda_1 a{}^{(-3)}%
 - \frac{3}{4}\, N_0 \kappa \pi_{\lambda 1}{}^{2} \lambda_0{}^{2} a{}^{(-3)}  -\nonumber \\
&&~~~~ \frac{3}{2}\, N_1 {}\pi_{\lambda0} \pi_{\lambda 1} \kappa \lambda_0{}^{2} a{}^{(-3)} - {N_1}^{i} {}\pi_{\lambda0} \lambda_0 {\partial}_{i}{N_1}\,  N_0{}^{-1} - S_0 \alpha N_0{}^{3} \pi_{\lambda 1}{}^{2} a{}^{3}  - \nonumber \\
&&~~~~ 3\, N_0 S_0 \alpha N_1{}^{2} {}\pi_{\lambda0}{}^{2} a{}^{3} - 2\, {}\pi_{\lambda0} \pi_{\lambda 1} S_1 \alpha N_0{}^{3} a{}^{3} - 3\, N_1 S_1 \alpha N_0{}^{2} {}\pi_{\lambda0}{}^{2} a{}^{3}  + \nonumber \\
&&~~~~\frac{1}{2}\, N_0 \beta {\delta}^{i j} {\partial}_{i}{\varphi_1}\,  {\partial}_{j}{\varphi_1}\,  a - N_0 \pi_{\lambda 1} {\delta}_{i j} \kappa \lambda_0 {\pi_1}^{i j} a{}^{-1} - N_1 {}\pi_{\lambda0} {\delta}_{i j} \kappa \lambda_0 {\pi_1}^{i j} a{}^{-1}  -\nonumber \\
&&~~~~ N_0 {}\pi_{\lambda0} {\delta}_{i j} \kappa \lambda_1 {\pi_1}^{i j} a{}^{-1} - N_0 \pi_{\lambda 1} {\delta}_{i j} \kappa \lambda_1 \pi_0{}^{i j} a{}^{-1} - N_1 {}\pi_{\lambda0} {\delta}_{i j} \kappa \lambda_1 \pi_0{}^{i j} a{}^{-1}  -\nonumber \\
&&~~~~ N_0 {\delta}_{i j} {\delta}_{k l} \kappa {\pi_1}^{i j} {\pi_1}^{k l} a - 2\, N_1 {\delta}_{i j} {\delta}_{k l} \kappa \pi_0{}^{i j} {\pi_1}^{k l} a%
 + 2\, N_0 {\delta}_{i j} {\delta}_{k l} \kappa {\pi_1}^{i k} {\pi_1}^{j l} a   +\nonumber \\
&&~~~~  \pi_{N1} \xi_1 - N_1 {}\pi_{\lambda0} \lambda_0 N_0{}^{(-2)} \xi_1 + {}\pi_{\lambda0} \lambda_0 N_0{}^{(-3)} N_1{}^{2} \xi_0 + \pi_{\lambda 1} \lambda_0 N_0{}^{-1} \xi_1  + \nonumber \\
&&~~~~ {}\pi_{\lambda0} \lambda_1 N_0{}^{-1} \xi_1 - N_1 {}\pi_{\lambda0} \lambda_1 N_0{}^{(-2)} \xi_0 + \pi_{\lambda 1} \lambda_1 N_0{}^{-1} \xi_0- {}\pi_{\lambda0} {}\pi_{\varphi0} N_1{}^{2} +\nonumber \\
&&~~~~ {}\pi_{\lambda0} \lambda_1 {\partial}_{i}{{N_1}^{i}}\,  +  2  {\delta}_{i j} {\partial}_{k}{{N_1}^{i}}\,  {\pi_1}^{j k} a{}^{2}- \frac{3}{2}\, N_1 \kappa \lambda_0 \lambda_1 {}\pi_{\lambda0}{}^{2} a{}^{(-3)}-\nonumber \\
&&~~~~6\, N_1 {}\pi_{\lambda0} \pi_{\lambda 1} S_0 \alpha N_0{}^{2} a{}^{3} + {\delta}^{i j} \lambda_0 {\partial}_{i}{N_1}\,  {\partial}_{j}{\varphi_1}\,  N_0{}^{-1} a{}^{(-2)}- N_1 \pi_{\lambda 1} {\delta}_{i j} \kappa \lambda_0 \pi_0{}^{i j} a{}^{-1}+\nonumber \\
&&~~~~ N_0 S_0 \alpha {\delta}^{i j} {\partial}_{i}{\varphi_1}\,  {\partial}_{j}{\varphi_1}\,  a- N_1 \pi_{\lambda 1} \lambda_0 N_0{}^{(-2)} \xi_0+ 4\, N_1 {\delta}_{i j} {\delta}_{k l} \kappa {{}\pi_0{}}^{i k} {\pi_1}^{j l} a
\end{eqnarray}
Since we have got the perturbed second order Hamiltonian, we can
obtained the field equations using Hamilton's equations. First order
Momentum constraint or the equation of motion of $N_1^{i}$ is given by
\begin{eqnarray}
 &&- 6\, \alpha H\,  {\partial}_{i}{\varphi_1}\,  \varphi_0^{\prime}\, {}^{2}  + \beta \varphi_0^{\prime}\,  {\partial}_{i}{\varphi_1}\,  a{}^{2} - 2\, \alpha {\partial}_{i}{\phi_1}\,  \varphi_0^{\prime}\, {}^{3} +  2\, \alpha {\partial}_{i}{\varphi_1^\prime}\,  \varphi_0^{\prime}\, {}^{2} - \nonumber \\
 &&~~~~~~~~2\, H\,  {\partial}_{i}{\phi_1}\,  \kappa{}^{-1} a^2 = 0.
\end{eqnarray}

Similarly, First order Hamiltonian constraint or the equation of
motion of $N_1$ is given by
\begin{eqnarray}
&&24\, \alpha H\,  \phi_1 \varphi_0^{\prime}\, {}^{3} a{}^{(-2)} - 18\, \alpha \varphi_1^{\prime}\,  H\,  \varphi_0^{\prime}\, {}^{2} a{}^{(-2)} + V_\varphi \,  a{}^{2} \varphi_1 - \beta \phi_1 \varphi_0^{\prime}\, {}^{2}  +\nonumber \\
 && 2\, \alpha {\delta}^{i j} {\partial}_{i j}{B_1}\,  \varphi_0^{\prime}\, {}^{3} a{}^{-2} + \beta \varphi_0^{\prime}\,  \varphi_1^{\prime} + 2\, \alpha {\delta}^{i j} {\partial}_{i j}{\varphi_1}\,  \varphi_0^{\prime}\, {}^{2} a{}^{-2} \nonumber \\
 &&+ 2\, {\delta}^{i j} H\,  {\partial}_{i j}{B_1}\,  \kappa{}^{-1} + 6\, \phi_1 \kappa{}^{-1} H = 0
\end{eqnarray}
and the equation of motion of $\varphi_1$ is given by

\begin{eqnarray}
&&- 18\, \alpha \phi_1 \varphi_0^{\prime}\, {}^{2} H^2 + 12\, \alpha \varphi_0^{\prime}\,  \varphi_1^{\prime}\,  H^2 + 18\, \alpha {\phi_1^\prime}\,  H\,  \varphi_0^{\prime}\,{}^{2} \nonumber \\
 && + 36\, \alpha \varphi_0^{\prime}\,  H\,  \varphi_0^{\prime \prime}\,  \phi_1  - 12\, \alpha \varphi_0^{\prime}\,  H\,  \varphi_1^{\prime \prime} - 12\, \alpha \varphi_1^{\prime}\,H\,  \varphi_0^{\prime \prime}\nonumber \\
 && + 18\, \alpha a^{\prime \prime}\,  \phi_1 \varphi_0^{\prime}\, {}^{2} a{}^{-1} - 12\, \alpha \varphi_0^{\prime}\,  \varphi_1^{\prime}\,  a^{\prime \prime}\,  a{}^{-1} - 2\, \beta \varphi_0^{\prime}\,  H\,  \phi_1 a^2 - \beta {\phi_1^\prime}\,  \varphi_0^{\prime}\,  a{}^{2}\nonumber \\
 && - \beta \varphi_0^{\prime \prime}\,  \phi_1 a{}^{2} + 6\, \alpha {\delta}^{i j} H\,  {\partial}_{i j}{B_1}\,  \varphi_0^{\prime}\, {}^{2}  + 4\, \alpha {\delta}^{i j} \varphi_0^{\prime}\,  \varphi_0^{\prime \prime}\,  {\partial}_{i j}{B_1}\,  + 2\, \beta \varphi_1^{\prime}\,  H\,  a^2\nonumber \\
 && + \beta \varphi_1^{\prime \prime}\,  a{}^{2} + 4\, \alpha {\delta}^{i j} \varphi_0^{\prime}\,  H\,  {\partial}_{i j}{\varphi_1} + 4\, \alpha {\delta}^{i j} \varphi_0^{\prime \prime}\,  {\partial}_{i j}{\varphi_1}\,  + 2\, \alpha {\delta}^{i j} {\partial}_{i j}{B_1^\prime}\,  \varphi_0^{\prime}\, {}^{2}\nonumber \\
 && + V_\varphi\,  \phi_1 a{}^{4}%
 + V_{\varphi \varphi}\,  a{}^{4} \varphi_1 - \beta {\delta}^{i j} \varphi_0^{\prime}\,  {\partial}_{i j}{B_1}\,  a{}^{2} + 2\, \alpha {\delta}^{i j} {\partial}_{i j}{\phi_1}\,  \varphi_0^{\prime}\, {}^{2} - \beta {\delta}^{i j} {\partial}_{i j}{\varphi_1}\,  a{}^{2} = 0.~~~~~~~~~
\end{eqnarray}
	\pagestyle{fancy}
\chapter{}\label{app.3}
\section{Inversion formulae of $X$ and $\pi_\varphi$ and $G(\gamma,
  \pi_\varphi, Y, \varphi)$ for different scalar field
  models}{\label{Inversion}}
\subsection{Canonical scalar field}{\label{InvCanonical}}
In case of canonical scalar field, $P(X, \varphi)$ is given by $$P(X, \varphi) = - X - V(\varphi).$$
Hence, using equation (\ref{piphigeneral}), we get
\begin{eqnarray}
\pi_\varphi &=& \sqrt{\gamma}\sqrt{- 2X + Y} \nonumber \\
\Rightarrow X &=& \frac{1}{2} Y - \frac{\pi_\varphi^2}{2 \gamma}
\end{eqnarray}
and $G(\gamma, \pi_\varphi, Y, \varphi)$ is given by the relation (\ref{G})
\begin{equation}
G(\gamma, \pi_\varphi, Y, \varphi) = -\frac{1}{2}\frac{\pi_\varphi^2}{\gamma} - \frac{1}{2}\,Y - V(\varphi).
\end{equation}

\subsection{Tachyonic field}
Tachyons are described by $$P(X, \varphi) = - V(\varphi) \sqrt{1 + 2X}.$$
Similarly, in case of Tachyons, we get
\begin{eqnarray}
X &=& \frac{\gamma \, V^2\, Y - \pi_\varphi^2}{2\,(\pi_\varphi^2 + \gamma\, V^2)} \\
G(\gamma, \pi_\varphi, Y, \varphi) &=& -\frac{1}{\sqrt{\gamma}}\,\sqrt{1 + Y} \sqrt{\pi_\varphi^2 + \gamma \,V^2}.
\end{eqnarray}

\subsection{DBI field}
For DBI field, $$P(X, \varphi) = -\frac{1}{f(\varphi)} \left(\sqrt{1 +
    2\,f(\varphi)\, X} - 1 \right)- V(\varphi)$$ which implies that,
\begin{eqnarray}
X &=& \frac{\gamma \, Y - \pi_\varphi^2}{2\,(\gamma + f\,\pi_\varphi^2 )} \\
G(\gamma, \pi_\varphi, Y, \varphi) &=& -\frac{1}{\gamma\,f(\varphi)}\sqrt{\left(\gamma + f(\varphi)\, Y\right) \left(f(\varphi)\, \pi_\varphi^2 + \gamma\right)} + \frac{1}{f(\varphi)} - V(\varphi)
\end{eqnarray}

\section{Langlois' approach for non-canonical scalar field}\label{Langlois-nonC}
Two decades back, Langlois' obtained a consistent Hamiltonian for
canonical scalar field\cite{Langlois1994}. In this section, we extend
the method to non-canonical scalar fields.

Following \cite{Langlois1994}, expressing background 3-metric
$\gamma_{0 i j} = e^{2 \alpha}$, it can be shown that the first order
perturbed Hamiltonian constraint takes the form

\begin{eqnarray}\label{LangHam}
  \mathcal{H}_{N1} &\equiv & - \frac{e^{3 \alpha}}{2 \kappa}\, \left[\,\gamma_{0}^{i k} \gamma_0^{j l} - \gamma_{0}^{i j} \gamma_0^{k l} \,\right] \partial_{i j} \gamma_{1 k l} - \frac{\kappa}{3} e^{- 3 \alpha} \,\pi_\alpha\, \gamma_{0ij}\, \pi_1^{i j} - \Big[\,\frac{\kappa}{72}\,e^{-3\alpha}\,\pi_\alpha^2 + \nonumber \\
  &&\frac{1}{2}\,e^{3\alpha}\, G \,\Big]\,\gamma_0^{i j}\, \gamma_{1 i j} - e^{3\alpha}\, G_{\pi_\varphi}\, \pi_{\varphi 1} - e^{3\alpha}\,G_{\varphi}\, \varphi_1 = 0,
\end{eqnarray}  
where $\pi_\alpha$ is the momentum corresponding to $\alpha$. Similarly, first order perturbed Momentum constraint becomes
\begin{equation}\label{LangMom}
\mathcal{H}_{1i} \equiv - 2\, \partial_k \gamma_{1i j}\, \pi_0^{j k} - 2\, \gamma_{0ij}\, \partial_k \pi_1^{j k} + \pi_0^{j k}\, \partial_i \gamma_{1jk} + \pi_{\varphi1}\, \partial_i \varphi_1 = 0.
\end{equation}
 
In momentum space, equations (\ref{LangHam}) and (\ref{LangMom}) becomes
\begin{eqnarray}\label{HamK}
  \mathcal{H}_{N1}(k) &\equiv & - \frac{e^{3 \alpha}}{2 \kappa}\, \left[\,\gamma_{0}^{i k} \gamma_0^{j l} - \gamma_{0}^{i j} \gamma_0^{k l} \,\right] k_i k_j \gamma_{1 k l} - \frac{\kappa}{3} e^{- 3 \alpha} \,\pi_\alpha\, \gamma_{0ij}\, \pi_1^{i j} - \Big[\,\frac{\kappa}{72}\,e^{-3\alpha}\,\pi_\alpha^2 + \nonumber \\
  &&\frac{1}{2}\,e^{3\alpha}\, G \,\Big]\,\gamma_0^{i j}\, \gamma_{1 i j} - e^{3\alpha}\, G_{\pi_\varphi}\, \pi_{\varphi 1} - e^{3\alpha}\,G_{\varphi}\, \varphi_1 = 0 \\
\label{MomK}
\mathcal{H}_{1i}(k) &\equiv & - 2\, k_k\, \gamma_{1i j}\, \pi_0^{j k} - 2\, \gamma_{0ij}\, k_k\, \pi_1^{j k} + \pi_0^{j k}\, k_i\, \gamma_{1jk} + \pi_{\varphi1}\, k_i\, \varphi_1 = 0
\end{eqnarray}

The scalar configuration variables are
\begin{equation}
  \gamma_1  = \frac{1}{3} \gamma_0^{i j} \, \gamma_{1ij}, \quad \gamma_2 = \frac{1}{2}\,\left[\frac{3 k^i \,k^j}{k^2}\,\gamma_{1ij} - \gamma_0^{i j} \, \gamma_{1ij}\right],
\end{equation}
which are associated with their conjugate momenta
\begin{equation}
  \pi^1 = \gamma_{0ij}\, \pi_1^{i j}, \quad \pi^2 = \frac{k_i k_j}{k^2}\, \pi_1^{i j} - \frac{1}{3} \gamma_{0ij} \pi_1^{i j}.
\end{equation}

Hence the energy constraint (\ref{HamK}) becomes
\begin{eqnarray}
  E &=& - \left[ \frac{1}{24} \kappa e^{-3\alpha}\, \pi_\alpha^2 + \frac{3}{2} e^{3\alpha}\, G\right]\, \gamma_1 - \frac{e^{3\alpha}}{\kappa}\,k^2 \,\gamma_1 + \frac{e^{3\alpha}}{3 \kappa}\, k^2 \, \gamma_2 \nonumber \\
  && -\frac{\kappa}{3} e^{-3\alpha}\, \pi_\alpha\, \pi^1 - e^{3\alpha}\, G_{\pi_\varphi}\, \pi_{\varphi1} - e^{3\alpha}\, G_{\varphi}\, \varphi_1 = 0.
\end{eqnarray}

Momentum constraint contains scalar and vector, both
modes. Contracting with $k^i$, we obtain the scalar part of Momentum
constraint
\begin{equation}
M \equiv \frac{1}{6} \pi_\alpha\, \gamma_1 - \frac{2}{9} \pi_\alpha \, \gamma_2 - \frac{2}{3}\pi^1 - 2 \pi^2 + \pi_\alpha\,\varphi_1 = 0
\end{equation}

In case of scalar phase-space, there exist two first class constraints, namely $E$ and $M$ and
\begin{eqnarray}\label{ham1}
E\left(\gamma_\alpha, \pi^\beta = \frac{\partial S}{\partial \gamma_\beta}\right) &=& 0 \\
\label{mom1}
M\left(\gamma_\alpha, \pi^\beta = \frac{\partial S}{\partial \gamma_\beta}\right) &=& 0
\end{eqnarray}
where $S$ is the quadratic generating function, given by
\begin{equation}
S = 	\frac{1}{2}A_{\alpha\beta}\, \gamma_\alpha\,\gamma_\beta+ B_\alpha\,\gamma_\alpha.
\end{equation}
$\alpha, \beta = 0, 1, 2, \gamma_0 = \varphi_1, \pi^0 =
\pi_{\varphi1}$ and $A_{\alpha\beta}$ are symmetric. Hence, equations
(\ref{ham1}) and (\ref{mom1}) become equations for $A_{\alpha\beta}$
and $B_\alpha$ with a polynomial form in $\gamma_\alpha$ and lead to
the following four equations for the Energy constraint:
\begin{eqnarray}
  &&- \left[\frac{1}{24}\kappa\, e^{-3\alpha}\, \pi_\alpha^2 + \frac{3}{2}e^{3\alpha} \,G\right] - \frac{e^{3\alpha}}{\kappa} k^2 - \frac{e^{-3\alpha}\, \kappa}{3}\pi_\alpha\, A_{11} - e^{3\alpha}\, G_{\pi_\varphi} \,A_{01} = 0 \\
  && \frac{e^{3\alpha}}{3\kappa}k^2 - \frac{\kappa}{3} e^{-3\alpha}\, \pi_\varphi \, A_{12} - e^{3\alpha}\,G_{\pi_\varphi}\,A_{02} = 0 \\
  && -\frac{\kappa}{3} e^{-3\alpha}\,\pi_\alpha\, A_{01} - e^{-3\alpha}\, G_{\pi_\varphi}\,A_{00}- e^{3\alpha}\,G_{\varphi} = 0 \\
  && -\frac{\kappa}{3} e^{-3\alpha}\, \pi_\alpha\,B_1- e^{3\alpha}\,G_{\pi_\varphi}\, B_0 = 0
\end{eqnarray}
 and for the Momentum constraint
\begin{eqnarray}
&& \frac{1}{6} \pi_\alpha - \frac{2}{3} A_{11} - 2 A_{21} = 0\\
&& -\frac{2}{9}\pi_\alpha - \frac{2}{3} A_{12}- 2 A_{22} = 0 \\
&& -\frac{2}{3}A_{10} - 2 A_{20} + \pi_\alpha = 0 \\
&& -\frac{2}{3}B_1 - 2 B_2 = 0
\end{eqnarray}
respectively.

The solutions for the $B_\alpha$ form a one-dimensional space and can be written as
\begin{equation}
B_0 = P, \quad B_1 = -\frac{3}{\kappa} e^{6\alpha}\,\frac{G_{\pi_\varphi}}{\pi_\alpha}\,B_0, \quad B_2 = \frac{e^{6\alpha}}{\kappa}\, \frac{G_{\pi_\varphi}}{\pi_\alpha}\,B_0
\end{equation}
where dependence of the $B_\alpha$ on the free parameter $P$ is chosen
for later convenience. $A_{\alpha\beta}$ are undetermined since there
are five out of six independent equations and one is background
equation. The additional condition is arbitrary and independent of any
physical change in the system. The quantity $P$ is the momentum in the
reduced phase-space and its conjugate coordinate is given by

\begin{equation}
Q = \frac{\partial S}{\partial P} =  \varphi_1 + \frac{e^{6\alpha}}{\kappa}\,\frac{G_{\pi_\varphi}}{\pi_\alpha}\, \left(\gamma_2 - 3\gamma_1\right)
\end{equation}
which coincides with gauge-invariant Mukhanov's variable. Other relations between old and new variables are given as
\begin{eqnarray}
\varphi_1 &=& Q + [\gamma_1, \gamma_2], \quad \pi_{\varphi1} = A_{00}\,Q + P + [\gamma_1, \gamma_2] \\
\pi^1 &=& A_{10}\,Q - \frac{3}{\kappa}\,e^{6\alpha}\,\frac{G_{\pi_\varphi}}{\pi_\alpha}\,P + [\gamma_1, \gamma_2], \quad \pi^2 = A_{20}\, Q + \frac{e^{6\alpha}}{\kappa}\,\frac{G_{\pi_\varphi}}{\pi_\alpha}\, P + [\gamma_1, \gamma_2]
\end{eqnarray}
where brackets contain all the terms with $\gamma_1$ or
$\gamma_2$. These are not written explicitly since they are `pure
gauge' and do not contribute to the `true' dynamics.

The second order expansion of the Energy constraint is given by
\begin{eqnarray}\label{b24}
  \mathcal{H}_{N2} &=& \frac{2\kappa}{\sqrt{\gamma}}\, \left(\gamma_{0ik} \gamma_{0jl} -\frac{1}{2} \gamma_{0ij} \gamma_{0kl}\right)\,\pi_1^{ij}\pi_1^{k l} - \frac{\sqrt{\gamma}}{2}\, G_{\pi_\varphi\pi_\varphi}\,\pi_{\varphi1}^2 \nonumber \\
  &&-\frac{\sqrt{\gamma}}{2}\,G_{\varphi\varphi}\,\varphi_1^2-\sqrt{\gamma}\,G_Y\,\gamma_0^{i j} \partial_i\varphi_1\partial_j \varphi_1-\sqrt{\gamma}\,G_{\varphi\pi_\varphi}\,\pi_{\varphi1}\,\varphi_1 + [\gamma_{i j}]
\end{eqnarray}
where $[\gamma_{i j}]$ collectively represents all the terms that
involve $\gamma_{1 i j}$. By choosing $N=1$ and $N^i = 0$ to simplify
calculations, the scalar part of the Hamiltonian is easily obtained
and is given by
\begin{eqnarray}\label{b25}
  H^s &=& \int d^3k \left\{N \mathcal{H}_{N} + N^i \mathcal{H}_i\right\} \nonumber \\
  &=& \int d^3k \Big\{\frac{2\kappa}{\sqrt{\gamma}}\,\left(-\frac{1}{6}\pi_1{}^2 - \frac{3}{2}\pi_2{}^2\right) - \frac{\sqrt{\gamma}}{2}\, G_{\pi_\varphi\pi_\varphi}\,\pi_{\varphi1}^2 \nonumber \\
  &&-\frac{\sqrt{\gamma}}{2}\,G_{\varphi\varphi}\,\varphi_1^2-\sqrt{\gamma}\,G_Y\,k^2 \varphi_1^2-\sqrt{\gamma}\,G_{\varphi\pi_\varphi}\,\pi_{\varphi1}\,\varphi_1 + [\gamma_1, \gamma_2]
  \Big\}.
\end{eqnarray}
Hence the gauge-invariant Hamiltonian is given by
\begin{eqnarray}
  H_{GI}^s &=& H^s + \{S, H_0\}_{Background} \\
  &=& \int d^3k\Bigg[ -\frac{\sqrt{\gamma}}{2}\,G_{\pi_\varphi\pi_\varphi}\,P^2 + \Big\{-\frac{\kappa}{3\sqrt{\gamma}}\,A_{10}^2 + \frac{3\kappa}{\sqrt{\gamma}}\,A_{20}^2 - \frac{\sqrt{\gamma}}{2}\,G_{\pi_\varphi\pi_\varphi}\,A_{00}^2 \nonumber \\
  && -\frac{\sqrt{\gamma}}{2}\,G_{\varphi\varphi}+ \sqrt{\gamma}\,G_Y\,k^2 + \frac{1}{2}\,\dot{A}_{00} - \sqrt{\gamma}\,G_{\varphi\pi_\varphi}\,A_{00}\Big\}Q^2 \nonumber \\
  && + \left\{\frac{2}{\sqrt{\gamma}}e^{6\alpha}\,\frac{G_{\pi_\varphi}}{\pi_\alpha}\,A_{10} + \frac{6}{\sqrt{\gamma}}\, e^{6\alpha}\, \frac{G_{\pi_\varphi}}{\pi_\alpha}\, A_{20} - \sqrt{\gamma}\, G_{\pi_\varphi\pi_\varphi}\,A_{00}\right\}\,P\,Q\Bigg]
\end{eqnarray}
where $\dot{A}_{00} = \{A_{00}, H_0\}_{Background}$. If we impose additional condition
\begin{equation}
\frac{2}{\sqrt{\gamma}}e^{6\alpha}\,\frac{G_{\pi_\varphi}}{\pi_\alpha}\,A_{10} + \frac{6}{\sqrt{\gamma}}\, e^{6\alpha}\, \frac{G_{\pi_\varphi}}{\pi_\alpha}\, A_{20} - \sqrt{\gamma}\, G_{\pi_\varphi\pi_\varphi}\,A_{00} = 0
\end{equation}
in order to cancel cross terms in the above Hamiltonian, we get the following solutions:
\begin{eqnarray}
A_{00} &=& \frac{3\, G_{\pi_\varphi}}{G_{\pi_\varphi\pi_\varphi}}\,\frac{\pi_{\varphi0}}{\pi_\alpha}, \quad A_{10} = -\frac{9 }{\kappa}\,e^{6\alpha}\, \frac{G_{\pi_\varphi}^2\,\pi_{\varphi0}}{G_{\pi_\varphi \pi_\varphi}\,\pi_\alpha^2} - \frac{3}{\kappa}\,e^{6\alpha}\,\frac{\pi_{\varphi0}}{\pi_\alpha}, \nonumber \\
 A_{20} &=& \frac{3}{\kappa}\,e^{6\alpha}\,\frac{G_{\pi_\varphi}^2\,\pi_{\varphi0}}{G_{\pi_\varphi \pi_\varphi}\,\pi_\alpha^2} +  \frac{1}{\kappa} \,e^{6\alpha}\,\frac{\pi_{\varphi0}}{\pi_\alpha} + \frac{1}{2}\, \pi_{\varphi0}.
\end{eqnarray}
Finally, the Hamiltonian takes the form
\begin{eqnarray}
H_{GI}^{s} &=& \int d^3 k \left\{-\frac{1}{2}\,e^{3\alpha}\,G_{\pi_\varphi\pi_\varphi}\,P^2 + \frac{1}{2}\, \left(X + 2\, e^{3\alpha}\,k^2\,G_Y\right)\,Q^2 \right\}, \quad \mbox{where} \\
X &\equiv & 9\,e^{3\alpha}\,\frac{G_{\pi_\varphi}{}^2\,\pi_{\varphi0}{}^2}{G_{\pi_\varphi\pi_\varphi}\pi_\alpha{}^2} - 6\,e^{3\alpha}\,\frac{G_{\varphi\pi_\varphi}\,G_{\pi_\varphi}\,\pi_{\varphi0}}{\pi_\alpha} + \dot{A}_{00} + \frac{3}{2}\,\kappa\,e^{-3\alpha}\,\pi_{\varphi0}^2 \nonumber \\
&&\qquad \qquad 6 \,e^{3\alpha}\,\frac{G_{\varphi}\,\pi_{\varphi0}}{\pi_\alpha} - e^{3\alpha}\,G_{\varphi\varphi} 
\end{eqnarray}
and the corresponding equation of motion becomes
\begin{eqnarray}
\ddot{Q} - 2\,k^2\, e^{6\alpha}\, G_{\pi_\varphi\pi_\varphi}\, G_Y\,Q + \left(3\,H - \frac{\dot{G}_{\pi_\varphi\pi_\varphi}}{G_{\pi_\varphi\pi_\varphi}}\right)\,\dot{Q} - e^{3\alpha}\,G_{\pi_\varphi\pi_\varphi}\,X\,Q = 0.
\end{eqnarray}

Note that, speed of sound
\begin{eqnarray}
c_s^2 &=& 2\, e^{6\alpha}\, G_{\pi_\varphi\pi_\varphi}\, G_Y \nonumber \\
&=& 2\, a^6\, G_{\pi_\varphi\pi_\varphi}\, G_Y.
\end{eqnarray}

Note that, in Langlois' approach, the time coordinate represents
cosmic time and hence the sound speed, according to our approach, is
given by $2\, a^4\, G_{\pi_\varphi\pi_\varphi}\, G_Y$. The discrepancy
arises due to the fact that, in Langlois' approach, $\gamma^{i
  j} \partial_{i j} \rightarrow - k^2$ (see eqs. (\ref{b24}) and
(\ref{b25})), where we have used $\delta^{i j} \partial_{i j}
\rightarrow - k^2$. Hence, extra $a^2$ factor appears in Langlois'
approach.
	\chapter{}\label{app4}
\section{Non-minimal coupling terms} \label{app:model construction}

All possible combinations of contraction between Vector fields and Riemann tensor are given below:
\begin{eqnarray}
\mathcal{S}_1 &=& D_1\,\int d^4x\, \sqrt{-g}\, {g}^{\mu \nu} {g}^{\alpha \beta} {g}^{\gamma \delta}\, R_{\mu \nu} \,{\nabla}_{\alpha}{A}_{\gamma}\, {\nabla}_{\beta}{A}_{\delta} \\
\mathcal{S}_2 &=& D_2\,\int d^4x\, \sqrt{-g}\,{g}^{\mu \alpha} {g}^{\nu \beta} {g}^{\gamma \delta}\, {R}_{\mu \nu}\, {\nabla}_{\alpha}{A}_{\gamma}\, {\nabla}_{\beta}{A}_{\delta}\\
\mathcal{S}_3 &=& D_3\,\int d^4x\, \sqrt{-g}\,{g}^{\mu \nu} {g}^{\alpha \beta} {g}^{\gamma \delta} \,{R}_{\mu \nu} \,{\nabla}_{\alpha}{A}_{\beta} \,{\nabla}_{\gamma}{A}_{\delta}\\
\mathcal{S}_4 &=& D_4\,\int d^4x\, \sqrt{-g}\,{g}^{\mu \nu} {g}^{\alpha \delta} {g}^{\gamma \beta}\, {R}_{\mu \nu} \,{\nabla}_{\alpha}{A}_{\beta}\, {\nabla}_{\gamma}{A}_{\delta}\\
\mathcal{S}_5 &=& D_5\,\int d^4x\, \sqrt{-g}\,{g}^{\mu \gamma} {g}^{\alpha \beta} {g}^{\nu \delta}\, {R}_{\mu \nu}\, {\nabla}_{\alpha}{A}_{\beta}\, {\nabla}_{\gamma}{A}_{\delta}\\
\mathcal{S}_{6} &=& D_{6}\,\int d^4x\, \sqrt{-g}\,{g}^{\mu \alpha} {g}^{\nu \delta} {g}^{\gamma \beta}\, {R}_{\mu \nu}\, {\nabla}_{\alpha}{A}_{\beta}\, {\nabla}_{\gamma}{A}_{\delta}\\
\mathcal{S}_{7} &=& D_{7}\,\int d^4x\, \sqrt{-g}\,{g}^{\mu \alpha} {g}^{\nu \beta} {g}^{\gamma \zeta} {g}^{\delta \eta} \,{R}_{\alpha \beta \gamma \delta}\, {\nabla}_{\mu}{A}_{\nu}\, {\nabla}_{\zeta}{A}_{\eta}\\
\mathcal{S}_{8} &=& D_{8}\,\int d^4x\, \sqrt{-g}\,{g}^{\mu \alpha} {g}^{\eta \beta} {g}^{\gamma \zeta} {g}^{\delta \nu} \,{R}_{\alpha \beta \gamma \delta}\, {\nabla}_{\mu}{A}_{\nu} \,{\nabla}_{\zeta}{A}_{\eta}\\
\mathcal{S}_{9} &=& D_{9}\,\int d^4x\, \sqrt{-g}\,{g}^{\alpha \beta} {g}^{\gamma \delta} {g}^{\mu \nu}\, {R}_{\alpha \beta}\, {R}_{\gamma \delta}\, {A}_{\mu} {A}_{\nu}\\
\mathcal{S}_{10} &=& D_{10}\,\int d^4x\, \sqrt{-g}\,{g}^{\alpha \beta} {g}^{\gamma \mu} {g}^{\delta \nu}\, {R}_{\alpha \beta}\, {R}_{\gamma \delta}\, {A}_{\mu} {A}_{\nu}\\
\mathcal{S}_{11} &=& D_{11}\,\int d^4x\, \sqrt{-g}\,{g}^{\alpha \gamma} {g}^{\beta \delta} {g}^{\mu \nu}\, {R}_{\alpha \beta}\, {R}_{\gamma \delta} \,{A}_{\mu} {A}_{\nu}\\
\mathcal{S}_{12} &=& D_{12}\,\int d^4x\, \sqrt{-g}\,{g}^{\alpha \gamma} {g}^{\beta \mu} {g}^{\delta \nu}\, {R}_{\alpha \beta}\, {R}_{\gamma \delta}\, {A}_{\mu} {A}_{\nu}\\
\end{eqnarray}

\section{Details of fixing the coefficients}
\label{app:Gaugefixing}

Evaluating the equations of motion for an arbitrary metric is hard and also non-transparent.  Hence, to calculate equations of motion and thus to fix the coefficients, we consider FRW background 
$$ds^2 = - N^2 d\eta^2 + a^2 d{\bf x}^2$$ where 
$N(\eta)$ is the Lapse function. The action becomes

{\small
\begin{eqnarray}
&&\mathcal{L}_{SV} = 4\, \delta^{i j} {A}_{i} {\partial}_{0 0}{{A}_{j}}\,  N{}^{(-4)} a^\prime{}^{2} a{}^{(-4)} - 2\, \delta^{i j} {A}_{i} {A}_{j}  {\delta}_{i j} N{}^{(-4)} a^\prime{}^{4} a{}^{(-6)} - 4\,  {\delta}^{i j}{A}_{i} {A}_{j}  a^{\prime \prime}\,  N{}^{(-4)} a^\prime{}^{2} a{}^{(-5)} - \nonumber \\
&&8\,\delta^{i j} {A}_{i}  {\partial}_{0}{{A}_{j}}\,  N^{\prime}\,  N{}^{(-5)} a^\prime{}^{2} a{}^{(-4)} + 4\,\delta^{i j} {A}_{i} {A}_{j}   N^{\prime}\,  N{}^{(-5)} a^\prime{}^{3} a{}^{(-5)} - 4\,  {\delta}^{i j} {\partial}_{0}{{A}_{i}}\,  a^{\prime}\,  {\partial}_{0 0}{{A}_{j}}\,  N{}^{(-4)} a{}^{(-3)} +\nonumber \\
 && 4\, \delta^{i j} {A}_{i}  {\partial}_{0}{{A}_{j}}\,  a^{\prime}\,  a^{\prime \prime}\,  N{}^{(-4)} a{}^{(-4)} + 2\, {\delta}^{i j} {\partial}_{0}{{A}_{i}}\,  {\partial}_{0}{{A}_{j}}\,  N{}^{(-4)} a^\prime{}^{2} a{}^{(-4)} + 4\,  {\delta}^{i j} {\partial}_{0}{{A}_{i}}\,  {\partial}_{0}{{A}_{j}}\,  N^{\prime}\,  a^{\prime}\,  N{}^{(-5)} a{}^{(-3)} + \nonumber \\
 &&8\,\delta^{i j} {A}_{i}  N^{\prime}\,  {\partial}_{j}{{A}_{0}}\,  N{}^{(-5)} a^\prime{}^{2} a{}^{(-4)} - 4\, \delta^{i j} {A}_{i}  a^{\prime}\,  {\partial}_{j}{{A}_{0}}\,  a^{\prime \prime}\,  N{}^{(-4)} a{}^{(-4)} - 4\,  {\delta}^{i j} {\partial}_{0}{{A}_{i}}\,  {\partial}_{j}{{A}_{0}}\,  N{}^{(-4)} a^\prime{}^{2} a{}^{(-4)} - \nonumber \\
 && 8\,  {\delta}^{i j} {\partial}_{0}{{A}_{i}}\,  N^{\prime}\,  a^{\prime}\,  {\partial}_{j}{{A}_{0}}\,  N{}^{(-5)} a{}^{(-3)} + 4\,  {\delta}^{i j} N^{\prime}\,  a^{\prime}\,  {\partial}_{i}{{A}_{0}}\,  {\partial}_{j}{{A}_{0}}\,  N{}^{(-5)} a{}^{(-3)} - 2\,  {\delta}^{i j} {\delta}^{k l} {\partial}_{0 i}{{A}_{k}}\,  {\partial}_{0 j}{{A}_{l}}\,  N{}^{(-2)} a{}^{(-4)} + \nonumber \\
 &&4\, {A}_{0}  {\delta}^{i j} N^{\prime}\,  a^{\prime}\,  {\partial}_{0 i}{{A}_{j}}\,  N{}^{(-5)} a{}^{(-3)} - 4\, {A}_{0}  {\delta}^{i j} a^{\prime \prime}\,  {\partial}_{0 i}{{A}_{j}}\,  N{}^{(-4)} a{}^{(-3)} + 6\,  {\delta}^{i j} {\delta}^{k l} a^{\prime}\,  {\partial}_{i}{{A}_{k}}\,  {\partial}_{0 j}{{A}_{l}}\,  N{}^{(-2)} a{}^{(-5)} + \nonumber \\
 &&6\,  {A}_{0}{}^{2} N{}^{(-8)} N^\prime{}^{2} a^\prime{}^{2} a{}^{(-2)}%
 - 12\,  N^{\prime}\,  a^{\prime}\,  a^{\prime \prime}\,  {A}_{0}{}^{2} N{}^{(-7)} a{}^{(-2)} + 6\,  {A}_{0}{}^{2} N{}^{(-6)} a^{\prime \prime}{}^{2} a{}^{(-2)} - \nonumber \\
 && 6\,  {\delta}^{i j} {\delta}^{k l} {\partial}_{i}{{A}_{k}}\,  {\partial}_{j}{{A}_{l}}\,  N{}^{(-2)} a^\prime{}^{2} a{}^{(-6)} - 4\, \delta^{i j} {A}_{i}  {\partial}_{0 j}{{A}_{0}}\,  N{}^{(-4)} a^\prime{}^{2} a{}^{(-4)} - 4\,  {\delta}^{i j} a^{\prime}\,  {\partial}_{i}{{A}_{0}}\,  {\partial}_{0 j}{{A}_{0}}\,  N{}^{(-4)} a{}^{(-3)} + \nonumber \\
 &&2\,  {\delta}^{i j} {\partial}_{i}{{A}_{0}}\,  {\partial}_{j}{{A}_{0}}\,  N{}^{(-4)} a^\prime{}^{2} a{}^{(-4)} + 4\,  {\delta}^{i j} {\partial}_{0}{{A}_{i}}\,  a^{\prime}\,  {\partial}_{0 j}{{A}_{0}}\,  N{}^{(-4)} a{}^{(-3)} -  {\delta}^{i j} {\delta}^{k l} {\partial}_{i k}{{A}_{0}}\,  {\partial}_{j l}{{A}_{0}}\,  N{}^{(-2)} a{}^{(-4)} - \nonumber \\
 &&4\, {A}_{0}  {\delta}^{i j} N^{\prime}\,  a^{\prime}\,  {\partial}_{i j}{{A}_{0}}\,  N{}^{(-5)} a{}^{(-3)} + 6\,  {\delta}^{i j} {\delta}^{k l} {\partial}_{i}{{A}_{k}}\,  {\partial}_{l}{{A}_{j}}\,  N{}^{(-2)} a^\prime{}^{2} a{}^{(-6)} +  {\delta}^{i j} {\delta}^{k l} {\delta}^{m n} {\partial}_{i k}{{A}_{m}}\,  {\partial}_{j l}{{A}_{n}}\,  a{}^{(-6)} - \nonumber \\
 &&2\, {A}^{j}  {\delta}^{k l} {\partial}_{k l}{{A}_{j}}\,  N{}^{(-2)} a^\prime{}^{2} a{}^{(-6)} + 2\, \delta^{i j} {A}_{i}  {\delta}^{k l} {\partial}_{j k}{{A}_{l}}\,  N{}^{(-2)} a^\prime{}^{2} a{}^{(-6)} +  {\delta}^{i j} {\delta}^{k l} {\partial}_{0 i}{{A}_{j}}\,  {\partial}_{0 k}{{A}_{l}}\,  N{}^{(-2)} a{}^{(-4)} + \nonumber \\
 &&2\,  {\delta}^{i j} {\delta}^{k l} {\partial}_{0 i}{{A}_{0}}\,  {\partial}_{j k}{{A}_{l}}\,  N{}^{(-2)} a{}^{(-4)} - 2\,  {\delta}^{i j} {\delta}^{k l} N^{\prime}\,  {\partial}_{i}{{A}_{0}}\,  {\partial}_{j k}{{A}_{l}}\,  N{}^{(-3)} a{}^{(-4)} -  {\delta}^{i j} {\delta}^{k l} {\delta}^{m n} {\partial}_{i k}{{A}_{j}}\,  {\partial}_{l m}{{A}_{n}}\,  a{}^{(-6)} - \nonumber \\
 &&2\,  {\delta}^{i j} {\delta}^{k l} {\partial}_{0 i}{{A}_{j}}\,  {\partial}_{k l}{{A}_{0}}\,  N{}^{(-2)} a{}^{(-4)} + 4\, {A}_{0}  {\delta}^{i j} a^{\prime \prime}\,  {\partial}_{i j}{{A}_{0}}\,  N{}^{(-4)} a{}^{(-3)} + 4\,  {\delta}^{i j} a^{\prime}\,  {\partial}_{i}{{A}_{0}}\,  {\partial}_{0 0}{{A}_{j}}\,  N{}^{(-4)} a{}^{(-3)}%
 - \nonumber \\
 &&2\,  {\delta}^{i j} {\delta}^{k l} {\partial}_{0 i}{{A}_{0}}\,  {\partial}_{k l}{{A}_{j}}\,  N{}^{(-2)} a{}^{(-4)} + 2\,  {\delta}^{i j} {\delta}^{k l} N^{\prime}\,  {\partial}_{i}{{A}_{0}}\,  {\partial}_{k l}{{A}_{j}}\,  N{}^{(-3)} a{}^{(-4)} - 2\, {\delta}^{i j} {\delta}^{k l} {\partial}_{0 0}{{A}_{i}}\,  {\partial}_{j k}{{A}_{l}}\,  N{}^{(-2)} a{}^{(-4)} +\nonumber \\
 && 2\, \delta^{i j}{A}_{i}  {\delta}^{k l} a^{\prime \prime}\,  {\partial}_{j k}{{A}_{l}}\,  N{}^{(-2)} a{}^{(-5)} + 2\,  {\delta}^{i j} {\delta}^{k l} {\partial}_{0}{{A}_{i}}\,  N^{\prime}\,  {\partial}_{j k}{{A}_{l}}\,  N{}^{(-3)} a{}^{(-4)} - 2\, \delta^{i j} {\delta}^{k l} {A}_{i}   N^{\prime}\,  a^{\prime}\,  {\partial}_{j k}{{A}_{l}}\,  N{}^{(-3)} a{}^{(-5)} + \nonumber \\
 && 2\,  {\delta}^{i j} {\delta}^{k l} {\delta}^{m n} {\partial}_{i j}{{A}_{k}}\,  {\partial}_{l m}{{A}_{n}}\,  a{}^{(-6)} + 2\,  {\delta}^{i j} {\delta}^{k l} {\partial}_{0 i}{{A}_{k}}\,  {\partial}_{j l}{{A}_{0}}\,  N{}^{(-2)} a{}^{(-4)} - 6\,  {\delta}^{i j} {\delta}^{k l} a^{\prime}\,  {\partial}_{i}{{A}_{k}}\,  {\partial}_{0 l}{{A}_{j}}\,  N{}^{(-2)} a{}^{(-5)} +\nonumber \\
 &&  {\delta}^{i j} {\delta}^{k l} {\partial}_{0 i}{{A}_{k}}\,  {\partial}_{0 l}{{A}_{j}}\,  N{}^{(-2)} a{}^{(-4)} -  {\delta}^{i j} {\delta}^{k l} {\delta}^{m n} {\partial}_{i k}{{A}_{m}}\,  {\partial}_{j n}{{A}_{l}}\,  a{}^{(-6)} + 2\,  {\delta}^{i j} {\delta}^{k l} {\partial}_{0 0}{{A}_{i}}\,  {\partial}_{k l}{{A}_{j}}\,  N{}^{(-2)} a{}^{(-4)} - \nonumber \\
 &&2\,\delta^{i j} {A}_{i}  {\delta}^{k l} a^{\prime \prime}\,  {\partial}_{k l}{{A}_{j}}\,  N{}^{(-2)} a{}^{(-5)} - 2\,  {\delta}^{i j} {\delta}^{k l} {\partial}_{0}{{A}_{i}}\,  N^{\prime}\,  {\partial}_{k l}{{A}_{j}}\,  N{}^{(-3)} a{}^{(-4)} + 2\,\delta^{i j} {\delta}^{k l} {A}_{i}   N^{\prime}\,  a^{\prime}\,  {\partial}_{k l}{{A}_{j}}\,  N{}^{(-3)} a{}^{(-5)} + \nonumber \\
 && {\delta}^{i j} {\delta}^{k l} {\partial}_{i j}{{A}_{0}}\,  {\partial}_{k l}{{A}_{0}}\,  N{}^{(-2)} a{}^{(-4)} -  {\delta}^{i j} {\delta}^{k l} {\delta}^{m n} {\partial}_{i j}{{A}_{k}}\,  {\partial}_{m n}{{A}_{l}}\,  a{}^{(-6)}
\end{eqnarray}}
{\small
\begin{eqnarray}
&& \mathcal{L}_1 = 6\,  a^{\prime \prime}\,  N{}^{(-6)} {\partial}_{0}{{A}_{0}}\, {}^{2} a{}^{-1} - 6\,  N^{\prime}\,  a^{\prime}\,  N{}^{(-7)} {\partial}_{0}{{A}_{0}}\, {}^{2} a{}^{-1} + 6\,  N{}^{(-6)} {\partial}_{0}{{A}_{0}}\, {}^{2} a^\prime{}^{2} a{}^{(-2)} - \nonumber \\
 &&6\,  {\delta}^{i j} {\partial}_{0}{{A}_{i}}\,  {\partial}_{0}{{A}_{j}}\,  a^{\prime \prime}\,  N{}^{(-4)} a{}^{(-3)} + 6\,  {\delta}^{i j} {\partial}_{0}{{A}_{i}}\,  {\partial}_{0}{{A}_{j}}\,  N^{\prime}\,  a^{\prime}\,  N{}^{(-5)} a{}^{(-3)} - 6\,  {\delta}^{i j} {\partial}_{0}{{A}_{i}}\,  {\partial}_{0}{{A}_{j}}\,  N{}^{(-4)} a^\prime{}^{2} a{}^{(-4)} - \nonumber \\
 &&6\,  {\delta}^{i j} {\partial}_{i}{{A}_{0}}\,  {\partial}_{j}{{A}_{0}}\,  a^{\prime \prime}\,  N{}^{(-4)} a{}^{(-3)} + 6\,  {\delta}^{i j} N^{\prime}\,  a^{\prime}\,  {\partial}_{i}{{A}_{0}}\,  {\partial}_{j}{{A}_{0}}\,  N{}^{(-5)} a{}^{(-3)} - 6\,  {\delta}^{i j} {\partial}_{i}{{A}_{0}}\,  {\partial}_{j}{{A}_{0}}\,  N{}^{(-4)} a^\prime{}^{2} a{}^{(-4)} + \nonumber \\
 &&6\,  {\delta}^{i j} {\delta}^{k l} {\partial}_{i}{{A}_{k}}\,  {\partial}_{j}{{A}_{l}}\,  a^{\prime \prime}\,  N{}^{(-2)} a{}^{(-5)} - 6\,  {\delta}^{i j} {\delta}^{k l} N^{\prime}\,  a^{\prime}\,  {\partial}_{i}{{A}_{k}}\,  {\partial}_{j}{{A}_{l}}\,  N{}^{(-3)} a{}^{(-5)} + 6\,  {\delta}^{i j} {\delta}^{k l} {\partial}_{i}{{A}_{k}}\,  {\partial}_{j}{{A}_{l}}\,  N{}^{(-2)} a^\prime{}^{2} a{}^{(-6)} - \nonumber \\
 &&12\, {A}_{0}  {\partial}_{0}{{A}_{0}}\,  N^{\prime}\,  a^{\prime \prime}\,  N{}^{(-7)} a{}^{-1} + 12\, {A}_{0}  {\partial}_{0}{{A}_{0}}\,  a^{\prime}\,  N{}^{(-8)} N^\prime{}^{2} a{}^{-1} - 12\, {A}_{0}  {\partial}_{0}{{A}_{0}}\,  N^{\prime}\,  N{}^{(-7)} a^\prime{}^{2} a{}^{(-2)} + \nonumber \\
 &&6\,   {\delta}^{j i} A_j{\partial}_{0}{{A}_{i}}\,  a^{\prime}\,  a^{\prime \prime}\,  N{}^{(-4)} a{}^{(-4)} - 6\,  {\delta}^{j i} A_j {\partial}_{0}{{A}_{i}}\,  N^{\prime}\,  N{}^{(-5)} a^\prime{}^{2} a{}^{(-4)} + 6\,  {\delta}^{j i} A_j {\partial}_{0}{{A}_{i}}\,  N{}^{(-4)} a^\prime{}^{3} a{}^{(-5)} + \nonumber \\
 &&6\,   {\delta}^{j i} A_j a^{\prime}\,  {\partial}_{i}{{A}_{0}}\,  a^{\prime \prime}\,  N{}^{(-4)} a{}^{(-4)}%
 - 6\, {A}_{j}  {\delta}^{j i} N^{\prime}\,  {\partial}_{i}{{A}_{0}}\,  N{}^{(-5)} a^\prime{}^{2} a{}^{(-4)} + 6\, {A}_{j}  {\delta}^{j i} {\partial}_{i}{{A}_{0}}\,  N{}^{(-4)} a^\prime{}^{3} a{}^{(-5)} - \nonumber \\
 &&6\, {A}_{0}  {\delta}^{i j} a^{\prime}\,  {\partial}_{i}{{A}_{j}}\,  a^{\prime \prime}\,  N{}^{(-4)} a{}^{(-4)} + 6\, {A}_{0}  {\delta}^{i j} N^{\prime}\,  {\partial}_{i}{{A}_{j}}\,  N{}^{(-5)} a^\prime{}^{2} a{}^{(-4)} - 6\, {A}_{0}  {\delta}^{i j} {\partial}_{i}{{A}_{j}}\,  N{}^{(-4)} a^\prime{}^{3} a{}^{(-5)} + \nonumber \\
 &&6\, {A}_{i}  {\delta}^{i j} {\partial}_{0}{{A}_{j}}\,  a^{\prime}\,  a^{\prime \prime}\,  N{}^{(-4)} a{}^{(-4)} - 6\, {A}_{i}  {\delta}^{i j} {\partial}_{0}{{A}_{j}}\,  N^{\prime}\,  N{}^{(-5)} a^\prime{}^{2} a{}^{(-4)} + 6\, {A}_{i}  {\delta}^{i j} {\partial}_{0}{{A}_{j}}\,  N{}^{(-4)} a^\prime{}^{3} a{}^{(-5)} + \nonumber \\
 &&6\,  a^{\prime \prime}\,  {A}_{0}{}^{2} N{}^{(-8)} N^\prime{}^{2} a{}^{-1} - 6\,  a^{\prime}\,  {A}_{0}{}^{2} N{}^{(-9)} N^\prime{}^{3} a{}^{-1} + 6\,  {A}_{0}{}^{2} N{}^{(-8)} N^\prime{}^{2} a^\prime{}^{2} a{}^{(-2)} - \nonumber \\
 &&12\, {A}_{i} {A}_{j}  {\delta}^{j i} a^{\prime \prime}\,  N{}^{(-4)} a^\prime{}^{2} a{}^{(-5)} + 12\, {A}_{i} {A}_{j}  {\delta}^{j i} N^{\prime}\,  N{}^{(-5)} a^\prime{}^{3} a{}^{(-5)} - 12\, {A}_{i} {A}_{j}  {\delta}^{j i} N{}^{(-4)} a^\prime{}^{4} a{}^{(-6)} + \nonumber \\
 &&6\, {A}_{i}  {\delta}^{i j} a^{\prime}\,  {\partial}_{j}{{A}_{0}}\,  a^{\prime \prime}\,  N{}^{(-4)} a{}^{(-4)} - 6\, {A}_{i}  {\delta}^{i j} N^{\prime}\,  {\partial}_{j}{{A}_{0}}\,  N{}^{(-5)} a^\prime{}^{2} a{}^{(-4)} + 6\, {A}_{i}  {\delta}^{i j} {\partial}_{j}{{A}_{0}}\,  N{}^{(-4)} a^\prime{}^{3} a{}^{(-5)} - \nonumber \\
 &&6\, {A}_{0}  {\delta}^{i j} a^{\prime}\,  {\partial}_{j}{{A}_{i}}\,  a^{\prime \prime}\,  N{}^{(-4)} a{}^{(-4)} + 6\, {A}_{0}  {\delta}^{i j} N^{\prime}\,  {\partial}_{j}{{A}_{i}}\,  N{}^{(-5)} a^\prime{}^{2} a{}^{(-4)} - 6\, {A}_{0}  {\delta}^{i j} {\partial}_{j}{{A}_{i}}\,  N{}^{(-4)} a^\prime{}^{3} a{}^{(-5)}%
 + \nonumber \\
 &&18\,  a^{\prime \prime}\,  {A}_{0}{}^{2} N{}^{(-6)} a^\prime{}^{2} a{}^{(-3)} - 18\,  N^{\prime}\,  {A}_{0}{}^{2} N{}^{(-7)} a^\prime{}^{3} a{}^{(-3)} + 18\, {A}_{0}{}^{2} N{}^{(-6)} a^\prime{}^{4} a{}^{(-4)} 
 \end{eqnarray}
 \begin{eqnarray}
&& \mathcal{L}_2 = 3\,  a^{\prime \prime}\,  N{}^{(-6)} {\partial}_{0}{{A}_{0}}\, {}^{2} a{}^{-1} - 3\,  N^{\prime}\,  a^{\prime}\,  N{}^{(-7)} {\partial}_{0}{{A}_{0}}\, {}^{2} a{}^{-1} - 3\,  {\delta}^{i j} {\partial}_{0}{{A}_{i}}\,  {\partial}_{0}{{A}_{j}}\,  a^{\prime \prime}\,  N{}^{(-4)} a{}^{(-3)} + \nonumber \\
&&3\,  {\delta}^{i j} {\partial}_{0}{{A}_{i}}\,  {\partial}_{0}{{A}_{j}}\,  N^{\prime}\,  a^{\prime}\,  N{}^{(-5)} a{}^{(-3)} +  {\delta}^{i j} N^{\prime}\,  a^{\prime}\,  {\partial}_{i}{{A}_{0}}\,  {\partial}_{j}{{A}_{0}}\,  N{}^{(-5)} a{}^{(-3)} - 2\,  {\delta}^{i j} {\partial}_{i}{{A}_{0}}\,  {\partial}_{j}{{A}_{0}}\,  N{}^{(-4)} a^\prime{}^{2} a{}^{(-4)} - \nonumber \\
&& {\delta}^{i j} {\partial}_{i}{{A}_{0}}\,  {\partial}_{j}{{A}_{0}}\,  a^{\prime \prime}\,  N{}^{(-4)} a{}^{(-3)} -  {\delta}^{i j} {\delta}^{k l} N^{\prime}\,  a^{\prime}\,  {\partial}_{i}{{A}_{l}}\,  {\partial}_{j}{{A}_{k}}\,  N{}^{(-3)} a{}^{(-5)} + 4\,  {\delta}^{i j} {\delta}^{k l} {\partial}_{i}{{A}_{l}}\,  {\partial}_{j}{{A}_{k}}\,  N{}^{(-2)} a^\prime{}^{2} a{}^{(-6)} +\nonumber \\
&&  {\delta}^{i j} {\delta}^{k l} {\partial}_{i}{{A}_{l}}\,  {\partial}_{j}{{A}_{k}}\,  a^{\prime \prime}\,  N{}^{(-2)} a{}^{(-5)} - 2\,  {\delta}^{i j} {\delta}^{k l} {\partial}_{i}{{A}_{k}}\,  {\partial}_{j}{{A}_{l}}\,  N{}^{(-2)} a^\prime{}^{2} a{}^{(-6)} - 6\, {A}_{0}  {\partial}_{0}{{A}_{0}}\,  N^{\prime}\,  a^{\prime \prime}\,  N{}^{(-7)} a{}^{-1} +\nonumber \\
&& 6\, {A}_{0}  {\partial}_{0}{{A}_{0}}\,  a^{\prime}\,  N{}^{(-8)} N^\prime{}^{2} a{}^{-1} + 3\, {A}_{j}  {\delta}^{j i} {\partial}_{0}{{A}_{i}}\,  a^{\prime}\,  a^{\prime \prime}\,  N{}^{(-4)} a{}^{(-4)} - 3\, {A}_{j}  {\delta}^{j i} {\partial}_{0}{{A}_{i}}\,  N^{\prime}\,  N{}^{(-5)} a^\prime{}^{2} a{}^{(-4)} - \nonumber \\
&&{A}_{j}  {\delta}^{j i} N^{\prime}\,  {\partial}_{i}{{A}_{0}}\,  N{}^{(-5)} a^\prime{}^{2} a{}^{(-4)} + 2\, {A}_{j}  {\delta}^{j i} {\partial}_{i}{{A}_{0}}\,  N{}^{(-4)} a^\prime{}^{3} a{}^{(-5)} + {A}_{j}  {\delta}^{j i} a^{\prime}\,  {\partial}_{i}{{A}_{0}}\,  a^{\prime \prime}\,  N{}^{(-4)} a{}^{(-4)} + \nonumber \\
&&{A}_{0}  {\delta}^{i j} N^{\prime}\,  {\partial}_{j}{{A}_{i}}\,  N{}^{(-5)} a^\prime{}^{2} a{}^{(-4)}%
 - 2\, {A}_{0}  {\delta}^{i j} {\partial}_{j}{{A}_{i}}\,  N{}^{(-4)} a^\prime{}^{3} a{}^{(-5)} - {A}_{0}  {\delta}^{i j} a^{\prime}\,  {\partial}_{j}{{A}_{i}}\,  a^{\prime \prime}\,  N{}^{(-4)} a{}^{(-4)} - \nonumber \\
&&2\, {A}_{0}  {\delta}^{i j} {\partial}_{i}{{A}_{j}}\,  N{}^{(-4)} a^\prime{}^{3} a{}^{(-5)} + 3\, {A}_{i}  {\delta}^{i j} {\partial}_{0}{{A}_{j}}\,  a^{\prime}\,  a^{\prime \prime}\,  N{}^{(-4)} a{}^{(-4)} - 3\, {A}_{i}  {\delta}^{i j} {\partial}_{0}{{A}_{j}}\,  N^{\prime}\,  N{}^{(-5)} a^\prime{}^{2} a{}^{(-4)} + \nonumber \\
&&3\,  a^{\prime \prime}\,  {A}_{0}{}^{2} N{}^{(-8)} N^\prime{}^{2} a{}^{-1} - 3\,  a^{\prime}\,  {A}_{0}{}^{2} N{}^{(-9)} N^\prime{}^{3} a{}^{-1} - 3\, {A}_{i} {A}_{j}  {\delta}^{j i} a^{\prime \prime}\,  N{}^{(-4)} a^\prime{}^{2} a{}^{(-5)} + \nonumber \\
&&3\, {A}_{i} {A}_{j}  {\delta}^{j i} N^{\prime}\,  N{}^{(-5)} a^\prime{}^{3} a{}^{(-5)} - {A}_{i}  {\delta}^{i j} N^{\prime}\,  {\partial}_{j}{{A}_{0}}\,  N{}^{(-5)} a^\prime{}^{2} a{}^{(-4)} + 2\, {A}_{i}  {\delta}^{i j} {\partial}_{j}{{A}_{0}}\,  N{}^{(-4)} a^\prime{}^{3} a{}^{(-5)} + \nonumber \\
&&{A}_{i}  {\delta}^{i j} a^{\prime}\,  {\partial}_{j}{{A}_{0}}\,  a^{\prime \prime}\,  N{}^{(-4)} a{}^{(-4)} + {A}_{i} {A}_{j}  {\delta}^{i j} N^{\prime}\,  N{}^{(-5)} a^\prime{}^{3} a{}^{(-5)} - 4\, {A}_{i} {A}_{j}  {\delta}^{i j} N{}^{(-4)} a^\prime{}^{4} a{}^{(-6)} - \nonumber \\
&&{A}_{i} {A}_{j}  {\delta}^{i j} a^{\prime \prime}\,  N{}^{(-4)} a^\prime{}^{2} a{}^{(-5)} + 2\, {A}_{i} {A}_{j}  {\delta}^{j i} N{}^{(-4)} a^\prime{}^{4} a{}^{(-6)} + {A}_{0}  {\delta}^{i j} N^{\prime}\,  {\partial}_{i}{{A}_{j}}\,  N{}^{(-5)} a^\prime{}^{2} a{}^{(-4)} - \nonumber \\
&&{A}_{0}  {\delta}^{i j} a^{\prime}\,  {\partial}_{i}{{A}_{j}}\,  a^{\prime \prime}\,  N{}^{(-4)} a{}^{(-4)} - 3\,  N^{\prime}\,  {A}_{0}{}^{2} N{}^{(-7)} a^\prime{}^{3} a{}^{(-3)} + 6\,  {A}_{0}{}^{2} N{}^{(-6)} a^\prime{}^{4} a{}^{(-4)}%
 + \nonumber \\
&&3\,  a^{\prime \prime}\,  {A}_{0}{}^{2} N{}^{(-6)} a^\prime{}^{2} a{}^{(-3)}
\end{eqnarray}
}
{\small
\begin{eqnarray}
&& \mathcal{L}_3 = 6\,  a^{\prime \prime}\,  N{}^{(-6)} {\partial}_{0}{{A}_{0}}\, {}^{2} a{}^{-1} - 6\,  N^{\prime}\,  a^{\prime}\,  N{}^{(-7)} {\partial}_{0}{{A}_{0}}\, {}^{2} a{}^{-1} + 6\,  N{}^{(-6)} {\partial}_{0}{{A}_{0}}\, {}^{2} a^\prime{}^{2} a{}^{(-2)} - \nonumber \\
&&12\,  {\delta}^{i j} {\partial}_{0}{{A}_{0}}\,  {\partial}_{i}{{A}_{j}}\,  a^{\prime \prime}\,  N{}^{(-4)} a{}^{(-3)} + 12\,  {\delta}^{i j} {\partial}_{0}{{A}_{0}}\,  N^{\prime}\,  a^{\prime}\,  {\partial}_{i}{{A}_{j}}\,  N{}^{(-5)} a{}^{(-3)} - 12\,  {\delta}^{i j} {\partial}_{0}{{A}_{0}}\,  {\partial}_{i}{{A}_{j}}\,  N{}^{(-4)} a^\prime{}^{2} a{}^{(-4)} + \nonumber \\
&&6\,  {\delta}^{i j} {\delta}^{k l} {\partial}_{i}{{A}_{j}}\,  {\partial}_{k}{{A}_{l}}\,  a^{\prime \prime}\,  N{}^{(-2)} a{}^{(-5)} - 6\,  {\delta}^{i j} {\delta}^{k l} N^{\prime}\,  a^{\prime}\,  {\partial}_{i}{{A}_{j}}\,  {\partial}_{k}{{A}_{l}}\,  N{}^{(-3)} a{}^{(-5)} + 6\,  {\delta}^{i j} {\delta}^{k l} {\partial}_{i}{{A}_{j}}\,  {\partial}_{k}{{A}_{l}}\,  N{}^{(-2)} a^\prime{}^{2} a{}^{(-6)} - \nonumber \\
&&12\, {A}_{0}  {\partial}_{0}{{A}_{0}}\,  N^{\prime}\,  a^{\prime \prime}\,  N{}^{(-7)} a{}^{-1} + 12\, {A}_{0}  {\partial}_{0}{{A}_{0}}\,  a^{\prime}\,  N{}^{(-8)} N^\prime{}^{2} a{}^{-1} - 48\, {A}_{0}  {\partial}_{0}{{A}_{0}}\,  N^{\prime}\,  N{}^{(-7)} a^\prime{}^{2} a{}^{(-2)} + \nonumber \\
&&12\, {A}_{0}  {\delta}^{i j} N^{\prime}\,  {\partial}_{i}{{A}_{j}}\,  a^{\prime \prime}\,  N{}^{(-5)} a{}^{(-3)} - 12\, {A}_{0}  {\delta}^{i j} a^{\prime}\,  {\partial}_{i}{{A}_{j}}\,  N{}^{(-6)} N^\prime{}^{2} a{}^{(-3)} + 48\, {A}_{0}  {\delta}^{i j} N^{\prime}\,  {\partial}_{i}{{A}_{j}}\,  N{}^{(-5)} a^\prime{}^{2} a{}^{(-4)} + \nonumber \\
&& 36\, {A}_{0}  {\partial}_{0}{{A}_{0}}\,  a^{\prime}\,  a^{\prime \prime}\,  N{}^{(-6)} a{}^{(-2)} + 36\, {A}_{0}  {\partial}_{0}{{A}_{0}}\,  N{}^{(-6)} a^\prime{}^{3} a{}^{(-3)} - 36\, {A}_{0}  {\delta}^{i j} a^{\prime}\,  {\partial}_{i}{{A}_{j}}\,  a^{\prime \prime}\,  N{}^{(-4)} a{}^{(-4)} - \nonumber \\
&&36\, {A}_{0}  {\delta}^{i j} {\partial}_{i}{{A}_{j}}\,  N{}^{(-4)} a^\prime{}^{3} a{}^{(-5)}%
 + 6\,  a^{\prime \prime}\,  {A}_{0}{}^{2} N{}^{(-8)} N^\prime{}^{2} a{}^{-1} - 6\,  a^{\prime}\,  {A}_{0}{}^{2} N{}^{(-9)} N^\prime{}^{3} a{}^{-1} + \nonumber \\
&&42\,  {A}_{0}{}^{2} N{}^{(-8)} N^\prime{}^{2} a^\prime{}^{2} a{}^{(-2)} - 36\,  N^{\prime}\,  a^{\prime}\,  a^{\prime \prime}\,  {A}_{0}{}^{2} N{}^{(-7)} a{}^{(-2)} - 90\,  N^{\prime}\,  {A}_{0}{}^{2} N{}^{(-7)} a^\prime{}^{3} a{}^{(-3)} + \nonumber \\
&&54\,  a^{\prime \prime}\,  {A}_{0}{}^{2} N{}^{(-6)} a^\prime{}^{2} a{}^{(-3)} + 54\,  {A}_{0}{}^{2} N{}^{(-6)} a^\prime{}^{4} a{}^{(-4)}
\end{eqnarray}
}
{\small
\begin{eqnarray}
&&\mathcal{L}_4 = 6\,  a^{\prime \prime}\,  N{}^{(-6)} {\partial}_{0}{{A}_{0}}\, {}^{2} a{}^{-1} - 6\,  N^{\prime}\,  a^{\prime}\,  N{}^{(-7)} {\partial}_{0}{{A}_{0}}\, {}^{2} a{}^{-1} + 6\,  N{}^{(-6)} {\partial}_{0}{{A}_{0}}\, {}^{2} a^\prime{}^{2} a{}^{(-2)} - \nonumber \\
&&12\,  {\delta}^{i j} {\partial}_{0}{{A}_{j}}\,  {\partial}_{i}{{A}_{0}}\,  a^{\prime \prime}\,  N{}^{(-4)} a{}^{(-3)} + 12\,  {\delta}^{i j} {\partial}_{0}{{A}_{j}}\,  N^{\prime}\,  a^{\prime}\,  {\partial}_{i}{{A}_{0}}\,  N{}^{(-5)} a{}^{(-3)} - 12\,  {\delta}^{i j} {\partial}_{0}{{A}_{j}}\,  {\partial}_{i}{{A}_{0}}\,  N{}^{(-4)} a^\prime{}^{2} a{}^{(-4)} + \nonumber \\
&&6\,  {\delta}^{i j} {\delta}^{k l} {\partial}_{i}{{A}_{l}}\,  {\partial}_{k}{{A}_{j}}\,  a^{\prime \prime}\,  N{}^{(-2)} a{}^{(-5)} - 6\,  {\delta}^{i j} {\delta}^{k l} N^{\prime}\,  a^{\prime}\,  {\partial}_{i}{{A}_{l}}\,  {\partial}_{k}{{A}_{j}}\,  N{}^{(-3)} a{}^{(-5)} + 6\,  {\delta}^{i j} {\delta}^{k l} {\partial}_{i}{{A}_{l}}\,  {\partial}_{k}{{A}_{j}}\,  N{}^{(-2)} a^\prime{}^{2} a{}^{(-6)} - \nonumber \\
&&12\, {A}_{0}  {\partial}_{0}{{A}_{0}}\,  N^{\prime}\,  a^{\prime \prime}\,  N{}^{(-7)} a{}^{-1} + 12\, {A}_{0}  {\partial}_{0}{{A}_{0}}\,  a^{\prime}\,  N{}^{(-8)} N^\prime{}^{2} a{}^{-1} - 12\, {A}_{0}  {\partial}_{0}{{A}_{0}}\,  N^{\prime}\,  N{}^{(-7)} a^\prime{}^{2} a{}^{(-2)} + \nonumber \\
&&6\, {A}_{j}  {\delta}^{j i} a^{\prime}\,  {\partial}_{i}{{A}_{0}}\,  a^{\prime \prime}\,  N{}^{(-4)} a{}^{(-4)} - 6\, {A}_{j}  {\delta}^{j i} N^{\prime}\,  {\partial}_{i}{{A}_{0}}\,  N{}^{(-5)} a^\prime{}^{2} a{}^{(-4)} + 6\, {A}_{j}  {\delta}^{j i} {\partial}_{i}{{A}_{0}}\,  N{}^{(-4)} a^\prime{}^{3} a{}^{(-5)} + \nonumber \\
&&6\, {A}_{j}  {\delta}^{j i} {\partial}_{0}{{A}_{i}}\,  a^{\prime}\,  a^{\prime \prime}\,  N{}^{(-4)} a{}^{(-4)} - 6\, {A}_{j}  {\delta}^{j i} {\partial}_{0}{{A}_{i}}\,  N^{\prime}\,  N{}^{(-5)} a^\prime{}^{2} a{}^{(-4)} + 6\, {A}_{j}  {\delta}^{j i} {\partial}_{0}{{A}_{i}}\,  N{}^{(-4)} a^\prime{}^{3} a{}^{(-5)} - \nonumber \\
&&12\, {A}_{0}  {\delta}^{i j} a^{\prime}\,  {\partial}_{i}{{A}_{j}}\,  a^{\prime \prime}\,  N{}^{(-4)} a{}^{(-4)}%
 + 12\, {A}_{0}  {\delta}^{i j} N^{\prime}\,  {\partial}_{i}{{A}_{j}}\,  N{}^{(-5)} a^\prime{}^{2} a{}^{(-4)} - 12\, {A}_{0}  {\delta}^{i j} {\partial}_{i}{{A}_{j}}\,  N{}^{(-4)} a^\prime{}^{3} a{}^{(-5)} + \nonumber \\
&&6\, {A}_{i}  {\delta}^{i j} a^{\prime}\,  {\partial}_{j}{{A}_{0}}\,  a^{\prime \prime}\,  N{}^{(-4)} a{}^{(-4)} - 6\, {A}_{i}  {\delta}^{i j} N^{\prime}\,  {\partial}_{j}{{A}_{0}}\,  N{}^{(-5)} a^\prime{}^{2} a{}^{(-4)} + 6\, {A}_{i}  {\delta}^{i j} {\partial}_{j}{{A}_{0}}\,  N{}^{(-4)} a^\prime{}^{3} a{}^{(-5)} + \nonumber \\
&&6\,  a^{\prime \prime}\,  {A}_{0}{}^{2} N{}^{(-8)} N^\prime{}^{2} a{}^{-1} - 6\,  a^{\prime}\,  {A}_{0}{}^{2} N{}^{(-9)} N^\prime{}^{3} a{}^{-1} + 6\,  {A}_{0}{}^{2} N{}^{(-8)} N^\prime{}^{2} a^\prime{}^{2} a{}^{(-2)} - \nonumber \\
&&6\, {A}_{i} {A}_{j}  {\delta}^{i j} a^{\prime \prime}\,  N{}^{(-4)} a^\prime{}^{2} a{}^{(-5)} + 6\, {A}_{i} {A}_{j}  {\delta}^{i j} N^{\prime}\,  N{}^{(-5)} a^\prime{}^{3} a{}^{(-5)} - 6\, {A}_{i} {A}_{j}  {\delta}^{i j} N{}^{(-4)} a^\prime{}^{4} a{}^{(-6)} + \nonumber \\
&&6\, {A}_{i}  {\delta}^{i j} {\partial}_{0}{{A}_{j}}\,  a^{\prime}\,  a^{\prime \prime}\,  N{}^{(-4)} a{}^{(-4)} - 6\, {A}_{i}  {\delta}^{i j} {\partial}_{0}{{A}_{j}}\,  N^{\prime}\,  N{}^{(-5)} a^\prime{}^{2} a{}^{(-4)} + 6\, {A}_{i}  {\delta}^{i j} {\partial}_{0}{{A}_{j}}\,  N{}^{(-4)} a^\prime{}^{3} a{}^{(-5)} - \nonumber \\
&&6\, {A}_{i} {A}_{j}  {\delta}^{j i} a^{\prime \prime}\,  N{}^{(-4)} a^\prime{}^{2} a{}^{(-5)} + 6\, {A}_{i} {A}_{j}  {\delta}^{j i} N^{\prime}\,  N{}^{(-5)} a^\prime{}^{3} a{}^{(-5)} - 6\, {A}_{i} {A}_{j}  {\delta}^{j i} N{}^{(-4)} a^\prime{}^{4} a{}^{(-6)} + \nonumber \\
&&18\,  a^{\prime \prime}\,  {A}_{0}{}^{2} N{}^{(-6)} a^\prime{}^{2} a{}^{(-3)} - 18\,  N^{\prime}\,  {A}_{0}{}^{2} N{}^{(-7)} a^\prime{}^{3} a{}^{(-3)} + 18\,  {A}_{0}{}^{2} N{}^{(-6)} a^\prime{}^{4} a{}^{(-4)}
\end{eqnarray}}
{\small
\begin{eqnarray}
&& \mathcal{L}_5 = 3\,  a^{\prime \prime}\,  N{}^{(-6)} {\partial}_{0}{{A}_{0}}\, {}^{2} a{}^{-1} - 3\,  N^{\prime}\,  a^{\prime}\,  N{}^{(-7)} {\partial}_{0}{{A}_{0}}\, {}^{2} a{}^{-1} +  {\delta}^{i j} {\partial}_{0}{{A}_{0}}\,  N^{\prime}\,  a^{\prime}\,  {\partial}_{j}{{A}_{i}}\,  N{}^{(-5)} a{}^{(-3)} - \nonumber \\
&&4\,  {\delta}^{i j} {\partial}_{0}{{A}_{0}}\,  {\partial}_{j}{{A}_{i}}\,  N{}^{(-4)} a^\prime{}^{2} a{}^{(-4)} -  {\delta}^{i j} {\partial}_{0}{{A}_{0}}\,  {\partial}_{j}{{A}_{i}}\,  a^{\prime \prime}\,  N{}^{(-4)} a{}^{(-3)} + 2\,  {\delta}^{i j} {\partial}_{0}{{A}_{0}}\,  {\partial}_{i}{{A}_{j}}\,  N{}^{(-4)} a^\prime{}^{2} a{}^{(-4)} - \nonumber \\
&&3\,  {\delta}^{i j} {\partial}_{0}{{A}_{0}}\,  {\partial}_{i}{{A}_{j}}\,  a^{\prime \prime}\,  N{}^{(-4)} a{}^{(-3)} + 3\,  {\delta}^{i j} {\partial}_{0}{{A}_{0}}\,  N^{\prime}\,  a^{\prime}\,  {\partial}_{i}{{A}_{j}}\,  N{}^{(-5)} a{}^{(-3)} -  {\delta}^{i j} {\delta}^{k l} N^{\prime}\,  a^{\prime}\,  {\partial}_{j}{{A}_{i}}\,  {\partial}_{k}{{A}_{l}}\,  N{}^{(-3)} a{}^{(-5)} +\nonumber \\
&& 4\,  {\delta}^{i j} {\delta}^{k l} {\partial}_{j}{{A}_{i}}\,  {\partial}_{k}{{A}_{l}}\,  N{}^{(-2)} a^\prime{}^{2} a{}^{(-6)} +  {\delta}^{i j} {\delta}^{k l} {\partial}_{j}{{A}_{i}}\,  {\partial}_{k}{{A}_{l}}\,  a^{\prime \prime}\,  N{}^{(-2)} a{}^{(-5)} - 2\,  {\delta}^{i j} {\delta}^{k l} {\partial}_{i}{{A}_{j}}\,  {\partial}_{k}{{A}_{l}}\,  N{}^{(-2)} a^\prime{}^{2} a{}^{(-6)} - \nonumber \\
&&6\, {A}_{0}  {\partial}_{0}{{A}_{0}}\,  N^{\prime}\,  a^{\prime \prime}\,  N{}^{(-7)} a{}^{-1} + 6\, {A}_{0}  {\partial}_{0}{{A}_{0}}\,  a^{\prime}\,  N{}^{(-8)} N^\prime{}^{2} a{}^{-1} + 3\, {A}_{0}  {\delta}^{i j} N^{\prime}\,  {\partial}_{i}{{A}_{j}}\,  a^{\prime \prime}\,  N{}^{(-5)} a{}^{(-3)} -\nonumber \\
&& 3\, {A}_{0}  {\delta}^{i j} a^{\prime}\,  {\partial}_{i}{{A}_{j}}\,  N{}^{(-6)} N^\prime{}^{2} a{}^{(-3)} - 12\, {A}_{0}  {\partial}_{0}{{A}_{0}}\,  N^{\prime}\,  N{}^{(-7)} a^\prime{}^{2} a{}^{(-2)} + 6\, {A}_{0}  {\partial}_{0}{{A}_{0}}\,  N{}^{(-6)} a^\prime{}^{3} a{}^{(-3)} + \nonumber \\
&&12\, {A}_{0}  {\partial}_{0}{{A}_{0}}\,  a^{\prime}\,  a^{\prime \prime}\,  N{}^{(-6)} a{}^{(-2)}%
 + {A}_{0}  {\delta}^{i j} N^{\prime}\,  {\partial}_{i}{{A}_{j}}\,  N{}^{(-5)} a^\prime{}^{2} a{}^{(-4)} - 3\, {A}_{0}  {\delta}^{i j} a^{\prime}\,  {\partial}_{i}{{A}_{j}}\,  a^{\prime \prime}\,  N{}^{(-4)} a{}^{(-4)} - \nonumber \\
&&{A}_{0}  {\delta}^{i j} a^{\prime}\,  {\partial}_{j}{{A}_{i}}\,  N{}^{(-6)} N^\prime{}^{2} a{}^{(-3)} + 7\, {A}_{0}  {\delta}^{i j} N^{\prime}\,  {\partial}_{j}{{A}_{i}}\,  N{}^{(-5)} a^\prime{}^{2} a{}^{(-4)} + {A}_{0}  {\delta}^{i j} N^{\prime}\,  {\partial}_{j}{{A}_{i}}\,  a^{\prime \prime}\,  N{}^{(-5)} a{}^{(-3)} + \nonumber \\
&&3\,  a^{\prime \prime}\,  {A}_{0}{}^{2} N{}^{(-8)} N^\prime{}^{2} a{}^{-1} - 3\,  a^{\prime}\,  {A}_{0}{}^{2} N{}^{(-9)} N^\prime{}^{3} a{}^{-1} + 12\,  {A}_{0}{}^{2} N{}^{(-8)} N^\prime{}^{2} a^\prime{}^{2} a{}^{(-2)} - \nonumber \\
&&15\,  N^{\prime}\,  {A}_{0}{}^{2} N{}^{(-7)} a^\prime{}^{3} a{}^{(-3)} - 12\,  N^{\prime}\,  a^{\prime}\,  a^{\prime \prime}\,  {A}_{0}{}^{2} N{}^{(-7)} a{}^{(-2)} - 12\, {A}_{0}  {\delta}^{i j} {\partial}_{j}{{A}_{i}}\,  N{}^{(-4)} a^\prime{}^{3} a{}^{(-5)} - \nonumber \\
&&3\, {A}_{0}  {\delta}^{i j} a^{\prime}\,  {\partial}_{j}{{A}_{i}}\,  a^{\prime \prime}\,  N{}^{(-4)} a{}^{(-4)} + 18\,  {A}_{0}{}^{2} N{}^{(-6)} a^\prime{}^{4} a{}^{(-4)} + 9\,  a^{\prime \prime}\,  {A}_{0}{}^{2} N{}^{(-6)} a^\prime{}^{2} a{}^{(-3)}
\end{eqnarray}
\begin{eqnarray}
&& \mathcal{L}_6 = 3\,  a^{\prime \prime}\,  N{}^{(-6)} {\partial}_{0}{{A}_{0}}\, {}^{2} a{}^{-1} - 3\,  N^{\prime}\,  a^{\prime}\,  N{}^{(-7)} {\partial}_{0}{{A}_{0}}\, {}^{2} a{}^{-1} - 3\,  {\delta}^{i j} {\partial}_{0}{{A}_{j}}\,  {\partial}_{i}{{A}_{0}}\,  a^{\prime \prime}\,  N{}^{(-4)} a{}^{(-3)} + \nonumber \\
&&3\,  {\delta}^{i j} {\partial}_{0}{{A}_{j}}\,  N^{\prime}\,  a^{\prime}\,  {\partial}_{i}{{A}_{0}}\,  N{}^{(-5)} a{}^{(-3)} +  {\delta}^{i j} {\partial}_{0}{{A}_{i}}\,  N^{\prime}\,  a^{\prime}\,  {\partial}_{j}{{A}_{0}}\,  N{}^{(-5)} a{}^{(-3)} - 4\,  {\delta}^{i j} {\partial}_{0}{{A}_{i}}\,  {\partial}_{j}{{A}_{0}}\,  N{}^{(-4)} a^\prime{}^{2} a{}^{(-4)} - \nonumber \\
&& {\delta}^{i j} {\partial}_{0}{{A}_{i}}\,  {\partial}_{j}{{A}_{0}}\,  a^{\prime \prime}\,  N{}^{(-4)} a{}^{(-3)} + 2\,  {\delta}^{i j} {\partial}_{0}{{A}_{j}}\,  {\partial}_{i}{{A}_{0}}\,  N{}^{(-4)} a^\prime{}^{2} a{}^{(-4)} -  {\delta}^{i j} {\delta}^{k l} N^{\prime}\,  a^{\prime}\,  {\partial}_{j}{{A}_{l}}\,  {\partial}_{k}{{A}_{i}}\,  N{}^{(-3)} a{}^{(-5)} + \nonumber \\
&&4\,  {\delta}^{i j} {\delta}^{k l} {\partial}_{j}{{A}_{l}}\,  {\partial}_{k}{{A}_{i}}\,  N{}^{(-2)} a^\prime{}^{2} a{}^{(-6)} +  {\delta}^{i j} {\delta}^{k l} {\partial}_{j}{{A}_{l}}\,  {\partial}_{k}{{A}_{i}}\,  a^{\prime \prime}\,  N{}^{(-2)} a{}^{(-5)} - 2\,  {\delta}^{i j} {\delta}^{k l} {\partial}_{i}{{A}_{l}}\,  {\partial}_{k}{{A}_{j}}\,  N{}^{(-2)} a^\prime{}^{2} a{}^{(-6)} - \nonumber \\
&&6\, {A}_{0}  {\partial}_{0}{{A}_{0}}\,  N^{\prime}\,  a^{\prime \prime}\,  N{}^{(-7)} a{}^{-1} + 6\, {A}_{0}  {\partial}_{0}{{A}_{0}}\,  a^{\prime}\,  N{}^{(-8)} N^\prime{}^{2} a{}^{-1} - {A}_{j}  {\delta}^{j i} N^{\prime}\,  {\partial}_{i}{{A}_{0}}\,  N{}^{(-5)} a^\prime{}^{2} a{}^{(-4)} + \nonumber \\
&&2\, {A}_{j}  {\delta}^{j i} {\partial}_{i}{{A}_{0}}\,  N{}^{(-4)} a^\prime{}^{3} a{}^{(-5)} + {A}_{j}  {\delta}^{j i} a^{\prime}\,  {\partial}_{i}{{A}_{0}}\,  a^{\prime \prime}\,  N{}^{(-4)} a{}^{(-4)} + 3\, {A}_{j}  {\delta}^{j i} {\partial}_{0}{{A}_{i}}\,  a^{\prime}\,  a^{\prime \prime}\,  N{}^{(-4)} a{}^{(-4)} - \nonumber \\
&&3\, {A}_{j}  {\delta}^{j i} {\partial}_{0}{{A}_{i}}\,  N^{\prime}\,  N{}^{(-5)} a^\prime{}^{2} a{}^{(-4)}%
 + {A}_{0}  {\delta}^{i j} N^{\prime}\,  {\partial}_{i}{{A}_{j}}\,  N{}^{(-5)} a^\prime{}^{2} a{}^{(-4)} - {A}_{0}  {\delta}^{i j} a^{\prime}\,  {\partial}_{i}{{A}_{j}}\,  a^{\prime \prime}\,  N{}^{(-4)} a{}^{(-4)} + \nonumber \\
&&3\, {A}_{i}  {\delta}^{i j} a^{\prime}\,  {\partial}_{j}{{A}_{0}}\,  a^{\prime \prime}\,  N{}^{(-4)} a{}^{(-4)} - 3\, {A}_{i}  {\delta}^{i j} N^{\prime}\,  {\partial}_{j}{{A}_{0}}\,  N{}^{(-5)} a^\prime{}^{2} a{}^{(-4)} + 3\,  a^{\prime \prime}\,  {A}_{0}{}^{2} N{}^{(-8)} N^\prime{}^{2} a{}^{-1} - \nonumber \\
&&3\,  a^{\prime}\,  {A}_{0}{}^{2} N{}^{(-9)} N^\prime{}^{3} a{}^{-1} - 4\, {A}_{i} {A}_{j}  {\delta}^{i j} a^{\prime \prime}\,  N{}^{(-4)} a^\prime{}^{2} a{}^{(-5)} + 4\, {A}_{i} {A}_{j}  {\delta}^{i j} N^{\prime}\,  N{}^{(-5)} a^\prime{}^{3} a{}^{(-5)} - \nonumber \\
&&{A}_{i}  {\delta}^{i j} {\partial}_{0}{{A}_{j}}\,  N^{\prime}\,  N{}^{(-5)} a^\prime{}^{2} a{}^{(-4)} + 2\, {A}_{i}  {\delta}^{i j} {\partial}_{0}{{A}_{j}}\,  N{}^{(-4)} a^\prime{}^{3} a{}^{(-5)} + {A}_{i}  {\delta}^{i j} {\partial}_{0}{{A}_{j}}\,  a^{\prime}\,  a^{\prime \prime}\,  N{}^{(-4)} a{}^{(-4)} - \nonumber \\
&&4\, {A}_{i} {A}_{j}  {\delta}^{i j} N{}^{(-4)} a^\prime{}^{4} a{}^{(-6)} + 2\, {A}_{i} {A}_{j}  {\delta}^{j i} N{}^{(-4)} a^\prime{}^{4} a{}^{(-6)} + {A}_{0}  {\delta}^{i j} N^{\prime}\,  {\partial}_{j}{{A}_{i}}\,  N{}^{(-5)} a^\prime{}^{2} a{}^{(-4)} - \nonumber \\
&&4\, {A}_{0}  {\delta}^{i j} {\partial}_{j}{{A}_{i}}\,  N{}^{(-4)} a^\prime{}^{3} a{}^{(-5)} - {A}_{0}  {\delta}^{i j} a^{\prime}\,  {\partial}_{j}{{A}_{i}}\,  a^{\prime \prime}\,  N{}^{(-4)} a{}^{(-4)} - 3\,  N^{\prime}\,  {A}_{0}{}^{2} N{}^{(-7)} a^\prime{}^{3} a{}^{(-3)} + \nonumber \\
&&6\,  {A}_{0}{}^{2} N{}^{(-6)} a^\prime{}^{4} a{}^{(-4)} + 3\,  a^{\prime \prime}\,  {A}_{0}{}^{2} N{}^{(-6)} a^\prime{}^{2} a{}^{(-3)}
\end{eqnarray}
\begin{eqnarray}
&& \mathcal{L}_7 = - {\delta}^{i j} {\partial}_{0}{{A}_{i}}\,  N^{\prime}\,  a^{\prime}\,  {\partial}_{j}{{A}_{0}}\,  N{}^{(-5)} a{}^{(-3)} + {\delta}^{i j} {\partial}_{0}{{A}_{i}}\,  {\partial}_{j}{{A}_{0}}\,  a^{\prime \prime}\,  N{}^{(-4)} a{}^{(-3)} + {\delta}^{i j} {\partial}_{0}{{A}_{i}}\,  {\partial}_{0}{{A}_{j}}\,  N^{\prime}\,  a^{\prime}\,  N{}^{(-5)} a{}^{(-3)} - \nonumber \\
&&{\delta}^{i j} {\partial}_{0}{{A}_{i}}\,  {\partial}_{0}{{A}_{j}}\,  a^{\prime \prime}\,  N{}^{(-4)} a{}^{(-3)} + {\delta}^{i j} N^{\prime}\,  a^{\prime}\,  {\partial}_{i}{{A}_{0}}\,  {\partial}_{j}{{A}_{0}}\,  N{}^{(-5)} a{}^{(-3)} - {\delta}^{i j} {\partial}_{i}{{A}_{0}}\,  {\partial}_{j}{{A}_{0}}\,  a^{\prime \prime}\,  N{}^{(-4)} a{}^{(-3)} - \nonumber \\
&&{\delta}^{i j} {\partial}_{0}{{A}_{j}}\,  N^{\prime}\,  a^{\prime}\,  {\partial}_{i}{{A}_{0}}\,  N{}^{(-5)} a{}^{(-3)} + {\delta}^{i j} {\partial}_{0}{{A}_{j}}\,  {\partial}_{i}{{A}_{0}}\,  a^{\prime \prime}\,  N{}^{(-4)} a{}^{(-3)} + {\delta}^{i j} {\delta}^{k l} {\partial}_{k}{{A}_{i}}\,  {\partial}_{l}{{A}_{j}}\,  N{}^{(-2)} a^\prime{}^{2} a{}^{(-6)} - \nonumber \\
&&{\delta}^{i j} {\delta}^{k l} {\partial}_{j}{{A}_{l}}\,  {\partial}_{k}{{A}_{i}}\,  N{}^{(-2)} a^\prime{}^{2} a{}^{(-6)}
\end{eqnarray}}
{\small
\begin{eqnarray}
&&\mathcal{L}_8 = - {\delta}^{i j} {\partial}_{0}{{A}_{0}}\,  N^{\prime}\,  a^{\prime}\,  {\partial}_{j}{{A}_{i}}\,  N{}^{(-5)} a{}^{(-3)} + {\delta}^{i j} {\partial}_{0}{{A}_{0}}\,  {\partial}_{j}{{A}_{i}}\,  a^{\prime \prime}\,  N{}^{(-4)} a{}^{(-3)} + {\delta}^{i j} {\partial}_{0}{{A}_{i}}\,  {\partial}_{0}{{A}_{j}}\,  N^{\prime}\,  a^{\prime}\,  N{}^{(-5)} a{}^{(-3)} -\nonumber \\
&& {\delta}^{i j} {\partial}_{0}{{A}_{i}}\,  {\partial}_{0}{{A}_{j}}\,  a^{\prime \prime}\,  N{}^{(-4)} a{}^{(-3)} + {\delta}^{i j} N^{\prime}\,  a^{\prime}\,  {\partial}_{i}{{A}_{0}}\,  {\partial}_{j}{{A}_{0}}\,  N{}^{(-5)} a{}^{(-3)} - {\delta}^{i j} {\partial}_{i}{{A}_{0}}\,  {\partial}_{j}{{A}_{0}}\,  a^{\prime \prime}\,  N{}^{(-4)} a{}^{(-3)} -\nonumber \\
&& {\delta}^{i j} {\partial}_{0}{{A}_{0}}\,  N^{\prime}\,  a^{\prime}\,  {\partial}_{i}{{A}_{j}}\,  N{}^{(-5)} a{}^{(-3)} + {\delta}^{i j} {\partial}_{0}{{A}_{0}}\,  {\partial}_{i}{{A}_{j}}\,  a^{\prime \prime}\,  N{}^{(-4)} a{}^{(-3)} + {\delta}^{i j} {\delta}^{k l} {\partial}_{k}{{A}_{j}}\,  {\partial}_{l}{{A}_{i}}\,  N{}^{(-2)} a^\prime{}^{2} a{}^{(-6)} -\nonumber \\
&& {\delta}^{i j} {\delta}^{k l} {\partial}_{j}{{A}_{i}}\,  {\partial}_{k}{{A}_{l}}\,  N{}^{(-2)} a^\prime{}^{2} a{}^{(-6)} - 3\, {A}_{i} {\delta}^{i j} {\partial}_{0}{{A}_{j}}\,  N^{\prime}\,  N{}^{(-5)} a^\prime{}^{2} a{}^{(-4)} + 2\, {A}_{i} {\delta}^{i j} {\partial}_{0}{{A}_{j}}\,  a^{\prime}\,  a^{\prime \prime}\,  N{}^{(-4)} a{}^{(-4)} + \nonumber \\
&&{A}_{j} {\delta}^{j i} {\partial}_{0}{{A}_{i}}\,  N^{\prime}\,  N{}^{(-5)} a^\prime{}^{2} a{}^{(-4)} + {A}_{0} {\delta}^{i j} a^{\prime}\,  {\partial}_{i}{{A}_{j}}\,  N{}^{(-6)} N^\prime{}^{2} a{}^{(-3)} - {A}_{0} {\delta}^{i j} N^{\prime}\,  {\partial}_{i}{{A}_{j}}\,  a^{\prime \prime}\,  N{}^{(-5)} a{}^{(-3)} - \nonumber \\
&&{A}_{j} {\delta}^{j i} N^{\prime}\,  {\partial}_{i}{{A}_{0}}\,  N{}^{(-5)} a^\prime{}^{2} a{}^{(-4)} + {A}_{j} {\delta}^{j i} a^{\prime}\,  {\partial}_{i}{{A}_{0}}\,  a^{\prime \prime}\,  N{}^{(-4)} a{}^{(-4)} + 6\, {A}_{0} {\partial}_{0}{{A}_{0}}\,  N^{\prime}\,  N{}^{(-7)} a^\prime{}^{2} a{}^{(-2)} - \nonumber \\
&&6\, {A}_{0} {\partial}_{0}{{A}_{0}}\,  a^{\prime}\,  a^{\prime \prime}\,  N{}^{(-6)} a{}^{(-2)}%
 + 2\, {A}_{0} {\delta}^{i j} {\partial}_{i}{{A}_{j}}\,  N{}^{(-4)} a^\prime{}^{3} a{}^{(-5)} + {A}_{0} {\delta}^{i j} a^{\prime}\,  {\partial}_{j}{{A}_{i}}\,  N{}^{(-6)} N^\prime{}^{2} a{}^{(-3)} - \nonumber \\
&&{A}_{0} {\delta}^{i j} N^{\prime}\,  {\partial}_{j}{{A}_{i}}\,  a^{\prime \prime}\,  N{}^{(-5)} a{}^{(-3)} + {A}_{i} {A}_{j} {\delta}^{i j} N^{\prime}\,  N{}^{(-5)} a^\prime{}^{3} a{}^{(-5)} - {A}_{i} {A}_{j} {\delta}^{i j} a^{\prime \prime}\,  N{}^{(-4)} a^\prime{}^{2} a{}^{(-5)} - \nonumber \\
&&6\, {A}_{0}{}^{2} N{}^{(-8)} N^\prime{}^{2} a^\prime{}^{2} a{}^{(-2)} + 6\, N^{\prime}\,  a^{\prime}\,  a^{\prime \prime}\,  {A}_{0}{}^{2} N{}^{(-7)} a{}^{(-2)} - {A}_{i} {\delta}^{i j} N^{\prime}\,  {\partial}_{j}{{A}_{0}}\,  N{}^{(-5)} a^\prime{}^{2} a{}^{(-4)} + \nonumber \\
&&{A}_{i} {\delta}^{i j} a^{\prime}\,  {\partial}_{j}{{A}_{0}}\,  a^{\prime \prime}\,  N{}^{(-4)} a{}^{(-4)} + {A}_{i} {A}_{j} {\delta}^{j i} N^{\prime}\,  N{}^{(-5)} a^\prime{}^{3} a{}^{(-5)} - {A}_{i} {A}_{j} {\delta}^{j i} a^{\prime \prime}\,  N{}^{(-4)} a^\prime{}^{2} a{}^{(-5)} + \nonumber \\
&&2\, {A}_{0} {\delta}^{i j} {\partial}_{j}{{A}_{i}}\,  N{}^{(-4)} a^\prime{}^{3} a{}^{(-5)} - 6 {A}_{0}{}^{2} N{}^{(-6)} a^\prime{}^{4} a{}^{(-4)}
\end{eqnarray}
\begin{eqnarray}
&&\mathcal{L}_9 = - 36\, {A}_{0}{}^{2} N{}^{(-6)} a^{\prime \prime}{}^{2} a{}^{(-2)} + 72\, N^{\prime}\,  a^{\prime}\,  a^{\prime \prime}\,  {A}_{0}{}^{2} N{}^{(-7)} a{}^{(-2)} - 36\, {A}_{0}{}^{2} N{}^{(-8)} N^\prime{}^{2} a^\prime{}^{2} a{}^{(-2)} + \nonumber \\
&&36\, {A}_{i} {A}_{j} {\delta}^{i j} N{}^{(-4)} a^{\prime \prime}{}^{2} a{}^{(-4)} - 72\, {A}_{i} {A}_{j} {\delta}^{i j} N^{\prime}\,  a^{\prime}\,  a^{\prime \prime}\,  N{}^{(-5)} a{}^{(-4)} + 36\, {A}_{i} {A}_{j} {\delta}^{i j} N{}^{(-6)} N^\prime{}^{2} a^\prime{}^{2} a{}^{(-4)} - \nonumber \\
&&72\, a^{\prime \prime}\,  {A}_{0}{}^{2} N{}^{(-6)} a^\prime{}^{2} a{}^{(-3)} + 72\, N^{\prime}\,  {A}_{0}{}^{2} N{}^{(-7)} a^\prime{}^{3} a{}^{(-3)} + 72\, {A}_{i} {A}_{j} {\delta}^{i j} a^{\prime \prime}\,  N{}^{(-4)} a^\prime{}^{2} a{}^{(-5)} - \nonumber \\
&&72\, {A}_{i} {A}_{j} {\delta}^{i j} N^{\prime}\,  N{}^{(-5)} a^\prime{}^{3} a{}^{(-5)} - 36\, {A}_{0}{}^{2} N{}^{(-6)} a^\prime{}^{4} a{}^{(-4)} + 36\, {A}_{i} {A}_{j} {\delta}^{i j} N{}^{(-4)} a^\prime{}^{4} a{}^{(-6)}
\end{eqnarray}
\begin{eqnarray}
&&\mathcal{L}_{10} = - 18\, {A}_{0}{}^{2} N{}^{(-6)} a^{\prime \prime}{}^{2} a{}^{(-2)} + 36\, N^{\prime}\,  a^{\prime}\,  a^{\prime \prime}\,  {A}_{0}{}^{2} N{}^{(-7)} a{}^{(-2)} - 18\, {A}_{0}{}^{2} N{}^{(-8)} N^\prime{}^{2} a^\prime{}^{2} a{}^{(-2)} - \nonumber \\
&&12\, {A}_{i} {A}_{j} {\delta}^{i j} N^{\prime}\,  a^{\prime}\,  a^{\prime \prime}\,  N{}^{(-5)} a{}^{(-4)} + 18\, {A}_{i} {A}_{j} {\delta}^{i j} a^{\prime \prime}\,  N{}^{(-4)} a^\prime{}^{2} a{}^{(-5)} + 6\, {A}_{i} {A}_{j} {\delta}^{i j} N{}^{(-4)} a^{\prime \prime}{}^{2} a{}^{(-4)} + \nonumber \\
&&6\, {A}_{i} {A}_{j} {\delta}^{i j} N{}^{(-6)} N^\prime{}^{2} a^\prime{}^{2} a{}^{(-4)} - 18\, {A}_{i} {A}_{j} {\delta}^{i j} N^{\prime}\,  N{}^{(-5)} a^\prime{}^{3} a{}^{(-5)} - 18\, a^{\prime \prime}\,  {A}_{0}{}^{2} N{}^{(-6)} a^\prime{}^{2} a{}^{(-3)} + \nonumber \\
&&18\, N^{\prime}\,  {A}_{0}{}^{2} N{}^{(-7)} a^\prime{}^{3} a{}^{(-3)} + 12\, {A}_{i} {A}_{j} {\delta}^{i j} N{}^{(-4)} a^\prime{}^{4} a{}^{(-6)} 
\end{eqnarray}
\begin{eqnarray}
&&\mathcal{L}_{11} = - 12\, {A}_{0}{}^{2} N{}^{(-6)} a^{\prime \prime}{}^{2} a{}^{(-2)} + 24\, N^{\prime}\,  a^{\prime}\,  a^{\prime \prime}\,  {A}_{0}{}^{2} N{}^{(-7)} a{}^{(-2)} - 12\, {A}_{0}{}^{2} N{}^{(-8)} N^\prime{}^{2} a^\prime{}^{2} a{}^{(-2)} + \nonumber \\
&&12\, {A}_{i} {A}_{j} {\delta}^{i j} N{}^{(-4)} a^{\prime \prime}{}^{2} a{}^{(-4)} - 24\, {A}_{i} {A}_{j} {\delta}^{i j} N^{\prime}\,  a^{\prime}\,  a^{\prime \prime}\,  N{}^{(-5)} a{}^{(-4)} + 12\, {A}_{i} {A}_{j} {\delta}^{i j} N{}^{(-6)} N^\prime{}^{2} a^\prime{}^{2} a{}^{(-4)} + \nonumber \\
&&12\, N^{\prime}\,  {A}_{0}{}^{2} N{}^{(-7)} a^\prime{}^{3} a{}^{(-3)} - 12\, {A}_{0}{}^{2} N{}^{(-6)} a^\prime{}^{4} a{}^{(-4)} - 12\, a^{\prime \prime}\,  {A}_{0}{}^{2} N{}^{(-6)} a^\prime{}^{2} a{}^{(-3)} - \nonumber \\
&&12\, {A}_{i} {A}_{j} {\delta}^{i j} N^{\prime}\,  N{}^{(-5)} a^\prime{}^{3} a{}^{(-5)} + 12\, {A}_{i} {A}_{j} {\delta}^{i j} N{}^{(-4)} a^\prime{}^{4} a{}^{(-6)} + 12\, {A}_{i} {A}_{j} {\delta}^{i j} a^{\prime \prime}\,  N{}^{(-4)} a^\prime{}^{2} a{}^{(-5)} 
\end{eqnarray}
\begin{eqnarray}
&&\mathcal{L}_{12} = - 9\, {A}_{0}{}^{2} N{}^{(-6)} a^{\prime \prime}{}^{2} a{}^{(-2)} + 18\, N^{\prime}\,  a^{\prime}\,  a^{\prime \prime}\,  {A}_{0}{}^{2} N{}^{(-7)} a{}^{(-2)} - 9\, {A}_{0}{}^{2} N{}^{(-8)} N^\prime{}^{2} a^\prime{}^{2} a{}^{(-2)} + \nonumber \\
&&{A}_{i} {A}_{j} {\delta}^{i j} N{}^{(-6)} N^\prime{}^{2} a^\prime{}^{2} a{}^{(-4)} - 4\, {A}_{i} {A}_{j} {\delta}^{i j} N^{\prime}\,  N{}^{(-5)} a^\prime{}^{3} a{}^{(-5)} - 2\, {A}_{i} {A}_{j} {\delta}^{i j} N^{\prime}\,  a^{\prime}\,  a^{\prime \prime}\,  N{}^{(-5)} a{}^{(-4)} + \nonumber \\
&&4\, {A}_{i} {A}_{j} {\delta}^{i j} N{}^{(-4)} a^\prime{}^{4} a{}^{(-6)} + 4\, {A}_{i} {A}_{j} {\delta}^{i j} a^{\prime \prime}\,  N{}^{(-4)} a^\prime{}^{2} a{}^{(-5)} + {A}_{i} {A}_{j} {\delta}^{i j} N{}^{(-4)} a^{\prime \prime}{}^{2} a{}^{(-4)}
\end{eqnarray}
}

where the action is defined as:
\begin{eqnarray}
\mathcal{S}_i &=&  D_i\int d^4 x \sqrt{-g}\,\mathcal{L}_i \nonumber \\
 &=& D_i\int d^4x\,N\,a^3\,\mathcal{L}_i
\end{eqnarray}

Varying these along with (\ref{VecGalAction}) w.r.t $A_{\mu}$ in the FRW background, and demanding that the action is gauge-invariant and that the equation of motion only contains derivatives upto second order, the coefficients $D_i$ are given below:
\begin{eqnarray}
&&D_2 = 2,~D_4 = - D_1 - D_3,~D_{6} = -2 - D_5, ~D_{10} = -\frac{1}{6}, 
\nonumber \\
& & D_{7} = - 4 - 6 D_1 - 6 D_3 - 2 D_5,~D_{8} = 6 D_3 + 2 D_5, ~ D_{12} = 1 \nonumber \\
&&  D_{9} = \frac{1}{12} - \frac{D_3}{2} - \frac{D_5}{12},~ D_{11} = - \frac{1}{4} + \frac{3 D_3}{2} + \frac{D_5}{4},~
\end{eqnarray}

\section{Spectrum of electric and $B.B^\prime$}\label{app:otherEner}

Electric part and $B.B^\prime$ part of the energy density at sound horizon become

\begin{eqnarray}
\mathcal{P}_E &=& - \frac{24\,D}{\pi\,c_s^{7+2\beta}}\,\mathcal{G}_1(\beta)\,H_*^4\,\left(\frac{k}{k_*}\right)^{2\beta + 8}, \beta \langle -\frac{5}{2},~~\mathcal{G}_1(\beta) = \frac{|C_1|^2}{2^{2\beta+3}\left(\Gamma(\beta + 5/2)\right)^2} \nonumber \\
&& - \frac{24\,D}{\pi\,c_s^{1-2\beta}}\,\mathcal{G}_2(\beta)\,H_*^4\,\left(\frac{k}{k_*}\right)^{2 - 2\beta}, \beta \rangle -\frac{5}{2},~~\mathcal{G}_2(\beta) = \frac{|C_2|^2}{2^{ -2\beta-3}\left(\Gamma( - \beta -1/2)\right)^2} \\
\mathcal{P}_{B.B^\prime} &=& -\frac{16\,D}{\pi\,c_s^{10+2\beta}} \mathcal{J}_1(\beta)\,H_*^4 \left(\frac{k}{k_*}\right)^{2\beta + 10}, \beta\langle -\frac{5}{2}, \mathcal{J}_1(\beta) = \frac{|C_1|^2}{2^{2\beta+4} (-\beta - 5/2)\,\left(\Gamma(\beta + 5/2)\right)^2} \nonumber \\
&&  - \frac{16\,D}{\pi\,c_s^{2 - 2 \beta}} \mathcal{J}_2(\beta)\,H_*^4 \left(\frac{ k}{k_*}\right)^{2 -2\beta}, \beta \rangle -\frac{5}{2}, \mathcal{J}_2(\beta) =  \frac{|C_2|^2}{2^{-2\beta - 4} (-\beta - 3/2)\,\left(\Gamma(-\beta - 3/2)\right)^2}
\end{eqnarray}

\section{Slow-roll inflation and spectrum of the energy densities}
\label{app:slow-roll}

In case of slow-roll inflation, the slow-roll parameters are defined as
\begin{eqnarray}
&&\epsilon_1 = -\frac{\dot{H}}{H^2},~~ \mbox{where dot used for cosmic time }t \nonumber \\
&&\epsilon_{n+1} = \frac{\dot{\epsilon}_n}{H\epsilon}, ~~n~\mbox{is natural numbers.}
\end{eqnarray}

In conformal coordinate, 
\begin{eqnarray}
&&\epsilon_1 - 1 = -\frac{\mathcal{H}^\prime}{\mathcal{H}^2} \\
\Rightarrow&& c_s \equiv \frac{\mathcal{H}^\prime}{\mathcal{H}^2} = 1 - \epsilon_1~~~~\mbox{(exact)} \\
\end{eqnarray}
and
\begin{eqnarray}
\frac{J^{\prime\prime}}{J} = \mathcal{H}^2\left(-\epsilon_1 + 2 \epsilon_1^2 - \epsilon_1\epsilon_2\right)~~~~\mbox{(exact)}
\end{eqnarray} 

In case of leading order slow-roll approximation with $\epsilon_n \ll 1$,  $\mathcal{H} = -\frac{1+\epsilon_1}{\eta}$, thus
\begin{eqnarray}
\frac{J^{\prime\prime}}{J} = \frac{\epsilon_1(\epsilon_2 - 1)}{\eta^2}
\end{eqnarray}

Hence, the equation for $\mathcal{A}_k$ becomes
\begin{eqnarray}
\mathcal{A}_k^{\prime\prime} + \left(c_s^2 k^2 - \frac{\epsilon_1(\epsilon_2 - 1)}{\eta^2}\right)\,\mathcal{A}_k = 0
\end{eqnarray}

Solution for the above equation using Bunch-Davies vacuum (see section (\ref{Sec.DynamicsInflation})) becomes
\begin{eqnarray}
\mathcal{A}_k = \frac{\pi}{4} \sqrt{-\eta} H^1_\nu (-c_s k \eta), ~~\nu = \frac{1}{2}\sqrt{1 + \epsilon_1(\epsilon_2 - 1)}
\end{eqnarray}

Using the above solution for $-k\eta \rightarrow 0$, at sound horizon, spectral energy density of the magnetic field becomes
\begin{eqnarray}
\mathcal{P}_B = \frac{8\,D}{\pi\,c_s^5} \,H_*^4 \left(\frac{k}{k_*}\right)^4
\end{eqnarray}

	\clearpage
	\thispagestyle{empty}
	\phantom{a}
	\vfill
	\vfill

\end{appendices}
\newpage
\thispagestyle{fancy}
\bigskip{}
\vspace{1cm} \cleardoublepage
\clearpage
\thispagestyle{empty}
\phantom{a}
\vfill
\vfill

\providecommand{\href}[2]{#2}\begingroup\raggedright\endgroup

\bibliographystyle{jcappub}
\bibliography{Mycollection}
\clearpage
\thispagestyle{empty}
\phantom{a}
\vfill
\vfill
\newpage
\thispagestyle{fancy}
\bigskip{}
\vspace{1cm} \cleardoublepage
\clearpage
\thispagestyle{empty}
\phantom{a}
\vfill
\vfill
\end{document}